# UNIVERSIDAD COMPLUTENSE DE MADRID

## FACULTAD DE CIENCIAS FÍSICAS

Materiales multifuncionales magnéticos amorfos y basados en grafeno: síntesis, caracterización y aplicaciones en apantallamiento electromagnético y monitorización avanzada de gases.

Multifunctional graphene-based and amorphous magnetic materials: synthesis, characterization, and applications in electromagnetic shielding and advanced gas sensing.

Autor: **Álvaro Peña Moreno**

Directores: **Pilar Marín Palacios**

**Maria del Carmen Horrillo Güemes**

Madrid, 2023

# UNIVERSIDAD COMPLUTENSE DE MADRID

## FACULTAD DE CIENCIAS FÍSICAS

# TESIS DOCTORAL

Materiales Multifuncionales Magnéticos Amorfos y Basados en Grafeno: Síntesis, Caracterización y Aplicaciones en Apantallamiento Electromagnético y Monitorización Avanzada de Gases./
Multifunctional Graphene-based and Amorphous Magnetic Materials: Synthesis, Characterization, and Applications in Electromagnetic Shielding and Advanced Gas Sensing.

MEMORIA PARA OPTAR AL GRADO DE DOCTOR

PRESENTADA POR

Álvaro Peña Moreno

DIRECTORAS

Pilar Marín Palacios
Maria del Carmen Horrillo Güemes

“Crystals are like people; it is the defects in them which tend to make them interesting”.

- Colin Humphreys

# Agradecimientos

Durante la carrera un profesor dijo sobre mi que era un buen alumno, pero no un buen estudiante. Más de diez años después de empezar mis estudios puedo decir que fue demasiado amable.

Primero, quiero expresar mi más sincero agradecimiento a mis directoras de tesis, Pilar y Mari Carmen. Pilar, tu dedicación a múltiples roles, incluyendo la dirección del Instituto de Magnetismo Aplicado, ha sido una fuente constante de inspiración. Aprecio tu apoyo inquebrantable y tu habilidad para equilibrar las necesidades del laboratorio con las de tu carrera docente.

A Mari Carmen, agradezco tu perspicacia y supervisión atenta, que me ha ayudado a dar forma a esta tesis desde su concepción hasta su conclusión. Tu capacidad para ofrecer orientación valiosa y relevante, a pesar de los retos, es un testimonio de tu compromiso con la excelencia académica y tu dedicación a tus estudiantes.

Deseo agradecer a mis compañeros investigadores del Instituto de Magnetismo Aplicado y del otro laboratorio. Cada uno de vosotros ha contribuido a mi éxito de alguna manera, y por eso estoy agradecido. Quiero hacer una mención especial a aquellos que han participado en los artículos y patentes incluidas en esta tesis.

También quiero expresar mi gratitud a los servicios administrativos y técnicos del Instituto de Magnetismo Aplicado, tanto a los de siempre como a las nuevas incorporaciones. Vuestro trabajo ha facilitado en gran medida el mío.

Agradezco profundamente a los revisores de la tesis por su tiempo y esfuerzo. Igualmente agradezco a los miembros del tribunal por aceptar la responsabilidad de juzgar este trabajo.

Agradezco de corazón a mi familia y amigos, quienes han estado a mi lado, apoyándome en los momentos más desafiantes. Vuestra confianza en mi capacidad para completar este trabajo me ha brindado la motivación necesaria para superar los obstáculos.

Por último, agradezco a la Universidad Complutense de Madrid y al Ministerio de Ciencia (y todos los nombres por los que ha pasado en los últimos años) por su apoyo financiero. Vuestra ayuda no solo ha hecho posible mi trabajo, sino que también ha contribuido a expandir el conocimiento en este campo.

# Index

## Resumen

Esta tesis doctoral presenta una investigación en detalle sobre materiales basados en grafeno (GBMs) y materiales magnéticos amorfos, centrándose en su producción, caracterización y posibles aplicaciones en el apantallamiento electromagnético y la detección de gases.

El estudio incluye el desarrollo de materiales y sus aplicaciones dentro del campo de la física del estado sólido. Lo hace explorando, con un enfoque eminentemente experimental, cuatro líneas de investigación principales: la producción de GBMs, el uso de GBMs y materiales magnéticos amorfos en aplicaciones de apantallamiento electromagnético, el uso de GBMs en sensores químico-resistivos de gases, y el uso de materiales magnéticos amorfos en sensores de gas basados en la resonancia magnetoelástica.

La investigación empieza explorando métodos escalables para la producción de GBMs, ya que, para los materiales magnéticos amorfos, la producción industrial ya está consolidada. Se presentan los GBMs y sus actuales métodos de producción, incluyendo sus ventajas y limitaciones, se estudia sistemáticamente la técnica de molienda de bolas utilizando grafito como precursor. Tras probar diferentes molinos de bolas y parámetros de molienda, la técnica se postula como prometedora para una producción a gran escala. De hecho, a partir de uno de los procesos de molienda se obtuvo grafeno mesoporoso de pocas capas (FLMG), una aglomeración de dominios de grafeno de pocas capas formando una estructura mesoporosa.

Aunque este método de producción de FLMG resulta ser económicamente viable, respetuoso con el medio ambiente y adecuado para la producción a escala industrial, la viabilidad comercial de este material está limitada por sus posibles aplicaciones. En este sentido, primero se investigó el uso de FLMG para aplicaciones de apantallamiento electromagnético.

Los materiales magnéticos amorfos, en particular los micróhilos magnéticos amorfos (MW), tienen una reputación bien afianzada como materiales de apantallamiento electromagnético (ESMs). Sin embargo, estos materiales están potencialmente limitados por su alta densidad relacionada con su composición metálica de base hierro. El uso combinado de FLMG y MW - ya sea por sustitución o adición -, resultó en ESMs con un rendimiento excepcional y parámetros de apantallamiento ajustables. Es más, la investigación sugiere un efecto sinérgico entre FLMG y MW que mejora el rendimiento más allá de lo que se esperaría a partir de sus componentes por separado.

También se investigó la aplicación de FLMG en sensores de gases para el control de la calidad del aire en general y la detección de $NO_2$ en particular. El dispositivo sensor basado en FLMG mostró un buen rendimiento con la capacidad de detectar $NO_2$ en bajas concentraciones (25 ppb), por debajo del umbral de toxicidad humana, de manera rápida y fiable. Además, una limitación tradicional de los sensores basados en grafeno, esto es, la lenta desorción del analito que conduce a una disminución del rendimiento durante la operación del sensor se abordó mediante irradiación UV. Gracias a una colaboración internacional, se estudiaron en profundidad los efectos de la irradiación UV en un grafeno multicapa crecido mediante CVD. Se encontró que, ajustando los parámetros de irradiación, pueden obtenerse una desorción rápida y completa y una mejora general del rendimiento en sensores de gas basados en grafeno.

Por último y no menos importante, también se ha investigado el uso de materiales magnéticos amorfos en aplicaciones de detección de gases. A partir del efecto magnetoelástico, es posible desarrollar un sensor sin contacto físico y sin fuentes de alimentación integradas, lo que lo hace particularmente interesante para aplicaciones biomédicas. A partir de un nuevo sistema capaz de monitorizar la resonancia magnetoelástica en tiempo real se desarrolló un sensor operado remotamente capaz de detectar diferentes biomarcadores relacionados con el diagnostico de enfermedades por análisis del aliento y, lo que es más importante, capaz de monitorizar la respiración real de una manera rápida y reproducible. Esta investigación también presenta y explora diferentes estrategias para mejorar el rendimiento de estos sensores.

En resumen, esta tesis se enfoca en las aplicaciones de apantallamiento electromagnético y detección de gases, y en ese contexto, realiza un estudio crítico del estado actual del arte, desarrolla materiales y técnicas, y presenta métodos innovadores para superar las limitaciones actuales. La calidad y novedad de esta tesis de doctorado están respaldadas por una serie de patentes (2) y publicaciones científicas (5) incluidas en los capítulos de la tesis.

# Summary

This doctoral thesis comprehensively investigates graphene-based materials (GBMs) and amorphous magnetic materials, focusing on their production, characterization, and potential applications in electromagnetic shielding and gas sensing.

The study is primarily concerned with the development of materials and their applications within the field of solid-state physics. It does so by exploring, with a predominantly experimental approach, four major research lines: the production of GBMs, the use of GBMs and amorphous magnetic materials in electromagnetic shielding applications, the use of GBMs in chemiresistive gas sensors, and the use of amorphous magnetic materials in magnetoelastic resonance-based gas sensors.

The research begins with the exploration of scalable methods for the production of GBMs, since for amorphous magnetic materials, industrial production has already been achieved. After revisiting GBMs and their current production methods, presenting their advantages and limitations, a ball milling approach using graphite as a precursor is systematically studied. After testing different ball mills and milling parameters, the technique shows promise for large-scale production. Interestingly, one of the milling processes produced few-layered mesoporous graphene (FLMG), an agglomeration of few-layered graphene domains forming a mesoporous structure.

Although the production method for FLMG proved to be cost-effective, environmentally friendly, and suitable for industrial-scale production, the commercial viability of this material is limited by its potential applications. In this regard, FLMG is first tested for its potential in electromagnetic shielding applications.

Amorphous magnetic materials, in particular, Fe-based amorphous magnetic microwires (MW), have a well-established reputation as electromagnetic shielding material (ESMs). However, they face potential challenges due to the high density related to their Fe-based metallic nature. This high density can limit potential applications where weight is a critical parameter. Using FLMG in combination with MW -by substitution or addition-, resulted in ESMs with exceptional performance and tuneable shielding parameters. Interestingly, the research suggests a synergistic effect between FLMG and MW that enhances performance beyond what would be expected from the individual components.

Next, the research focused on the use of FLMG in gas sensors, specifically for air quality control and $NO_2$ detection. The FLMG-based gas sensor device exhibited good performance with the capability of detecting $NO_2$ below the human toxicity threshold (25 ppb) in a fast and accurate manner. Furthermore, a traditional limitation of graphene-based sensors, slow analyte desorption leading to decreasing performance during the sensor's operation, is assessed by means of UV irradiation. Thanks to an international collaboration the effects of UV irradiation were comprehensively studied in a CVD-grown multilayer graphene. It was found that careful tuning of the irradiation parameters leads to fast and complete desorption and overall increased sensing performance in graphene-based gas sensors.

Finally, the application of amorphous magnetic materials in gas sensing applications is also explored. Interestingly, by exploiting the magnetoelastic phenomenon, it is possible to develop a sensor without physical contact and integrated power sources, being of great interest for biomedical-related applications. A novel system to monitor the magnetoelastic resonance in real time allowed the development of a remotely operated sensor capable of detecting different breath analysis-related biomarkers and, importantly, monitoring real breath exhalation events in a fast and reproducible manner. The research also presented and explored different strategies for enhancing the performance of these sensors.

In summary, this thesis conducts a critical review of the current state of the art, develops materials and techniques, and presents innovative methods to overcome current limitations within the context of electromagnetic shielding and gas detection applications. The quality and novelty of this PhD thesis are backed up by a series of patents (2) and scientific publications (5) included throughout the thesis chapters.

# Chapter 0: General introduction.

*Chapter introduction; Graphene-based materials; Amorphous magnetic materials; Electromagnetic Shielding; Gas sensors; Research gap; Objectives and methodology; Thesis structure; References.*

## 0.1 Chapter introduction.

In this introductory chapter, a general background is provided to establish the foundation for the subsequent chapters of the thesis. This background encompasses an overview of the materials and research fields that will be further discussed, emphasizing their significance in the context of the research objectives. The information presented here sets the context for understanding the overall research objectives and scope of this thesis, without delving into specific details that will be covered in the following chapters.

Over recent years, the research landscape on graphene synthesis, amorphous magnetic materials, microwave-absorbing materials, and gas-sensing technologies has experienced remarkable progress. This thesis focuses on integrating these materials and technologies to develop innovative solutions for addressing challenges associated with electromagnetic shielding and advanced gas sensing, emphasising practical applications and commercial viability.

Multifunctional materials, as denoted in the title of the thesis, are a class of materials that possess two or more distinct functional properties, enabling them to perform multiple tasks simultaneously. These materials are designed to offer various advantages in different applications by integrating multiple functionalities into a single material system. In this thesis, the functionality of two materials, or rather two families of materials, has been studied for two distinct applications. Both graphene-based materials and amorphous magnetic materials have been evaluated for their potential in microwave absorbance and gas-sensing applications.

## 0.2 Graphene-based materials.

The first family of materials studied in this thesis is graphene-based materials (GBMs) [1,2]. This family encompasses materials derived from graphene, the two-dimensional allotrope of carbon, consisting of carbon atoms bonded with sp2 covalent bonds in a hexagonal lattice and a dangling electron per atom at the basal planes [3].

Notably, GBMs demonstrate an array of unique properties derived from nanoscale physics, such as exceptional electrical and thermal conductivity, high mechanical strength, and remarkable optical properties. Furthermore, GBMs often exhibit diverse physicochemical characteristics among themselves, making their research particularly fascinating [4]. One notable example may be the distinct difference in oxidation degrees seen between graphene oxide, where all dangling bonds are fully saturated with oxygen or oxygen-containing groups, and graphene [5]. Reduced graphene oxide falls somewhere in the middle, effectively spanning the full spectrum of oxidation degrees [6]. This leads to a fascinating scenario where, within the same family of materials, some can exhibit high hydrophilicity while others can be extremely hydrophobic, or any level of these properties in between [5,7,8].

These attributes can be exploited simultaneously, rendering GBMs attractive for a wide range of applications, including electronics, energy storage, and sensing technologies [9–12]. Several current commercial applications demonstrate the multifunctionality of GBMs as a critical component:

- Smart/advanced clothing: Manufacturers claim that sportswear featuring GBMs outperforms traditional clothing materials. In this context, GBMs' remarkable thermal conductivity, strength, and antibacterial properties are harnessed to create high-performance athletic wear [13,14]. In footwear, GBMs have been incorporated into shoe soles to enhance energy return and durability [15,16]. In protective gear, GBMs offer a lightweight and robust alternative to ceramic bulletproof vests [17–19].
- Electronics: Owing to their exceptional thermal and mechanical properties, GBMs have gained significant popularity in heat dissipation applications for smartphones [20–22]. Additionally, they contribute to the development of batteries with increased capacity and extended lifetimes [23,24].
- Structural applications: GBMs can serve multiple functions as structural reinforcements, enhancing mechanical properties, fire resistance, and corrosion resistance while reducing weight. These characteristics make them ideal for construction and vehicles [25–29].

It is worth noting that in the vast majority of applications, GBMs are not used as a standalone material, but as part of composites that benefits from synergies between GBMs and other materials. The immense potential of GBMs has spurred significant interest in both academia and industry to explore novel applications and expand the existing ones. As researchers continue to unveil the unique characteristics of GBMs, their potential applications in various fields are expected to grow exponentially.

The research presented in this thesis aims to further explore the capabilities of GBMs in electromagnetic shielding and advanced gas sensing applications, potentially unlocking new possibilities, and further expanding the range of applications for these versatile materials.

### *0.3 Amorphous Magnetic Materials.*

The second family of materials investigated in this thesis are amorphous magnetic materials. The focus on this family of materials is not solely due to their historical significance in the research conducted at the *Instituto de Magnetismo Aplicado (IMA)*, a research institute at the *Universidad Complutense de Madrid (UCM)* where this thesis has primarily been conducted. Instead, the emphasis on these materials is motivated by their intrinsic properties and the promising potential they exhibit in various applications. Specifically, amorphous magnetic microwires (MWs) serve as an outstanding representative of this family of materials for the ensuing discussion. Another form of amorphous magnetic materials, microribbons, would be later explored.

MWs consist of a metallic alloy nucleus and borosilicate glass or Pyrex-like cover (see Figure 0. 1). They are produced by Taylor-Ulitovsky technique, which consists of a melt spinning with ultrafast cooling [30,31]. Briefly, the metallic alloy is prepared in a crucible and is then transferred to a glass tube with a protective atmosphere. Using inductive heating, the alloy and the glass are molten (Figure 0. 2a to c), and at the first step, they are manually pulled and threaded onto the spinning plate (Figure 0. 2d to f). Then, as the plate spins, the molten material is automatically pulled and winded, the wire is air- or water-cooled during this process. The spinning rate is a critical parameter in this technique as it controls the thickness of the microwire, which affects the cooling process [32,33]. It is important to note that this technique can also be used to produce amorphous alloys in various shapes, such as the micro ribbons that will be introduced in a subsequent chapter.

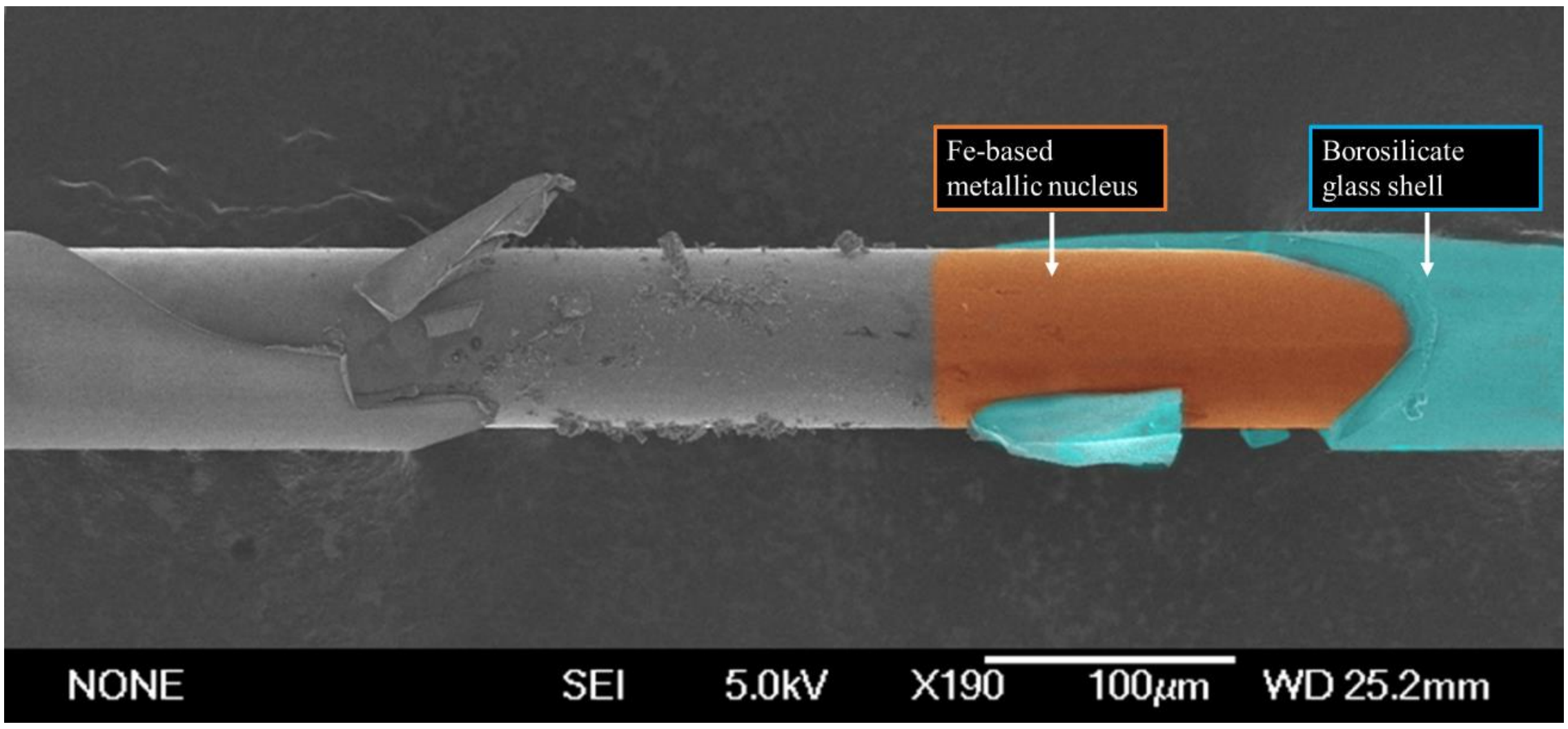

*Figure 0. 1 SEM image of a microwire, obtained through the Taylor-Ulitovsky technique, showing a cracked glass shell that reveals the metallic nucleus (the right part of the image has been coloured to highlight each material).*

Ideally, the microwire will be thin enough to promote ultra-fast cooling dynamics that prevent the crystallization of the metallic nucleus, which solidifies into an amorphous or metastable phase [34]. However, the microwire must be thick enough to retain some mechanical resistance and withstand the pulling stresses. The speed of the spinning plate is usually controlled with a feedback system that continuously monitors the wire thickness. Altogether, the process resembles the one to produce optic fibre and relies on the flexibility given by the thinness of an otherwise rigid or brittle material [35].

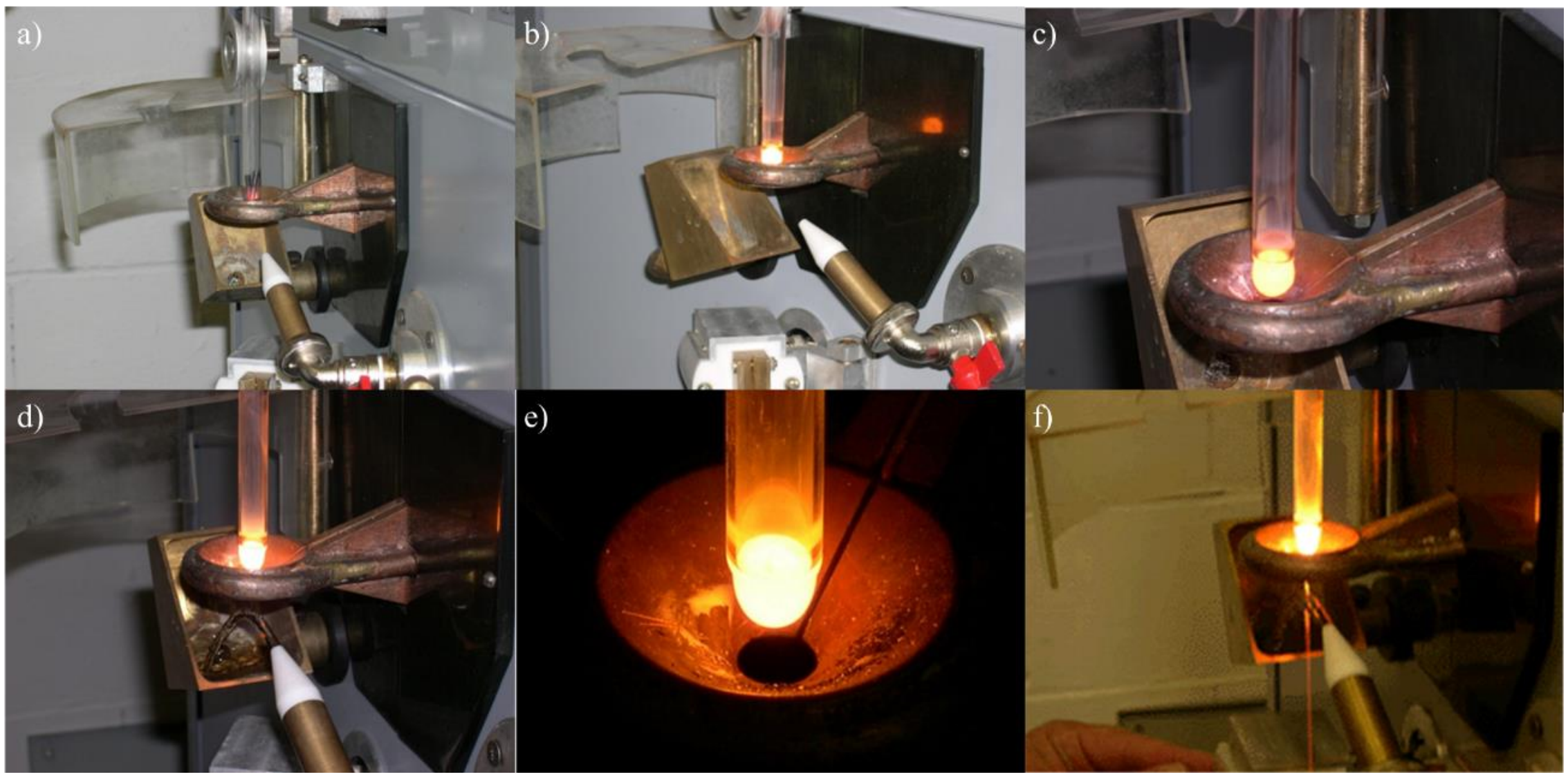

*Figure 0. 2 Illustration of the Taylor-Ulitovsky technique. Rep. from* [32].

This technique allows the fabrication of different alloys, mainly Fe- and Co-based alloys [32,36]. The production is quasi-continuous, and it is possible to obtain reels containing MWs several hundreds of meters long. However, some rules must be kept in mind when designing the metal alloy for the nucleus [32,33]: (1) The metal components should not react with glass, as is the case for Al or Zr; (2) The oxidation of the molten alloy should be prevented and highly reactive metals (such as Nb or Mg) should be avoided if possible; (3) the melting temperature of the alloy should be lower than the working temperature of the glass; (4) the thermal expansion coefficient of the metal alloy should be similar to that of the glass.

The magnetic hysteresis loop is used to evaluate the magnetic behaviour of MWs. This loop is a graphical representation of the magnetic behaviour of a material when subjected to a varying external magnetic field (see Figure 0.3). The loop depicts the relationship between the magnetic field strength (H), expressed in Oe, and the resulting magnetization (M), expressed in emu/g (CGS units are often used in magnetism [37]).

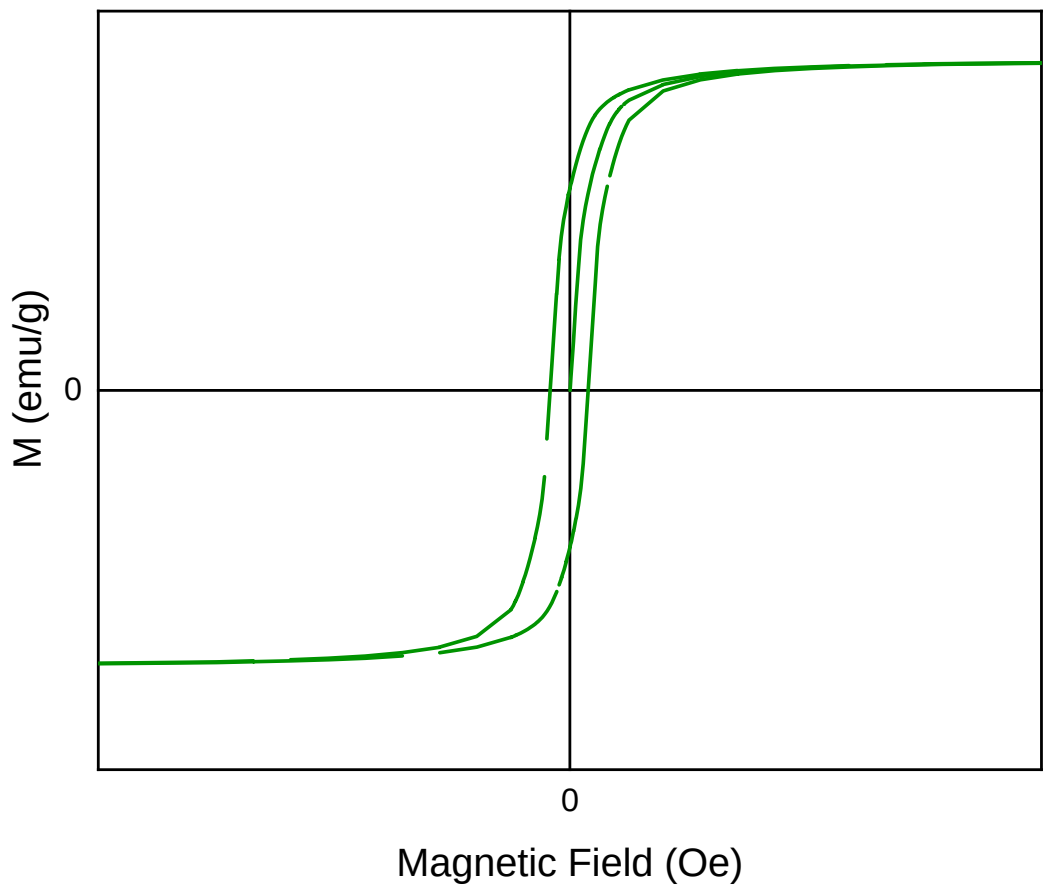

*Figure 0. 3 Typical hysteresis loop for a ferromagnet. The discontinuities are a measure artefact of the VSM at IMA.*

When a ferromagnetic material -such is the case for the Fe-based amorphous alloys- is exposed to an external magnetic field, it becomes magnetized. As the magnetic field is increased, the magnetization of the material increases until it reaches a saturation point ($M_s$). Initially, this increase is linear, and its slope provides the magnetic susceptibility ($\chi$).

When the magnetic field is reduced and eventually reversed, the magnetization decreases but does not return to zero immediately. This lag is called "hysteresis", i.e., the material retains some residual magnetization, called "remanent magnetization," even after the external magnetic field is removed.

To completely demagnetize the material, a magnetic field in the opposite direction, known as the "coercive field ($H_k$)" must be applied. By cycling through increasing and decreasing magnetic fields, the material traces a closed loop on the graph, which is the magnetic hysteresis loop.

Due to the low coercivity and low remanent magnetization, amorphous magnetic materials are considered soft magnets, which means they can be easily magnetized and demagnetized. They are often used on transformers, inductors, electric motors, and generators [38].

Interestingly, amorphous magnetic materials often exhibit magnetostrictive behaviour. Magnetostrictive or magnetoelastic materials exhibit a strong coupling between their magnetic and mechanical properties. On the one hand, magnetization can induce mechanical stress, leading to physical strain. This phenomenon, known as the magnetoelastic, inverse magnetostrictive, or Joule effect, is attributed to magnetization rotation in response to an applied magnetic field that alters the arrangement of magnetic domains, causing the material to elongate or contract [39–41]. Strain magnitudes, measured in parts per million, can reach up to thousands of ppm [42–44]. On the other hand, when a material experiences mechanical strain or stress, its magnetic properties change through the magnetostrictive or Villari effect, which occurs due to distortion in the atoms that affect magnetic anisotropy [45].

Magnetic anisotropy is the directional dependence of a material's magnetic properties. It arises from the preferential alignment of magnetic moments along specific axes, leading to a variation in magnetic behaviour depending on the direction of the applied magnetic field. In other words, the existence of magnetic anisotropy lowers or increases the energetic cost of magnetization.

Magnetic anisotropy can be classified into several types, with magnetocrystalline anisotropy being a prominent example. Magnetocrystalline anisotropy arises when magnetic moments become constrained to align along specific crystallographic directions [46]. However, due to the inherent disorder of the amorphous state, the anisotropy governing these materials is predominantly influenced by magnetoelastic and shape anisotropies. Among these, magnetoelastic anisotropy plays a more dominant role in determining the anisotropy of amorphous microwires, due to the stronger coupling between the magnetic and elastic properties of the material, compared to the anisotropy arising from their shape [47].

As Fe-based amorphous materials exhibit great magnetostrictive effect, the stress field within the can affect the easy magnetization axis distribution. These residual stresses originate during the manufacturing process and are due to the different thermal expansion coefficients between the glass and the metal and due to the stretching of the MW during the process [36].

When magnetizing a material, there is an energetic cost directly related to the stresses within the material according to the magnetoelastic anisotropy constant ($K_\sigma$):

$$\boldsymbol{K_\sigma = \frac{3}{2}\lambda_s\sigma} \qquad \textit{(Eq. 0. 1)}$$

Where $\lambda_s$ is the magnetostrictive coefficient (positive for Fe-based MWs), and σ is the stress components. Thus, magnetoelastic anisotropy dictates that as-cast Fe-based MWs should present the ease of magnetization along the MW's length. In the most outer region of the nucleus, a circumferential compressive stress may induce radial magnetic domains if the thickness of the wire allows them, energetically, to exist [47]. Furthermore, as the cover radius is reduced relative to the nucleus, the intensity of the stress field is reduced, modifying the contribution of the magnetoelastic anisotropy [36].

It is essential to consider the magnetic domain structure and the difference in behaviour between positive and negative magnetostrictive microwires. Positive magnetostrictive microwires tend to exhibit magnetic domains with their magnetization aligned parallel to the wire's axis due to the magnetostrictive effect, while negative magnetostrictive microwires typically show domains with a radial and circular orientation [48]. The magnetic behaviour of microwires is determined by this type of domain structure (see Chapter 5).

Regarding shape anisotropy, the aspect ratio, i.e., the ratio between radius and length, will also determine the axis of easy magnetization. At a constant length, a thinner nucleus will give a higher aspect ratio and force the magnetization to occur, preferably in the longitudinal axis. When magnetizing a material, there is an energetic cost directly related to its shape according to the shape or magnetostatic anisotropy constant ($K_{shape}$):

$$\boldsymbol{K_{shape} = \frac{1}{2}(N_a - N_c)\mu_0 M^2} \qquad \textit{(Eq. 0. 2)}$$

Where $N_a$ is the length of the short direction (in this case, the MW's diameter) and is $N_c$ the length of the long direction (MW's length). Briefly, shape anisotropy dictates that the higher the shape ratio, the more energetically favourable it is to magnetize the material in the long direction.

In the absence of other contributions, the magnetoelastic and magnetostriction anisotropies favour a single magnetic domain across the longitudinal axis in Fe-rich microwires with a positive magnetostriction constant. This leads to magnetic bi-stability where the magnetic domain can be in either direction of the MW's longitudinal axis, and the magnetization process to change it will be characterized by a single large Barkhausen jump, as seen in Figure 0. 4.

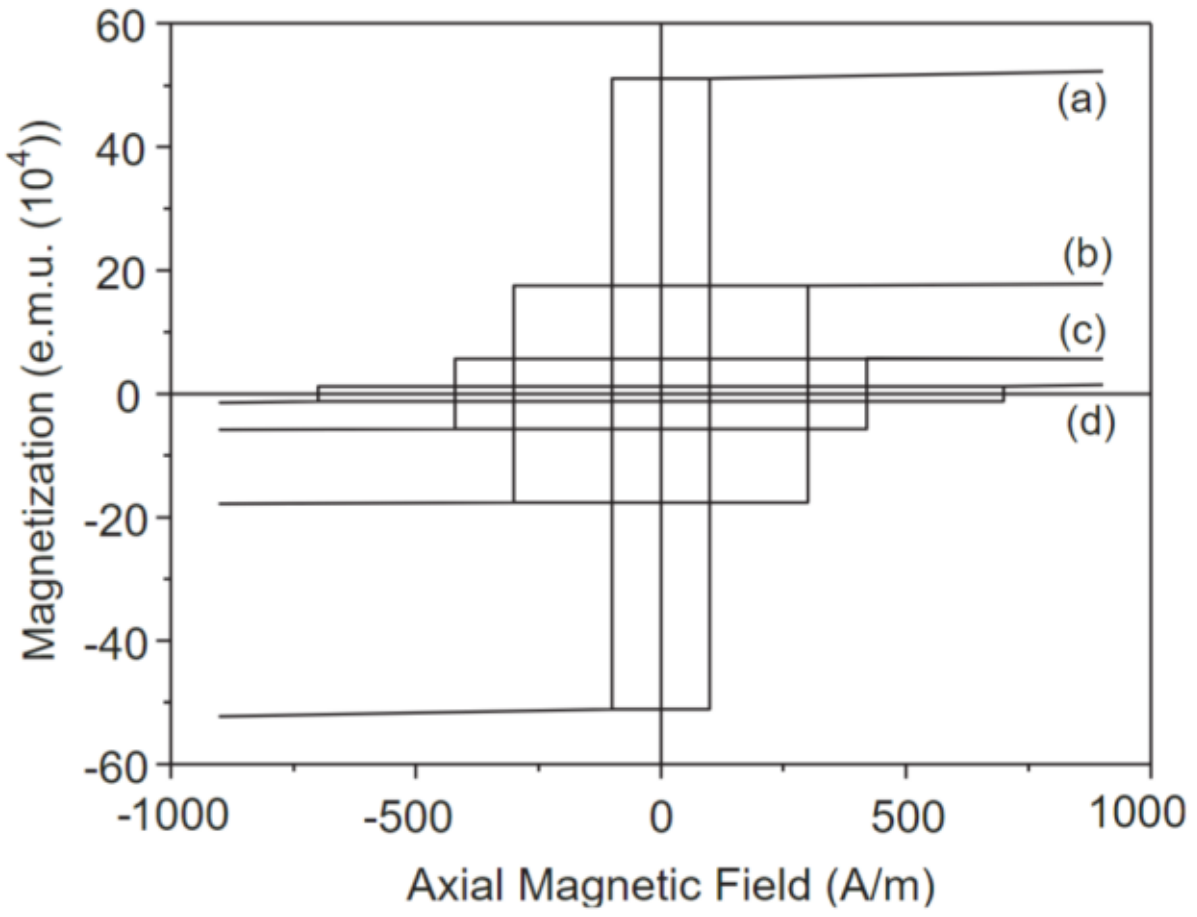

*Figure 0. 4 Hysteresis loops for bi-stable $Fe_{89}Si_3B_1C_3Mn_4$ microwires. Reproduced from* [49].

In contrast, the relatively large thickness of MWs as those used in Chapter 4, allows a different distribution of the magnetic domains, with axial and radial domains that can rotate towards the longitudinal axis, making them particularly useful for magnetoelastic-based applications.

The microstructure of microwires with a FeSiB composition is influenced by thermal annealing, which modifies the microstructure and, in turn, affects their magnetic and electrical properties. Causing a loss of the bi-stable magnetization process. The addition of Cu and Nb elements helps control the processes during thermal annealing, further enhancing the ability to tailor these properties [50–52]. A more detailed explanation of these processes will be provided in a subsequent chapter.

As for the microwires geometry, i.e., the inner and outer diameter and the ratio between them, it mainly affects the magnetic anisotropy and the solidification process (the thicker the wire, the slower the cooling and thus the lower the chance of getting a non-homogeneous amorphous state) [49].

### *0.4 Electromagnetic shielding.*

The growing demand for microwave-based applications, such as wireless communication, 5G, IoT, and radar systems, has increased the need for microwave absorbance technologies. These technologies help mitigate electromagnetic interference, ensuring the proper functioning of electronic devices and systems.

Electromagnetic shielding is particularly important in military applications for stealth purposes by reducing radar visibility. Materials used for microwave absorption, known as electromagnetic shielding materials (ESMs), work by interacting with incident radiation to prevent its transmission and reflection through various mechanisms. They mainly consist of magnetic materials, dielectric materials, or a combination of both.

At first, it was hypothesized that the alternating magnetic component of the GHz signal would induce ferromagnetic resonance (FMR) within the MW [49]. Thus, their electromagnetic shielding performance would be mainly due to magnetic losses [53], adding MW to the list of other magnetic ESMs such as magnetic metals (Fe or Co) [54,55], magnetic alloys (CoNi, FeCo, NiFe) [56–58], and magnetic oxides ($Fe_3O_4$, $Fe_2O_3$, $CoFe_2O_4$) [59–61]. Magnetic ESMs are generally characterized by a high magnetic saturation and low coercivity, i.e., soft magnets. The low coercivity rises from the need to have a material easily magnetizable by the relatively low intensity of the magnetic component in the incident wave. The high saturation on the other hand allows for a large number of magnetic moments to interact with the wave and generate a hysteresis loop with a large surface.

However, the electric permittivity, $\varepsilon_r$, also contributes to microwave absorption through the interaction with the alternating electric field component. Since the electrical conductivity, and thus $\varepsilon_r$, is also affected by the MW's characteristics, possible dielectric loss mechanisms need to be considered. Tests comparing the Fe-based amorphous MW with copper microwires revealed a loss mechanism through a dipolar antenna-like attenuation [62,63]. Briefly, the electric component of the microwave excited a dipolar excitation along the MW with a resonant frequency dependent on the MW's length and conductivity.

The latest advancements in this field suggest the shielding mechanism of MWs may be related to the destructive interference of the microwave. In this case, MWs influence the $\varepsilon_r$ of the composite material, explained through an effective medium standard model, in such a way that extremely thin ESMs can be obtained [64].

As outlined in Chapter 3, numerous uncertainties and knowledge gaps persist concerning the mechanisms of electromagnetic shielding and the formulation of descriptive or predictive models. Despite these challenges, microwires have emerged as promising contenders for the development of ESMs. Research carried out at the *IMA* has significantly contributed to this understanding, paving the way for ongoing applications and commercialization of this technology in military and stealth-related fields through a spin-off company, Micromag 2000, S.L. [65,66].

On the other hand, GBMs, with their unique properties, have emerged as promising candidates for electromagnetic shielding applications. Furthermore, the multifunctionality of GBMs as a component of ESMs may include enhancing mechanical properties, corrosion resistance, thermal stability, and flame retardancy [67–69]. Several studies have demonstrated the potential effectiveness of GBMs in electromagnetic shielding applications [70,71], but they are yet to reach commercial implantation.

## *0.5 Gas sensors.*

Just as electromagnetic pollution can be dealt with using ESMs, gas pollution is dealt with using gas sensors. More precisely, accurate gas monitoring helps identify sources of pollution, track changes in gas concentrations over time and allows for timely intervention and effective mitigation strategies. In the medical field, breath analysis is a non-invasive and user-friendly gas-sensing application used as a diagnostic method that can swiftly detect biomarkers related to health conditions, offering valuable insights into a patient's well-being.

Gas sensors play a crucial role in detecting and addressing gas pollution in various settings. Different gas sensing technologies have been developed, the most relevant or common include electrochemical, optical, surface acoustic wave (SAW), and chemiresistive sensors. Electrochemical sensors use an electrolyte and electrodes to create an electrochemical reaction that detects the presence of a specific gas [72–74]. Optical sensors include infrared (IR) sensors that analyse the absorption or transmission spectra of gases, and laser-based sensors that measure the absorption or scattering of light [75,76]. SAW sensors use acoustic waves that propagate through a material to detect gas molecules, and are particularly attractive due to their small size, low cost, and fast response time [77,78]. Finally, chemiresistive sensors change the material's electrical resistance upon interaction with the target gas [79–83]. Each of these technologies has its advantages and limitations and may be suited for the detection of specific gases.

Within the chemiresistive sensors, GBMs have shown tremendous potential in gas sensing applications. These materials, which include graphene and other graphene-based materials, offer several advantages over traditional gas sensors. With high sensitivity, fast response time, and room temperature operation, graphene-based gas sensors can detect a wide range of gases, such as nitrogen dioxide, carbon monoxide, sulphur dioxide, and ammonia [84–86]. Additionally, these sensors can operate with low power consumption, enhancing their long-term stability and making them suitable for various real-world applications. However, these devices present their own challenges, as is discussed in Chapter 4.

Magnetoelastic materials offer another promising approach to gas sensing applications. These materials, used in magnetoelastic resonance-based gas sensors, are particularly appealing for biomedical applications due to their remote operation and the lack of an integrated power source. They operate under externally applied magnetic fields, which enables easy miniaturization and portability of the sensors. Magnetoelastic materials are sensitive to a range of parameters affecting their mechanical and magnetic properties, making them versatile for detecting different gases and environmental factors [87–91].

In summary, both graphene-based materials and magnetoelastic materials offer unique benefits for gas sensing applications.

### *0.6 Research gap.*

Throughout the thesis, a primary cross-cutting theme will be the transfer of technology. This will manifest in various ways, including standardization challenges, the commercialization or industrialization of processes and materials, and other barriers that hinder the practical realization of the technologies presented in this thesis. The thesis will examine and highlight these challenges and offer assessments wherever feasible.

The standardization challenges are observed and discussed in the thesis for GBMs and amorphous magnetic materials, from the controversial use of the term “graphene” to the lack of standardized methodology to evaluate the performance of different materials in electromagnetic shielding and gas sensing. This challenge is an obvious result of faulty communication between industrial needs and the researcher’s capabilities. The understanding between these two agents is crucial for successful technological development.

For instance, the materials within the GBM family are radically different in terms of physicochemical characteristics and performance in various applications. Understanding these differences may be the key to seeing the long-anticipated graphene technological revolution.

Additionally, navigating the regulatory landscape associated with the introduction of new materials in various sectors, such as healthcare, aerospace, and automotive, can be a challenge that requires close collaboration between researchers, manufacturers, and regulatory agencies.

Another crucial agent for technological development is the general public, i.e., the potential customers and users of these technologies. The public, being non-experts in materials science, may be influenced by hype or concerns surrounding new materials and applications. Promoting public awareness and market acceptance of these technologies can contribute to driving demand, stimulating investment, and supporting the growth of the topics discussed in the thesis.

Bridging the gap between research and commercial applications of GBMs requires a concerted effort from researchers, manufacturers, policymakers, and other stakeholders. By addressing the challenges associated with scalability, cost, material optimization, standardization, and market acceptance, it is possible to unlock the full potential of GBMs and amorphous magnetic materials and revolutionize various industries with their unique properties and multi-functionality.

### *0.7 Research Objectives and Methodology.*

This thesis aims to advance the progress in technological fields with huge commercial potential by (1) developing scalable methods for GBMs, (2) enhancing electromagnetic shielding performance of microwire materials using GBMs, (3) developing and improving graphene-based gas sensors, and (4) creating magnetoelastic resonance-based sensors.

To achieve these objectives, the research presented in this thesis covers every aspect of materials science, including synthesis, characterization through various techniques, comprehensive discussion of results, application development and finally, communication of the results.

The characteristics of the materials are comprehensively examined using a range of analytical methods, while the effectiveness of the devices developed is assessed through meticulous experimentation and in-depth analysis.

Despite the experimental nature of this work, a significant effort has been paid to include theoretical and analytical discussions to support the research, all with an applied approach and commercial viability in mind.

## *0.8 Thesis structure.*

This thesis is organized into eight chapters, including the present one and an annexe. Chapters 2, 3, 4, and 5 are monographic in nature, each focusing on a specific research topic and including a state-of-the-art review. A brief description of the content and motivation for each chapter is included below.

### *0.8.1 Chapter 1: Experimental Techniques.*

In the first chapter, the experimental techniques and equipment generally available that have been used throughout are presented and described, providing technical details and examples when necessary.

It is worth noting that the scope of this chapter is to methods generally accessible to scientific researchers and equipment that are commercially available. For instance, the thesis often includes the development of custom-made experimental techniques or devices that are thus excluded from this chapter.

### *0.8.2 Chapter 2: Pursuing a Scalable Synthesis of Graphene-Based Materials.*

In the second chapter, a comprehensive revision of the synthesis of graphene and graphene-based materials will be presented.

From the first studies analysing the physicochemical characteristics to the state-of-the-art production methods, this chapter dwells on the controversial terminology around “graphene” and the differences within the GBM family, the effect it has had on the commercialization of its potential applications and presents a detailed experimental work about graphene-based materials synthesis.

This experimental work will mainly focus on the ball-milling technique. Several conditions, parameters and variations on the technique have been tested, obtaining interesting results that pave the way to large-scale production of graphene-based materials.

Interestingly, the result of one of these ball-milling technique variations has been called Few-Layered Mesoporous Graphene (FLMG). This material is comprised of a mesoporous agglomeration of few-layered graphene domains.

FLMG represents a change in the paradigm of high-quality graphene production, i.e., the presence of structural defects in the material becomes a desirable feature instead of a negative attribute. Briefly, by focusing on production quantity instead of structural quality, ball-milling may offer a realistic option to meet the huge and increasing demand for graphene-based material for certain applications, such as functional or structural composite reinforcements or energy-related applications [92–98]. In particular, FLMG has been tested in microwave absorbance and gas sensing applications during this thesis, yielding interesting results that are presented in subsequent chapters.

The chapter will also introduce other synthesis techniques and will further elaborate on chemical vapour deposition. Explaining the work done on this method for the development of chemiresistive graphene-based gas sensors is presented in the second part of Chapter 4.

### *0.8.3 Chapter 3: Practical Contexts and Synergy in Microwave Absorbing Materials.*

Chapter 3 delves into microwave absorbance applications, specifically focusing on shielding electromagnetic radiation within the 2-18 GHz frequency range. This range is particularly relevant for applications such as wireless communication, radar systems, and medical applications [99,100]. The chapter originates from the long-standing expertise at the *IMA* in using MW for microwave absorption and pushes this know-how even further by introducing graphene-based materials into the mix.

The chapter highlights the discrepancies between the popular theoretical approach based on transmission line theory and the heuristic, empirical approach used in this study. An analytical argument is provided, questioning the applicability of the theoretical approach, and explaining the lack of commercial success of materials with outstanding theoretical performances reported in the literature.

A heuristic approach, which is a problem-solving method that involves using practical, intuitive, and common-sense strategies to find solutions rather than a formal or theoretical understanding of the problem, is then employed for characterizing microwave-absorbing materials. By utilizing free-space measurements (FSM), this approach emulates real-world conditions.

This chapter then explores the use of FLMG in ESMs and evaluates their performance using the free-space measurement (FSM) method. The experiments demonstrate that FLMG, when combined with MW, can produce ultrathin composite ESMs with tuneable microwave absorbance properties, showing potential in electromagnetic interference shielding and radar stealth applications.

### *0.8.4 Chapter 4: Staying on Track with Graphene Gas Sensor Breakthroughs.*

Chapter 4 reflects the expertise on chemiresistive gas sensors, and other types of gas sensing technologies, from the SENSAVAN research group at the *Instituto de Tecnologías Físicas y de la Información* (*ITEFI*) from *Consejo Superior de Investigaciones Cientificas* (*CSIC*) and focus on the use of graphene-based gas sensors towards gas pollutants monitoring.

After a comprehensive revision of gas pollutants, the operation principle of graphene-based sensors will be discussed. The limitations of these sensors and alternatives to boost their performance will also be studied. The fundamentals of gas sensor evaluation will be explained, dealing with the relevant figures of merit and experiment design to understand the difficulties of direct comparison within the state-of-the-art examples.

The use of FLMG on gas sensing applications will be studied through experimental work demonstrating its potential as a low-cost and large-scale producible technology for gas pollution ($NO_2$) monitoring.

In a second work, attention was directed to the use of UV irradiation to overcome the main drawbacks of graphene-based gas sensors compared to their more commercially accepted alternatives. This work resulted from the collaboration with the group of Sten Vollebregt at TU Delft and benefitted from their great expertise on CVD-grown GBMs. The use of continuous irradiation of UV during the device's operation resulted in increased sensitivity and response, and recovery times, greatly enhancing the performance of the sensor.

The extraordinary research possibilities of working with different materials, analytes, UV irradiation conditions, and sensing evaluation protocols highlighted the importance of maintaining focus during research. Striking a balance between exploration and sustaining a clear direction is essential to ensure the efficient progress of projects and the development of innovative solutions, particularly, to address the pressing challenges in gas sensing and air quality monitoring.

In sum, this chapter investigates the use of graphene-based materials, specifically FLMG and MLG, for gas sensing applications, with a focus on $NO_2$ detection. The research highlights the potential of graphene-based materials in gas-sensing applications. It contributes to understanding the sensing mechanisms, paving the way for advanced gas sensors to monitor air quality and protect public health.

### *0.8.5 Chapter 5: Integrating Magnetism and Gas Sensing for Novel Applications.*

The last experimental chapter joins the knowledge of *IMA* and *ITEFI* to create a new research line for either Institute, the use of magnetostrictive material on gas sensing applications.

Although this line is not strictly new in the literature, the first examples can be traced to the early 2000s [87,88,101], and scarce examples have been recently reported. It is my understanding that this scarcity is due to the technological limitations of the measuring setups that monitor the response of magnetostrictive materials towards fast-occurring reactions.

This problem has been assessed by developing a real-time characterization system that is carefully detailed in the chapter.

The system was then validated by developing a gas sensor focused on breath analysis applications. Breath analysis, as well as other biomedical applications, can greatly benefit from the remote operation nature of magnetostrictive sensors.

Finally, a comprehensive experimental study on the optimization of magnetostrictive materials is presented to further enhance the performance of this type of sensor.

The work presented in this chapter is expected to bring a significant jump forward in the field of magnetostrictive sensors.

### *0.8.6 Closing Remarks: Synthesis and Reflections - Concluding the Thesis Journey,*

In this concluding chapter, the key findings, implications, and contributions of the research are reflected upon, acknowledging advancements and limitations. The main results are summarized and their relevance to the existing literature is discussed. The implications of these findings for the scientific community, potential applications, and future research directions are then examined. By recognizing the limitations of the work, areas, where further investigation is needed, are also explored.

### *0.8.7 Scientific production overview and Annex 1.*

In Annex I, the relevant published works resulting from the thesis are presented and summarized, including scientific papers and patents. Moreover, contributions to different scientific conferences are briefly mentioned.

Finally, unpublished related works, such as bachelor's and master's degree theses, are also covered. Although there is no formal recognition of collaboration with these students, their work has had an impact on this thesis. As a result, it is acknowledged in this additional chapter.

## *0.9 References.*

# Chapter 1: Equipment and techniques.

*Chapter Introduction; Mills; Electronics; Physicochemical characterization techniques; References.*

## 1.1 Chapter Introduction.

Chapter 1 provides an elaborated glossary of the relevant equipment and characterization techniques used in this work. The chapter includes definitions and references to aid in understanding the present work and guide future studies. The chapter begins by introducing the ball mills used for graphene synthesis in Chapter 2 and their specifications, followed by the microstructural characterization techniques used throughout all chapters. Finally, electronic devices and related techniques presented in Chapters 3-5 are presented.

## 1.2 Mechanical Mills.

In this work, mechanical ball mills were used to investigate an industrial-scale suitable technique for graphene-based materials (GBMs) synthesis. The details of this work are presented in *Chapter 2: Graphene Synthesis*. Briefly, three mills with different characteristics were used, planetary, oscillatory, and high-energy mill. It was found that the characteristics of the mill greatly affect the product material, but overall, it was found that milling can produce GBMs from graphite through exfoliation and is a potential technique for large-scale GBM production. These mills belong to *Instituto de Magnetismo Aplicado (IMA)*, a research institute at the *Universidad Complutense de Madrid (UCM)*

### 1.2.1 Planetary mill (Retsch PM400).

The planetary ball mill (see Figure 1. 1a) consists of four milling jars that rotate both in place and eccentrically within a larger disc, resulting in a characteristic planetary movement that includes rotational and translational motion (see Figure 1. 1b). This movement creates shear forces that are favourable for exfoliation of stacked materials like graphite during the process, as opposed to less effective impact forces. Studies have shown that this configuration promotes exfoliation in an effective way [1,2]. When setting up the mill, it is important to balance the jars to reduce anomalous centrifugal forces. This is done by setting the jars in pairs with similar weights in opposite configurations, like the care taken for other types of equipment with spinning components.

The PM400 can run pre-set milling sequences from 30 to 400 rpm, with 1 rpm increments, and from 1 minute to 99 hours, with 1-minute increments, and it can be purchased for around 20.000 € [3]. The PM400 could be considered a mill suited for long-time mild-energy millings [4].

This equipment was used for dry and wet ball-milling tests using stainless steel 250 ml jars and balls of the same material with diameters of 10 and 6.25 mm. Other works with this mill at *IMA* have previously been reported in [4–7].

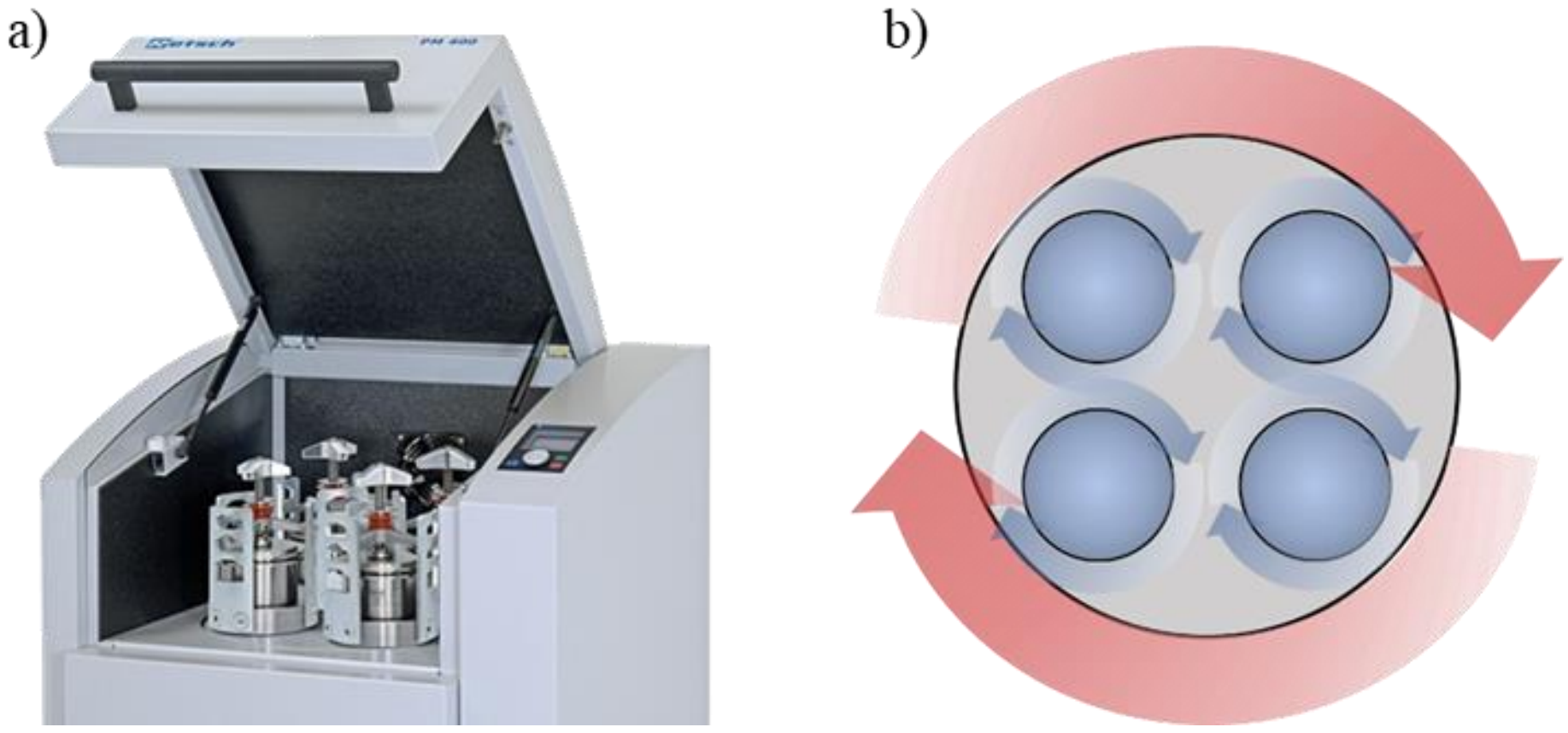

*Figure 1. 1 a) The PM400 mill and b) the planetary movement of the milling jars.*

### 1.2.2 Oscillatory mill (Retsch MM400).

This oscillatory mill (see Figure 1. 2a) has an oscillating or swinging movement of the two milling jars (see Figure 1. 2b). The ball inside the jars moves from side to side, impacting against the materials and the walls of the jar. This movement generates great impact forces, whereas neglectable shear forces should be expected. When setting up the mill, it is essential to balance the jars using similar weights in both milling positions. Each milling arm moves in opposite directions, and unbalanced jars may cause vibrations with adverse effects on the equipment and the milling process. Additionally, the milling of graphite at high frequencies may cause fractures in the inner coating of the jars due to its high lubricity. The hardness of the milling instruments should thus be carefully considered.

The MM400 can run pre-set milling sequences from 3 to 30 Hz (180 to 1800 rpm), with 0.1 Hz (6 rpm) increments, and from 1 minute to 99 minutes, with 1-minute increments [8]. It can be purchased for around 10.000 €. Contrary to the PM400, the MM400 could be considered a mill suited for short bursts of high-energy impact-based millings [4].

This equipment was used for dry ball-milling tests using WC 50 ml jars and balls of the same material with a diameter of 15 mm.

This equipment was used during this thesis to investigate the dry ball-milling of graphite, and a material called few-layered mesoporous graphene (FLMG) was obtained [9]. Other works with this mill at *IMA* have previously been reported in [4,10].

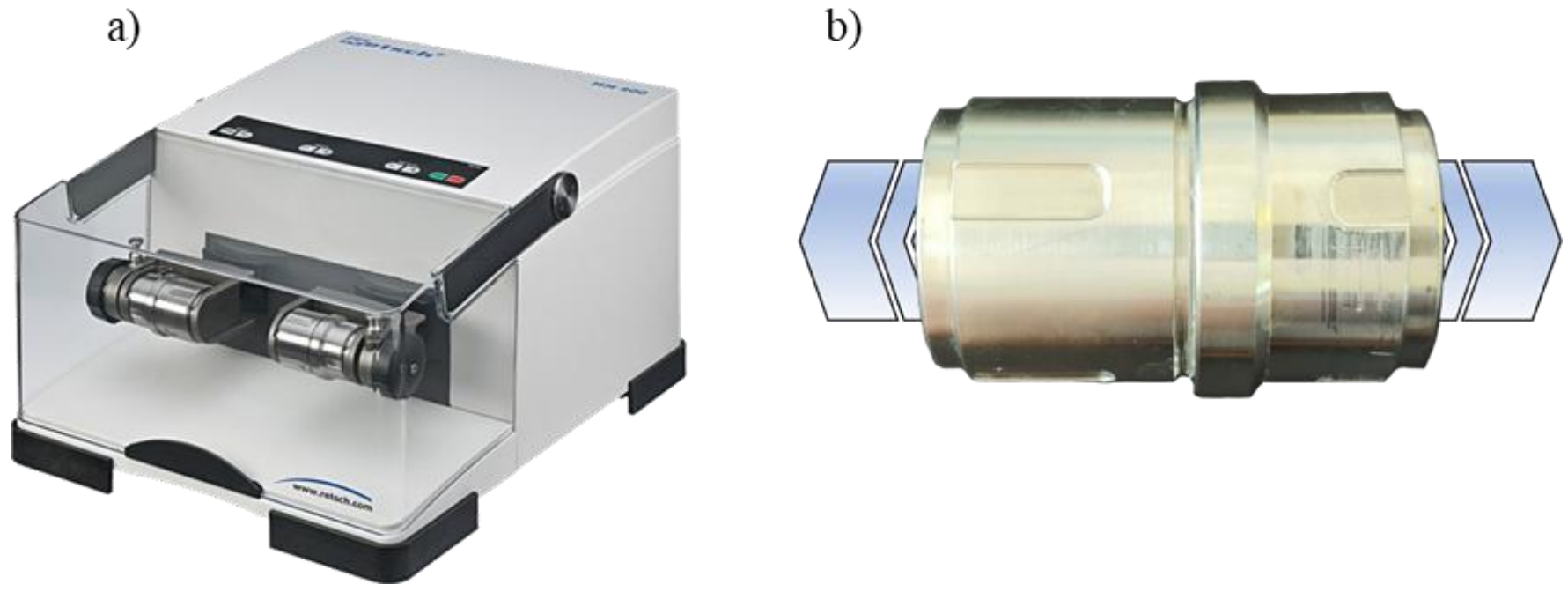

*Figure 1. 2 a) The MM400 mill and b) the swinging movement of the milling jars.*

### *1.2.3 High-energy mill (Retsch Emax).*

This high-energy mill (see Figure 1. 3a) has an eccentric circular movement of the jar (see Figure 1. 3b). It resembles the movement of the planetary mill if the jars stayed still, and the disk was much smaller. The ball inside the jars moves circularly, rubbing the material against the jar's walls. This movement generates mainly shear forces. Similarly to the MM400, the Emax could be considered a mill suited for short bursts of high-energy milling but providing mainly shear force instead of impact force.

The Emax can run advanced pre-set milling sequences from 300 to 2000 rpm with 100 rpm increments and from 1 min to 99 h with 1-second increments. In addition, the mill includes liquid-cooled sleeves for the jar that provide temperature control during the milling, and temperature can be controlled from 0 to 119 ºC with 1 ºC increments [11]. For contextualization, it can be purchased for around 50.000 €.

It was used with 50 ml WC jars and ~600 balls of the same material with 3 mm in diameter. With one standard jar and one atmosphere-controllable jar, i.e., a jar that allows the modification of its atmosphere through connectors in its lid.

This equipment was used during the thesis to investigate the dry ball-milling of graphite under different atmospheres and to compare these results with the ones obtained with other mills. Other works using this mill have previously been reported in [12].

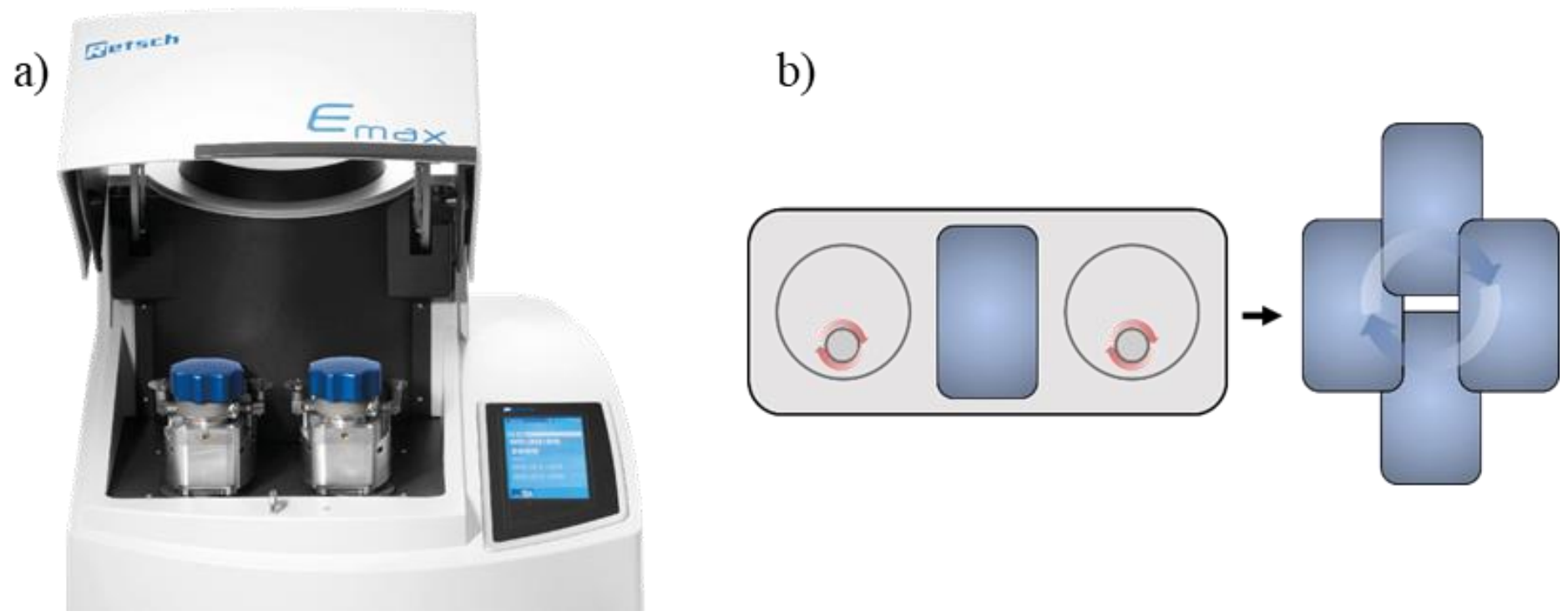

*Figure 1. 3 a) The Emax mill and b) the eccentric movement of the milling jars.*

### *1.3 Electro- and magnetic characterization.*

Some crucial electronic-related characterization techniques have been omitted from this section. These are the techniques related to magnetoelastic resonance characterization. The reason is that the development of these techniques constitutes core experimental work with custom-made components and is therefore presented in great detail in *Chapter 5*.

#### *1.3.1 Electromagnetic Shielding.*

In the electromagnetic shielding characterization experiment, a free-space measurement approach was taken. Briefly, an anechoic chamber is used to create an environment free of electromagnetic reflections and interference. The chamber is lined with electromagnetic radiation absorbers, including the plates, to eliminate border effects and ensure accurate measurements. The experimental setup consists of two EMCO 3160-07 horn antennas with an operating range between 800 MHz and 19 GHz, positioned under far-field conditions of the absorbing optimized painting. Under these conditions, the transmission of electromagnetic waves can be considered plane waves, providing a more controlled environment for testing [10,13,14].

An Agilent E8362B PNA Vector Network Analyser (VNA) is employed to control and analyse the electromagnetic signal in the microwave absorbance characterization experiment. The VNA is a specialized test instrument that measures the complex scattering parameters (S-parameters) of a device under test (DUT), characterizing how the DUT interacts with and responds to electromagnetic waves.

In this experimental setup (see Figure 1. 4), the scattering coefficient S21 is measured directly by the network analyser. The S21 parameter is one of the four fundamental S-parameters (S11, S12, S21, and S22) and represents the transmission from port 1 (emitting antenna) to port 2 (receiving antenna). Specifically, S21 quantifies the ratio of the output signal at port 2 to the input signal at port 1 in terms of both magnitude and phase.

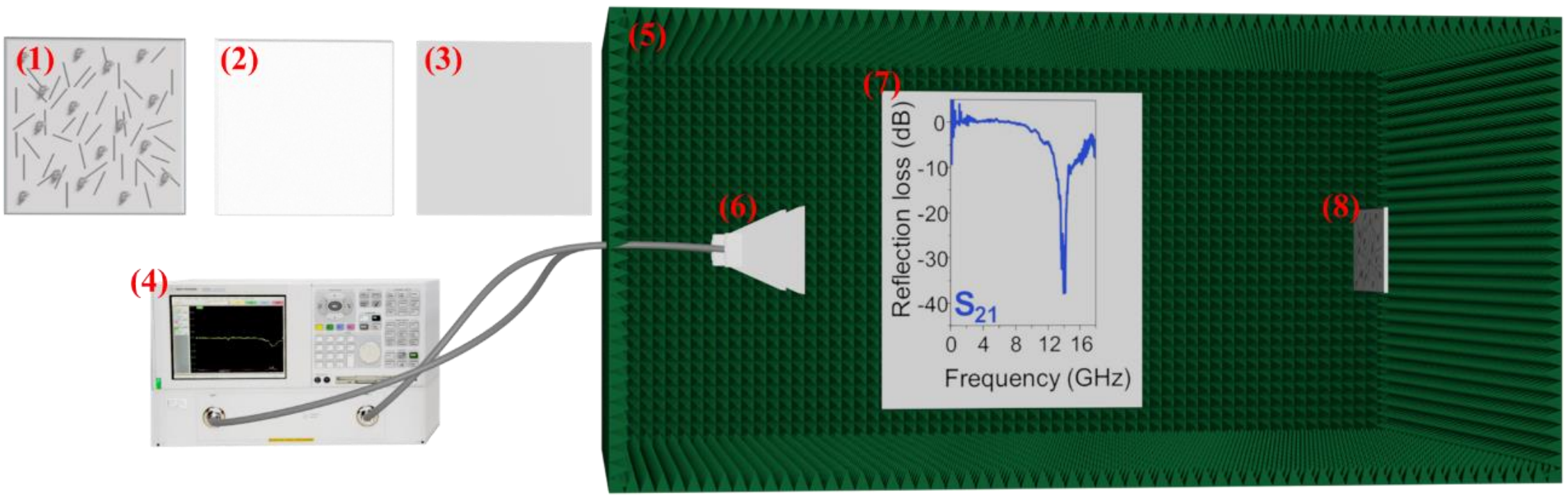

*Figure 1. 4 Illustration of the setup, including the paint-based composite (1), the dielectric substrate (2), the metal backplate (3), the VNA (4), the anechoic chamber (5), the antenna with a 2-port configuration (6), a representation of the measure (7) and the sample (8).*

By measuring S21, the experiment evaluates the amount of electromagnetic signal that is reflected by the metal-backed sample, providing insight into the sample's microwave absorption properties. The sample preparation is further described in Chapter 3.

Before the measurements, the network analyser is calibrated considering free-space measurement (FSM) conditions to guarantee accurate results. The FSM calibration is done by acquiring a background measurement, i.e., with the fully prepared setup except for the sample. The background measurement is then mathematically extracted from the subsequent measurements. This calibration considers the reflection and losses throughout the cables and within the chamber and is essential to eliminate systematic errors in the measurements and ensure that the VNA provides reliable data.

It is important to note that any deviations within the setup from the calibrated configuration will induce artefacts and alternation to the measure. This deviation may go from the orientation of the sample holder, with modifies the reflection plane of the sample respecting the antenna, to the position and curvature of the cables.

#### *1.3.1.1 Optimization of the FSM Technique.*

One notable example of these deviations was investigated during the thesis. Traditionally, the samples were attached to the sample holder using several metallic clips (see Figure 1. 5a). Since the metallic surfaces act as reflection planes, the placement and amount of these clips affect the measurement by creating conditions different to the calibrated ones. Furthermore, with these metallic clips, the plane orientation is affected by the thickness of the sample, meaning that even if the amount and placement of the clips are unaltered, a new calibration is needed for each sample thickness. To solve this problem, plastic (PLA) clips were fabricated using a 3D printer (see Figure 1. 5b).

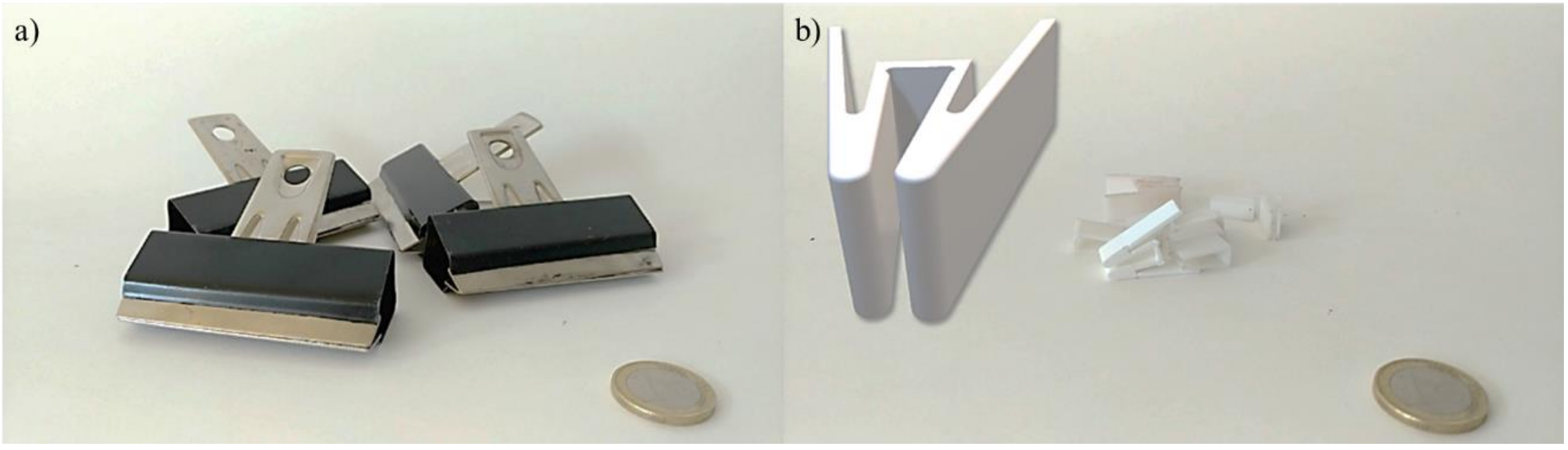

*Figure 1. 5 Photography of (a) the metallic clips and (b) the plastic clips, including the 3D model.*

An experiment was designed to measure the RL of a dummy non-absorbent sample, with metallic and plastic clips in different configurations. The results are presented in Figure 1. 6.

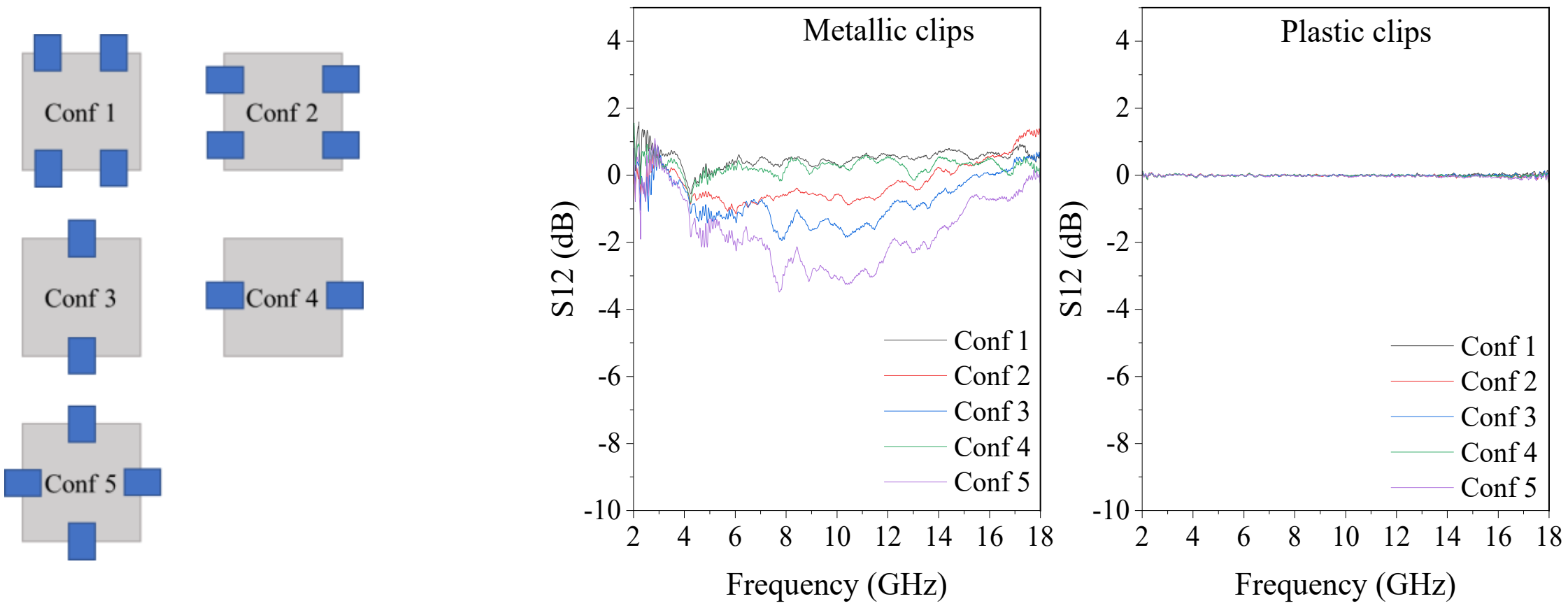

*Figure 1. 6 Different arrangements of clips and the resulting RL spectra with a non-absorbing sample.*

The effect of the arrangement of metallic clips has a clear effect on the RL measurements, as previously discussed since the reflections due to the metallic surfaces of the clips differ from the calibrated conditions, these are manifested in the RL measurement, both in terms of the shape of the RL curve and intensity values. Notably, these differences are not manifested when using plastic clips.

It could be argued that these effects may become unnoticeable when using a real active sample, in which the contribution of the electromagnetic shielding material (EMS) outweighs the contribution of the metallic clips. With that in mind, the test was repeated with a well-known microwire-containing sample that is often used as a reference sample.

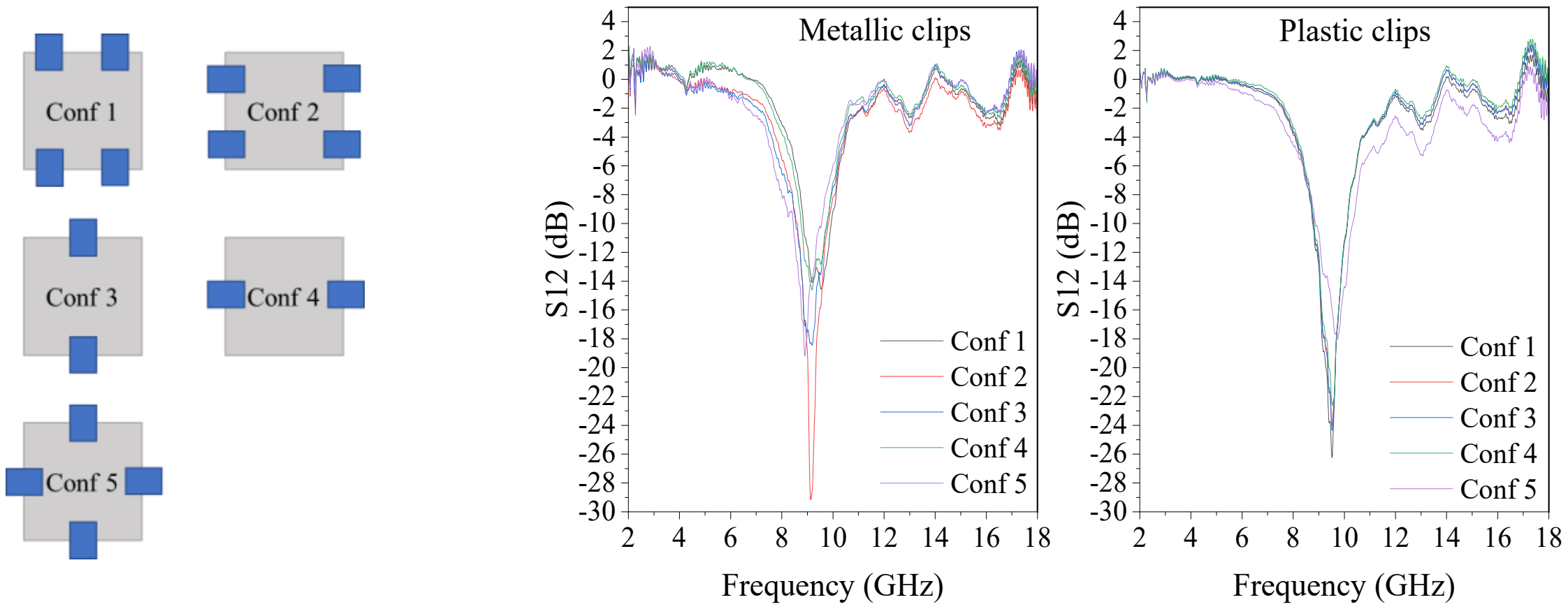

*Figure 1. 7 Different arrangements of clips and the resulting RL spectra with absorbing sample.*

As observed in Figure 1. 7, the arrangement of the metallic clips has a significant effect on an EMS sample. In particular, the frequency of the minimum shifts between 8.80 GHz and 9.20 GHz, and the value of this minimum shifts between -13.0 dB and -29.1 dB. With the plastic clips, the variations in configurations 1 to 4 the minimum sits at 9.53 GHz with a value between -22.6 dB and -26.2 dB. The plastic clips in configuration 5 are noticeably different, although this could be related to an experimental error of different nature.

Consequently, all the RL measurements presented in this thesis were done using these plastic clips. Furthermore, the calibration step was periodically repeated during the measurement after each change of sample or conditions.

The microwave characterization was done at *IMA*. Prior works using this technique have been reported in [10,14].

### *1.3.2 Electrical Resistance Monitoring.*

The measuring of electrical resistance or conductivity has been extensively used as an auxiliary technique throughout most of the experimental works involving electronic components. More notable, it is a core technique for evaluating chemiresistive sensors such as those presented in Chapter 4.

Regarding these sensors, the electrical resistance has been measured with a digital multimeter, often controlled remotely with a computer typically using a two-probe configuration. Briefly, the multimeter sends a small current through the material being measured and measures the voltage drop across it. The resistance can then be calculated using Ohm's law (R=V/I).

Several multimeters have been used, like the Fluke 8842A, the Keithley 2001, or the Keithley 2400. There are no significant operational differences between them other than the drivers for the acquisition software integration or their accuracy. Electrical resistance characterization was done at *IMA* and at the SENSAVAN's lab at the *Instituto de Tecnologías Físicas y de la Información* (*ITEFI*) from *Consejo Superior de Investigaciones Científicas* (*CSIC*), Spain.

### *1.3.3 Vibrating Sample Magnetometer.*

The magnetic characterization presented in the form of hysteresis loops was performed using a vibrating sample magnetometer (VSM, PPSM from Quantum Design). The VSM applies a magnetic field to a small sample attached to a cantilever, which is set into a state of mechanical oscillation (see Figure 1. 8). By measuring the magnetization of the sample as the external magnetic field is varied, the VSM provides high sensitivity and accuracy in determining the magnetic moment, susceptibility, and coercivity of the material.

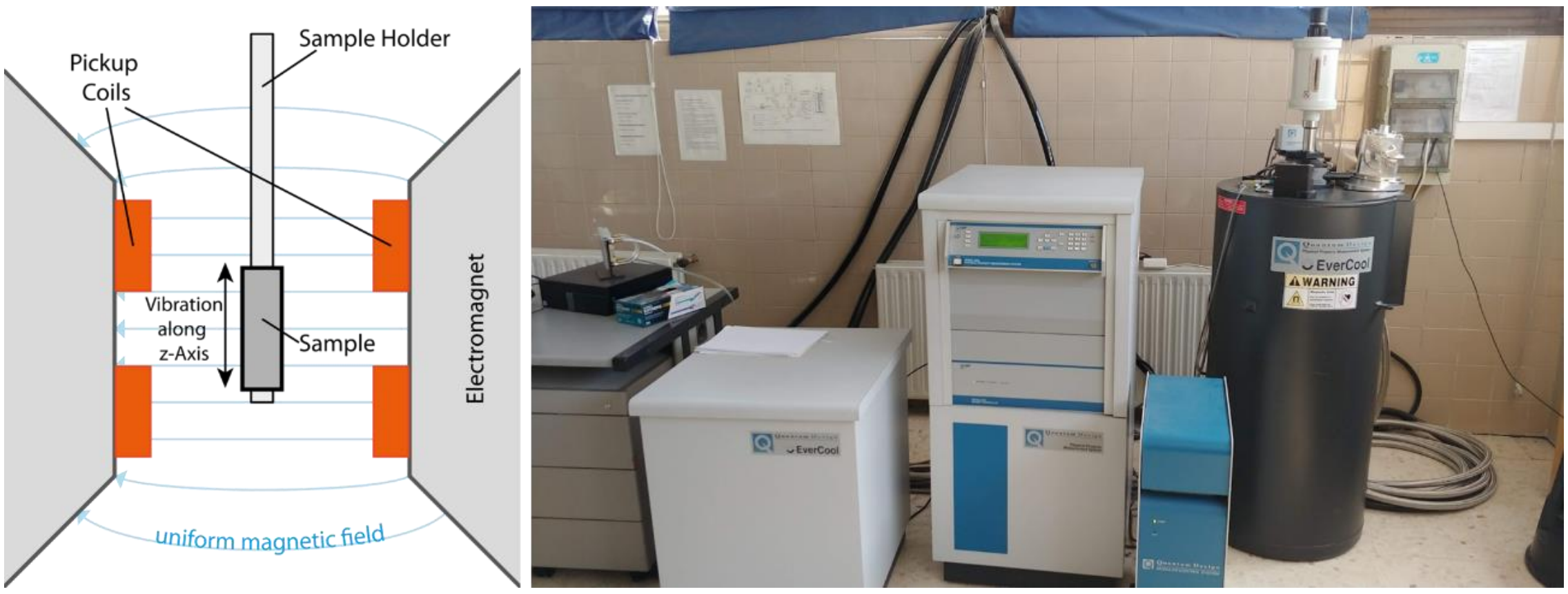

*Figure 1. 8 Representation of a VSM (left) and photography of the PPSM from Quantum Design at Instituto de Magnetismo Aplicado (right).*

This specific VSM incorporates a set of superconductive coils that are immersed in liquid helium, allowing for magnetic fields up to 7 T in the sample space, although such intense fields were rarely used in our soft magnetic materials. Despite its high accuracy, with magnetic field accuracy of ~0.5 Oe and magnetization accuracy of $10^{-5}$ emu, the VSM is limited by its sample size, which is approximately 1 X 2.2 X 0.4 mm for the thin-film sample holder and 4x4x4 mm for the pill-shaped holder.

Overall, the VSM is a valuable tool for characterizing the magnetic properties of materials, providing high sensitivity and accuracy for analysis. In this study, we utilized the VSM to obtain hysteresis loops that allowed us to characterize the magnetic properties of our materials. The VMS characterization was carried out at the *IMA*, Spain. Previous uses of this VSM have been reported in [15,16].

## *1.4 Physicochemical characterization techniques.*

The following physicochemical characterization techniques were used to investigate the properties of the materials synthesized or applied in this thesis. Each technique offers a distinct perspective into the materials' physical and chemical characteristics, providing insights into their morphology, structure, and properties. The specific application of each technique to the materials studied is provided as relevant examples when needed. The combination of these techniques offers a comprehensive understanding of the materials' physicochemical properties.

### *1.4.1 X-ray diffraction.*

X-ray diffraction is a technique that uses X-rays to investigate the crystal structure of a material. When a crystalline sample is exposed to an X-ray beam, the atoms in the sample diffract the X-rays. The resulting diffraction pattern consists of angles and intensity that are unique to each type of crystal. The pattern displays the diffraction angle, theta, versus intensity and typically includes a series of maxima that arise from various periodic arrangements, as described by Bragg's Law (Equation 1.1) [17]. By analysing the scattered X-ray pattern, it is possible to determine the arrangement of atoms in the crystal and identify the type of crystal present in the sample. While the X-ray diffraction technique does not provide information on the chemical nature of the sample itself, it is possible to determine the composition and phase of a material by comparing its spectra with reference data (see Table 1. 1 for an example).

$$2d\sin\theta = n\lambda \qquad \textit{(Eq. 1. 2)}$$

Where $d$ is the interplanar distance, $n$ is an integer number referring to the diffraction order, and $\lambda$ is the X-ray wavelength used.

*Table 1. 1 Calculated pattern for graphite. Extracted from* [18].

| Peaks | hkl | d spacing (Å) | 2θ angle | Relative intensity |
|---|---|---|---|---|
| 1 | 002 | 3.3553 | 26.543 | 100 |
| 2 | 100 | 2.1319 | 42.360 | 3.5 |
| 3 | 101 | 2.0318 | 44.555 | 16.7 |
| 4 | 102 | 1.7994 | 50.689 | 3.0 |
| 5 | 004 | 1.6777 | 54.661 | 5.4 |
| 6 | 103 | 1.5433 | 59.882 | 4.5 |
| 7 | 104 | 1.3184 | 71.498 | 0.7 |
| 8 | 110 | 1.2309 | 77.480 | 5.0 |
| 9 | 112 | 1.1556 | 83.605 | 7.3 |
| 10 | 105 | 1.1358 | 85.400 | 1.0 |

The Scherrer equation (Equation 1.2) is a widely used method to determine the size of crystalline domains or crystallites from X-ray diffraction data. The equation is based on the relationship between the width of diffraction peaks in the X-ray diffraction pattern and the size of crystallites in the sample [19]. The equation relates the crystallite size D to the full width at half maximum intensity (FWHM) of the diffraction peak. The constant K in the equation depends on the shape of the crystallites, which can be approximated as spherical, cubic, or hexagonal, depending on the material and the crystalline structure.

$$D = \frac{K\lambda}{FWHM\cos\theta} \qquad \textit{(Eq. 1. 3)}$$

Regarding its use for graphene characterization, the regular arrangement of the carbon atoms in graphene gives rise to a characteristic diffraction pattern. Furthermore, this pattern is different from that of graphite or oxidated graphene, making this technique helpful in identifying these materials [20,21]. The maximum located at 26.5º corresponds to an interplanar distance of ~0.33 nm, the typical spacing between graphene sheets in graphite or other GBMs. On the other hand, the maximum located at 10º corresponds to an interplanar distance of 0.9 nm, roughly the thickness of an oxidated graphene sheet [22]. The diffraction pattern can also provide information about the particle size and crystalline domain size, providing some insight into the lateral size of the graphene sheet and the presence of defects.

During the GBM synthesis reported in Chapter 2, XRD was extensively used as a first characterization technique. It allowed for preliminary discussions on the choice of milling parameters, being a powerful and cost-effective technique. The equipment used was a PANalytical X'Pert PRO MPD with a Cu source (λ= 1.54056 Å). The spectra were obtained at the Unidad de Difracción de Rayos X from *UCM*, Spain.

### *1.4.2 X-ray fluorescence.*

X-ray fluorescence (XRF) is an elemental analysis technique in which the sample is irradiated with X-rays, and the radiation is absorbed by the sample's atoms, which subsequently causes the emission of secondary X-rays with a characteristic energy that is specific to the element [23]. The XRF spectrometer measures the intensity of the emitted X-rays as a function of their energy. This data is used to identify the elements present in the sample and their relative concentrations. The data is typically analysed using software that compares the measured X-ray intensities to a reference database of known X-ray energies for each element.

Regarding graphene characterization, XRF provides an elemental composition of graphene samples thanks to its high sensitivity to a wide range of elements, including light elements such as hydrogen and carbon. It can be used to measure trace amounts of these elements [24]. Therefore, XRF can also be used to determine the oxygen content and C/O ratio of GBMs or the presence of trace contaminants.

XRF was used to corroborate the presence of contaminants in the ball-milled samples presented in Chapter 2. It should be noted that as opposed to inductively coupled plasma mass spectrometry (ICP-MS), also used for elemental analysis, XRF is a less-destructive technique. It is not strictly non-destructive since the sample is still irradiated with X-rays, and therefore, there is some level of interaction between the X-rays and the sample. This can cause minor changes in the sample, especially if it is sensitive to radiation. The data used in this thesis were provided by the Unidad de Difracción de Rayos X from *UCM*, Spain.

### *1.4.3 Electron Microscopies Techniques.*

Scanning electron microscopy (SEM) is a technique that uses a focused beam of electrons to produce detailed images of the surface of a sample. The electrons interact with the atoms in the sample, and the beam is reflected towards the detector. This technique is particularly useful in studying the arrangement of particles and measuring the size and shape of nanoparticles, integrated circuits and other structures.

SEM is useful for studying morphology below optical resolution and has been used to analyse nanofibers, microwires and GBMs throughout this research. For instance, under SEM graphene sheets appear thin and translucent with folds and wrinkles (see Figure 1. 9). However, SEM has no chemical sensitivity to distinguish between some GBMs; for example, rGO, GO, and graphene will have a similar appearance when observed with this technique [25].

Transmission electron microscopy (TEM) uses a beam of electrons to produce detailed images of the internal structure of a sample. The electrons are transmitted through the sample, and the resulting signals are used to create a high-resolution image of the sample's internal structure. TEM can be used to study the shape, size, and composition of individual atoms or molecules in a sample. Moreover, the analysis of diffraction patterns, either by direct measurement or by using methods based on Fourier transformation, can provide information on the chemical aspects of the sample [26].

Regarding graphene, TEM allows the study of the arrangement of the carbon atoms in the graphene lattice and to measure the size and shape of the graphene flakes. In addition, TEM can be used to study the distribution of defects in the graphene lattice, such as vacancies or impurities, and to identify the presence of other materials in the graphene sample [26].

For both electron microscopy techniques, additional modules can be incorporated for supplementary characterization, such as energy dispersive x-ray spectroscopy (EDS) for elemental analysis or electron energy loss spectroscopy (EELS) that provides information about the electronic structure.

The main differences between SEM and TEM lie in their penetration capabilities. SEM has limited penetration within samples and provides information mainly on the sample's surface, while TEM completely crosses the sample. Consequently, SEM is best suited for bulky samples, while TEM requires thinner samples.

SEM was utilized in Chapter 2 to investigate the particle morphology of the milled samples. As this technique is capable of producing a relatively large number of images in one session, it is also commonly used to obtain size distributions through statistical treatments. Additionally, SEM proved useful in accurately measuring the thickness and diameters of microwires or FLMG particles.

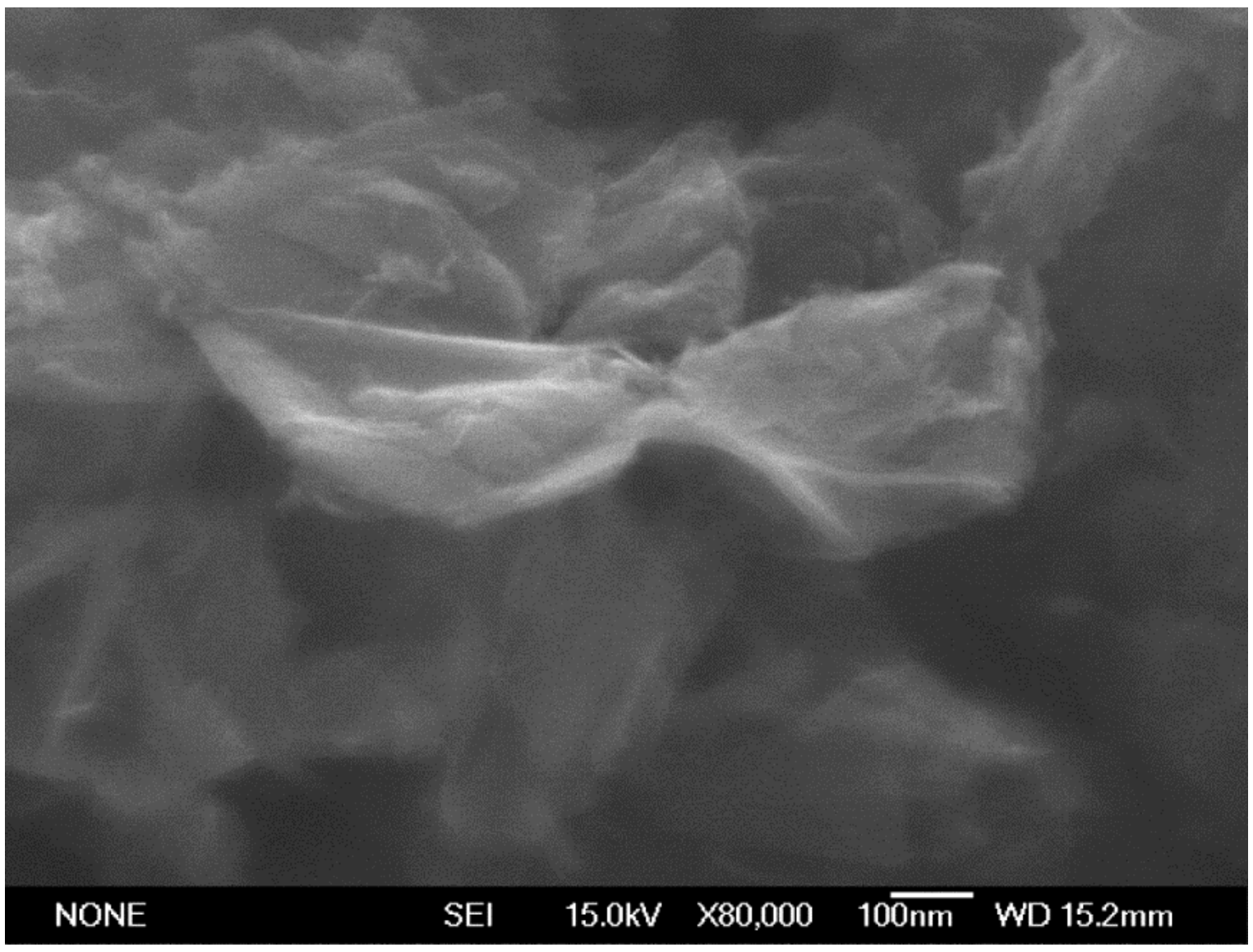

*Figure 1. 9 Typical SEM images for GBMs, particularly graphene found in a 60-min milled sample in the oscillatory mill.*

However, an important limitation of the SEM technique is that the samples must be conductive to avoid static charging from the electron beam. If the sample is not conductive, the static charge can deflect the beam, resulting in artefacts in the image. During the course of this work, the microwire samples, which were covered with non-conductive Pyrex glass, and the functional coatings for biosensors, made from polymeric nanofibers, exhibited this phenomenon (see Figure 1.10 ).

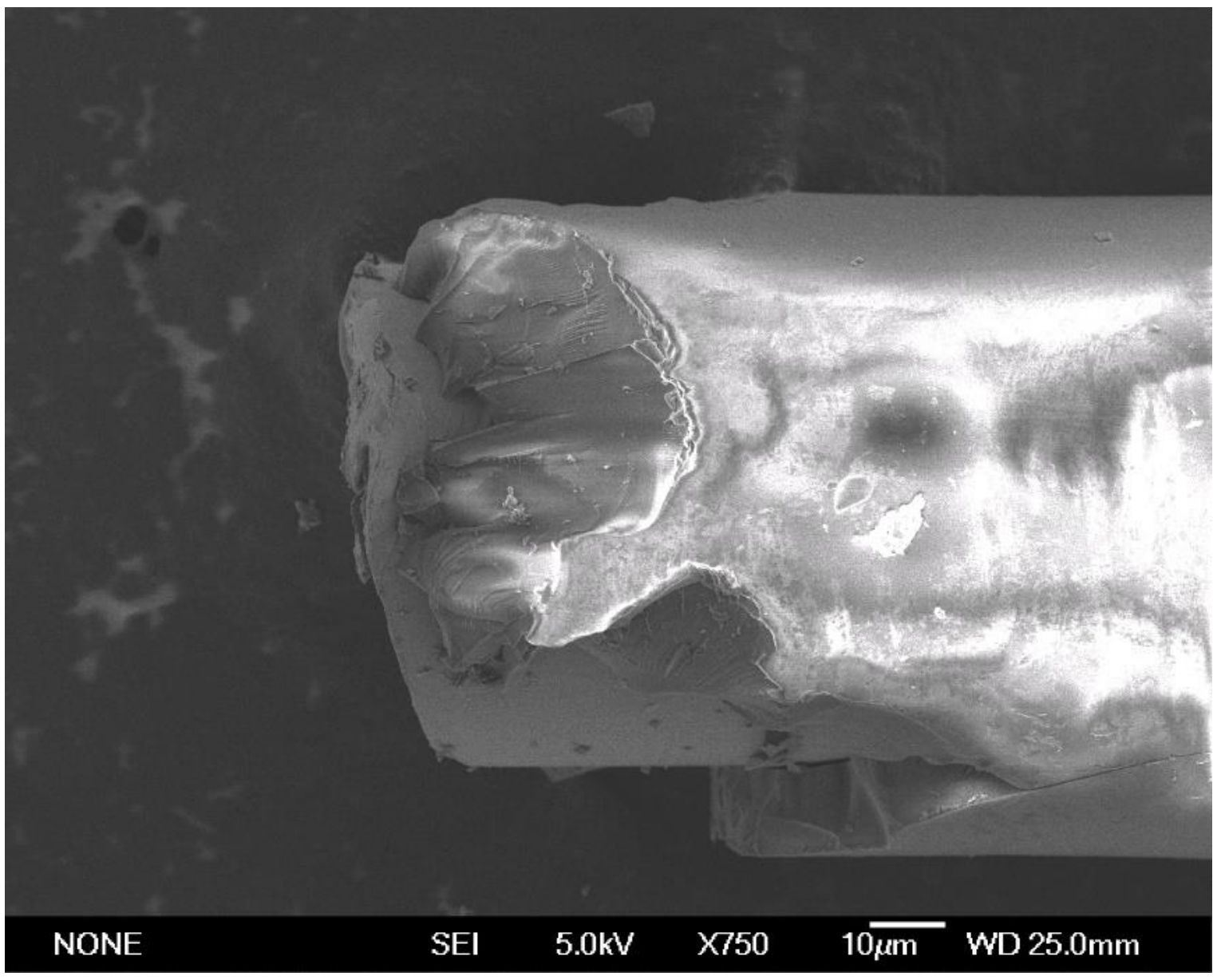

*Figure 1. 10 Fe-based and Pyrex coated microwire. Note how the shell presents heterogeneous brightness due to the electrostatic charging.*

For samples where this could be a problem, a thin coating of gold or other conductive material can be deposited over the sample to promote the discharge of the electron beam. For instance, a gold coating was applied to the nanofiber deposits for the biosensor to avoid the static charging of the non-conductive polymers (Figure 1. 11), note how the deposit is unnoticeable under a few magnifications. Only at high magnifications, the deposit would be observed as a scaled surface.

SEM (JEOL JSM 6335F and JSM 7600F) and TEM (JEOL JEM-3000F) images were obtained at Centro Nacional de Microscopía Electronica (CNME), Spain.

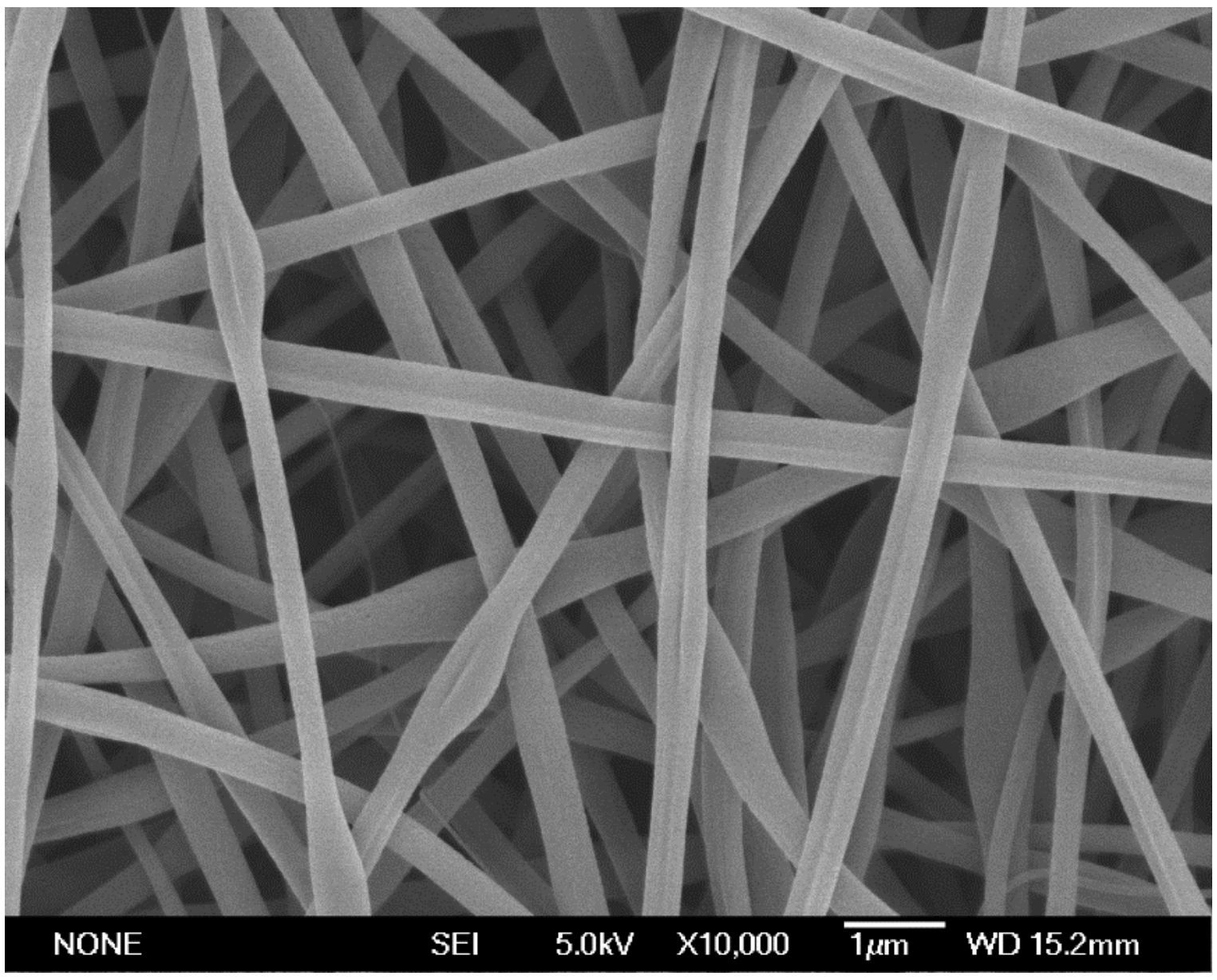

*Figure 1. 11 Gold-coated electrospun polymeric nanofiber deposit.*

### *1.4.4 Raman and IR Spectroscopy.*

Raman spectroscopy is a technique that uses the inelastic scattering of light to study a molecule's vibrational, rotational, and other low-frequency modes. The technique is based on the Raman effect, which is the inelastic scattering of light by a molecule when it changes its vibrational energy. The change in the vibrational energy of the molecule causes a change in the frequency of the scattered light, and this change in frequency, called the Raman shift, is unique for each type of molecule. The spectra are presented as Raman shift versus intensity and can provide information about the chemical nature of the sample [27].

In the case of graphene, the vibrational modes of the carbon atoms in the material give rise to characteristic peaks in the Raman spectrum. The peaks are typically observed at wavelengths corresponding to the vibrational modes of the sp2-bonded carbon atoms in the graphene lattice (see Figure 1. 12). The position and shape of the peaks can provide information about the quality and structure of the graphene. For example, the position and shape of the D and D' bands in the Raman spectrum of graphene can be used to determine the number and type of defects in the graphene lattice. The intensity of the G peak can be used to measure the size and shape of the graphene flakes. Finally, the intensity of the 2D band can be related to the number of graphene layers [28–30].

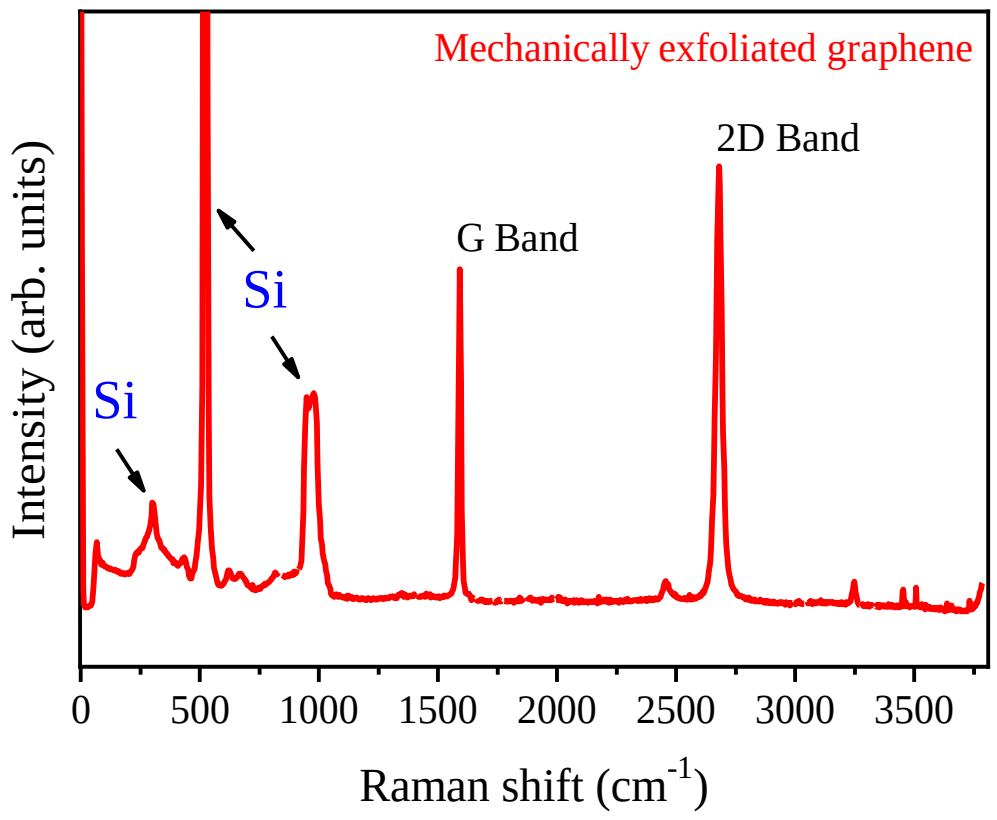

*Figure 1. 12 Typical Raman spectra of a graphene sample, obtained from a mechanically exfoliated sample used in Chapter 4. The silicon substrate contributes to the spectra with its characteristic peaks.*

Raman spectroscopy was performed at Instituto de Cerámica y Vidrio (CSIC), Spain, using a WITec ALPHA 300RA (Nd:YAG laser light source of 532 nm in p-polarization).

Infrared absorbance spectroscopy (IRAS) is a technique that uses the absorption of infrared light to study the molecular structure of a material. When a sample is irradiated with infrared light, molecules inside it absorb specific frequencies of IR radiation, which correspond to the vibrational modes of the bonds within the molecule. This absorbance (or transmittance) is then plotted as a function of wavelength to produce an IR absorbance spectrum with characteristic absorption bands. The absorption bands are unique for each type of molecule and can be used to identify and quantify the chemical nature of the sample [31].

Regarding graphene analysis, the absorption bands can also provide information about the quality and structure of the graphene, but more importantly, about the presence of carboxyl, hydroxyl, epoxy, or other oxygenated functional groups that may rise from discontinuities of the sp2 lattice, and evidence the nature and degree of an oxidated GBM [32].

The IRAS characterization of the ball-milled samples involved mixing the GBM powder with KBr, with no absorptions at IR wavelengths, at a ratio of approximately 1:200 using a mortar, then compressing the mixture to obtain a thin disk with a homogeneous dispersion of the sample. The mixture was then compressed in a vacuum die to obtain a thin disk with a homogeneous dispersion of the sample. IRAS was performed at Facultad de Ciencias Físicas (UCM), Spain, using a Bruker VERTEX 80v instrument, fully evacuated to reduce atmospheric perturbations, with a DGTS detector.

Both infrared absorbance and Raman spectroscopy can detect the vibrational modes of the molecules in a sample. In infrared spectroscopy, the vibrational modes cause light absorption at specific wavelengths. In contrast, in Raman spectroscopy, the vibrational modes cause a change in the wavelength of the scattered light. The main difference between the two techniques is that infrared spectroscopy is sensitive to the vibrations of the bonds between atoms, and Raman spectroscopy is sensitive to the vibrations of the atoms within a lattice. This means that infrared spectroscopy is often used to study the functional groups in a molecule, whereas Raman spectroscopy is often used to study the vibrational modes of the molecule as a whole. Both techniques can provide valuable information about the structure and properties of a sample, but they are sensitive to different aspects of the molecular structure.

### *1.4.5 UV-vis Absorption.*

This technique uses the absorption of ultraviolet (UV) and visible (vis) light to study the electronic structure of a material. The technique is based on the principle that the electrons in a material can absorb energy from incident light. This absorbed energy can excite the electrons from their ground state to a higher energy level. The energy of the absorbed light is quantized, and the absorption is typically strongest at specific wavelengths, called absorption bands. The absorption bands are unique for each type of material and can be used to identify the material in the sample [33].

In the case of graphene, the material's electronic structure gives rise to characteristic absorption bands depending on whether the material is oxidized. For example, the $\pi$- $\pi^*$ electronic promotion, where the $\pi$ electron is promoted to a conjugated state, is excited by approximately 260 nm absorptions. This promotion is typical of non-oxidized materials [21]. Conversely, GO presents a decreased or broken $\pi$-conjugated network, requiring more energy for the previous transition and causing the absorption band to blueshift to 235 nm [34,35].

UV-vis absorption was performed with a Shimadzu UV-1603 double-beam spectrophotometer, coupled with an integrating sphere, located at Facultad de Ciencias Físicas from *UCM*, Spain.

### *1.4.6 Atomic Force Microscopy.*

Atomic Force Microscopy (AFM) is a high-resolution imaging technique used to study the topography, properties, and behaviour of surfaces at the nanoscale level. In AFM, a small probe (a cantilever with a sharp tip) is scanned over the surface of the sample, while the interaction between the tip and the surface is monitored and used to generate a three-dimensional image of the surface. A laser is projected into the tip of the cantilevers which reflects it to a photodetector. The reflection of the laser thus mimics the surface and is used to create a three-dimensional map (see Figure 1. 13).

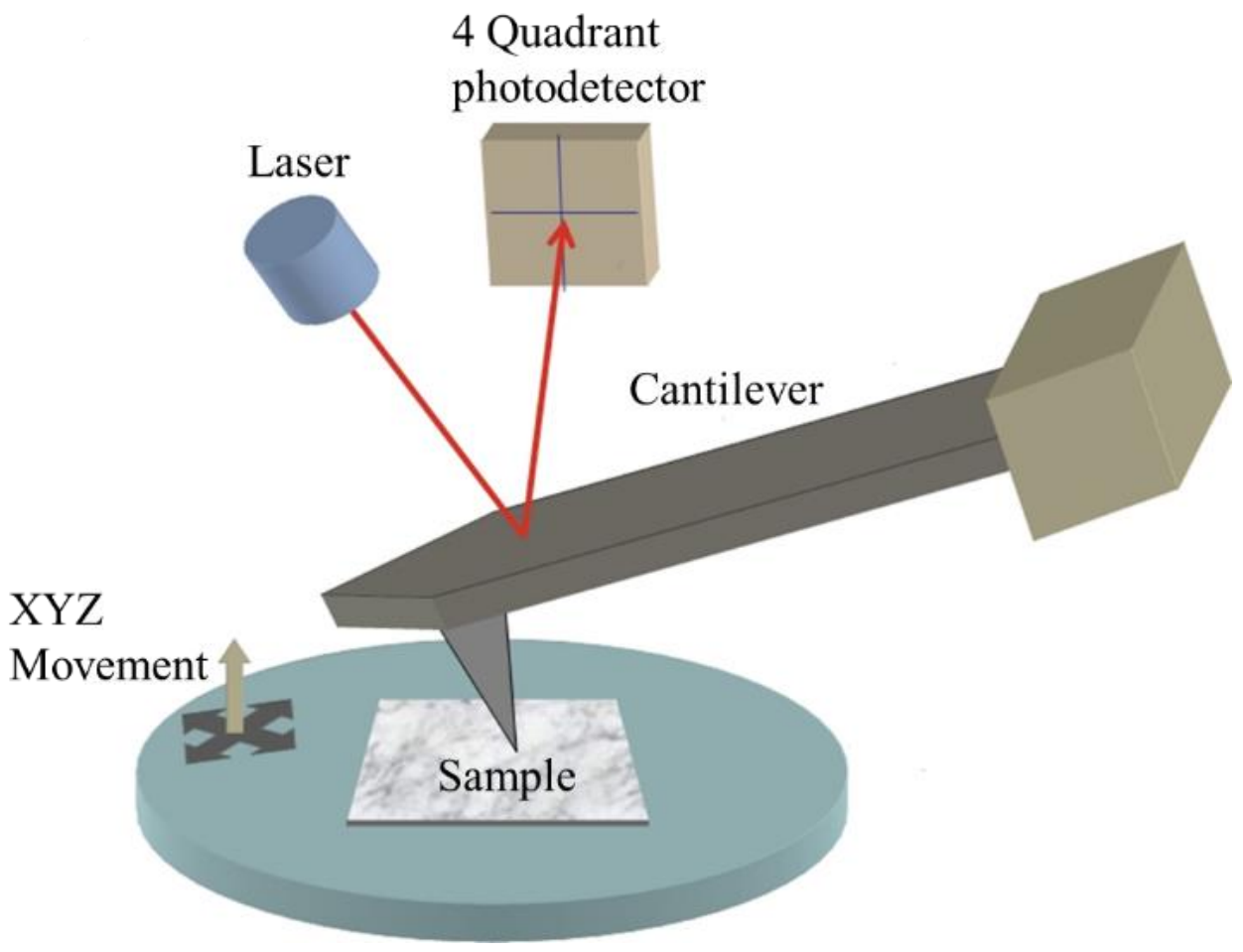

*Figure 1. 13 Schematic representation of the AFM technique.*

The AFM was used to characterize the surface roughness of multilayer graphene gas sensors and to verify the surface state after UV irradiation [36]. In particular, a Nanotec scanning probe microscopy Cervantes's system controlled by WSxM software was used [37]. The probe was a Nanosensors PPP-FMR with a resonance frequency of 85.5 kHz. The measurements were carried out in amplitude modulation mode, with the phase-lock-loop (PLL) feedback on. The analysis was done at Laboratorio de Microscopía de Campo Cercano from Instituto de Ciencia de Materiales from *Consejo Superior de Investigaciones Científicas* (*CSIC*), Spain with the support of Agustina Asenjo.

### *1.4.7 Differential Scanning Calorimeter.*

The DSC measures both the heat flow into and out of a sample, as well as the change in weight of the sample as it is subjected to a controlled temperature program. The instrument allows for the precise measurement of various thermal properties such as specific heat, enthalpy, and glass transition temperature, as well as providing information on changes in mass or weight during the thermal process.

During the DSC scan, a small sample is placed into a crucible and heated at a controlled rate, while the heat flow into or out of the sample and the sample weight is recorded. The difference in heat flow between the sample and a reference material is measured, providing information about the thermal behaviour of the sample, as well as any changes in mass or weight that may occur during the heating or cooling process.

In our study, the DSC scan was performed over a temperature range of 18-800 Cº, with a heating rate of 5 Cº/min. The resulting data allowed us to determine the phase transition temperatures of the amorphous magnetic alloys to design thermal annealing treatments. The DSC characterization was performed at CAI de Ciencias de la Tierra y Arqueometría from *UCM*, Spain.

## *1.5 References.*

# Chapter 2: Graphene-based materials synthesis.

*Chapter Introduction; The history of "graphene"; Graphene-based materials; Top-down methods and ball-milling; Bottom-up methods and CVD; Conclusions; References.*

## *2.1 Chapter introduction.*

One should begin this chapter by defining the terms of its title, which will be the focus of the discussion and the experimental work presented. The term "graphene-based materials" refers to a family of materials derived from or related to graphene, a single-atom-thick honeycomb-like structure of carbon atoms with hybridized sp2 covalent bonds and an unpaired electron per atom (the fourth valence electron of carbon) either above or below the basal plane, as depicted in Figure 2. 1. The term "synthesis" refers to the production of these materials through physical or chemical reactions, carried out artificially and with specific purposes in mind.

The definition and use of the term "graphene" have often been controversial, and revisiting the origins of the material can provide valuable context for understanding graphene-based materials [1]. From its initial discovery, graphene has inspired the development of various graphene-based materials with different structures and properties, extending its potential applications and impact. In this chapter, I will explore the synthesis of these materials, drawing from the experimental work conducted to produce a range of graphene-based materials with diverse characteristics and applications.

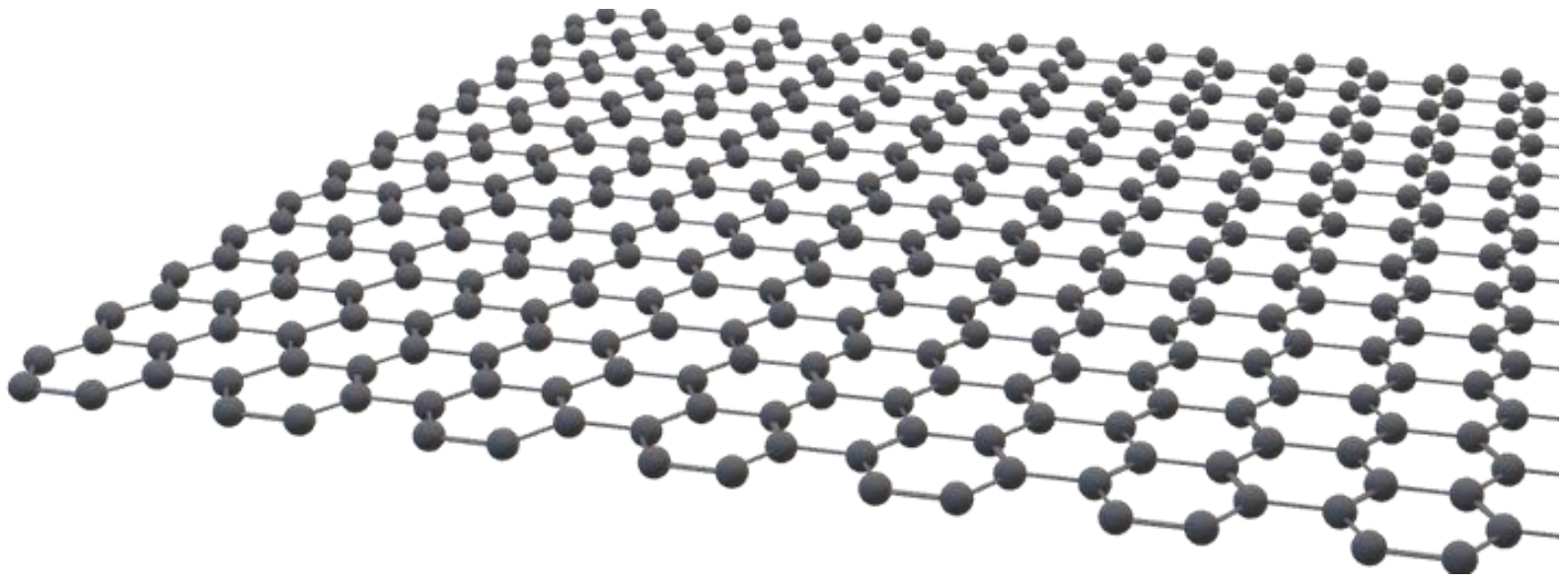

*Figure 2. 1 Representation of pristine graphene according to its strict definition.*

## *2.2 The history of "graphene".*

The first reference for the definition of graphene may be traced to the 19th century, particularly in 1860. Graphite was found to have a lamellar structure composed of extremely thin sheets, although not yet known how thin, after exfoliation through an oxidation process [2], [3]. B. C. Brodie called this material graphon, which we now know as graphene oxide (GO). In 1840, C. Schafhaeutl described the oxidation method, later used by B. C. Brodie, while investigating graphite precipitates from molten iron alloys [4], [5]. However, it is unclear whether C. Schafhaeutl perceived the lamellar structure as B. C. Brodie, which is the matter at hand.

The honeycomb, or hexagonal, disposition of the carbon atoms in the oxidized form of graphene was reported in the 1960s using X-ray diffraction patterns [6]. Around the same time, GO was reduced by Hans Peter Boehm et al. [7]. The material they obtained had a vast specific surface and the presence of Carboxyl- and Hydroxyl- groups. Although they did not formally name it, it is known as reduced graphene oxide (rGO) by today's standards. This rGO greatly resembles the ones reported nowadays (see Figure 2. 2). Along with graphite oxidation, the reduction of GO will receive attention later in this discussion, as these techniques are still relevant.

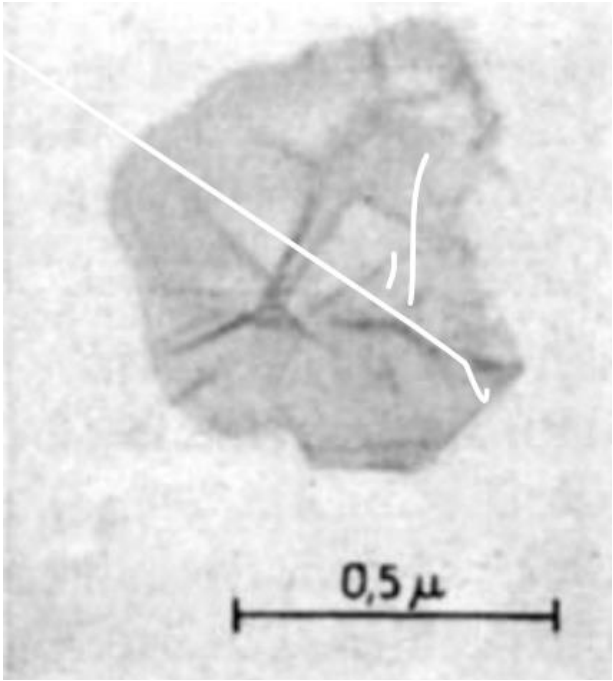

*Figure 2. 2 Graphite oxide after chemical reduction. Reproduced* from Ref. [7]

H. P. Boehm, a German chemist who devoted his scientific career to graphene research until his recent passing away, was also responsible for the name "graphene". In 1986, Boehm et al. suggested the ending "-ene" on graphite to refer to a single carbon atom layer [8]. The IUPAC later adopted the term as a single carbon layer of the graphite structure [9]. The IUPAC also collected the definitions for GO and rGO. However surprising it may seem; we are far from understanding today's controversy regarding "graphene".

The unique electronic properties and sp2 hybridization in graphene have been anticipated since 1947. Unfortunately, the work from P. R. Wallace [10] was limited to a theoretical analysis, as pristine, non-oxidized graphene was yet to be obtained. However, considerable advancements in this regard have been made, both theoretical and experimental [11]. Still, only a small part of the electronic properties will be relevant throughout this document, so I will only provide a few details in this chapter.

The next milestone in the definition of graphene, and the key to understanding its controversy, occurred in 2004 when Andre Geim and Konstantin Novoselov obtained single-layer graphene by repeatedly peeling graphite with a scotch tape (Figure 2. 3), i.e., mechanical exfoliation, leading to an outburst in experimental research on graphene [12].

With graphene finally available for experimental research, its definition was expanded with comprehensive characterization, such as:

- A high electron density of ~$2x10^{11}$ $cm^{-2}$, and carrier mobility of 200.000 $cm^2$/ V s with low-temperature dependence, which is the highest mobility ever reported on a semiconductor material [13], [14].
- High thermal conductivity above 3000 W $nK^{-1}$ [15], [16].
- Specific surface of 2630 m2/g [17].
- Strength of 130 GPa and an Elastic modulus of 1 TPa [18].

Many of these works were led by A. Geim and K. Novoselov, granting them the Physics Nobel Prize in 2010 for "ground-breaking experiments regarding the two-dimensional material graphene" [19].

From this point on, the popularity of graphene, the greatly anticipated material with outstanding properties, exploded outside the scientific community [20]. Graphene became omnipresent in public media, often accompanied by the word revolutionary [21] or miraculous [22]. Undoubtedly generating huge commercial interest for a myriad of applications.

However, the hype at that moment greatly exceeded the technological possibilities of the field [23], [24]. To illustrate, employing the methodology of exfoliating graphene flakes to fortify the concrete infrastructure of a skyscraper exceeding 250,000 tons [25] would be far from feasible. Therefore, the research on new synthesis methods to meet the industry's requirements, i.e., scalable methods, was promoted mainly by public funding, from relatively small local or national funding to the gigantic Graphene Flagship consortium, and has relentlessly provided new methods and, consequently, commercial applications for graphene.

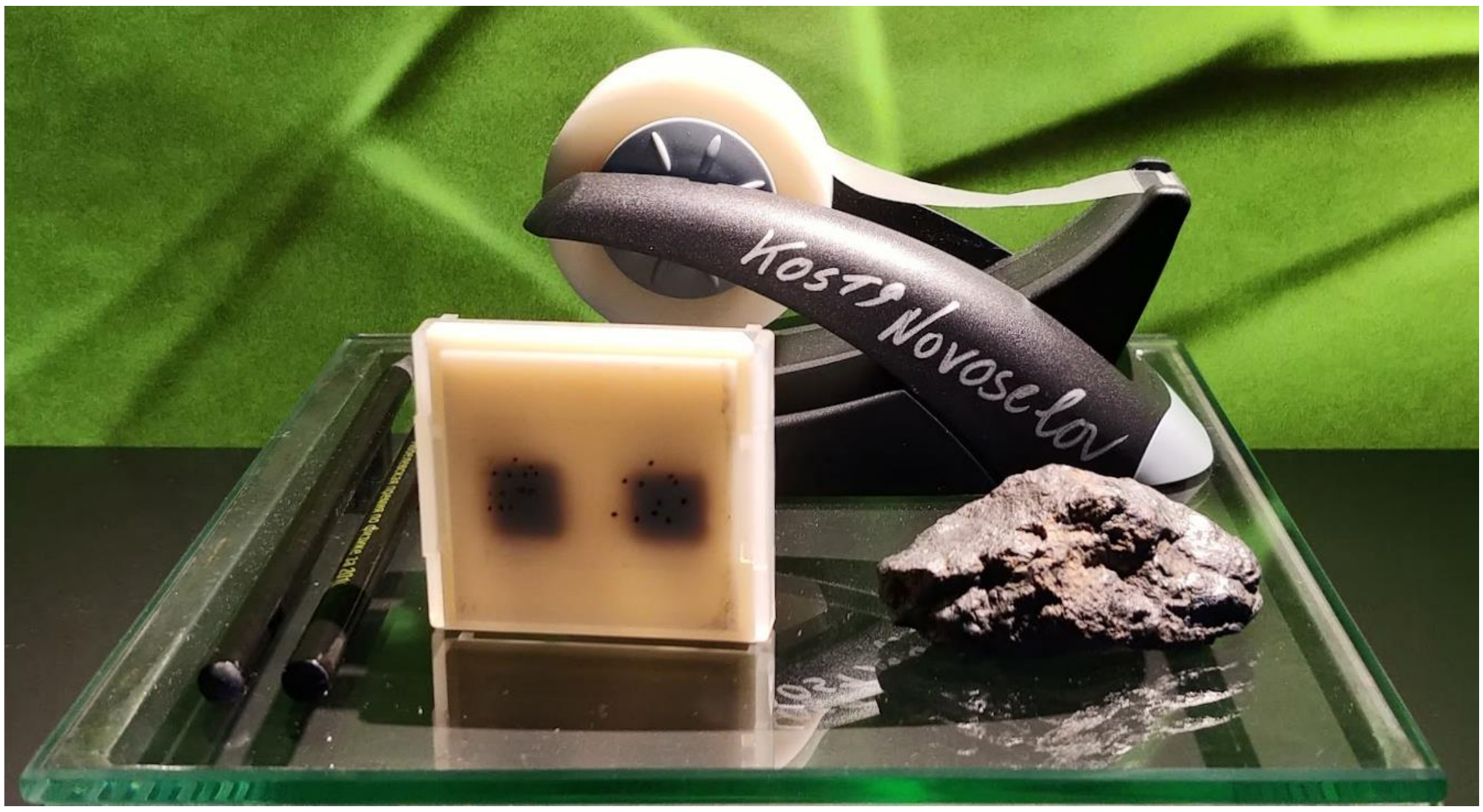

*Figure 2. 3 The "Scotch tape" from K. Novoselov, found at the Nobel Prize Museum in Stockholm, Sweden, where prize winners are asked to leave a memento from their awarded research.*

### *2.3 Graphene-based materials.*

Nonetheless, due to practical considerations or even intentional misappropriation, the term "graphene" has been and continues to be frequently utilized in commercial contexts to refer to a variety of different materials. An examination of commercial products marketed as graphene disclosed that the majority were not, in fact, pristine single-layer graphene [26]. Conversely, these materials belonged to a broad family of materials with some common characteristics called graphene-based materials (GBMs), sometimes called graphene-related materials (GRMs), two of which, GO and rGO, have already been presented [27].

GBMs can be defined from graphene, i.e., the single layer is the construction block of these materials. Graphene is chemically modified, stacked, or re-arranged to build the most popular GBMs, such as multi-layer graphene (MLG), graphene nanoplatelets (GNPs), and even carbon nanotubes (CNTs), among many more. GBMs have been given names as they were first reported, and the lack of standardization has caused many incoherencies in different research fields. This becomes a much bigger issue when the technology is to be transferred to commercialization, especially after "graphene" gained so much mystique and appeal. To solve it, the International Organization for Standardization (ISO) reported a standard for names and definitions, "ISO 80004-13:2017: Graphene and related two-dimensional (2D) materials" whereas the Graphene Flagship proposes a classification strategy based on three easy-to-measure and quantifiable material characteristics to avoid naming convention [28], [29]. In addition, several efforts are being made to register GBMs under the European Chemicals Agency (ECHA) REACH regulation, a *must* to commercialize large quantities of materials in European territory [30], [31]. Along the same line, the Graphene Council launched the "Verified Graphene Producer" program to help customers identify reputable graphene producers that have added themselves to the standardization efforts.

One of the consequences of the misspelt terminology is the lack of standardization and subsequent inconsistency of the research results. Probably the most notable case of this inconsistency is found in biological applications where lateral size, chemistry, and layer stacking affect the toxicity and antimicrobial activity of the material [32]–[36]. For example, it would be easy to find an article reporting how GO's biocompatibility, in terms of cellular viability, can be >80% [37], between 60-80 % [38] or <60% [39]. Although GO is usually considered non-toxic, it is understandable how these inconsistencies can negatively affect the public perception and commercial applications of GBMs. A recent event (2021) regarding graphene-enhanced face masks during the CoViD-19 pandemic constitutes a perfect example of this [40]–[42].

The synthesis method dramatically affects the physicochemical characteristics of GBMs [20]. As for other nanomaterials, GBM synthesis can be divided into two approaches: bottom-up methods and top-down methods [43], [44]. An oversimplified description suggests that the top-down methods achieve quantity over quality, whereas bottom-up methods achieve quality over quantity. The truth is, of course, more complex, and a brief description of some relevant methods from each approach is required. Two methods, however, will be explained in more detail as they will be relevant in the following chapters. These methods are ball-milling and chemical vapour deposition (CVD).

### *2.4 Top-down methods.*

Top-down refers to the methods in which the precursor is a bulk material, and the product is a nanomaterial. Hence the process is based on a scale reduction [45].

Regarding graphene synthesis, the precursor in top-down methods is usually graphite, exceptionally low-crystallinity carbons [46], [47] and nano-sized carbon materials (sometimes other GBMs) [48]–[50] have been used. The product's characteristics are highly influenced by those from the precursor [51]–[53]. The lateral sheet size, crystallinity or purity tend to decrease, or at least be maintained, during the top-down processes, although some post-processing can recover these characteristics [54], [55].

One of the most popular GBM synthesis methods in the scientific literature is the oxidation of graphite to obtain GO, and its posterior reduction, to obtain rGO. The oxidation method originated from the work of Brodie, as previously discussed. It was later improved by L. Staudenmaler [56]–[58] and later by W. S. Hummers Jr. et al. [59] Nevertheless, these methods produce slightly different oxides [60], [61]. Such improvements to the oxidation method have been mainly focused on its simplification and the use of less hazardous reactants, which is still the focus of current research leading to several variations of the so-called modified Hummers method [62]–[65].

Briefly, the method consists in mixing graphite with sulfuric acid and potassium permanganate ($KMnO_4$) under constant vigorous stirring. When the oxidation is complete, $H_2O_2$ is added to stop the reaction. The GO is finally obtained after comprehensive washing. The oxidation can be strongly exothermic and violent, and extreme care should be paid for. The evolution of the process can be visualized through colour changes in the solution [66].

This GO can be later reduced, with chemical or thermal [63] methods, to obtain rGO. Alternative reduction methods have also been proposed to make the process more environmentally friendly and scalable, such as green chemicals [67]–[69] or light [70]. From my experience before this thesis, ascorbic acid [67] not only renders good results as a reducer but is also very welcomed in processes related to GBM's biomedical applications. Overall, rGO, either due to its ease of preparation or its well-balanced properties, is probably the most popular GBM.

Liquid phase exfoliation (LPE) is based on intercalation methods to exfoliate graphite in a liquid medium using a combination of solvents, surfactants, and energy in a variety of forms, mainly sonication [71]–[73] and shear [74], [75]). Energy is given to overcome the van der Waals attraction between the graphene sheets, and the solvents and surfactants are used to prevent re-stacking and obtain a stable liquid dispersion [76], [77]. This method possesses advantages such as high production, high crystalline quality, large sheet size (highly dependent on the form of energy used [71], [78], [79]) and ease of implementation. Regarding the drawbacks, it produces a high dispersion on lateral size and thickness, often requiring further processing, whereas surfactants' use saturates the sheet's surface, preventing its use in some applications [80]. LPE is generally regarded as a technique well suited to produce GBMs in liquid suspensions to be used in electronic inks, conductive coatings, or composite fillers [81], [82]

The Scotch tape was used to obtain pristine graphene for the first time [12]. This method consists of a mechanical exfoliation using highly oriented pyrolytic graphite as a precursor and an adhesive tape, hence its name, to cleave the layers [45], [83]. First, graphite flakes are rubbed off against a surface. Once a flake has adhered to the tape, it is repeatedly cleaved to obtain thinner flakes until the desired thickness is achieved. The material is then transferred to a suitable substrate. This method, while capable of producing single-layer graphene and being applicable to other 2D materials, suffers from significant scalability and reproducibility constraints, confining its utility largely to the lab scale. [84].

The concluding top-down methodology to be examined within this chapter is ball-milling, which is based on the mechanical exfoliation and constitutes the core experimental research contribution to this chapter.

For ball-milling, a container is filled with milling instruments, such as balls made of steel or ceramic, and the precursor material to be ground. Additionally, the milling can occur in the presence of reactive or mediums. The container is then submitted to a movement, causing the balls to collide with the material, reducing its size through impact and friction. Additionally, the energy transferred through such movement can promote blending [85], [86] or chemical reactions [87]–[90]. The duration of the milling process, the size and type of media, and the material being milled can all be varied to achieve the desired outcome. It has been traditionally used for ceramic and metallic materials, but it has recently been adapted for GBM synthesis with promising results [91]–[94].

Regarding GBMs synthesis, the balls during the milling process generate impact and shear forces, i.e., forces normal and parallel to the stacking order. The former induces stresses in the π-π (stacking) bonds and the σ-σ bonds (structural), leading to both exfoliation and crystalline loss. In contrast, the latter overcome the Van der Waals interaction between layers and exfoliates the sheets more efficiently and less destructively [55], [95]–[97].

Ball-milling techniques can be divided into wet [98]–[100] or dry ball-milling [50], [92], depending on whether the milling is produced within a liquid medium or not, respectively. Furthermore, mill assisting agents can be added to either type of ball-milling to prevent re-stacking or favour the shear versus impact effects [55], [93], [99], [101]. Nevertheless, the choice of medium and additives to the milling process can affect the product's composition, which should be carefully considered for its intended application [102], [103].

Wet ball-milling is characterized by using a liquid medium during the milling that can help the exfoliation and dispersion of the sheets [104], [105]. The product is obtained as a liquid dispersion that can be washed and later dried off (vacuum filtering or freeze drying are popular options) or used directly depending on the intended application.

### *2.4.1 Wet ball milling in the planetary mill.*

In the present work, wet ball milling has been briefly tested. The precursor is pristine graphite flake (Natural, -325 Mesh, 99,8 %_ Metal basis from Alfa Aesar), and the mill used is a planetary ball mill (PM400 Retsch GmbH). The planetary ball mill has a characteristic rotational and translational movement of the jars that favours shear forces over impact forces which, as previously stated, should promote exfoliation during the process [55], [106], [107].

We have used 250 mL stainless steel jars and 100 balls of the same material with 10 mm in diameter and a weight of 4 g each. All jars contained 4 g of graphite to build a powder/ball weight ratio of 1:100, and the liquid medium was created using distilled water, Deoxycholate (SDC), $C_{24}H_{39}NaO_4$, as a surfactant and ammonia, $NH_3$, to adjust the pH to 9 [108], both from ITW Reagents. Four jars were prepared with different conditions, including one without liquid medium for reference as presented in Table 2. 1 and milled simultaneously. During the milling sequences, interruptions were made each 2 hours to control the temperature of the milling jars and ensure it did not exceed 50ºC. The sequence was resumed after the jars were thermalized to room temperature.

As a safety note for future users of this technique, sample D exhibited some sparkling without ignition when the jar was opened after the milling, similar to other reports [103]. The milling jars need to be let cooled down and opened under a controlled environment.

*Table 2. 1 Milling conditions used in the planetary mill for testing wet ball milling GBMs synthesis.*

| Name | Material | Solvent | Surfactant |
|---|---|---|---|
| A | Graphite (4 g) | Water + $NH_3$ | None |
| B | Graphite (4 g) | Water | SDC (0.4 g) |
| C | Graphite (4 g) | Water + $NH_3$ | SDC (0.4 g) |
| D | Graphite (4 g) | None (dry) | None |

The samples were milled for 12 h at 200 rpm. Samples A, B, and C were washed with distilled water and vacuum filtered to obtain a dry powder. The resulting products were investigated with X-ray diffraction (XRD) spectra (Figure 2. 4).

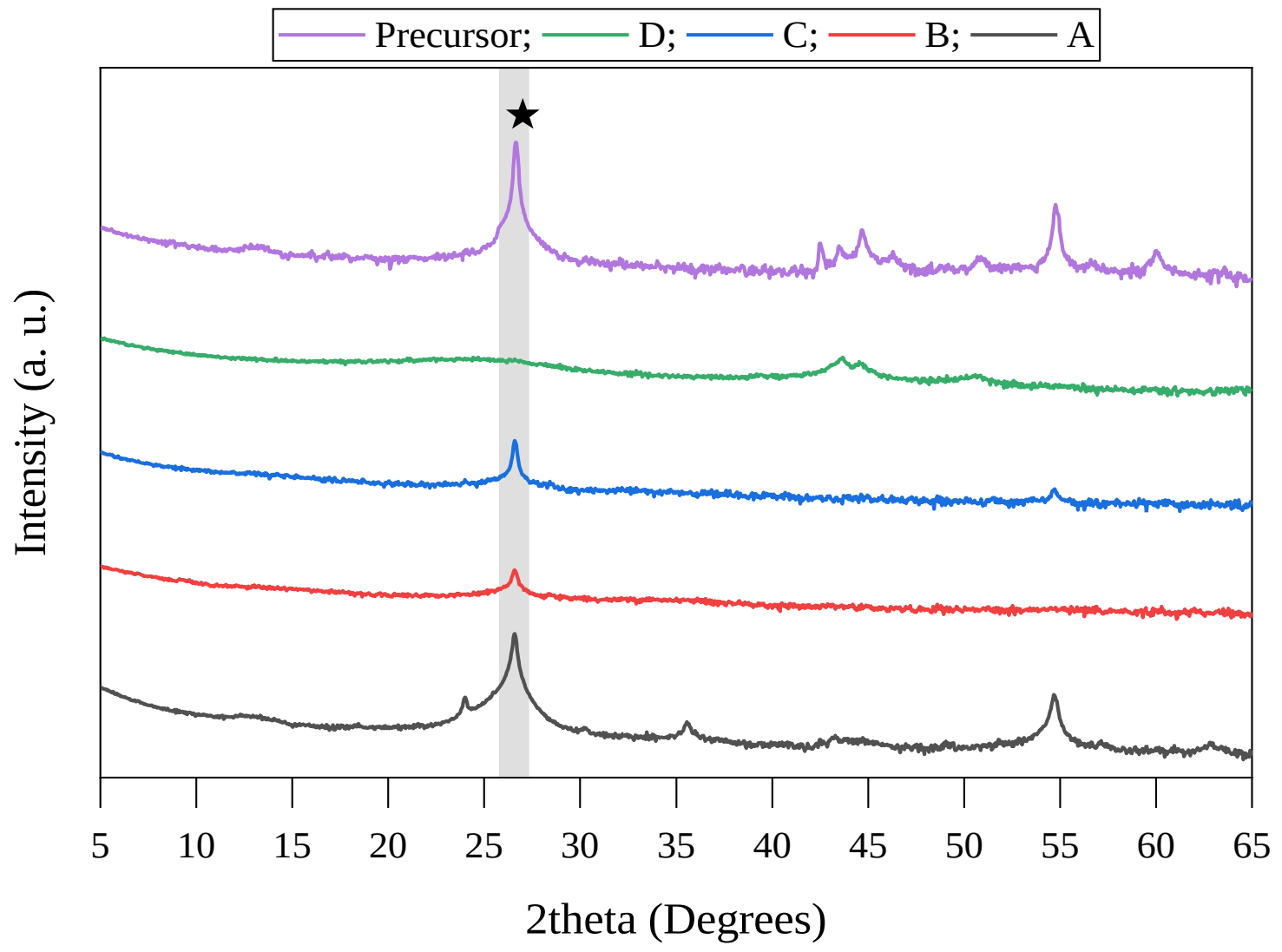

*Figure 2. 4 XRD spectra for a 12 h milling test in the planetary mill. The precursor refers to the graphite as purchased, and samples A-D refer to different milling conditions.*

The analysis of the results, shown in Figure 2. 4, is focalised on the maximum located at 26,5 º within the shaded line (★). This peak corresponds to the (002) plane, which is related to the stacking direction of the graphite layers. A decrease in the intensity of this peak can be associated with an increase in the exfoliation degree of the samples [93], [103], [109].

All milled samples exhibit some differences compared to the precursor graphite, indicating that milling has affected their structure. Sample A is the most similar to the precursor, exhibiting a lesser degree of exfoliation. In addition, sample A exhibits an additional peak at around 24 º which can be related to graphite nitrate resulting from the intercalation of nitrogen atoms [110].

Samples C and B are significantly different from sample A, highlighting the crucial role played by the surfactant in the wet milling process. Furthermore, the combination of surfactant with the modified pH of sample B has led to the highest exfoliation degree among the wet-milled samples.

In contrast, the dry-milled sample D shows the greatest difference in terms of exfoliation degree, indicating a more efficient exfoliation process compared to the wet milling methods. The absence of a liquid medium in the dry milling process not only simplifies the post-processing steps but also potentially makes it more suitable for large-scale synthesis of highly exfoliated graphene materials.

Based on the preliminary conclusions drawn from the XRD analysis, the presence of a liquid milling medium, as demonstrated by samples A, B, and C, may not be the most suitable approach for the large-scale synthesis of highly exfoliated graphene-based materials.

An alternative wet ball-milling technique was tested following the method proposed by F.M. Casallas Caiçedo et al. [87], which exploits the technique's capabilities to induce chemical modifications on the material simultaneously. This method uses potassium perchlorate ($KClO_4$) as a reactive, distilled water medium and graphite as the precursor. Using planetary milling, the PM100, a smaller version of the PM400 used in the previous test, Casallas Caiçedo et al. obtained a stable GO suspension. The oxidation generates functional groups in the basal plane of the sheets. These groups are usually hydrophilic, which makes GO easily dispersed in water as opposed to non-oxidized graphene [111]. Therefore, no surfactants are needed since the oxidated functional groups are sufficient to prevent re-stacking.

The method was then tested with a few adjustments. Using the PM400 with 250 ml jars, 12 samples were obtained following the conditions presented in Table 2. 2. All milling conditions were tested with and without $KClO_4$. In addition, two samples, #01 and #02, consisted of a mixture of the components without ball-milling.

Two sets of milling balls with different sizes were used (10 mm and 6.25 mm in diameter), as this parameter is expected to modify the interaction and force delivery to the material [55]. In particular, the transferred energy ($E_t$) is related to the ball by [112], [113]:

$$E_t = kN_bE_b;$$
$$E_b = \frac{1}{2}m_b v_b^2 \phi_b \qquad \text{(Eq. 2. 1)}$$

Where k is a constant related to the mill and milling type, $N_b$ is the number of balls and $E_b$ is the energy transferred by each ball, which is in turn related to their mass, $m_b$, velocity, $v_b$, and a constant related to the filling ratio of the jar, $\varphi_b$.

In addition, samples #10 to #12 used increasing amounts of the oxidizing agent, expecting to see a gradual increase in the oxidation degree of the samples. Interruptions every 2 hours were programmed analogously to the previous test.

*Table 2. 2 Milling conditions used in the planetary mill for wet ball-milling GO synthesis.*

| Name | Milling frequency | Milling time | Graphite | $KClO_4$ | Ball ratio | Ball diameter |
|---|---|---|---|---|---|---|
| #01 | 0 rpm | 12 h | 4 g | 6.04 g | 20:1 | - |
| #02 | 0 rpm | 12 h | 4 g | 0 | 20:1 | - |
| #03 | 250 rpm | 12 h | 4 g | 6.04 g | 20:1 | 10 mm |
| #04 | 250 rpm | 12 h | 4 g | 0 | 20:1 | 10 mm |
| #05 | 250 rpm | 24 h | 4 g | 6.04 g | 20:1 | 10 mm |
| #06 | 250 rpm | 24 h | 4 g | 0 | 20:1 | 10 mm |
| #07 | 500 rpm | 12 h | 4 g | 6.04 g | 20:1 | 10 mm |
| #08 | 500 rpm | 12 h | 4 g | 0 | 20:1 | 10 mm |
| #09 | 250 rpm | 12 h | 4 g | 0 g | 20:1 | 6.25 mm |
| #10 | 250 rpm | 12 h | 4 g | 2.01 g | 20:1 | 6.25 mm |
| #11 | 250 rpm | 12 h | 4 g | 4.02 g | 20:1 | 6.25 mm |
| #12 | 250 rpm | 12 h | 4 g | 6.04 g | 20:1 | 6.25 mm |

The prepared graphene-based material samples were washed with distilled water and vacuum filtered. Following the washing process, the samples were redispersed in distilled water using sonication. The samples were visually examined evaluating the colour and stability of the resulting dispersions as key indicators of the presence of graphene oxide.

GO, and unoxidized forms of GBMs exhibit different properties when it comes to stability in solution. GO is hydrophilic due to the presence of oxygen-containing functional groups, such as hydroxyl, epoxy, and carboxyl groups, on its basal planes and edges. These groups not only provide the necessary sites for hydrogen bonding with water molecules but also impart a negative charge to GO sheets, which allows them to be easily dispersed in polar solvents like water [114]–[116]. In addition, GO solutions exhibit a characteristic yellow-brown colour [117]–[120].

The suspension stability decreases with the decrease in oxidation degree, resulting in a hydrophobic and chemically inert surface that does not readily interact with polar solvents such as water. Consequently, unoxidized GBMs are poorly soluble in most solvents. They tend to form aggregates or precipitates in solution due to their strong van der Waals forces and $\pi$-$\pi$ interactions between individual sheets [115], [116], [121]. These materials exhibit a deep black colour solution when dispersed that rapidly turns into a clear liquid with the precipitated GBM [117]–[120].

Despite the variations, no evidence of oxidation was observed in any sample, regardless of the milling time, milling frequency, ball size or use of $KClO_4$ (see Figure 2. 5). All samples consisted of suspended black particles in the water solution, more stable than the non-milled graphite (#01 and #02).

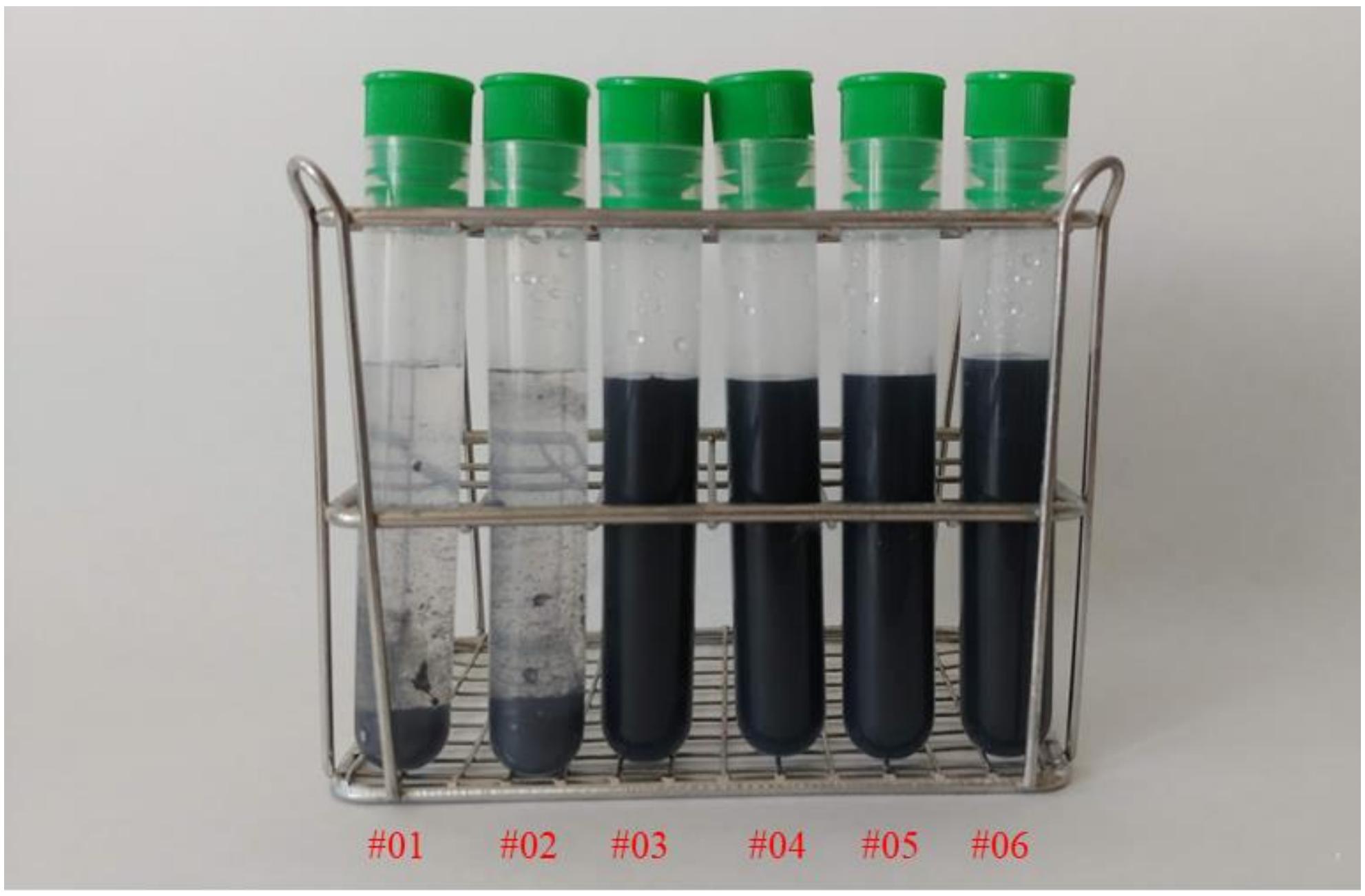

*Figure 2. 5 Photograph of the milled wet ball-milling GO synthesis test samples in water solution after 24h.*

As shown in Figure 2. 6 samples #01 and #02 were not submitted to the milling process and exhibited a strong (002) peak similar to that of graphite. However, sample #01 exhibits a variety of unidentified diffraction peaks not related to graphite or other components involved in the milling process.

As for the milled samples, sample #03 shows a slight contribution of $KClO_4$ to the diffraction spectra (●) [122], and a slight contribution of $Fe_3O_4$ (❄) [123], probably originating from the reaction of the oxidant agent with the milling instruments, which is more notable in sample #04.

From these preliminary results, it is yet unclear the origin of the possible contamination on sample #01 and the presence of iron oxide in the sample milled without an oxidant agent (#04). Nevertheless, no presence of GO is detected in either of the characterized samples, in close agreement with the visual examination (the presence of GO in a XRD spectra can be observed in Figure 2. 7).

This is an ongoing experiment and a comprehensive discussion and analysis of the results have not been conducted at the time of writing this thesis.

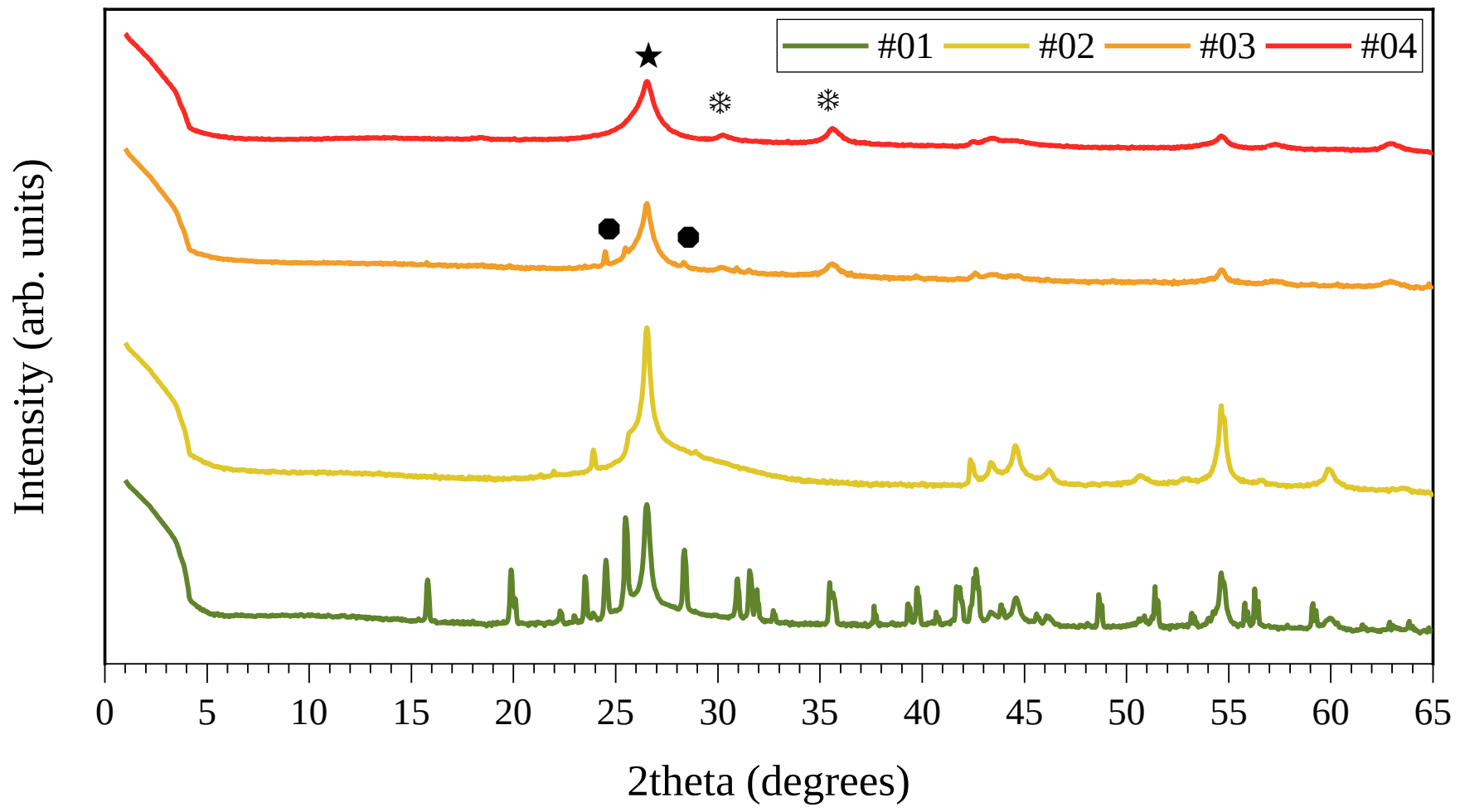

*Figure 2. 6 XRD spectra for the wet ball-milling GO synthesis test in the planetary mill. It shows non-milled samples with (#01) and without (#04) $KClO_4$ and samples milled for 100 min with (#03) and without (#04) $KClO_4$.*

Dry ball-milling is an interesting alternative, not only because the lack of liquid medium may simplify the synthesis process but also, according to results presented in Figure 2. 4, it promotes exfoliation to a greater extent than wet ball-milling, timewise.

### *2.4.2 Dry ball milling in the planetary mill.*

The first test on the dry ball-milling technique was done with a planetary mill, following the conditions set for sample D (see Table 2. 1) using a milling frequency of 200 rpm and milling times from 3 to 21 h. The milling results were examined using XRD (see Figure 2. 7). As previously noted, the intensity of the maximum located at 26.5º can be used to study the evolution of the exfoliation process during milling. Accordingly, the stacking contribution to the spectra is significantly reduced with increasing milling time, with the most noticeable changes occurring during the first 9 h and less noticeable changes occurring during further milling. These results reflect that the milling process reduces the crystallinity in terms of exfoliation up to a point after which excess milling has a lesser effect on the characteristics of the material, consistent with the further milling results presented in this chapter and with previous reports [124].

In addition, the maximum located at 44.5º (■) corresponds to a group of planes with a component in the crystalline stacking order and an in-plane component, i.e., 3R (101), 2H (101), and 3R (012) (left to right) [125]. Therefore, it roughly provides some information regarding the lateral size and crystallinity of the graphene sheets. Furthermore, comparing the evolution of the intensities of both maximums reveal that the exfoliation evolves faster than the reduction of the in-plane crystallinity with the milling time, which agrees with the description previously provided for the planetary mill and its tendency to generate shear forces against impact forces.

Lastly, the precursor spectra show a slight maximum located at 13º (●), which could be related to the contribution of GO (001) [126]–[128]. Oxidation could be naturally present near the edges of the sheets within the graphite precursor. This maximum is similarly observed during the first 3 hours of milling but not for higher milling times.

Although the test attained interesting results and confirmed the viability of exfoliation of graphite through planetary ball-milling, it still needs to be brought to completion. It is now presented as a potential line for future work.

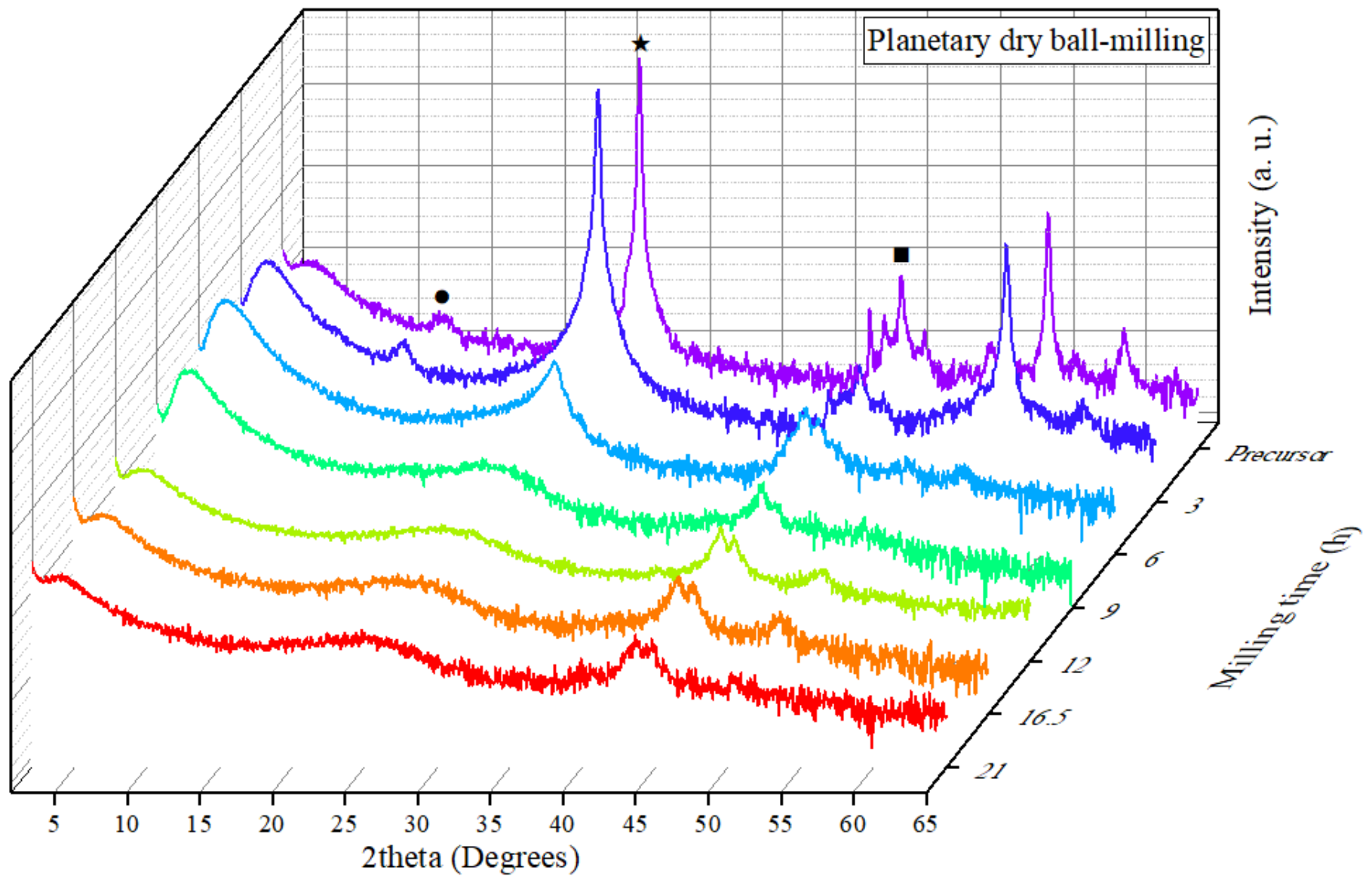

*Figure 2. 7 XRD spectra for the dry ball-milling test in the planetary mill. The precursor refers to the graphite as purchased.*

### *2.4.3 Dry ball milling in the oscillatory mill.*

Next, dry ball milling was tested in a ball mill with significantly different characteristics. The test was done using the same precursor as the previous test and an oscillator ball mill (MM400 Retsch GmbH), called a multipurpose mill by its manufacturer. This ball mill has an oscillating or swinging movement of the jars. This movement generates great impact forces, whereas neglectable shear forces should be expected.

The test was done using 25 mL WC jars and a single ball of the same material, 15 mm in diameter, with a weight of 25 g. All jars contained 1 g of graphite to build a powder/ball weight ratio of 1:25. No additives were used. The milling was done at a rate of 25 Hz (1500 rpm) with a milling time between 20 and 300 min. The milling results were examined using XRD (see Figure 2. 7). During the milling sequences, interruptions were made each 15-30 minutes to control the temperature of the milling jars and ensure it did not exceed 50ºC. The sequence was resumed after the jars were thermalized to room temperature [129], [130].

#### *2.4.3.1 Microstructural characterization.*

Attending the same criteria as previously, as shown in Figure 2. 8, the exfoliation process during the milling is time-dependent. A gradual decrease in the intensity of the maximum located at 26.5º was observed as the milling time increased. In contrast to the planetary mill, the decrease in the intensity of the maximum located at 44.5º was more accused, indicating a decrease of the in-plane crystallinity coinciding with the exfoliation process. No noticeable changes occurred for milling times beyond 200 min regarding these two maximums; however, the appearance of new maximums at approximately 31.5º, 35.6º, and 48,3º (▲) were a sign of WC contamination (planes (001), (100), and (101) respectively) of the samples from the milling instruments [131].

These results confirm how the forces induced by this kind of mill differ from those of the planetary mill. In essence, the oscillatory mill induced impact forces that overcame 1) the van der Waals forces responsible for the stacking of sheets and 2) the covalent bonds between carbon atoms within the layer, i.e., inducing structural defects and lateral size reduction.

Furthermore, the changes in the precursor material take a shorter time when milling in the oscillatory mill compared to the planetary. Therefore, we could consider the oscillatory mill as a harsher, more destructive mill from the differences in the milling rate and the forces induced.

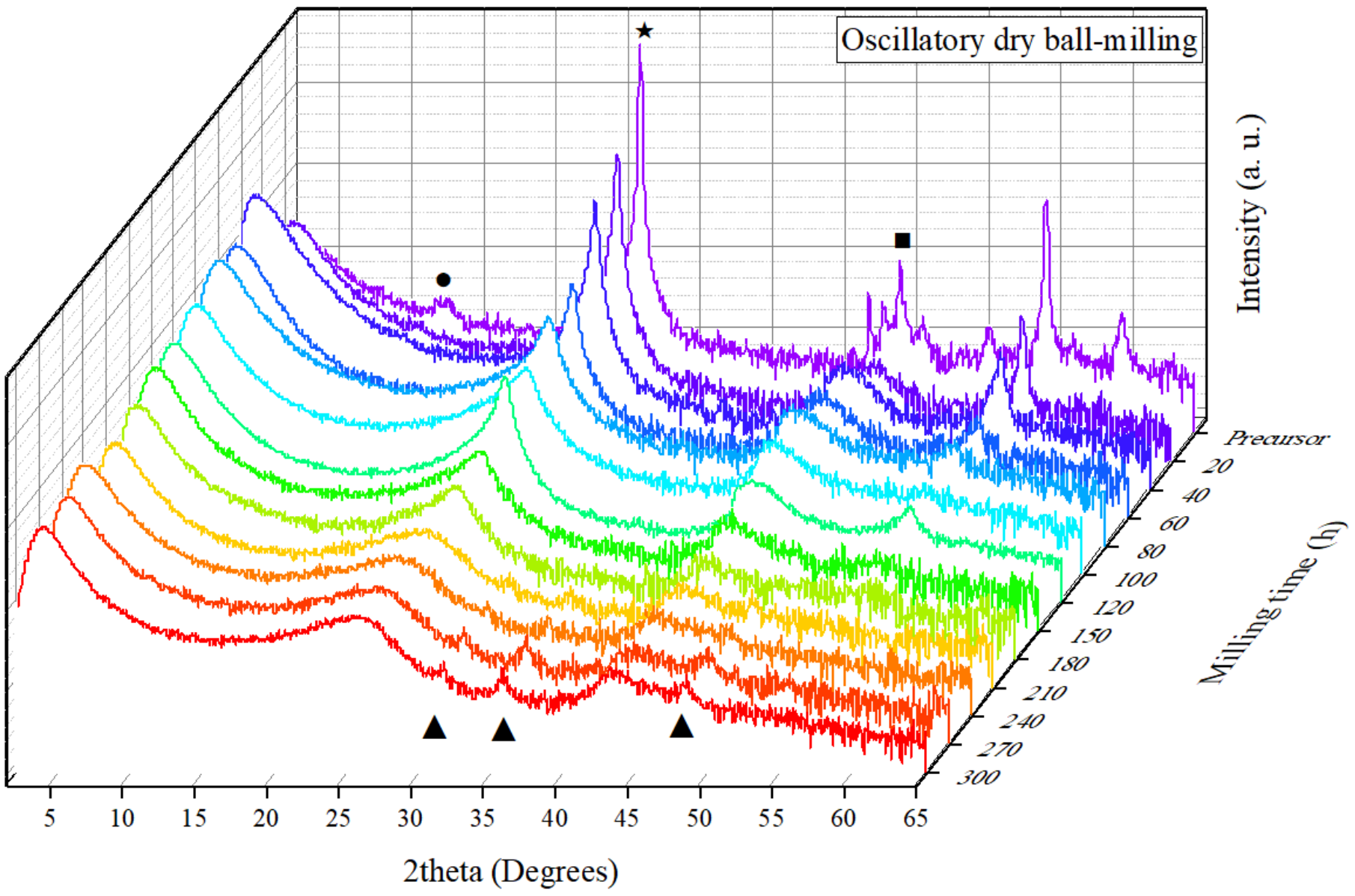

*Figure 2. 8 XRD spectra for the dry ball-milling test in the oscillatory mill. The precursor refers to the graphite as purchased.*

Additionally, Figure 2. 8 shows the rise of a low-angle maximum (located at around 3 - 4º) with a strong presence, evidencing the formation of a large order periodicity not observed during the planetary mill. From the Braggs law equation [132] such a maximum should have its origin in relatively high interplanar distances. This contribution could be due to the presence of mesostructures with sizes of a few nanometers with non-uniform organizations, as suggested by the width of these peaks [133]–[135].

Extensive structural changes were observed when the milling was initiated, substantially impacting the XRD patterns. At 20 min, the contribution of mesostructures increased in relative intensity and was accompanied by a loss of sharpness of the (002) maximum. Two trends can be observed regarding the evolution of the relative intensity of the mesostructure peak to those of the (002) as the milling progresses (see Figure 2. 9a). On the one hand, the relative intensity of the low-angle maximum increased linearly with milling time up to 8.55 times up to 240 min milling time. Thus, a massive generation of mesostructures was produced at this point (Region I, mesostructures generation in Figure 2. 9a). Eventually, a stabilization of the mesostructures generation was reached after 240 min of milling, leading to the establishment of a stabilization region (Region II in Figure 2. 9a) [130].

On the other hand, the resulting width of the mesostructures (w) was calculated from Bragg's Law and is presented in Figure 2. 9b. The starting graphite (0 min) already showed a certain degree of with a w=28.38(1) Å and decreased progressively up to 18.49(6) Å after 240 min (Figure 2. 9b). The analysis showed that contraction of the mesostructures occurred as they were generated, establishing the mesostructure compression range (Region I in Figure 2. 9b). Afterward, an inflexion point was observed at 240 min where the trend reverses, giving rise to a mesostructure enlargement up to 23.23(5) Å at 300 min (Region II in Figure 2. 9b). Consequently, the averaged width increased when the generation of mesostructures stopped. The stabilization and enlargement phenomena may be due to a loss of exfoliation efficiency due to a high degree of agglomeration from 240 min onwards, as later observed in the SEM images (Figure 2. 10b3-4) [130]. A highly porous mesostructure provides a huge specific surface compared to solid particles, improves the interaction of the material with gaseous components for catalysis or sensing applications [95], [135] and promotes multiple reflections [136], [137].

The samples obtained during this test were further examined to determine their nature.

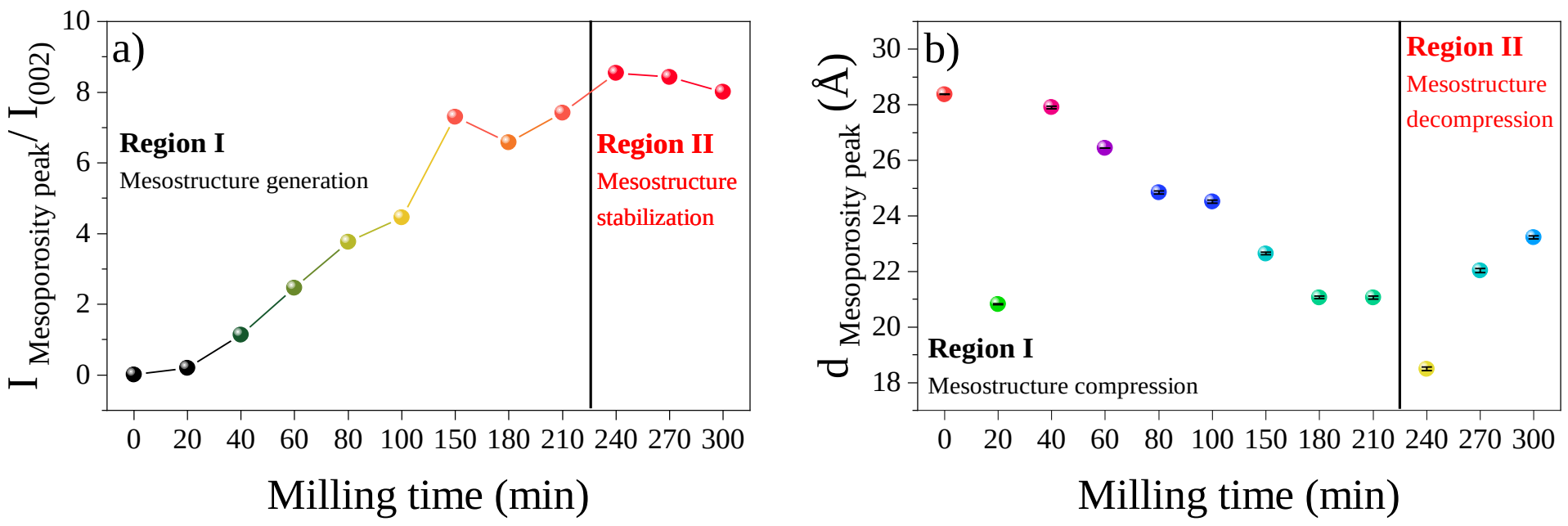

*Figure 2. 9 a) Ratio of the intensities between the mesoporosity peak and the (002) maximum (unitless) for increasing milling times. Two trends can be observed, Region I, where the ratio's value rises and Region II, where the ratio's value plateaus; b) and evolution of the interplanar distance (calculated from Braggs law) for increasing milling times. Again, two trends can be distinguished, Region I, where "d" falls and Region II, where "d" rises. As in other figures, 0 min refers to the precursor material.*

### 2.4.3.2 Morphological characterization.

The morphological properties of the raw graphite and milled samples between 20 and 300 min were studied by SEM measurements. As representative analysis, the samples with milling times 0 , 40, 150, and 240 min are shown in Figure 2. 10a1-a4, respectively. The raw graphite (0 min) was observed to have a flake shape with lateral dimensions between 2 and 50 µm and thicknesses less than 100 nm (Figure 2. 10a1, in close agreement with the manufacturer's description). In addition, the flakes showed pronounced and abrupt surfaces that could be considered of high crystalline quality [130], [135]. The morphology and size of the flakes decreased with milling time, while agglomeration signs could be observed after 40 minutes but were more noticeable after 150 minutes of milling (Figure 2. 10a3-a4).

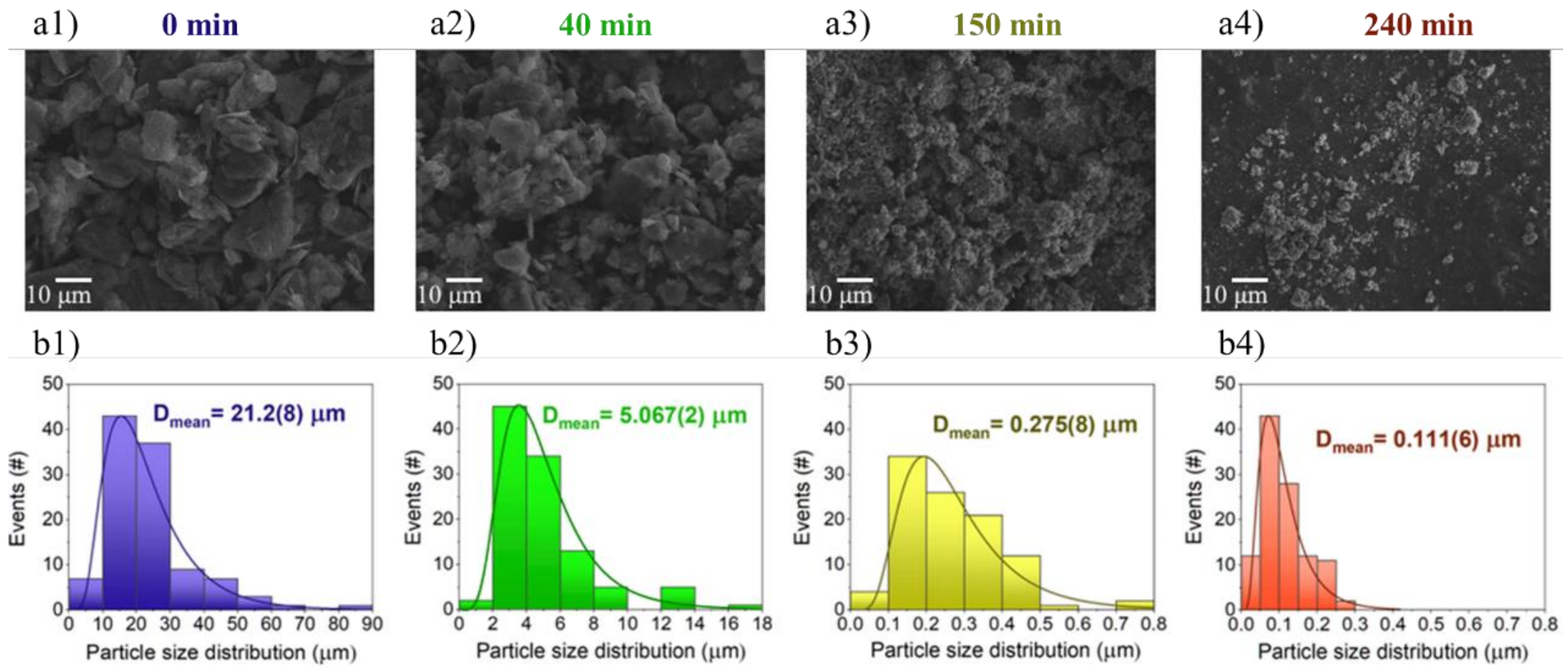

*Figure 2. 10 (a1-a4) SEM images and (b1-b4) particle size distributions of the samples 0 min, 40 min, 150 min, and 240 min.*

Additionally, particle size distributions were obtained for samples prepared at 0, 40, 150, and 240 min (Figure 2. 10b1-b4). The studied distributions followed a log-normal function with mean values ranging from 21.2(8) µm (0 min) to 0.111(6) mm (240 min), showing the typical trend when effective milling is performed [138], [139]. Furthermore, the progressive decrease of the particle size is associated with a narrowing of the distribution curves with milling since the full-width high maximum (FWHM) for the 0 min sample is 22.4(6) µm and for the 240 min sample is 0.145(9) µm. Therefore, the milling effectively reduced particle size relatively quickly, obtaining narrower and more homogeneous size distributions with milling.

Using transmission electron microscopy (TEM), images with a higher resolution were obtained. Figure 2. 11 shows samples for the precursor (a), a 100-min milled sample (b) and a 240-min milled sample (c). All figures show particles with crystalline domains periodic stacking. In this set of images, the number of layers at each stacking decreases with the milling time, from around 50 layers for the precursor to around 20 layers at 100 min and around 5 layers at 240 min.

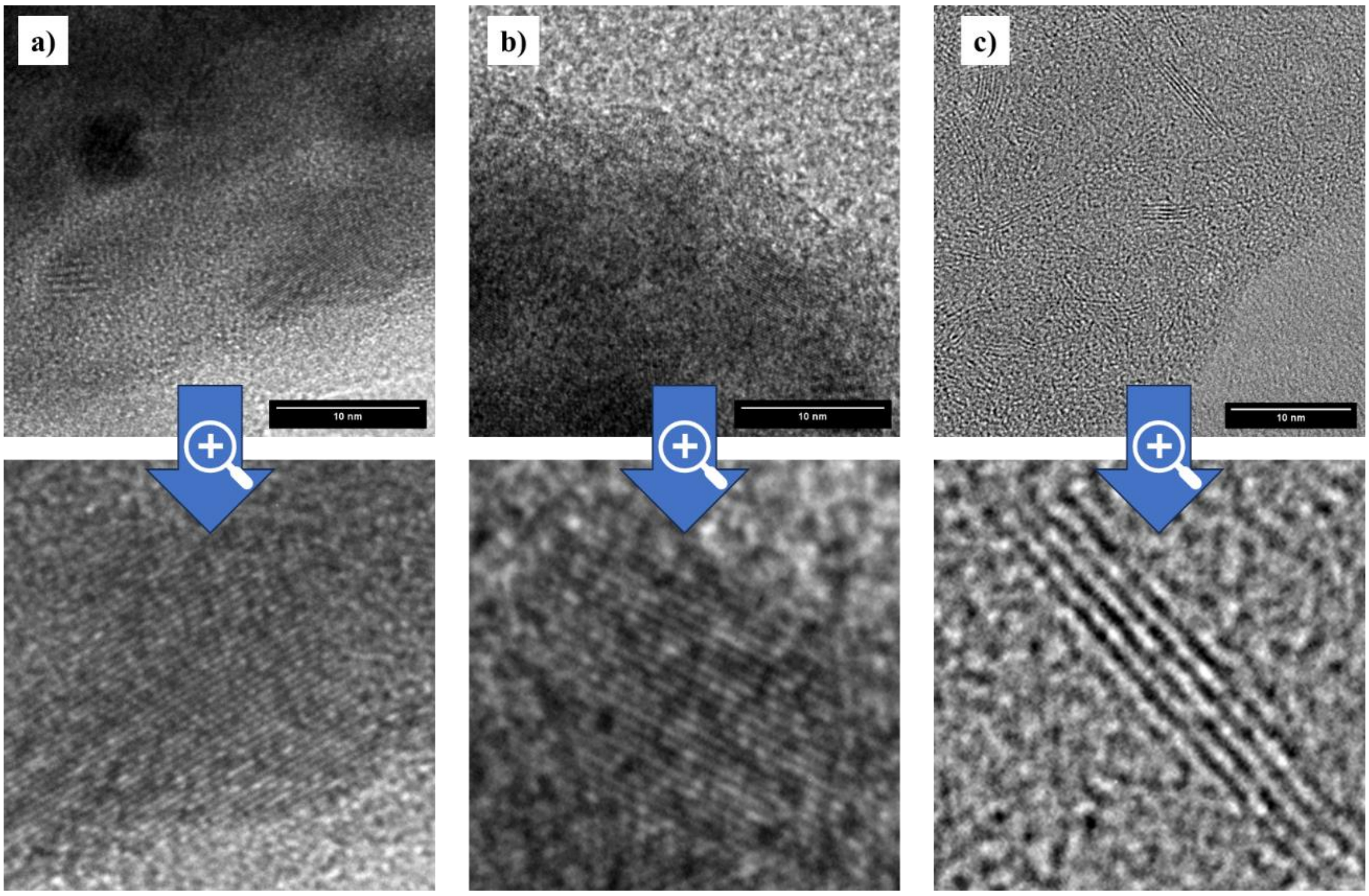

*Figure 2. 11 TEM images of a) the precursor and milled samples during b) 100 min and c) 240 min. The scale bar represents 10 nm.*

The quantification of layers using TEM is extremely sensitive to the chances of finding a stacking oriented perpendicular to the electron beam. The possibility of capturing a region not representative of the whole sample should not be neglected. Bearing this in mind, TEM provided additional information about the milled samples that, within the context provided by the rest of the techniques, further confirmed that exfoliation of the precursor occurs and is dependent on the milling time.

*2.4.3.3 Defects characterization.*

Regarding the structural quality of the milling product, Raman characterization was performed. The Raman spectra show a series of maximum related to the vibrational modes in the graphene lattice, the most relevant for graphene-based materials analysis are indicated over the results in Figure 2. 12. The G-band, located around 1580 $cm^{-1}$, is related to in-plane vibrations of the sp2 hybridized carbon atoms. The D band, located around 1350 $cm^{-1}$, and the D' band, located around 1620 $cm^{-1}$, come from vibration near defects or graphene edges. Finally, the 2D band (often referred to as G'), located around 2700 $cm^{-1}$, is a second-order band related to layer stacking [140], [141].

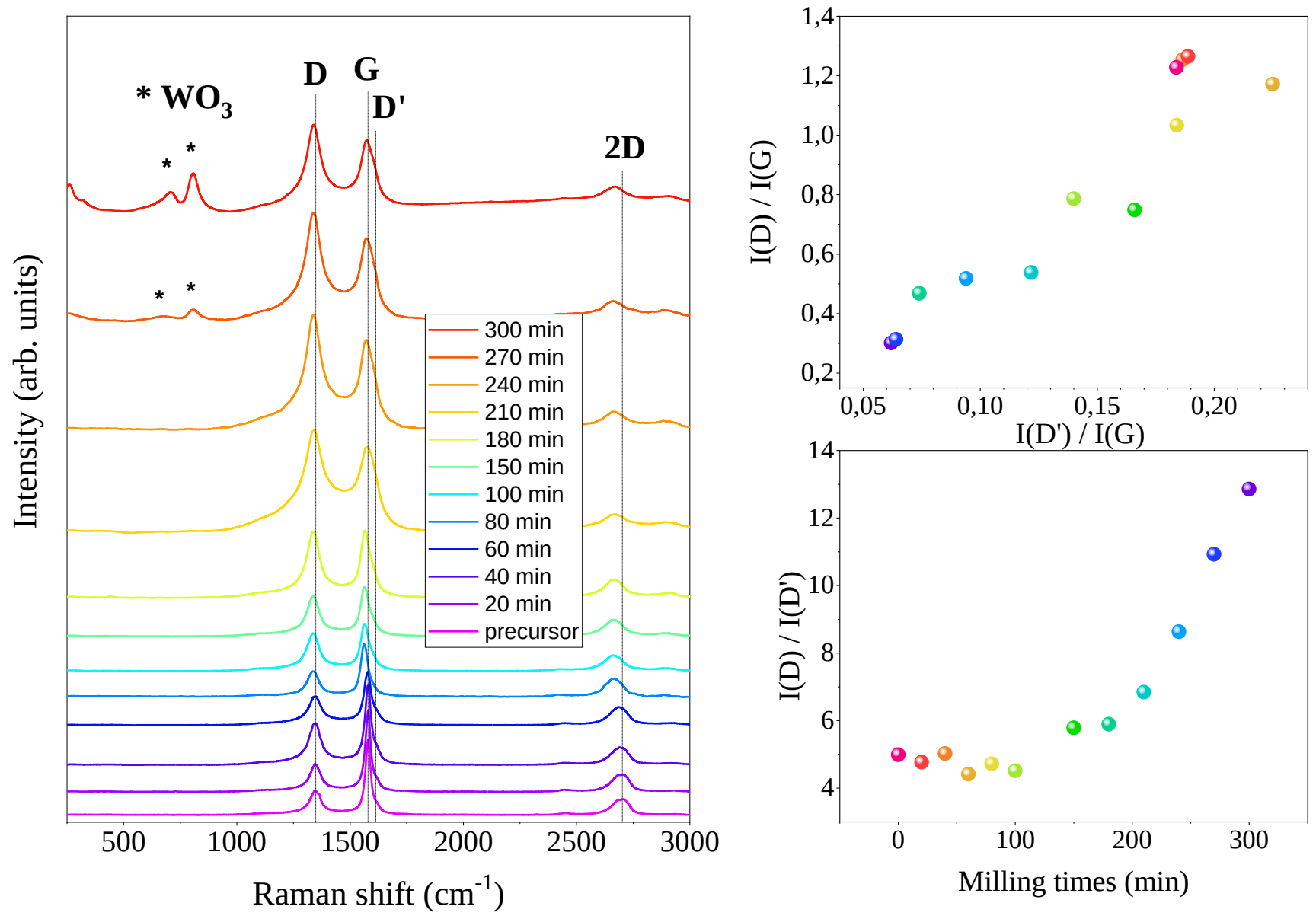

*Figure 2. 12 a) Evolution of the averaged Raman spectra, b) I(D)/I(G), and c) I (D)/I(D') ratios for the precursor graphite flakes and the milled samples between 20 and 300 min.*

In this line, the analysis between the intensity (I) of the D/G bands (I(D)/I(G) ratio) provides information related to the amount of defects, whereas the I(D)/I(D') ratio is similarly related to the nature of the defects (i.e., vacancies, grain boundaries, and sp3 bonding) [142]. For this purpose, Lorentzian fits were made to the bands obtained from the average spectra in each region corresponding to each sample (Figure 2. 12b). Starting with the precursor graphite flakes (0 min sample), Raman signal displays a certain degree of defects since the ratio I(D)/I(G) has a value of 0.31(6). When milling begins, there is a gradual increase of the I(D)/I(G) ratio with the milling time, up to 210 min with a value of 1.23(2), indicating a high degree of disorder. After 240 min, a stabilization of defect generation is observed with similar ratio values. Therefore, if the results shown in Figure 2. 9a are considered, the formation and stabilization of the defects are accompanied by the generation and stabilization of mesostructures (Region I and Region II, respectively) [130], [135].

As for the I(D)/I(D') ratio, it is known that the defects are predominantly related to grain boundaries when this ratio is close to ~3.5, to vacancies when it is close to ~7, and to sp3 bonding when it is close to ~13 [142]. To unveil the nature of the defects generated by high-energy milling, Figure 2. 12c shows the evolution of the ratio I(D)/I(D') as a function of milling time. The results obtained from the average Raman spectra indicate no significant variations up to 100 min of milling with values from 4.99(2) (0 min) to 4.51(4) (100 min) values. Thus, the nature of the defects is predominantly due to grain boundaries. However, from 150 min onwards, there is an intense trend change from 5.78(4) (150 min) to 12.86(5) (300 min), indicating a clear transition of the nature of the defects from vacancies to sp3 bonding. Interestingly, the average I(D)/I(D') ratio is equal to 6.84(8) for 210 min and equal to 12.86(3) for 300 min. Therefore, this milling enables tuning the nature of defects formed with remarkable selectivity. It could represent a breakthrough in technological applications, such as gas sensors or microwave absorbance, where defects and their nature can modify the material's response [135], [143], [144].

For the 270- and 300-min samples, additional Raman bands emerge, corresponding to the formation of $WO_3$ (indicated with asterisks in Figure 2. 12a) [145]. Therefore, the WC contribution observed in the diffraction patterns (Figure 2. 8), could have a certain degree of W oxidation, and the contamination-free limit would be marked up to 240 min of milling according to the Raman spectroscopy results.

### *2.4.3.4 Chemical Characterization.*

In the synthesis of a graphene-based material, it is important to study the oxidation level and the presence of oxygenated functional groups, as these factors can impact the material's properties and potential applications.

UV-vis spectroscopy was employed to understand the chemical nature of the product of the ball-milling (see Figure 2. 13). The absorption spectra of the samples presented a maximum located at around 257(3) nm, originated from π-π*, characteristic of non-oxidized GBMs [70], [127], [146], [147]. Noteworthy, the location of the maximum is largely unaltered after 300 min despite the milling process being carried out under normal atmospheric conditions. Due to the programmed interruptions of the milling sequences and the temperature control, extensive oxidative reactions affecting the material's composition may have been avoided [129], [130], [135].

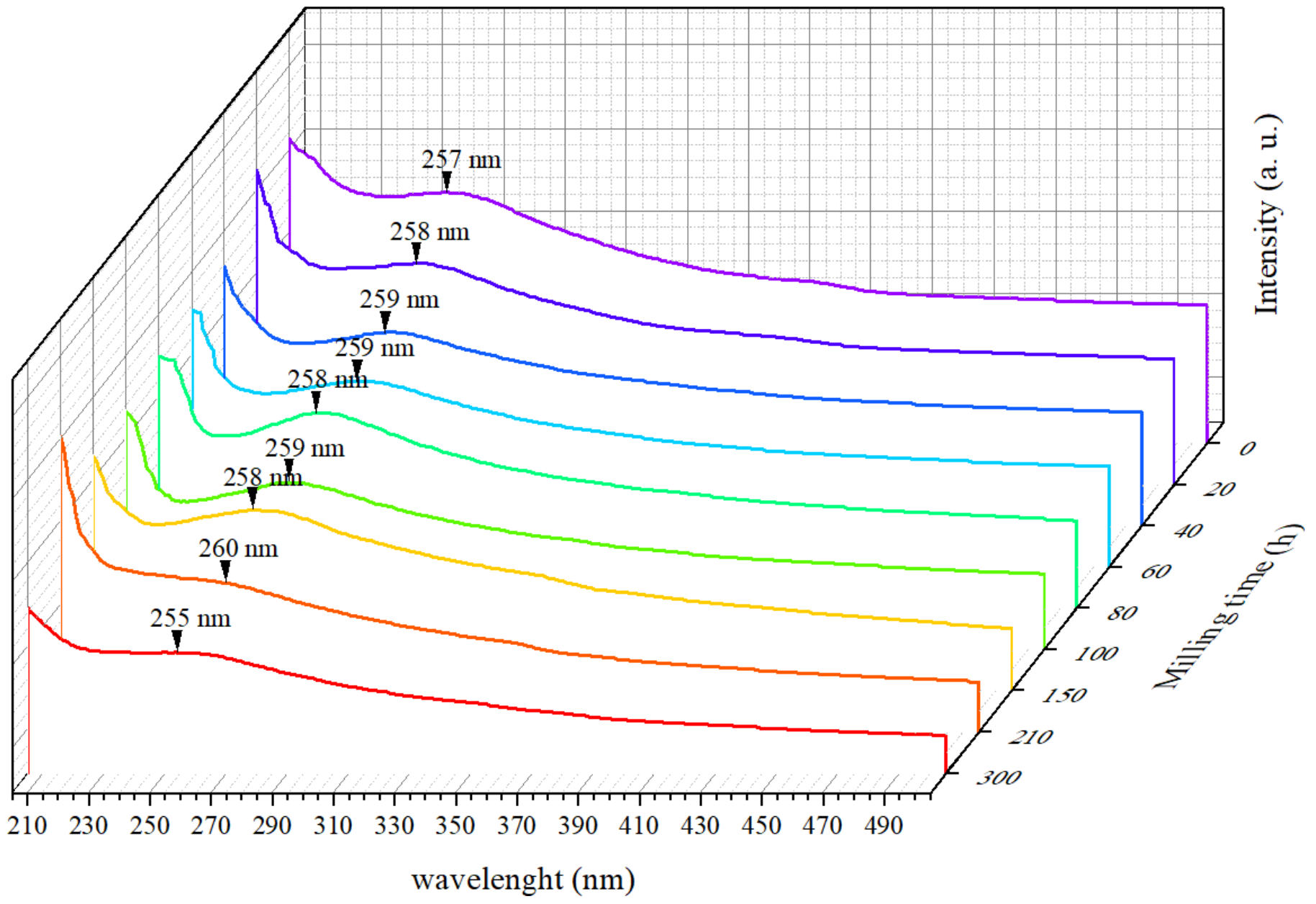

*Figure 2. 13 Ultraviolet-visible absorption spectra for the milled samples.*

It is worth noting that the presence of oxygenated functional groups to some degree is expected during the synthesis process. This is because the formation of defects in the material can lead to dangling bonds, creating highly reactive sites that are susceptible to reactions with oxygen or other elements. Oxygen can be present in various forms in graphene oxide (GO), reduced graphene oxide (rGO), or other graphene-based materials (GBMs), including hydroxyl (C-OH), ketonic species (C=O), carboxyl groups (COOH), or epoxide (C-O-C) structures [148].

The presence of such ligands was investigated using infrared absorption spectroscopy (IRAS) (Figure 2. 14). When comparing the IR bands of a synthesized material to those found in the literature, it is important to keep in mind that the exact positions and relative intensities of different modes can be significantly influenced by the local environment. Moreover, noticeable shifts in frequencies can occur, complicating direct comparisons with published data [149]. The spectra analysed in Figure 2. 14 can be divided into three regions [54]: (I) 3700 – 2700 $cm^{-1}$: CHx and OH stretching modes from hydrocarbons, either adsorbed in the powdered material or from functional groups attached to the lattice. Also, these bands may originate from water; (II) 1800-1500 $cm^{-1}$: C=O stretching vibrations and C−C lattice vibrations; and (III) 1500-600 $cm^{-1}$: deformations from CHx and stretching from C-O groups. The three spectral zones show a large number of organic ligands present in the flake graphite precursor, and no significant changes of the ligand nature are detected [129], [130], [135].

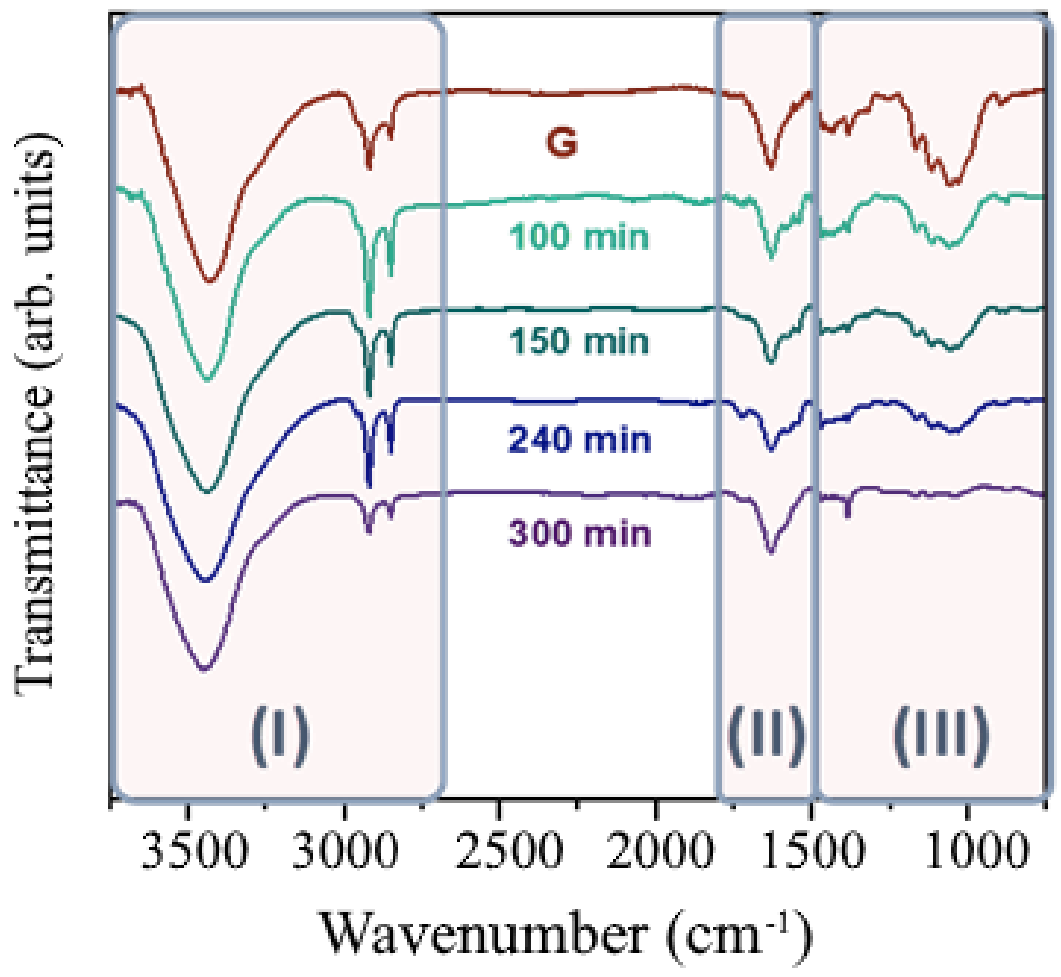

*Figure 2. 14 Infrared absorbance spectra for the milled samples. Rep. from* [130].

In sum, the dry ball-milling of graphite under the previously described conditions led to obtaining a nanomaterial in the form of mesoporous agglomeration of few-layered graphene sheets stackings with a high degree of defects and minor signs of oxidation. Said material was named few-layer mesoporous graphene (FLMG) after such characteristics [129], [130], [135]. To investigate the relevance of this material, FLMG was tested in applications where the mesostructure and the high specific surface derived from it, along with the presence or structural defects, are desired features, such as microwave absorbance and gas sensing, that will be presented in the following chapters. Regarding synthesis, FLMG proved to be a material obtainable through a cost-effective and large-scalable dry ball-milling technique.

#### *2.4.3.5 Final remarks.*

Finally, considering the patentability of the method, modifications of some parameters of the milling method were also studied. For instance, the previous test was repeated, first using two milling balls instead of one and second, with a lower milling frequency. For instance, the previous test was repeated, first using two milling balls instead of one and second, with a lower milling frequency.

For the former, two WC milling balls were placed inside the jar, and the quantity of graphite was doubled to maintain the powder/ ball weight ratio. The milling was done at 1500 rpm for 120 minutes. The XRD results, presented in Figure 2. 15, reveal severe contamination of the samples in the two balls case, probably due to the impact of one ball with another, and was thus discarded.

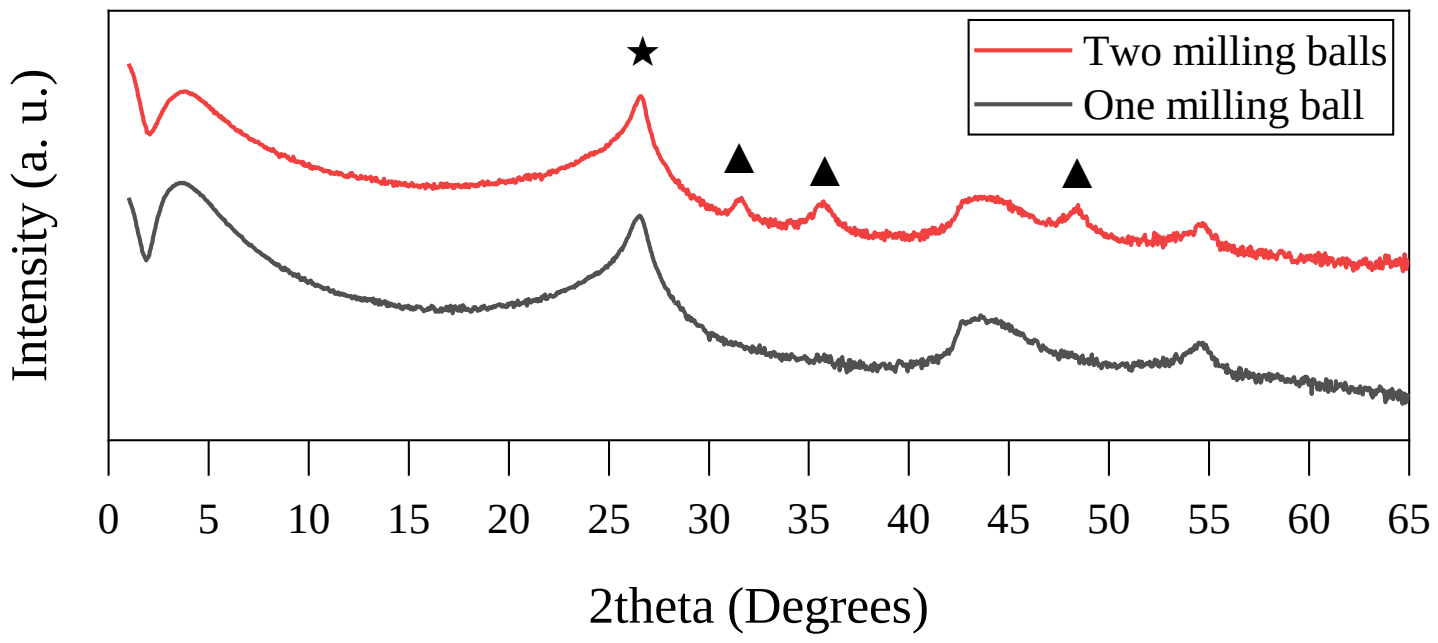

*Figure 2. 15 DRX results of 120-min milled samples using one or two milling balls* [129].

For the latter, two milling frequencies were tested. The rest of the conditions were maintained from the original experiment, and the milling time was set to 240 min. The DRX results, presented in Figure 2. 16, show that the decrease in milling frequency effectively slows the exfoliation process. The reduced exfoliation rate may be paired with a milder process similar to the results observed for the planetary milling, and, subsequently, the exfoliation takes process with lesser damage to the in-plane crystalline structure. This possibility, however, was not further explored during the present work.

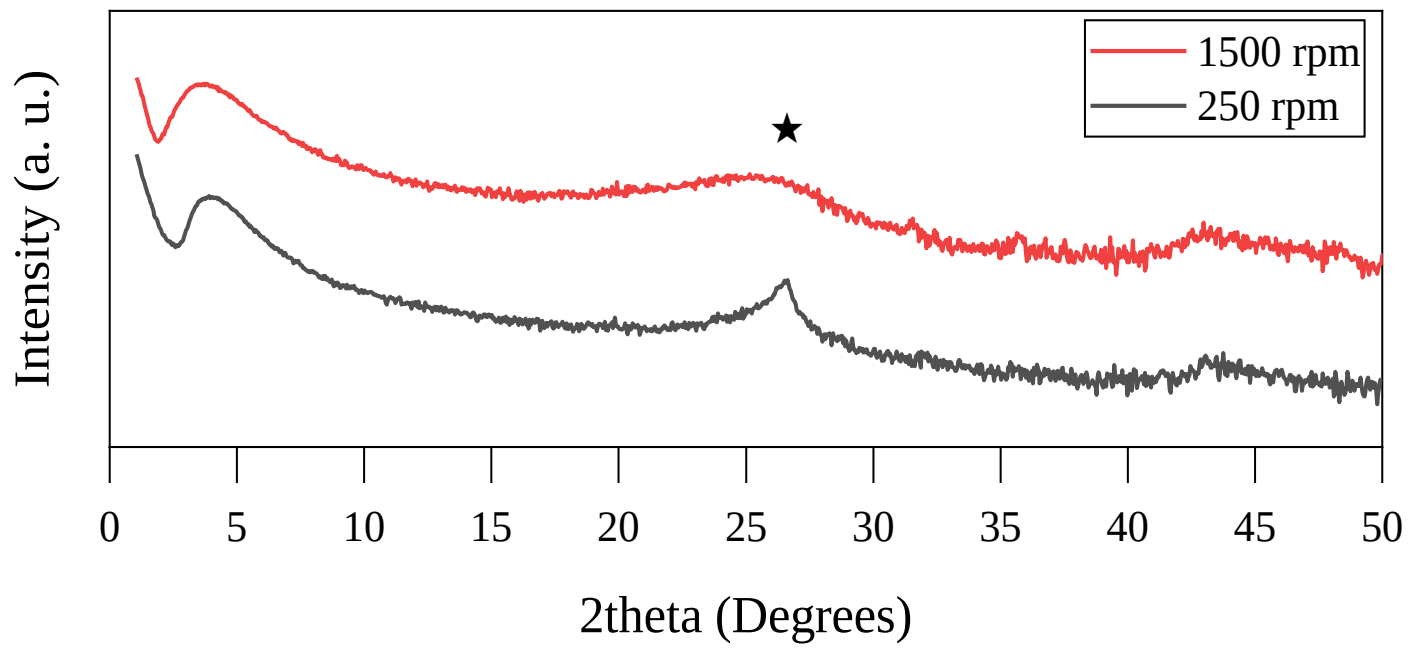

*Figure 2. 16 DRX results of a 240 min mill using milling frequencies of 250 and 1500 rpm* [129].

This work was published as a patent, ES2779151B2 "Obtención a gran escala en un solo paso y a temperature ambiente de material compuesto por pocas láminas de grafeno con un alto grado de defectos mediante molienda mecánica seca oscilatoria de alta energía" (please note that since it was presented at the Spanish patent office it was accordingly written in Spanish). The patent contained a comprehensive state-of-the-art revision and a detailed explanation of the method, as usual in this type of publication, followed by the claims, the critical points protected by the patent. The claims and high-quality versions of the figures are presented in Annex 1.

### *2.4.4 Dry ball milling in the high-energy mill.*

Lastly, a third mill was tested, a high-energy mill. This ball mill, the Emax, has a characteristic circular eccentric movement of the jars. This movement is expected to shear forces from the friction between the balls and the jar walls but far more energetic than in the case of the PM400.

The test was done using 50 mL WC jars and around 600 balls of the same material with 3 mm in diameter and a weight of 180 mg each. All jars contained 1 g of graphite to build a powder/ball weight ratio of 1:100. No additives were used. The milling was done at a rate of 25 Hz (1500 rpm) with a milling time between 20 and 100 min. During the milling sequences, interruptions were made each 4 minutes because an accumulation of material at the slits around the lid was observed, the jars were opened, and the accumulated material was removed and mixed again to prevent heterogeneous milling.

The milling results were examined using XRD, the spectra presented in Figure 2. 17, show how the decrease in the 26.5º peak intensity occurs more rapidly than in any of the previous milling tests. The low-angle maximum, previously attributed to the existence of mesostructures, is gradually reduced until it becomes no longer relevant, differing greatly from the milling done in the oscillatory mill.

Finally, the appearance and growth of the three maximums previously attributed to WC indicate that a huge degree of contamination is coming into the sample from the milling instruments. The first indicators of this contamination are noticeable from 60 min milling time.

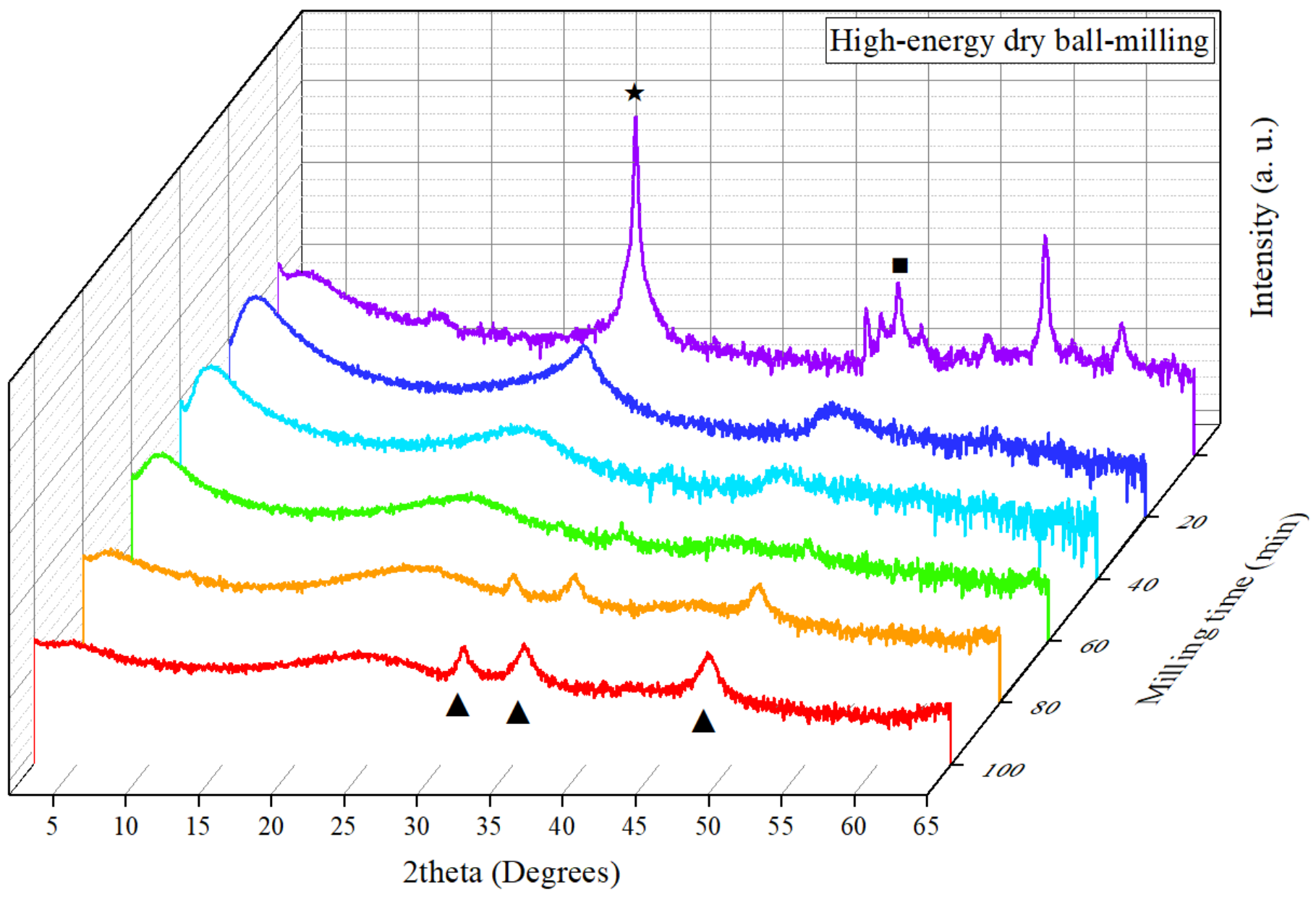

*Figure 2. 17 XRD spectra for the samples milled in the E-max.*

Since the contamination signs are the greatest observed in any milling technique, further investigation was required. X-ray fluoroscopy (XRF) was done in the same set of samples (see Figure 2. 18). In close agreement with the XRD results, the samples exhibited a rapidly increasing concentration of tungsten up to 5.0(1) % of the total weight after a 100 min mill. This W contamination is solely attributed to the milling instruments since no other sources of this element were present during the milling. It is also worth noting that the 20 and 40 min milled samples exhibit a relatively high purity (99.8(1) %) that matches the manufacturer's description of the precursor material. Such purity drops down to 93.6(1) % for higher milling times, as also noted in Figure 2. 17. The presence of oxygen in the sample increases slowly during the milling and as previously discussed, is a common phenomenon due to the generation of defects under an oxygen-containing atmosphere, even more considering that the jars were opened each interruption. The presence of other elements is neglectable in the samples.

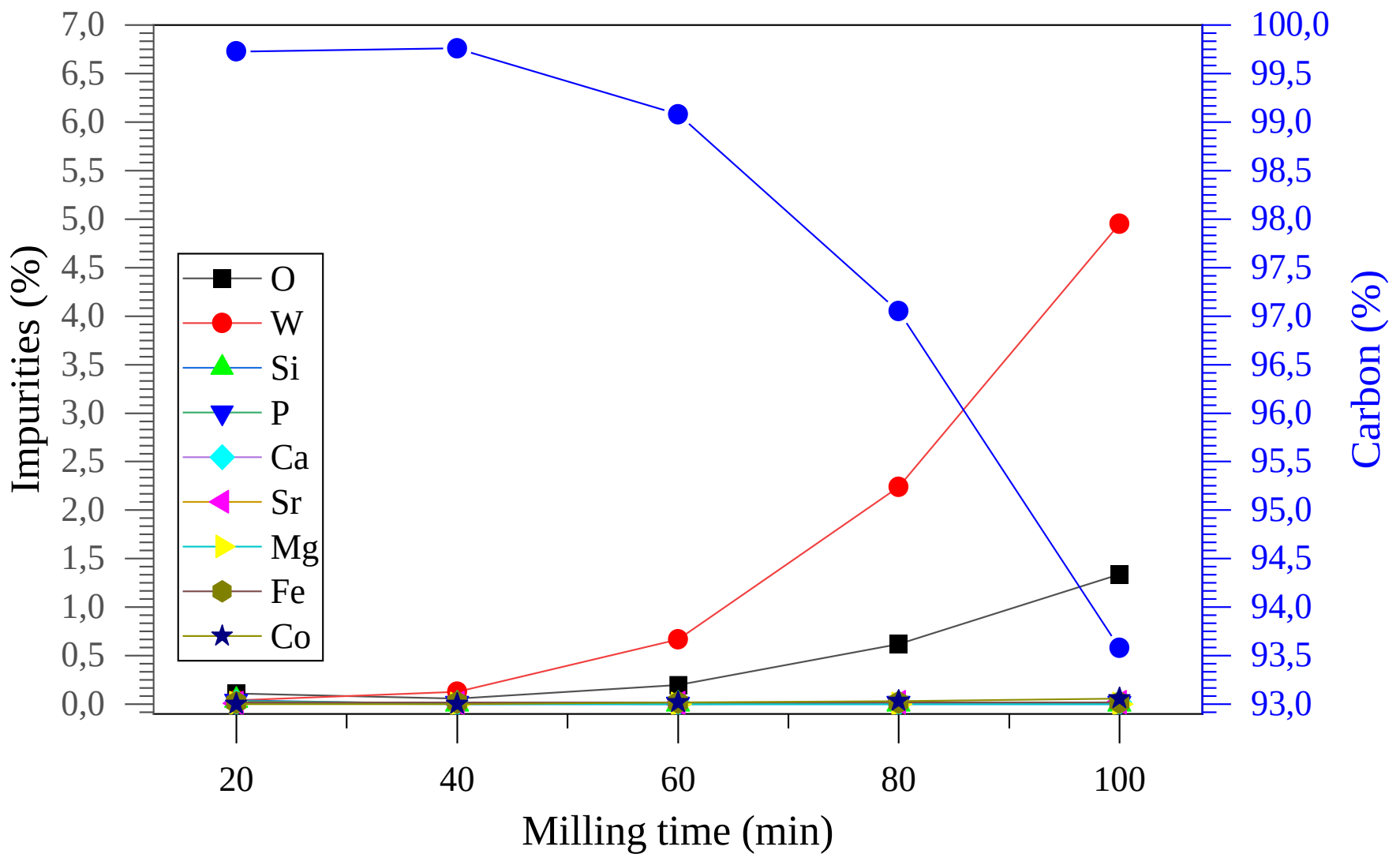

*Figure 2. 18 XRF elemental analysis of the samples high-energy milled samples*

To modulate the exfoliation and contamination dynamics, variations were introduced in the experiment and the milling was repeated at 8.3 and 16,7 Hz (500 and 1000 rpm). The samples were characterized using XRD.

The milling at 500 and 1000 rpm, analysed in Figure 2. 19a and Figure 2. 19b, respectively, show similar results in the obtained samples in terms of the exfoliation degree that greatly differ from those obtained at 1500 rpm. Figure 2. 19 shows that the milling frequency significantly affects the exfoliation degree. However, the abrupt differences observed during the high-energy mill suggest that this dependence is not linear, and a possible threshold may need to be overcome to attain effective exfoliation. The contamination of these samples is greatly reduced compared to those observed in Figure 2. 17, with only slight signs of WC, observed in the DRX spectra for 1000 rpm milling frequency.

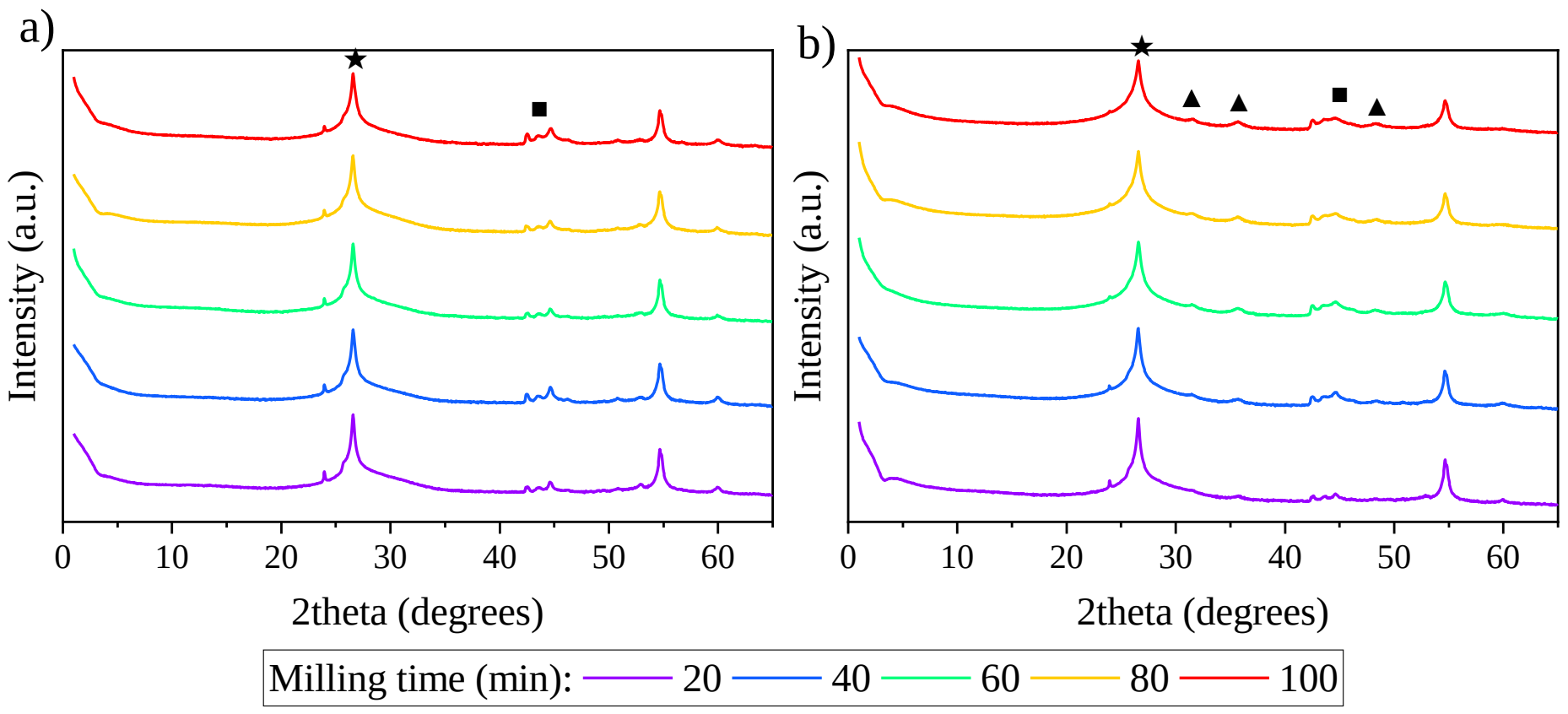

*Figure 2. 19 XRD spectra of the samples milled at a) 500 rpm and b) 1000 rpm using the high-energy mill.*

The results obtained testing the high-energy mill adverted that possibly due to the mill characteristics or the milling conditions, contamination from the milling instruments would always be present in a relevant amount before effective exfoliation of graphite could be obtained. Thus, the technique was no further explored in these preliminary results. However, later inspection of the milling instruments during cleaning revealed a fractured milling ball with exposed edges. It is unclear if the ball fractured during the milling or if it was a manufacturing defect, the fact that no other balls presented this problem suggests the latter. Nevertheless, it is highly possible that this ball originated the WC contamination, and as such, it would not be an inherent problem of the milling process.

### *2.4.5 Ball milling summary.*

To summarize the ball-milling technique, several experiments were designed and tested to explore ball-milling use for graphite exfoliation to obtain GBMs. The tests included millings using liquid and dry mediums, three milling equipment with different characteristics, a planetary mill, an oscillatory mill and a high-energy mill, and variations in the milling parameters (time, frequency, balls…). These tests revealed that ball-milling is a suitable technique for GBM synthesis, as evidenced by examples previously reported as presented in Table 2. 3, and that several physicochemical characteristics of the resulting GBM can be tuned through the careful determination of the milling parameters.

*Table 2. 3 Comparative table for ball milling GBM synthesis found in the literature. * The milling product was not fully characterized as previously stated, preliminary results suggest a form of exfoliated graphite.*

| Precursor | Milling-type | Additives | Milling parameters | Product | Application | Ref. |
|---|---|---|---|---|---|---|
| Graphite | Planetary | Potassium perchlorate | Liquid medium (water); 12 h; 300 rpm; ratio 1:20 | GO | - | [87] |
| Graphite | Planetary | Dry ice | Dry; 48 h; 500 rpm | Edge-selectively carboxylated graphene | - | [89] |
| | | Dry ice | | | | [103] |
| | | Sulphur trioxide | | Edge-Selectively Functionalized GNPs | Electrocatalysis | |
| | | Dry ice and sulphur trioxide mixture | | | | |
| | | $Cl_2$, $Br_2$ or $I_2$ | | Edge-selectively halogenated GNPs | Energy storage | [150] |
| | | Antimonium | | Sb-doped GNPs | Electrocatalysis | [151] |
| | | Sulphur | | S-functionalized GNPs | Energy storage | [152] |
| Expanded graphite | Planetary | - | Liquid medium (ethanol); 100 h; 450 rpm; ratio 1:40 | Defective GBM | - | [106] |
| Graphite | Planetary | Melamine | 30 min; 100 rpm | FLG | - | [153] |
| | | Triazine/ benzene derivates | | | - | [93] |
| Graphite nanosheets | Planetary | - | Liquid medium (DMF); 30 h; 300 rpm | Graphene | - | [104] |
| Graphene | Planetary | Polystyrene | Liquid medium (DMF); a few hours; 300 rpm | Polystyrene functionalized graphene | Conductive nanocomposites | [90] |
| GO | Planetary | - | 800 rpm; ratio 1:20 | rGO | - | [50] |
| Graphite | Planetary | Oxalic acid dihydrate | 20 h | Graphene | - | [55] |
| Graphite | Planetary | Direct yellow 50 | 2h; ratio 1:20 | GNPs solution | Conductive ink | [154] |
| | | Nuclear fast red | | GNPs poorly dispersable | - | |
| | | Methyl orange | | | - | |
| | | Methyl blue | | Graphite or non dispersable GNPs | - | |
| | | Alizarin red S | | | - | |
| | | Sodium dodecyl sulphate | | | - | |
| Graphite | Planetary | 1-pyrene carboxylic acid | Liquid medium (methanol and water); 30h; 300 rpm | FLG | Energy storage | [155] |
| Graphite | Planetary | - | 3-12 h; 250 rpm; ratio 1:100 | * | - | This work |
| | Oscillatory | - | 100-210 min; 1500 rpm; ratio 1:25 | FLMG | Gas sensors and microwave absorbance | This work [130], [135] |
| | High-energy | - | 20-100 min; 1500 rpm; ratio 1:100 | * | - | This work |

All milling tests were performed with the same precursor, a natural graphitic powder that is available and affordable and achieved exfoliation without additives or signs of oxidation. The oscillatory mill was the most explored ball-milling technique during this test. However, the planetary and high-energy milling yielded interesting results that should be further explored in future work.

The differences observed from each milling technique are outlined in Figure 2. 20. The planetary mill had a mild effect on the exfoliation degree and required more extensive time to obtain similar results to the other mills. However, the effect of the mild shear forces that originated from the planetary movement indicated that the exfoliation was related to less basal (sp2) crystallinity loss than other mills. On the other hand, the oscillatory and high-energy mill exhibited effective exfoliation at relatively low times. For these mills, characterized by the harshness of the process, contamination from the milling times was observed and should be kept in mind depending on the posterior application of the GBM. In this regard, HCl has been previously suggested to remove the metal impurities [55].

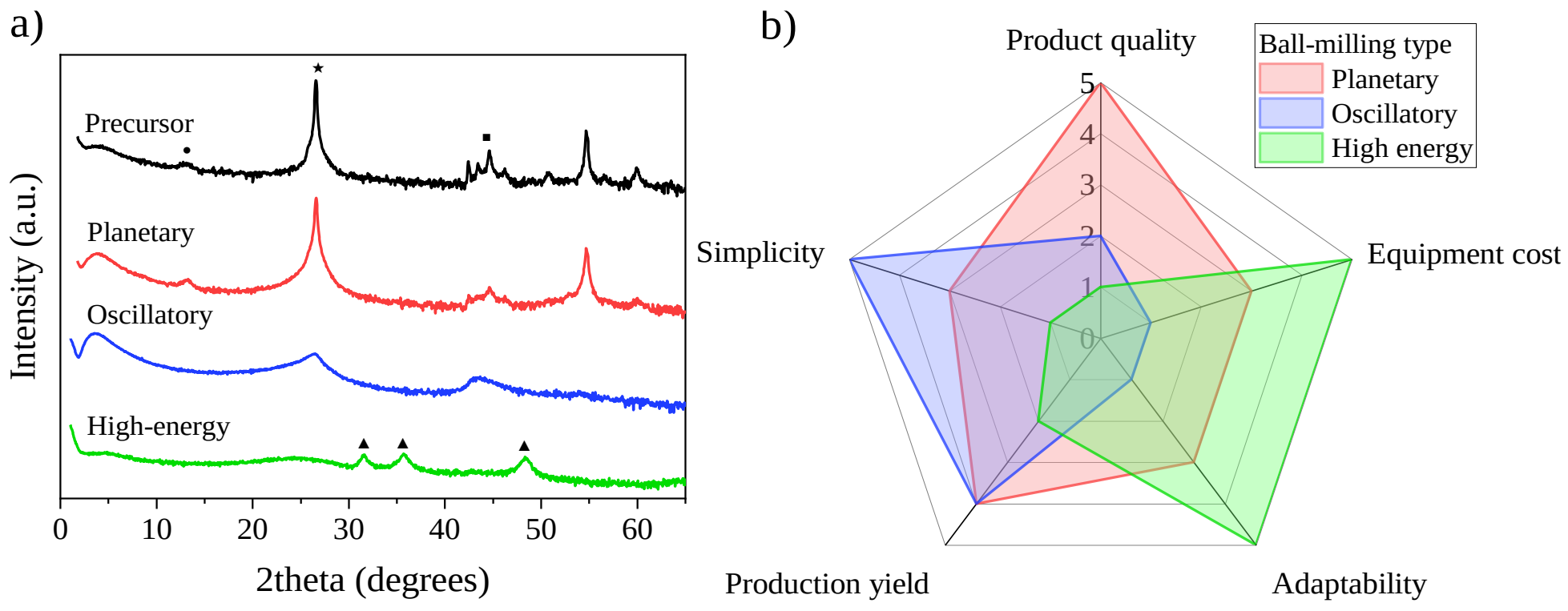

*Figure 2. 20 a) DRX comparison between graphite precursor, a 180 min dry milled sample in the planetary mill, a 100 min milled sample in the oscillatory mill and a 100 min milled sample in the high-energy mill; and b) Comparison between different milling techniques from the realized experiments. Where equipment costs refer to the acquisition and maintenance costs of the mills, production yield refers to the amount of material produced (timewise), adaptability refers to the capacity of the milling to be done under different conditions, product quality refers to the defectivity of the product and simplicity refers to the ease of preparation, operation, and post-production to obtain the material. The scale goes from 1-5, where 5 is the highest and 1 is the lowest, in arbitrary relative values.*

As previously introduced, top-down methods could be oversimplified to quantity over quality, which, far from being a negative aspect, makes them ideally suited for applications like composite reinforcement, energy (batteries or capacitors) or functional coatings [86], [90], [92], [94], [128], [156], [157].

The forthcoming chapters aim to assess the effectiveness of the milled product, FLMG, in applications such as electromagnetic shielding and gas sensing. Should FLMG display promising performance in these areas, it would present an enticing combination of efficacy and affordability. This combination could greatly accelerate the market uptake of graphene-based materials, bypassing the commercial challenges often tied to costlier or more complex alternatives.

## *2.5* Bottom-up Methods.

Bottom-up methods for graphene synthesis involve building up the graphene atom by atom and layer by layer from smaller precursors rather than starting with a bulk material as the top-down methods previously described. These techniques are generally more controllable and can produce high-quality graphene, but they tend to be slower and less scalable than top-down methods [44]. The principal bottom-up methods for graphene synthesis are epitaxial growth and chemical vapour deposition (CVD).

### *2.5.1 Epitaxial Growth.*

Epitaxial growth of graphene can produce single layers or FLG on the surface of a substrate. The first findings in this regard date back to 1962, when it was found that the sublimation of Si in a silicon carbide substrate left behind a carbon-rich layer [158]. Similar findings were later reported confirming this phenomenon [159]. Silicon carbide has been used ever since as a precursor for this method. Briefly, the substrate is heated under vacuum conditions, typically between 800 and 2000 ºC depending on the vacuum pressure, so that the silicon is sublimated and the carbon atoms re arrange themselves to form a graphene layer or FLG [160], [161]. The graphene layers formed this way are grown in a specific orientation with respect to the underlying SiC substrate.

The quality of the epitaxial graphene that is produced depends on the quality of the SiC substrate and the growth conditions used. Nevertheless, high-quality, single-layer graphene that has good electrical and mechanical properties can be obtained using this method. Epitaxial graphene is mainly used in electronic applications such as transistors, photodetectors, or gas sensors, among others [162]–[165].

### *2.5.2 Chemical Vapor Deposition.*

CVD is a process used to produce thin films or bidimensional materials on a substrate by the chemical reaction of gas with the substrate surface [166]. In the case of graphene, CVD is used to grow single-layer or few-layer graphene on a substrate such as copper, silicon carbide, or nickel. The precursor works for this technique can be traced back to 1968, when researchers found a strong affinity between carbon and a platinum substrate when studying the chemisorption of carbon-containing gases (CO, $C_2H_2$, and $C_2H_4$) [167]. These results were later interpreted to found that the gases decomposed in the surface and the carbon atoms adsorbed in the metal substrate, that acted as a catalyst causing the atoms to form "graphite sheets", i.e., a graphene layer [168], [169].

Several variations of this method can be found in literature, where different precursors, substrates, temperatures, and pressure are reported. Typically, organic gases such as methane and benzene are used as a carbon source, but virtually any carbon-rich product may be used as a precursor [170], [171]. The flow rates and ratios of the precursor gases have a significant impact on the quality of the resulting material, as these parameters control the supply of carbon atoms, the removal of unwanted carbon species from the surface, and the overall growth process [172].

As for the substrates, Cu and Ni are the most popular option, whereas Co is typically used on metal substrates as a catalyst. Cu is typically used for single layer graphene synthesis whereas Ni is used for MLG, meanwhile, Co can help to promote the decomposition of the hydrocarbon gas and the subsequent formation of graphene on the substrate [173]. The potential presence of cobalt as an impurity in graphene should be considered in such cases [174]. Before starting the CVD process, it is essential to clean the substrate thoroughly to remove any contaminants, such as dust, grease, or oxide layers, which can cause defects in the resulting graphene layer or lead to non-uniform coverage [175]. Common substrate cleaning methods include ultrasonic cleaning, rinsing with solvents like acetone, isopropanol, or deionized water, and using a mild acid or base to remove oxide layers. After cleaning, the substrate should be dried and loaded into the CVD reactor chamber without contacting any contaminants.

Precise temperature control is crucial during the CVD process, as it impacts the decomposition of precursor gases, the formation of graphene, and the overall quality of the product. Briefly, the temperature controls the kinetics of nucleation of crystal originated from the supersaturation of carbon atoms and their growth rate, it is thus essential to monitor the temperature and adjust it accordingly based on the specific requirements of the synthesis process [176]. The temperature range for graphene synthesis is typically set around 900-1000°C.

Furthermore, roll-to-roll production using Cu substrates receives huge interest thanks to its ability to obtain large sheets in an almost continuous production [177], [178]. Catalyst-free and low-temperature use are also key features for large-scale GBMs production [179].

Once the graphene has been produced on the substrate, it is necessary to transfer it to a substrate suitable for the intended application. Typically, a layer of PMMA is deposited on top of the graphene, and the Cu or Ni substrate is etched off using $FeCl_3$. Subsequently, the polymer is dissolved in acetone or another solvent [180]. However, the etching processes and manipulation of the graphene layer can damage, wrinkle, or introduce impurities in the graphene flake.

CVD-grown graphene is used for multiple electronic applications, including THz applications and gas sensors [181]. In addition, the recent field of twistronics also uses CVD-grown graphene, which is folded and twisted to obtain a slightly rotated BLG with interesting electronic properties [182], [183].

#### *2.5.1.1 CVD grown multilayer graphene.*

Devices presented in following chapter, i.e., the gas sensors using CVD-grown multi-layer graphene (MLG), were produced using a transfer-free method developed by Sten Vollebregt et al. [184]. First, Mo is sputtered over a $SiO_2$/Si wafer. During this step, Mo can be patterned so that graphene will grow in desired locations. The CVD process occurs at around 1000ºC and 25 mbar using Ar/$H_2$/$CH_4$ as a carbon source. After the CVD deposition of graphene, the Mo catalyst is rapidly etched using a phosphoric acid solution, after which the wafers are rinsed and spin-dried. As the distance between the graphene and $SiO_2$ after the Mo etching is relatively small, the graphene directly sticks to the $SiO_2$ without floating away during the etching and rinsing. Finally, the device is built by depositing Cr/Au (10/50 nm) electrical contacts using a lift-off process [174]. This method has been successfully used to produce various graphene-based electronic devices [185]–[187].

In this work, different temperatures between 850 ºC and 935 ºC were tested. By synthetizing MLG at these temperatures, different physicochemical properties are expected, which in turn may influence the sensing performance of the material [187]. Initially, these differences were interpreted as the density of defects originated during the growth process and characterized using the I(D)/I(G) and I(2D)/I(G) ratios from the Raman spectra [184], [187].

As shown in Figure 2. 21, a negative relation was found between the density of defects (I(D)/I(G) ratio) and the growth temperature, in close agreement with previous reports [188]. In addition, the evolution of the I(2D)/I(G) ratio suggests that higher temperatures result in the transformation from a material more similar to graphite to one more similar to MLG [187]. Overall, this indicates that higher growth temperatures lead to improved crystallinity and a more uniform graphene structure. Nevertheless, it is possible that limiting the analysis to the relative intensity ratios overlooks other characteristics such as the nature of these defects.

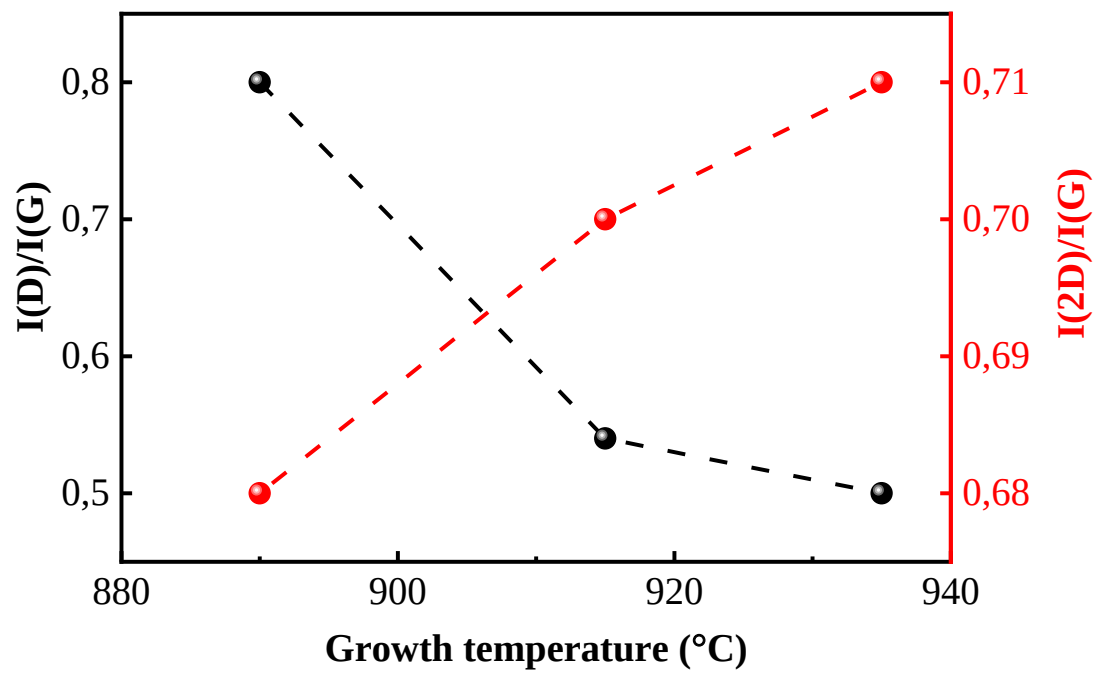

*Figure 2. 21 I(D)/I(G) and I(2D)/I(G) ratios for samples grown at different temperatures.*

Currently, research on the impact of growth temperature in graphene synthesis is centred on a more in-depth analysis of Raman spectra to identify the nature of defects. This deeper understanding can provide valuable insights into how these defects may affect the sensing performance of graphene-based materials.

A comprehensive Raman analysis was conducted on the devices, covering a large area as depicted in Figure 2. 22. The devices were probed in an area that included the Si/$SiO_2$ substrate, the Au electrical contact, and the MLG strip (see Figure 2. 22a-c). The analysis involved scanning the sample while acquiring a spectrum for each point, with the laser spot area being approximately 500 nm in size. These spectra can then be averaged or analysed independently to study possible inhomogeneities in the sample.

In the optical images presented in Figure 2. 22a-c, a first notable difference between the samples can be observed. The MLG grown at the higher temperature (935 ºC) exhibits a granulated aspect resulting from the existence of islands or pores probably originated from the terrace construction mechanisms of the graphene layers. Although it is not clear how these structures may affect the properties of the material, they have not been observed in MLGs grown at lower temperatures.

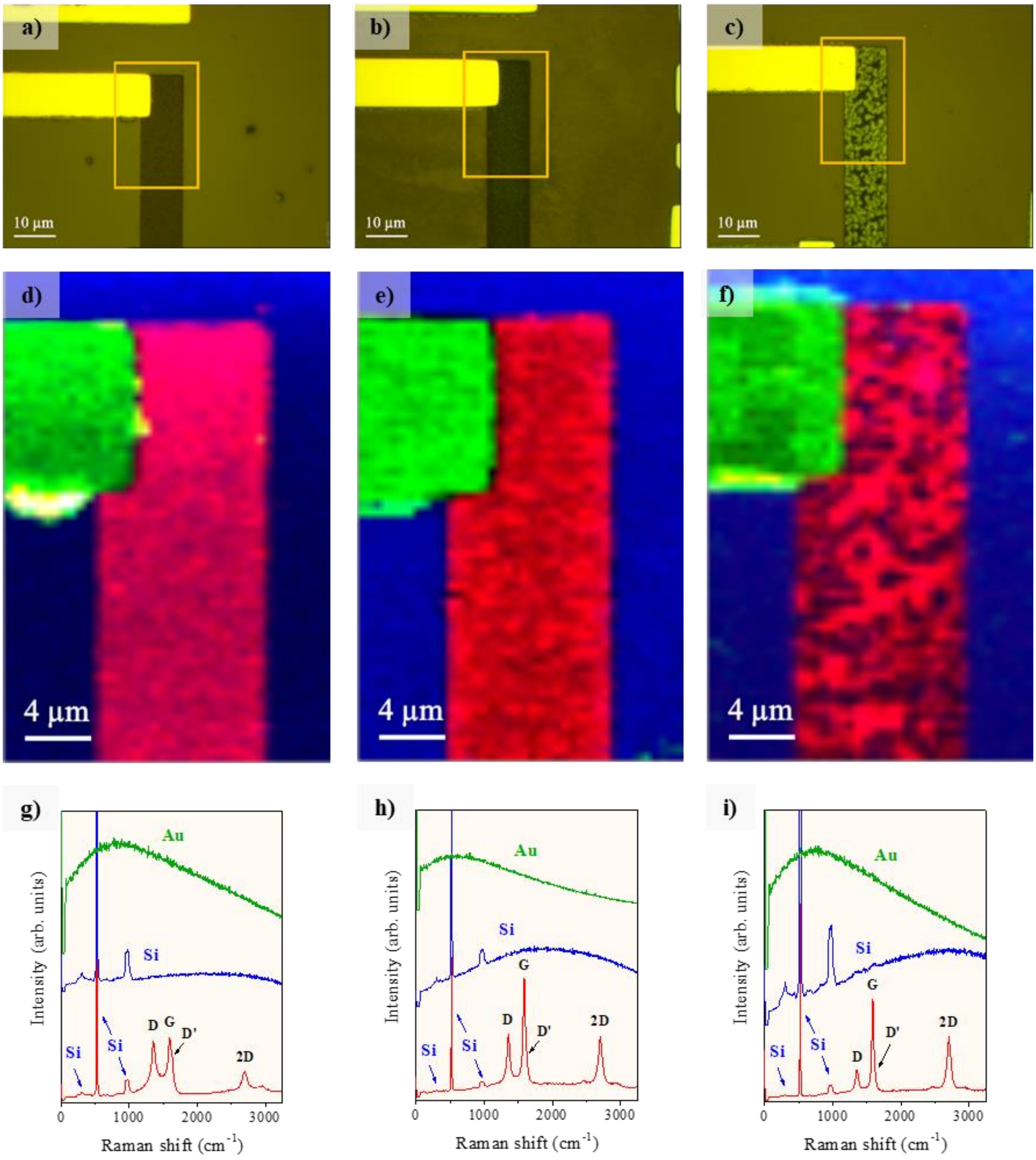

*Figure 2. 22 (a)-(c) Optical images of the devices with a marked are (yellow rectangle) where the Raman analysis is performed; (d)-(f) Raman intensity images in the XY plane the integration spectral area chosen are 50-450 cm$^{-1}$, 500-560 cm$^{-1}$, and 1560-1600 cm$^{-1}$, for the gold contacts (red), Si/SiO2 substrate (blue), and the MLG stripes respectively (green); (g)-(i) Average Raman spectra calculated from the Raman intensity images. Each column relates to a growth temperature, left to 850 C°, centre to 890 C°, and right to 935 C°.*

The Raman spectra, displayed in Figure 2. 22g-i, feature the characteristic bands for GBMs (red line) previously discussed, as well as those originating from the substrate. These additional bands are present in the MLG spectra due to the penetration depth of the Raman technique. The D, G, and 2D bands are clearly identifiable; however, it appears that there may be additional contributions contained within these bands. To study these contributions, the spectra is deconvoluted to identify and measure additional bands that provide further information about the material (see Figure 2. 23).

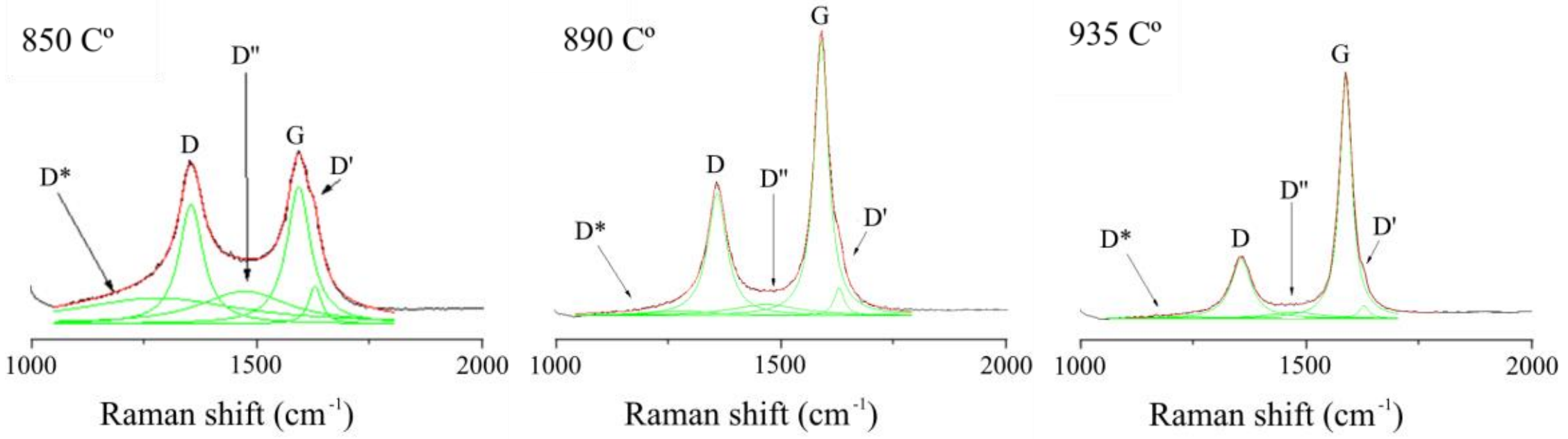

*Figure 2. 23 Deconvoluted Raman spectra of CVD-grown MLG materials synthesized at various temperatures.*

The D* band is located at 1050-1200 $cm^{-1}$ and originates from sp3 bonds in diamond-like carbons but can also be associated with vibrations of carbon atoms constrained by oxygen-containing groups [189]. In this case, the contribution of the D* band (I(D*)/I(G)) decreases with the increasing growth temperature.

Meanwhile, the D'' band is located at 1500-1550 $cm^{-1}$ and is related to an amorphous carbon phase [190]. For the MLG samples, the contributions of the D'' band (I(D'')/I(G)) decreases with the increasing growth temperature.

These preliminary results have not yet been discussed, and the mechanisms through which the D* and D" bands affect the material and its outcome in different applications are not fully understood. However, these findings confirm that differences exist in the materials grown by CVD at different temperatures beyond the conventional I(D)/I(G) and I(2D)/I(G) ratio analyses. The analysis of the D* and D" bands, along with the formation of islands or pores, may unveil crucial information about properties of the material.

## *2.6 Conclusions.*

As it has been presented, substantial efforts were made to obtain graphene and its definition. However, even greater efforts have been made, and would still need to be made, to keep the terminology and unambiguously distinguish between graphene-based materials (GBMs). The definition that has been generally accepted could be "a carbon-based nanomaterial that exploits the nanophysics phenomena". This definition deviates from the scientific one, but we must understand that graphene has escaped from the research laboratories and integrated itself into parts of society no one could have anticipated.

In materials science, no material fits all purposes. There is no *perfect* material, although there are materials that depending on the application, are more suited than others. Contrary to the expectations of some, graphene is no exception. The broad spectrum of physicochemical characteristics of GBMs provides us with materials that are an excellent fit for some applications while entirely inadequate for others, rendering us with the exceptional puzzle of pairing materials and applications and, in some cases, synthesis methods.

Comprehensive work exploring the ball-milling technique has been presented. The combination of scalable technique, graphite as a precursor and no additional reactives have yielded promising results that could lead to the obtention of GBM at an industrial scale and thus facilitate their use in commercial applications. The novelty of the present work was included in this chapter with a scalable ball-milling synthesis method of a GBM called few-layered mesoporous graphene and the suggestions of some applications that will be expanded in the following chapter.

The path of graphene has been long, as it can be traced back more than a century, and fascinating, as no other material has generated such interest in our era. Regardless of the unknown challenges that may lie ahead, thanks to the new synthesis methods, graphene has never been closer to delivering its promises.

## *2.7 References.*

# Chapter 3: Electromagnetic Shielding Materials.

*Chapter Introduction; The Transmission Line Approach; The λ/4 Approach; Free-Space Measurement Approach; Amorphous Magnetic Microwires in ESMs; Graphene-Based Materials in ESMs; Use of FLMG In ESMs; Conclusions; References.*

### *3.1 Chapter Introduction.*

In recent years, there has been a significant increase in microwave-based applications, i.e., applications that rely on electromagnetic radiation sources, especially in the range of 3-30 GHz. This is due, in part, to the development of relatively new technologies that rely on wireless communication, such as mobile phones and Wi-Fi, but also due to the projected expansion of applications based on 5G and the Internet of Things (IoT). The use of microwaves for radar and medical applications has also increased in recent years.

The increase in microwave sources is accompanied by the risks of electromagnetic interference, i.e., the effects these radiations can have over the normal operation of electronic devices. This explains the need for electromagnetic shielding to protect against the harmful effects of these sources. The ability of electronic devices and systems to function correctly in their electromagnetic environment without malfunction is referred to as electromagnetic compatibility (EMC). Ensuring EMC is important because it can help prevent equipment failure, data loss, and other problems caused by electromagnetic interference.

Health risk associated with electromagnetic pollution is usually claimed as a motivation for electromagnetic shielding technologies, e.g., electromagnetic absorbent contact lenses to prevent eye diseases like cataracts [1]. However, there does not seem to be strong evidence relating microwave radiation to health issues. There is still much that is not known about the potential health effects of microwave radiation, and more research is needed to understand the potential risks and how to minimize them fully. In the meantime, it is generally recommended that people are not exposed to microwave radiation beyond the justifiable need.

Finally, electromagnetic shielding is a crucial technology for military-related applications. It is typically used to absorb incoming Radar signals, located in the X-band (8-12 GHz, IEEE standard). Microwave absorption reduces the reflected signal and lowers the radar cross-section, i.e., the apparent size of an object under radar systems [2].

Materials capable of absorbing microwave radiation are commonly used as coatings and face important challenges towards a successful application. Not only do they have to present strong absorption in a certain frequency band or bands, but they also must be lightweight, thin, with decent mechanical and corrosion resistance properties, and overall suitable for large-scale use or good integrability into existing coatings.

Electromagnetic shielding materials (ESMs) include magnetic materials [3–5], carbon-based materials (GBMs and other such as carbon fibres) [6,7], and more popularly, composite materials including the previous [8–11]. A recent and interesting trend uses nanomaterials with special structures where the electromagnetic shielding comes greatly determined by their geometry [12–14].

There are different approaches employed in understanding microwave absorption and the associated characterization techniques [15]. In this chapter three approaches based on transmission/ reflection methods will be discussed, the transmission line approach, the quarter-wavelength approach and free-space measurement. Although the experimental work has been based on latter, the transmission line approach is here comprehensively revisited due to its widespread acceptance.

### *3.2 The Transmission Line Approach.*

The transmission line approach explains how the incident radiation is absorbed by different mechanisms intrinsic to the ESM including dielectric loss, magnetic loss or multiple reflection or scattering processes.

In particular, dielectric loss mechanisms interact with the electric component of the microwave and transform its energy into heat, similar to the operation of a microwave oven. They include conduction loss and polarization loss. The latter is due to the excitation of dipoles, i.e., bound charges localized around defects, interfaces, or residual groups, that become polarized at frequencies higher than their relaxation time. Ionic and electron polarization can absorb an electromagnetic wave similarly, but they occur at higher frequencies ($10^3$-$10^6$ GHz) [6,16].

The dielectric loss mechanisms can be studied using Cole-Cole semicircles. Each relaxation may appear as one semicircle when ε' vs ε'' (the real and imaginary parts of the electric permittivity) are plotted [17]. In the example presented in Figure 3. 1, five semicircles can be observed for the carbon black + barium

titanate/ polyvinylidene fluoride (PVDF) composite (a), whereas three semicircles are observed for the carbon black/ PVDF composite (b). In this example, the authors suggest that the two additional semicircles observed in Figure 3. 1a compared to Figure 3. 1b are due to relaxation processes within the barium titanate including its interfacial relaxation mechanisms [18].

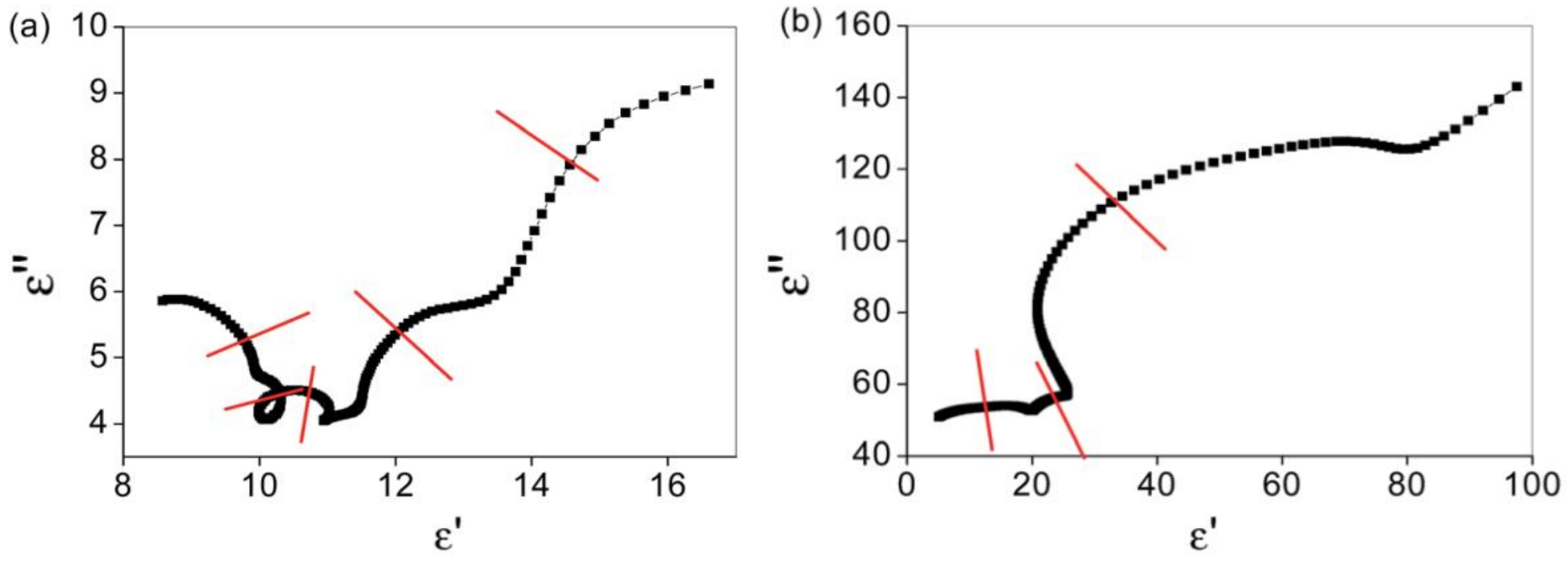

*Figure 3. 1 Cole-Cole semicircles for a) a carbon black + barium titanate / PVDF composite and b) a carbon black / PVDF composite. Adapted from* [18].

Meanwhile, magnetic loss mechanisms include ferromagnetic resonance, eddy current loss and exchange loss. In addition, domain wall resonance and magnetic hysteresis can cause magnetic loss, but these mechanisms are rarely seen in common electromagnetic shielding applications. Domain wall resonance occurs only at low frequencies (<2 GHz), and magnetic hysteresis is only observed under strong applied magnetic fields [19,20].

Finally, multiple reflection mechanisms also play an important role in ESM's operation. The incident wave is reflected within the ESM, increasing the propagation paths, and trapping the microwave. Although these mechanisms do not absorb electromagnetic energy *per se,* they promote dielectric and magnetic loss mechanisms' actuation, greatly improving the wave's absorption.

The amount of energy absorbed by the material depends on the relative complex permittivity ($\varepsilon_r = \varepsilon' - j\varepsilon''$) and permeability ($\mu_r = \mu' - j\mu''$), which is a measure of the material's ability to allow the passage of electromagnetic waves. Where the real parts (ε' and μ') are related to the storage capability of the incident energy, and the imaginary parts (ε'' and μ'') are related to the loss component and the mechanisms previously presented. Materials with high imaginary components are called lossy materials, and regarding electromagnetic shielding, they are characterized by the reflection loss (RL) parameter [16,21].

$$RL = 20\log\left[\frac{Z_{in} - Z_0}{Z_{in} + Z_0}\right] \qquad \textit{(Eq. 3. 1)}$$

Where $Z_{in}$ is the input impedance of a metal-backed electromagnetic shielding layer given by [22,23]:

$$Z_{in} = Z_0\sqrt{\frac{\mu_r}{\varepsilon_r}}tanh\left[j\left(\frac{2f\pi d}{c}\right)\sqrt{\mu_r\varepsilon_r}\right] \qquad \textit{(Eq. 3. 2)}$$

Where $Z_0$ is the characteristic impedance of free space (377 Ohm) [7], *f* is the frequency, *d* is the thickness of the absorbing layer, and c is the velocity of light.

RL is expressed in dB and is used to evaluate the microwave absorption efficiency of ESMs through two parameters. The first, is the minimum RL value at a specific frequency. Since RL is measured in a logarithmic scale, an RL equal to -10 dB represents the absorption of 90 % of the incident's wave intensity, an RL equal to -20 dB, 99% absorption, and so on. Thus, it could be safely said that RL values between -30 and -40 dB (99.9 and 99.99 % absorption) absorb the whole intensity of the incident wave, and neglectable interference should be expected. The absorbance is presented as RL vs frequency, and the spectra are typically characterized by absorption peaks, i.e., relatively sharp decreases in the RL value.

The second parameter is the effective absorbance bandwidth (EAB), the frequency range with RL equal to or less than -10 dB (≥ 90% absorption), i.e., the range of the spectrum that is suffering a significant reduction on the incident's wave intensity [19,24].

Additionally, the impedances of the medium at the interface, i.e., of air and the ESMs, need to be similar so that the incident wave is not reflected due to the mismatch. This is called impedance matching and occurs when the impedance matching characteristic ($Z= |Z_{in}/Z_0|$) is close or equal to 1 [16]. If the impedance is not matched, most of the incident electromagnetic waves will be reflected, resulting in poor RL, even if the dielectric and magnetic loss are good. Furthermore, impedance mismatching plays an important role in multiple scattering mechanisms, as the difference in impedance of two components in a material can promote reflection at their interface [6,25].

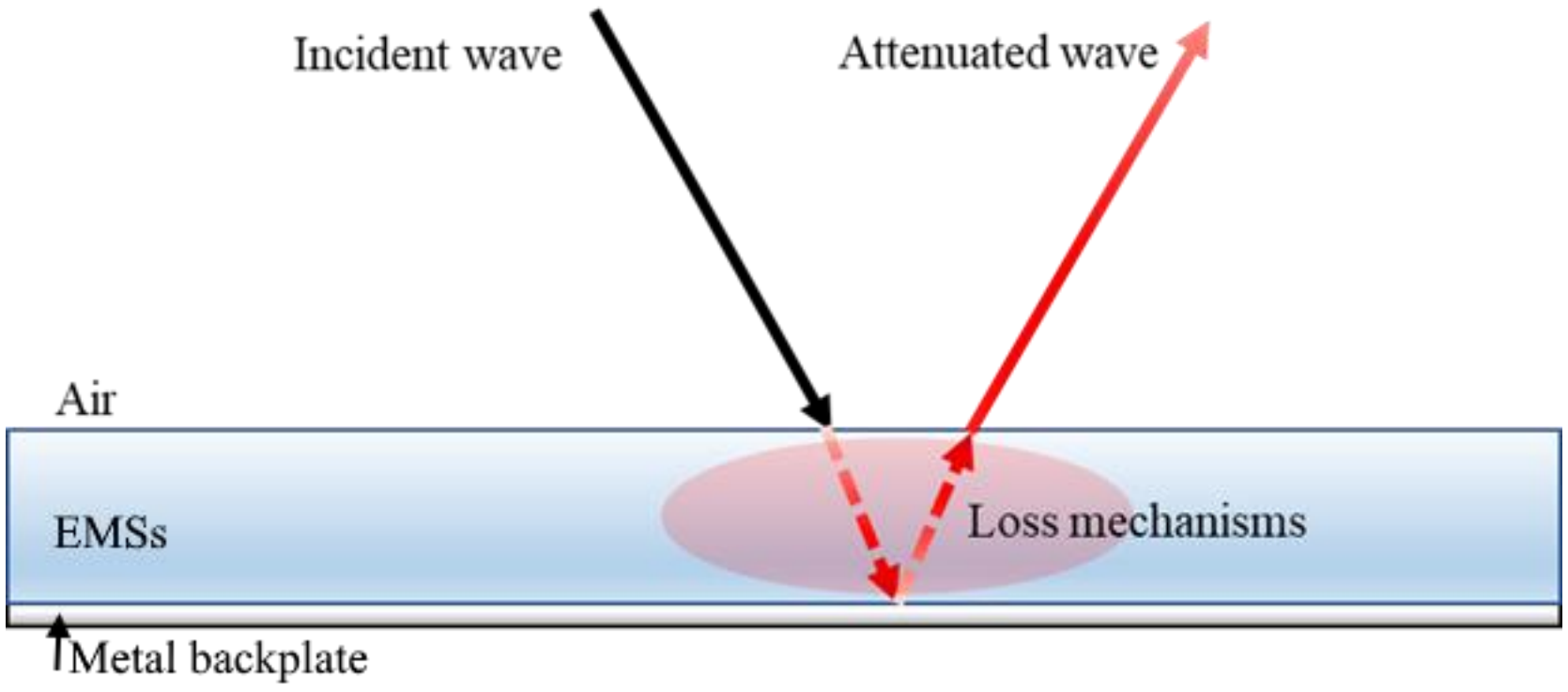

*Figure 3. 2 Schematic representation for the transmission line model.*

Note that since $Z_{in}=Z_0$, the impedance of the ESM and air respectively, there is no Air/ ESM interface for the incident wave and the only reflection is produced at the imposed metal backplate.

The RL is obtained indirectly from $\mu_r$ and $\varepsilon_r$ applying Equations 3.1 and 3.2. The $\mu_r$ and $\varepsilon_r$ as a function of f are obtained experimentally under a frequency sweep using a coaxial transmission line (see Figure 3. 3a) or a waveguide (see Figure 3. 3b) on a two-port setup from the scattering parameters, S11 and S12, using the Nicholson-Ross-Weir technique [26,27]. It is worth to note that the characterization of $\mu_r$ and $\varepsilon_r$ present limitations related to the sample thickness, depending on the used setup [28].

The samples in these studies possess a fixed geometry that is determined by the waveguide or coaxial sample holder. For instance, the WR90 waveguide for X-band frequencies uses samples with a rectangular shape (~ 23 x 10 x 1 mm) whereas the coaxial setup uses toroidal samples with typical dimensions of 7 mm outer diameter, 3 mm inner diameter [31,32].

The samples are typically composed of composites, wherein the lossy material is embedded within a polymeric or paraffin matrix, typically at a loading of 40-60% [33,34] and seldom below 15%. To calculate the RL, the thickness parameter is simulated, which may differ from the thickness of the measured sample, with several thickness values employed in the calculation, an optimal thickness is typically found that leads to the best return loss absorbing performance [35].

This ESM characterization approach and setup presents a series of advantages. The sample holders are standardized, and the measuring systems are available commercially. Due to the small size of the samples, small quantities of material are required. Finally, its wide acceptance in the scientific community allows direct comparison of results between researchers.

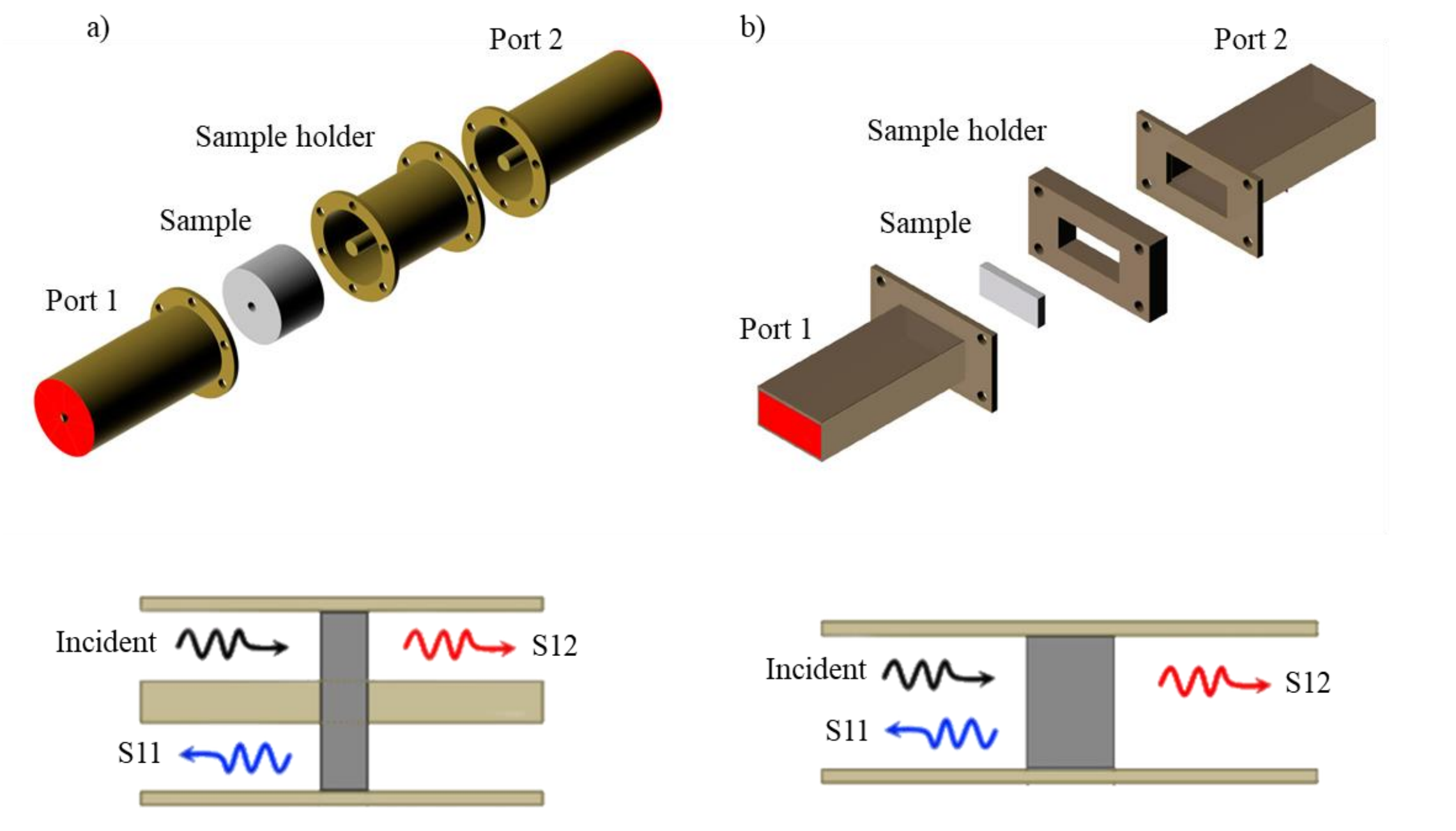

*Figure 3. 3 (a) Coaxial transmission line setup and (b) waveguide setup on two-port configurations. Adapted from* [29,30].

The Table 3. 1 presented in this study provides a selective survey of the literature on ESMs and includes examples of different types of materials, such as magnetic materials, graphene-based materials (GBMs), and GBM-hybrids. These examples demonstrate impressive microwave absorption properties of ESMs, with thicknesses in the range of a few millimetres, extraordinary absorptions up to ~ -60 dB in the range of microwaves up to ~20 GHz, and broad absorption bands covering a wide frequency range. Notably, all the electromagnetic shielding characterizations in the presented examples are based on the transmission line model, which employs experimentally obtained $\varepsilon_r$ and $\mu_r$. Despite the promising results, the application of ESMs in commercial electromagnetic shielding technologies remains limited.

*Table 3. 1 Comparative table of examples found in the literature based on the transmission line model. The materials include magnetic lossy, dielectric lossy, GBMs and hybrid ESMs.*

| Material | Method | Thickness (mm) | RL (dB) | Frequency (GHz) | EAB (GHz) | Ref. |
|---|---|---|---|---|---|---|
| Ni-Co coated PAN textiles | Waveguide | 2 | -47.7 | 14.9 | 7.7 | [3] |
| Fe nanoparticles | Coaxial | 1.7 | -47.0 | 9.6 | 3.1 | [5] |
| Hierarchical Co particles | Coaxial | 1.5 | -54.9 | 14.1 | 1.6 | [20] |
| Nano chain-like Fe arrays | Coaxial | 1.9 | -57.3 | 9.5 | 11.5 | [36] |
| Fe3O4@PEDOT microspheres | Coaxial | 4 | -29.1 | 9.5 | 4.9 | [37] |
| $CoFe_2O_4$ NPs | Waveguide | 2 | -55 | 9.3 | 2 | [38] |
| Co-doped $MnO_2$ | Coaxial | 5 | -24.2 | 4.2 | 1.32 | [39] |
| FeCo nanoplates | Coaxial | 1.8 | -43.2 | 8.2 | 2.9 | [40] |
| $Fe_7Co_3$ microparticles | Coaxial | 1.55 | -53.4 | 14.3 | 6.6 | [41] |
| Hollow magnetite nanospheres | Coaxial | 6.9 | -42.4 | 2.1 | 1 | [42] |
| Ni (Fe) flakes | Coaxial | 2 | -11.8 | 2.0 | 0.6 | [43] |
| GO/ Polypyrrole nanorods aerogel | Coaxial | 3.0 | -51.4 | 6.9 | 3.3 | [25] |
| Expanded graphite/ PANI | Coaxial | 3.5 | -17.1 | 10.3 | 5.0 | [32] |
| GO@CSA-PANi | Coaxial | 2.4 | -48.1 | 13.3 | 5.3 | [44] |
| rGO | Coaxial | 2.0 | -6.9 | 7 | 0 | [45] |
| Graphene micro-popcorns / paraffin | Coaxial | 2.5 | -45.3 | 11.9 | 3.5 | [46] |
| Graphene foam | Coaxial | 10.0 | -28.3 | 12.3 | 13.9 | [47] |
| rGO/ Polypyrrole aerogel | Coaxial | 3.0 | -53.3 | 12.8 | 6.4 | [48] |
| Magnetized N-doped graphene | Coaxial | 4.5 | -13.42 | 8.17 | 2.50 | [49] |
| CuS/ $NiFe_2O_4$ decorated graphene | Coaxial | 2.5 | -54.5 | 11.4 | 4.5 | [8] |
| rGO/Co3O4 | Coaxial | 3.3 | -43.7 | 13.8 | 2.7 | [11] |
| CoNi/ rGO aerogels | Coaxial | 0.8 | -53.3 | 16.2 | 3.5 | [19] |
| $\gamma$-$Fe_2O_3$/ rGO | Waveguide | 2.5 | -59.65 | 10.09 | 3 | [24] |
| rGO/Co | Waveguide | 2.2 | -68.1 | 13.8 | 7.1 | [33] |
| CoFe/rGO | Coaxial | 2.5 | -41.9 | 8.8 | 3.79 | [34] |
| Ni/ Graphene | Coaxial | 5.0 | -17.2 | 3.5 | 1.0 | [50] |
| Fe3O4/ porous graphene | Coaxial | 6.1 | -53.0 | 5.4 | 2.6 | [51] |

*The exact RL, Frequency (of the maximum RL), and EAB were not provided in some references, thus they were dug out from the RL-f curves using a plot digitizer app (https://apps.automeris.io/wpd/). The focus has been paid to the best material combination in each report within the 2-18 GHz range.*

### *3.2.1 The Limitation of The Transmission Line Model.*

Despite its popularity in the literature and industry, the transmission line approach has a fundamental limitation. Specifically, Equations 3.1 and 3.2 may not apply to the system depicted in Figure 3. 2, i.e., rendering the calculated RL values unreliable as will now be discussed.

The transmission line theory contains the following parameters [52]:

- Reflection coefficient ($\Gamma$): A complex number that describes the phase and magnitude of the reflected wave relative to the incident wave.
- Specific reflection coefficient ($R_M$): A dimensionless quantity that indicates the fraction of the incident wave that is reflected at the interface.
- Return loss ($L_R$): A measure given in dB of the amount of power reflected from the load to the source.
- And the scattering parameter $S_{11}$: A complex number that describes the phase and amplitude of the reflected wave relative to the incident wave at the input port. $S_{11}$ is the reflection coefficient of the first port (usually the input) of the network when the second port (usually the output) is terminated with a perfect load.

These parameters seem equivalent at first glance, and indeed they present some similarities as they are related to the reflection of electromagnetic waves. However, they differ in their specific definitions and applications and should not be mistaken, generally, with each other or with RL [53,54].

Figure 3. 4 depicts a model of the transmission line with an ESM sample. The model has two interfaces, $\chi_l$ and $\chi_{l+d}$, where d is the thickness of the sample. The superscript - or + refers to the side of the interface. V refers to the voltage of each wave, the incident ($V^+(\chi_l^-)$), the transmitted ($V^+(\chi_l^+)$), and so on following the notation just described. $Z_L$ is the impedance of the transmission line (or free space) and $Z_M$ is the transmission of the ESM defined from its intrinsic characteristics ($\mu_{r,M}$ ; $\varepsilon_{r,M}$) [55].

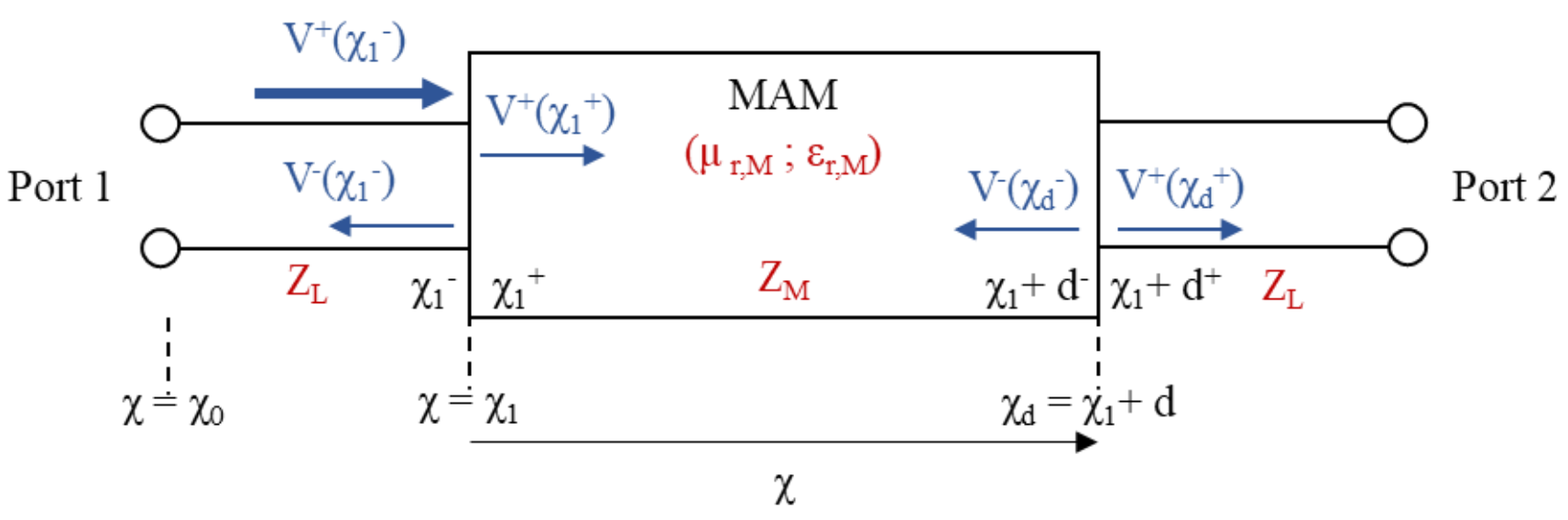

*Figure 3. 4 Model of the transmission line with an ESM sample.*

For the isolated ESM interfaces, Γ can be calculated from the assumption that $R_M$, the ratio of the waves going forward ($V^+$) and backwards ($V^-$), derived from Kirchhoff's law, is symmetrical at the interfaces as follow [56–58]:

$$R_M(\chi_1^-) = R_M(\chi_1 + d^+);$$

$$\left.\frac{V^-(\chi_1^-)}{V^+(\chi_1^-)}\right|_{V^-(\chi_1^+)=0} = \left.\frac{V^+(\chi_1 + d^+)}{V^-(\chi_1 + d^+)}\right|_{V^+(\chi_1+d^-)=0}; \qquad \textit{(Eq. 3. 3)}$$

$$\Gamma(\chi_1^-)|_{V^-(\chi_1^+)=0} = \Gamma(\chi_1 + d^+)|_{V^+(\chi_1+d^-)=0} = \frac{Z_M - Z_L}{Z_M + Z_L}$$

Equation 3.3. then states that the interfaces at either side of the ESM, treated as isolated, behave similarly. The same applies to the inner side of the interfaces, with a change of sign:

$$\Gamma(\chi_1^+)|_{V^+(\chi_1^-)=0} = \Gamma(\chi_1 + d^-)|_{V^-(\chi_1+d^+)=0} = -\frac{Z_M - Z_L}{Z_M + Z_L};$$

$$-\frac{Z_M - Z_L}{Z_M + Z_L} = \frac{Z_M - Z_L}{Z_M + Z_L}\, e^{-i\pi} = R_M(\chi_1^-)e^{-i\pi} \qquad \textit{(Eq. 3. 4)}$$

Then, $L_R$, defined as the ratio of the power of the incident wave to the power of the reflected wave, can be defined by Γ as [56–58]:

$$L_R(\chi_1^-) = 10\, log_{10}\left[\frac{P_{In}(\chi_1^-)}{P_R(\chi_1^-)}\right] = 10\, log_{10}\left[\frac{V^+(\chi_1^-)}{V^-(\chi_1^-)}\right]^2 = 20\, log_{10}\left[\frac{1}{\Gamma(\chi_1^-)}\right];$$

$$L_R(\chi_1^-) = 20\, log_{10}\left[\frac{1}{S_{11}(\chi_1^-)}\right] = -20\, log_{10}\left[\frac{Z_{in}(\chi_1^-) - Z_L}{Z_{in}(\chi_1^-) + Z_L}\right] \geq 0 \qquad \textit{(Eq. 3. 5)}$$

From Equation 3.5, $L_R$ from the ESM acting as a load in the circuit can be obtained from $S_{11}$. $L_R$ for an isolated interface can be obtained from $R_M$. The difference between treating the system as a load in the circuit or as an isolated interface comes from what happens at the rest of interfaces.

If $\chi_1$ is isolated, only beams 1, 2, and 3 are considered and Γ, for beams 1 and 2, is given by $R_M$ (see Figure 3. 3). If the whole ESM is considered as a part of the circuit, the waves transmitted waves will partially reflect at $\chi_{1+d}$ and be partially transmitted at $\chi_1$. Meaning that beams 5, 8…will contribute to Γ at $\chi_1$.

Note how Equation 3.5 resembles Equation 3.2, where $L_R$ (-20log$_{10}$[($Z_{in}(\chi_1^-)$-$Z_L$)/($Z_{in}(\chi_1^-)$+$Z_L$)]) could be mistaken with -RL. However, calculating RL from the assumption of an isolated interface does not correctly characterize the power dissipation of the material [53–55]. While $L_R$ is valid, and constant, from $\chi = \chi_1$ to $\chi$=0, RL should describe the range between $\chi = \chi_1$+d to $\chi = \chi_1$, i.e., it is dependent on the sample thickness. The isolated interface assumption to describe the whole system is only valid if a) the sample thickness is infinite or b) the absorption along the material is so strong that completely extinguish the wave.

Furthermore, in an isolated interface, the energy conservation is preserved, which is obviously not the case for ESMs [56–58].

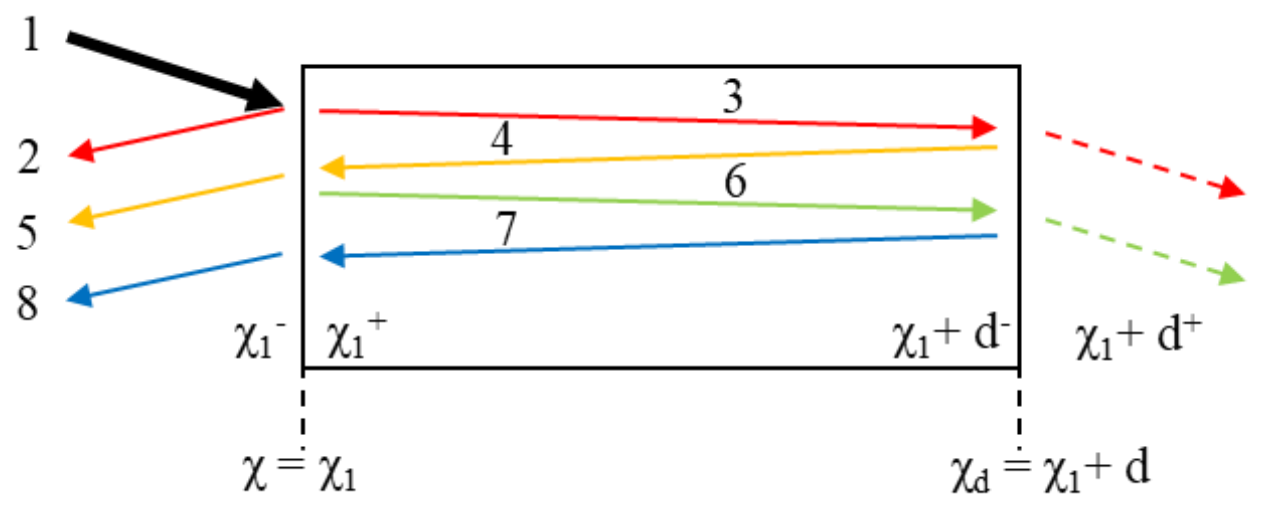

*Figure 3. 3 Multiple reflections occurring at interfaces when considering the ESM as a load in the circuit. $V^-(\chi_1^-)$ includes beams 2, 5, 8…*

There is another incorrect assumption in Equations 3.1 and 3.2 related to the concept of input impedance.

The input impedance at x (Port 1< x< $x_1$ or $x_1$+d< x<Port 2) can be defined as:

$$Z_{in(\chi)} = \frac{V_{(x)}}{I_{(x)}} = \frac{V^+_{(x)} + V^-_{(x)}}{I^+_{(x)} - I^-_{(x)}}; \ldots$$

$$Z_{in(\chi)} = Z_L \frac{1 + {V^-_{(x_1^-)}}/{V^+_{(x_1^-)}}}{1 - {V^-_{(x_1^-)}}/{V^+_{(x_1^-)}}} = Z_L \frac{1 + \Gamma_{(x_1^-)}}{1 - \Gamma_{(x_1^-)}} \quad \textit{(Eq. 3. 6)}$$

The first line is obtained simply by applying Kirchhoff's law to the interface [59].

The $Z_{in(x)}$ is then equivalent to the impedance at x (any x) in reflection measurement, meaning that at x= $x_1$, $Z_{in(x=x1)}$ is the impedance of the ESM ($Z_M$). It can be expressed for Γ when the ESM act as part of the circuit as:

$$\Gamma_{(x_1^-)} = \frac{Z_{in(x_1^-)} - Z_M}{Z_{in(x_1^-)} + Z_M} \quad \textit{(Eq. 3. 7)}$$

If the sample has infinite thickness and the system can be observed as an isolated interface, then $Z_{in}(x_1)$ is still equivalent to $Z_M$, but Equation 3.6 needs to be rewritten in terms of $R_M$:

$$Z_{in(x_1^-)} = Z_L \frac{1 + R_{M(x_1^-)}}{1 - R_{M(x_1^-)}} = Z_L \frac{1 + \frac{Z_M - Z_L}{Z_M + Z_L}}{1 - \frac{Z_M - Z_L}{Z_M + Z_L}} = Z_M \quad \textit{(Eq. 3. 8)}$$

The input impedance within the ESM ($x_1$ < x < $x_1$+d) can be defined similarly to Equation 3.6 as [59]:

$$Z_{in(\chi)} = \frac{V_{(x)}}{I_{(x)}} = \frac{V^+_{(x)} + V^-_{(x)}}{I^+_{(x)} - I^-_{(x)}}; \ldots$$

$$Z_{in(\chi)} = Z_M \frac{1 + \Gamma_{(\chi_1+d^-)-x}}{1 - \Gamma_{(\chi_1+d^-)-x}} \quad \textit{(Eq. 3.9)}$$

Or in terms of Γ analogously to Equation 3.7:

$$\Gamma_{(\chi_1+d^-)} = \frac{Z_{in(\chi_1+d^-)} - Z_M}{Z_{in(\chi_1+d^-)} + Z_M} \quad \textit{(Eq. 3.10)}$$

The difference between Equation 3.6 and Equation 3.9 is that the former describes the input impedance the transmission line and the latter the one in the material. Note how $Z_L$ is replaced by $Z_M$ in Equation 3.9, the correct version of Equation 3.2 should be derived from these conditions.

If the sample has an infinite thickness and the system can be considered and isolated interface, then the input impedance at ($x = x_1+d^-$) is:

$$Z_{in(\chi_1+d^-)} = Z_M \frac{1 + R_{M(x_1+d^-)}}{1 - R_{M(x_1+d^-)}} = Z_M \frac{1 + \frac{Z_L - Z_M}{Z_L + Z_M}}{1 - \frac{Z_L - Z_M}{Z_L + Z_M}} = Z_L \quad \textit{(Eq. 3. 11)}$$

Meaning that for the isolated surface within the material, $x_1+d^-$, the input impedance is the impedance for the line whereas for the isolated surface at the line, $x_1^-$, the input impedance is that of the material. As seen for a wave propagating from port 1, the input impedance is thus not the one where the wave is, but that of the medium it is about to interact with.

In sum, at $x_1+d^-$, where Equation 3.1 calls $Z_{in}$ the input impedance of the material, it is mistaking it with $Z_L$ [55,59]. Then, the popular description of Equation 3.1 would only be valid for the first interface, $x_1^-$. However, we have also seen how Equation 3.1 is not valid at $x_1^-$ because it treats the first interface as isolated instead of part of a material [56–58]. Whereas this assumption is true for infinite thickness or absolute absorbance, this is rarely a real-world situation. In other words, the concept of RL can only be applied to a film with a metal backplate, and RL is a characteristic of the device and not of a material.

As a side note, if the input impedance used in the transmission line approach is that of an ESM backed by a metal, with infinite $\varepsilon_r$, that the impedance is zero, then for Equation 3.11, $Z_{in}(x_1+d^-) = 0$ and Γ, from Equation 3.10, equals -1. Which is the formal expression of the total reflection from the metal backplate.

### *3.3 The λ/4 Approach.*

When dealing with a metal back plated film it is crucial to consider the quarter-wavelength model, which is related to destructive interference in the context of electromagnetic wave propagation. Strictly, it is not based on absorption mechanisms.

Briefly, this approach explains that when the microwave hits the material, part of it will be reflected (R1), and the rest will be transmitted. The transmitted part will eventually be reflected (R2) after travelling the material through a different path that has changed its phase related to the initial partial reflection (see Figure 3. 4). Thus, when the two waves meet outside the material, they will interfere. If the difference of phase is exactly a quarter of their wavelength, the interference will be destructive, and the waves will cancel each other. The difference in phase will depend on the incidence angle, the thickness of the material, the frequency, and, of course, on $\mu_r$ and $\varepsilon_r$.

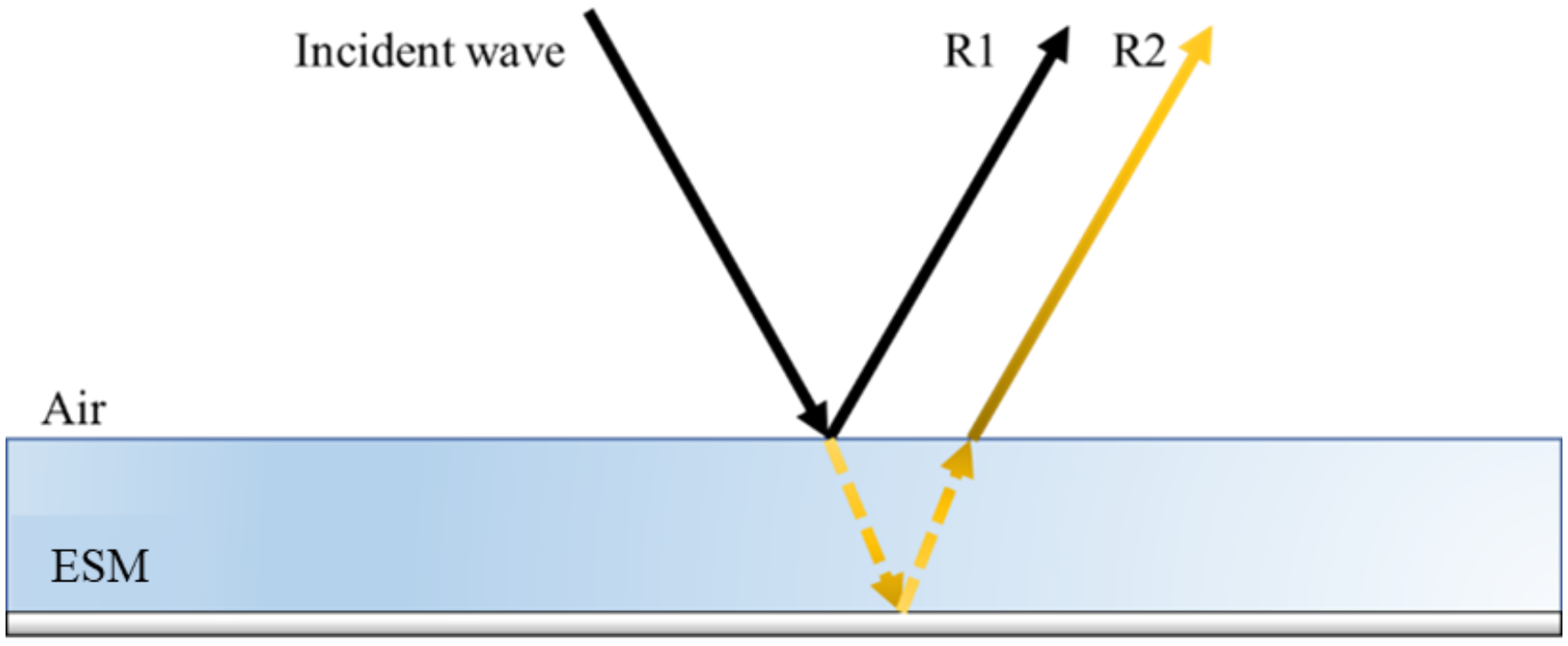

*Figure 3. 4 Schematic representation for the quarter-wavelength model.*

This phenomenon has been largely exploited for various applications such as filters, antennas, acoustics, or more popularly anti-reflective coatings.

Note how whereas the transmission line theoretical approach imposed that the incident wave is transmitted with minimum reflection, the quarter-wavelength approach requires that the incident wave is partially reflected. Ideally, R1 and R2, as described in Figure 3. 4, having similar intensities.

For a normal incidence, the thickness at which the optimum destructive interference occurs is given by [16,60]:

$$d = \frac{nc}{4f\sqrt{\varepsilon_r \mu_r}} \quad \text{(Eq. 3. 12)}$$

Where n is an uneven integer number (1, 3, 5…), i.e., there is a periodic condition at which the interference may be constructive, destructive or something in between. As in previous equations, f is the frequency of minimum RL. This method has been used to complement experimental work in ref [61] as well as in previous works at (*IMA*) [4].

It should be noted that some examples apply the quarter-wavelength approach to the results obtained through the transmission line approach [34]. While both models may be valid, the absorption through lossy behaviour and the absorption though wave extinction, it is important to acknowledge that they cannot be treated as interchangeable parts of the same system. Rather, it is essential to consider the underlying assumptions and implications of each approach carefully and to see them as distinct and complementary approaches to understanding microwave absorption.

Further discussion on the validity of these methods is beyond the scope of this thesis. Instead, this research has been based in an empirical approach, with the results and conclusions based on experimental data. Therefore, while the theoretical models may be of interest from a scientific standpoint, they have not played a significant role in the analysis of the work presented in this chapter.

### *3.4 Free-Space Measurement Approach.*

Finally, free-space measurement (FSM) is a non-contact method able to perform measurements over a wide bandwidth. Its operating conditions imitate the potential ESM applications (See Figure 3. 5). An emitter antenna radiates the electromagnetic wave towards the sample, backed by a metallic plate and reflects to a receiver antenna. Some FSM configurations may be based in the transmission of the wave by removing the metal backplate and placing the receiver antenna behind the sample [62,63]. The scattering parameter, i.e., a ratio of powers of the emitted signal ($P_0$) and the reflected signal ($P_r$), $S_{11}$ or $S_{12}$, can be obtained depending on whether the emitter antenna is the same as the receiver antenna or not. In a typical two-antennas configuration, $S_{12}$ is defined by Eq. 3.13 [7,64,65].

$$S_{12} = 10 \log \frac{P_r}{P_0} \quad \text{(Eq 3. 13)}$$

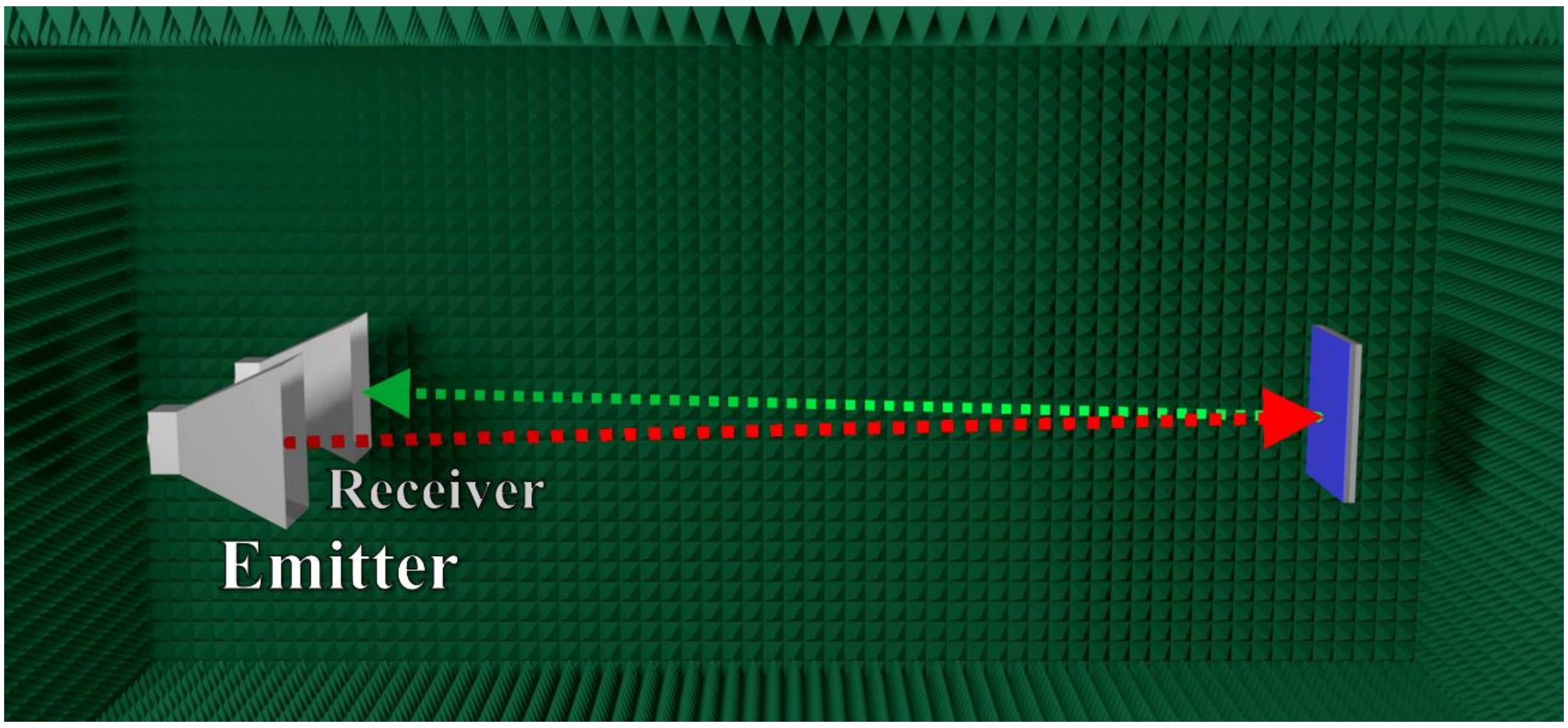

*Figure 3. 5 In this work, FSM measurements were taken inside an anechoic chamber to reduce external interferences, echoes, or reflection artefacts. The red line represents the incident wave and the green one represents the reflected wave one it has crossed the ESM.*

$S_{12}$, expressed in dB, in a configuration such as described in Figure 3. 5 can indeed be related to RL. Since the wave can only be reflected, that which is not reflected has been lost to its interaction with the material.

However, there is one assumption in this setup, the path of the wave must be as depicted in Figure 3. 5. The sample needs to be oriented so that the incident angle is similar to the reflected angle. Otherwise, a significant percentage of the microwave signal will be scattered towards the chamber walls and will not reach the receiver antenna. Such issue is usually solved through a time-consuming trial-and-error process.

Furthermore, FSM requires a relatively large quantity of sample that is sometimes not possible depending on the tested material, especially for nanomaterials or complex metaestructures. During the work presented in this chapter, each sample required a quantity of nano- and micromaterials in the range of grams.

As a characterization technique, FSM does not dwell into the absorption mechanisms. It can accurately describe the electromagnetic shielding performance in an ESM-based device operating under real-world conditions while the underlying mechanisms may be due to dielectric or magnetic losses, destructive interference, or a combination of both.

During the state-of-the-art revision of this field, we found no established correlations in the literature between the results obtained using the transmission line approach with coaxial or waveguide methods and the results obtained using FSM. Due to the shortcomings of the transmission line approach, any comparison of results obtained using this approach with those obtained by FSM should be carefully considered.

### *3.5 Amorphous Magnetic Microwires in ESMs.*

Magnetic ESMs include magnetic metals (Fe, Co, Ni), magnetic alloys (CoNi, FeCo, NiFe), and magnetic oxides ($Fe_3O_4$, $Fe_2O_3$, $CoFe_2O_4$) (see Table 3. 1).

Amorphous magnetic microwires (MWs) consist of a metallic magnetic alloy nucleus and a borosilicate glass or Pyrex-like cover (see Figure 3. 6). They are made with the Taylor-Ulitovsky technique, which consists of a melt spinning with ultrafast cooling [66,67].

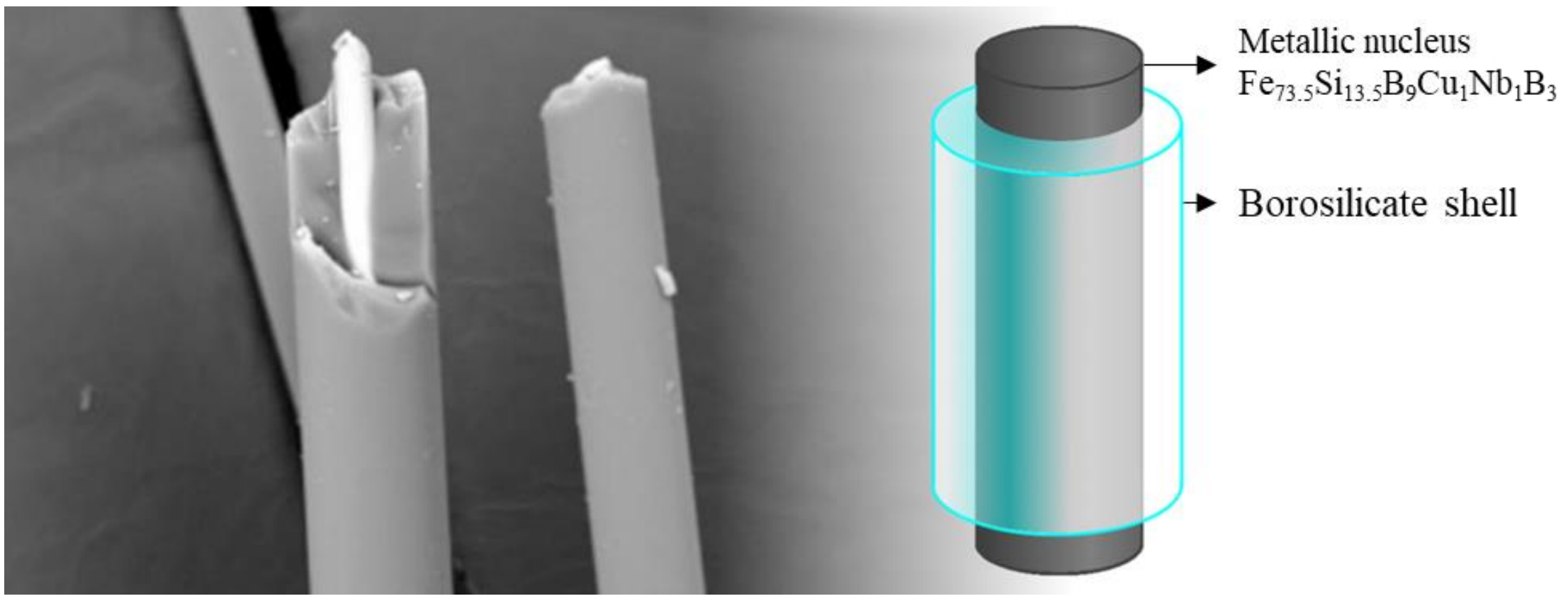

*Figure 3. 6 Illustration of a microwire obtained through the Taylor-Ulitovsky technique.*

The introductory chapter of this thesis provided an overview of magnetic microwires, including their synthesis methods, the research conducted by our group, and the discovery that their electromagnetic shielding performance is primarily influenced by dielectric losses and the modification of the electric permittivity of the embedding matrix, rather than magnetic losses.

The Fe-based amorphous MWs would thus have a microwave absorption capability related to both magnetic and dielectric losses. However, to what extent one mechanism relates to the other or contributes to the total RL remains an open question.

Recent works modelling the MWs behaviour as purely dielectric has led to a reasonable fitting with the experimental results, according to these, the magnetic loss contribution would be neglectable, at least for those sets of samples and that range of frequency. In particular, the microwave absorbent behaviour of composite with metallic inclusions can be modelled through an effective permittivity parameter ($\varepsilon_{eff}$) using the Maxwell-Garnett mixing model [68,69]:

$$\varepsilon_{eff} = \varepsilon_{host} + \frac{\frac{1}{3}\sum_{i=1}^{n} V_{f,i}(\varepsilon_{inclusion} - \varepsilon_{host})\sum_{k=1}^{3}\frac{\varepsilon_{host}}{\varepsilon_{host} + N_{z\,i,k}(\varepsilon_{inclusion} - \varepsilon_{host})}}{1 - \frac{1}{3}\sum_{i=1}^{n} V_{f,i}(\varepsilon_{inclusion} - \varepsilon_{host})\sum_{k=1}^{3}\frac{\varepsilon_{host}}{\varepsilon_{host} + N_{z\,i,k}(\varepsilon_{inclusion} - \varepsilon_{host})}} \quad \text{(Eq. 3. 14)}$$

Where $\varepsilon_{host}$ is the permittivity of the matrix, or host, $V_{f,i}$ is the volume fraction for *i* components and $N_z$ is the depolarizing factor [65]:

$$N_z = (d/l)^2 Ln(l/d) \quad \text{(Eq. 3. 15)}$$

Where *d* is the diameter of the NW and *l* its length. In sum, when designing amorphous microwires for electromagnetic shielding applications, the filling ratio, length, and inner and outer diameter will influence the microwave absorbent behaviour of the composite, a feature of huge interest in ESM development.

For the case of amorphous NW, Eq. 3. 14 can be simplified as [4]:

$$\varepsilon_{eff} = \varepsilon_{host}\left(1 + \frac{V_{f,i}}{3N_z}\right) \quad \text{(Eq. 3. 16)}$$

The current leading hypothesis is that the modification of the real part of ε is the primary factor contributing to the electromagnetic shielding by allowing the use of thin ESM coatings through wave extinction related to the quarter-wavelength model.

For the case of a thin amorphous microwire-based ESM coating, with thickness $t$, over a dielectric substrate, with thickness $\Delta$, and a metal backplate (see Figure 3. 7), Eq. 3.12 can be derived to:

$$f = \frac{c}{4\sqrt{\varepsilon_{eff} t \Delta}} \qquad \text{(Eq. 3. 17)}$$

It should be noted that magnetic loss contributions were not considered in this study due to the assumption of a magnetic permeability of 1. The use of a dielectric substrate served two primary purposes: firstly, to simulate a substrate that would exist in a real-world condition, such as a base coating or a dielectric structural material; and secondly, to fulfil an experimental requirement. Due to the necessarily thin nature of the paint-based ESM coating, $t$ is often a fixed value. However, by modifying $\Delta$, it is possible to tune the frequency value to within our measuring range, enabling observation of reduced reflectivity.

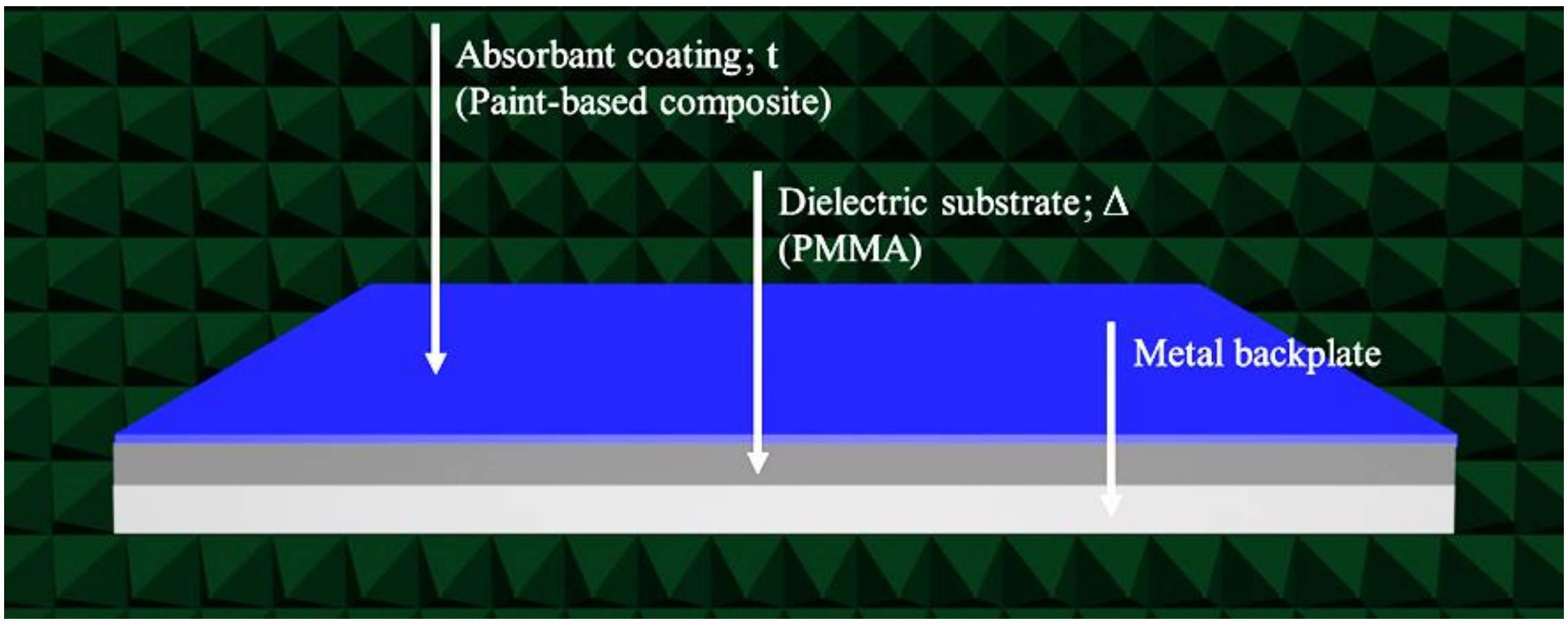

*Figure 3. 7 Illustration of the sample for FSM electromagnetic shielding characterization.*

Despite the good performance and significant advances in magnetic ESMs, their intrinsic high density, poor corrosion resistance and aggregation issues are still major drawbacks for some applications [19].

### 3.6 Graphene-Based Materials in ESMs.

Carbon-based ESMs include GBMs and other forms of carbon, such as carbon fibres, nanofibers, and carbon black, have recently attracted significant attention in electromagnetic shielding applications. These materials not only exhibit good RL properties through dielectric loss, but also demonstrate remarkable multifunctionality when employed as composite reinforcers [70,71].

For example, the use of multiwall carbon nanotubes (MWCNT) enhanced the mechanical properties so that the ESM may also be used as a structural material substituting aluminium or composite components [72,73]. Other examples may include microwave absorbent and corrosion-resistant GBM-based coatings [44], which rely on the comprehensively studied field of corrosion protection GBM-based coatings [74,75]. In addition, the thermal stability, and flame-retardant capabilities of GBM have also attracted great interest in multifunctional coating development [76–78].

As previously presented, dielectric loss mechanisms heavily rely on the polarization of charges bound to defects or functional groups [79]. This explains why GBMs like rGO perform better ESM than pristine graphene, CNTs or graphite [45].

As demonstrated in the previous chapter, graphene-based materials (GBMs) have exhibited remarkable performance-to-weight ratios. The use of GBMs facilitates the development of ESMs that are thinner and lighter in comparison to conventional materials, such as metals or ceramics [80]. This low thickness is crucial, as it enables the expanded application range of ESMs when utilized as coatings by reducing potential thickness and weight-related limitations.

Typically, ultrathin ESMs are considered to have a thickness of less than 1 mm [81–83]. However, it is important to note that there is currently no standardized definition for the term, and the ultimate thickness limits for ESMs will depend on the specific application requirements. Table 3. 1 illustrates that the majority of high performing ESMs possess a thickness that ranges from several millimetres to centimetres. However, it is important to exercise caution in drawing conclusions from the transmission line model.

Nevertheless, in the literature, it is widely believed that pristine graphene and other graphene-based materials (GBMs) with relatively high permittivity and low permeability exhibit poor electromagnetic shielding performance due to the difficulty in achieving impedance matching conditions [6,49,84]. To address this issue, two solutions have been proposed, which may be used in combination.

The first is the use of GBMs in a structured format; mesoporous carbon structures improve microwave absorption due to reduced effective permittivity and subsequent facilitation of impedance matching [84,85]. In addition, the structured format, which includes foams, networks, cavities, multi-layered interfaces, multi-shell, hierarchal, and heterostructures, promote multi-scattering and interfacial polarization [13,14]. Briefly, these structures create a "maze" for the incoming microwaves. As the microwaves try to pass through the material, they are repeatedly reflected by the walls of the pores. With each reflection, a portion of the microwave energy may be absorbed, leading to an overall increase in the microwave absorption of the material. Works compressing graphene foams have reported significant changes in the microwave absorption performance, suggesting that the geometry of these structures greatly affects the absorption mechanisms [47,79].

The second is the hybridization of GBMs with magnetic lossy materials to approximate the complex relative permittivity and permeability values and improve the impedance matching [6,16]. Furthermore, the hybridization can bring synergetic effects like additional interface polarization losses and the loss mechanisms of the magnetic materials [6]. It has been reported that electromagnetic shielding performance can be dramatically enhanced by adding heterogeneous or magnetic components to GBMs [8,19,24,49].

Despite exhibiting exceptional electromagnetic shielding performance, the widespread application of GBMs as ESMs is hindered by the lack of commercial availability. Highly complex materials with low production yields and high costs, or materials that are not suited for real-world conditions, may not be desirable for general ESM applications. As discussed in the previous chapter, ball-milling can be employed as a potential solution to this challenge.

## *3.7 Use Of FLMG In ESMs.*

From the points presented so far in this section, FLMG, a mesoporous aggregation of few-layered graphene domains with a strong presence of defects and some functional groups that are obtained through ball-milling graphite, could be an ideal candidate for electromagnetic shielding applications (see Chapter 2).

In the context of evaluating the application of FLMG on microwave absorption, paint-based composited were prepared with different compositions using FLMG, amorphous magnetic microwires, or a combination of both. The paint-based composited were then deposited on a non-absorbent substrate made of poly-methyl-methacrylate (PMMA) with different thickness. The substrate is also used to emulate common applications where a dielectric substrate is present such as base paint or epoxy. Finally, a metal backplate was used to ensure total reflection conditions, as described in the second section of the chapter. The samples were then evaluated using the FSM method previously described with the setup placed inside an anechoic chamber.

### *3.7.1 Test 1: Substitution of MWs with FLMG.*

For the first test [61], the aim was to study the substitution role of FLMG on MW-based ESMs. Samples were prepared by combining MWs and FLMG in different proportions to a fixed total filling ration of 2.4 %wt. In addition, samples with each on the active components separated were prepared according to Table 3. 2.

*Table 3. 2 Description of samples for Test 1.*

| Sample | FLMG content (%wt.) | MWs content (%wt.) | Total filling ratio (%wt.) |
|---|---|---|---|
| #01 | 0 | 0 | 0 |
| #02 | 0.4 | 0 | 0.4 |
| #03 | 0.8 | 0 | 0.8 |
| #04 | 1.2 | 0 | 1.2 |
| #05 | 1.6 | 0 | 1.6 |
| #06 | 0 | 0.8 | 0.8 |
| #07 | 0 | 1.2 | 1.2 |
| #08 | 0 | 1.6 | 1.6 |
| #09 | 0 | 2 | 2 |
| #10 | 0.4 | 2 | 2.4 |
| #11 | 0.8 | 1.6 | 2.4 |
| #12 | 1.2 | 1.2 | 2.4 |
| #13 | 1.6 | 0.8 | 2.4 |

FLMG was prepared following the instructions presented in Chapter 2 for the oscillatory ball-mill. Briefly, using graphite (Natural, -325 Mesh, 99,8 %_ Metal basis from Alfa Aesar) as a precursor, 25 Hz milling frequency, a weight ratio of 1:25 and 240 min milling time [86]. Between 0.2 and 0.8 g of FLMG were used for each sample. A sample containing only the precursor material was also prepared.

The MWs consisted of an amorphous magnetic nucleus, with a composition of $Fe_{73.5}Si_{13.5}B_9Cu_1Nb_3$, and a borosilicate glass cover. The nucleus has a diameter of 4.7 µm, and the total diameter was 15.8 µm. As for the length, the MWs were cut to approximately 2 mm. These MWs were prepared by the Taylor-Ulitosky technique by a third party upon custom requirements. Between 0.4 and 1 g of microwires were used for each sample.

The composites were prepared using 50 g of paint and thoroughly dispersing the samples until a homogeneous mixture was achieved. The paint-based composites were then deposited on a PMMA substrate as dielectric substrate sized 25 x 25 cm. The resulting thickness of the paint was 220 µm. The thickness of the substrate was 1.45 mm, selected after comprehensive tests for the prepared samples.

First, the non-absorbent characteristics of the commercial paint and PMMA substrate (sample #01) were tested. The results, shown in Figure 3. 8 show a flat line across the whole frequency range with an RL value of approximately 0 dB. However, the absorption curve is not perfectly flat. An analysis of several RL measurements of the paint-only sample revealed a typical RMS noise of < 0.1 dB/GHz and a maximum deviation from 0 dB of ~ 2.5 dB across the whole spectrum. This noise and variation fall within the expected experimental error associated with the used FSM setup.

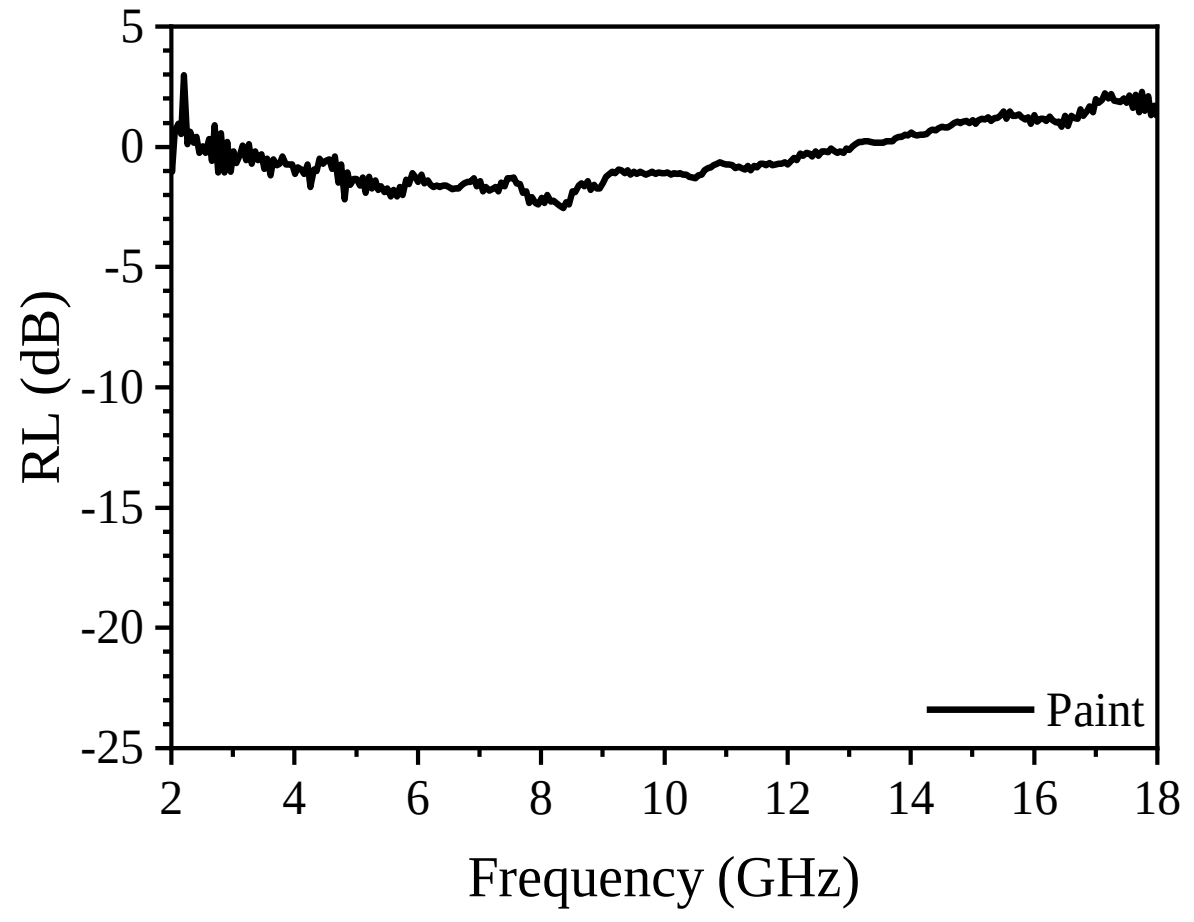

*Figure 3. 8 RL spectra for sample #01 (paint only).*

#### *3.7.1.1 FLMG-Containing Samples.*

Then, the individual contribution of FLMG with increasing filling ratio (0.4 to 1.6 wt.%) was evaluated with samples #02-#05. A sample containing 1.6 wt.% of graphite is added to the study for comparison. Results are shown in Figure 3. 9.

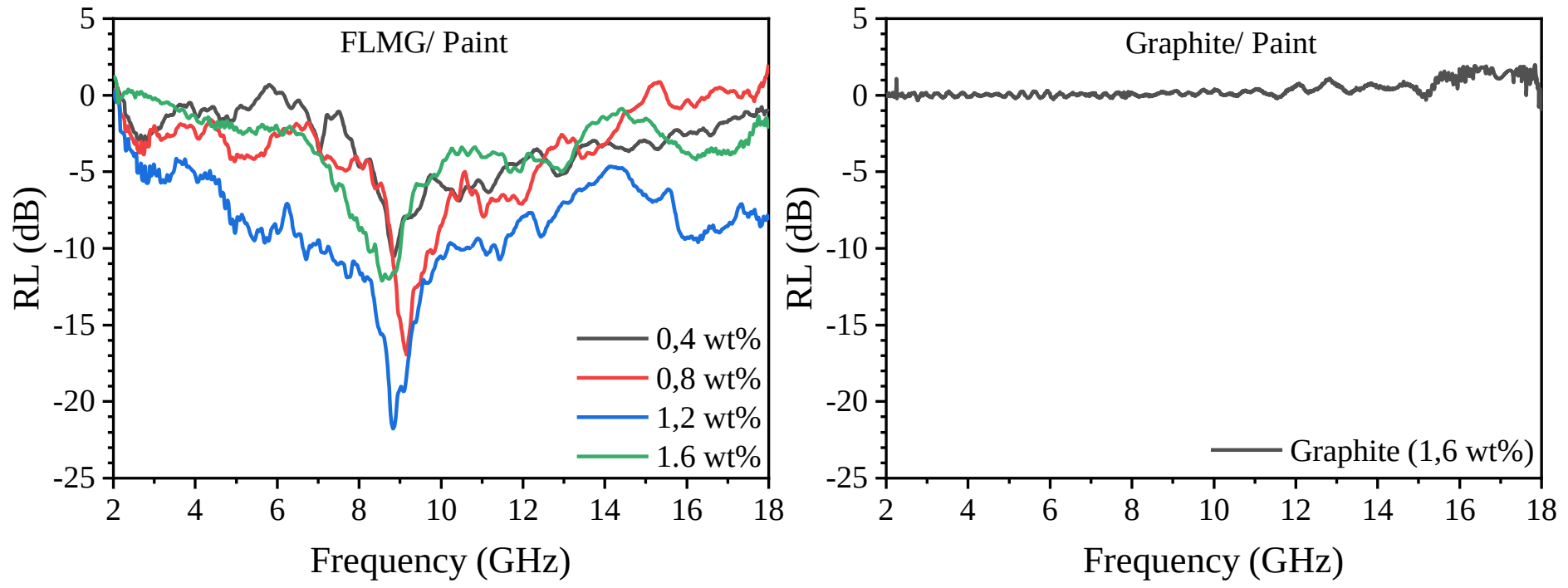

*Figure 3. 9 RL spectra for a) FLMG/ Paint ESMs, with different FLMG filling ratios, and b) Graphite/ Paint ESM with a 1.6 wt.% filling ratio.*

For the FLMG-containing samples, the RL spectra exhibit a minimum (maximum absorbance) at around 9 GHz, with the position of the peak remaining largely unaltered with the different FLMG concentrations.

The maximum absorbance ranges from -10 dB to -21,8 dB with increasing FLMG concentration. The maximum absorbance was observed with the samples containing 1.2 wt.% and 0.8 wt.% of FLMG. A noticeable difference in electromagnetic shielding behaviour was found between the FLMG samples and the sample containing 1.6 wt.% graphite, the latter showing no significant electromagnetic shielding, thereby highlighting the crucial role of defects, particle size reduction, and mesoporous structures, or a combination of these in designing graphene-based ESMs [87,88].

Meanwhile, EAB achieves a maximum value of ~ GHz for the sample containing 1.2 wt.% of FLMG, a notable value that covers almost the complete X-band.

The enhanced microwave absorption properties of FLMG can be attributed to increased dielectric losses, predominantly stemming from multiple reflections, dipolar interactions, and interfacial polarizations, all of which are enhanced due to the defects and mesostructures present in the FLMG.

These results successfully demonstrate that the milling process described in Chapter 2 can transform non-absorbent graphite, into FLMG capable of absorbing up to 99% of the incoming microwave radiation with extremely low loadings, doing so in a cost-effective and potentially scalable manner. Furthermore, the electromagnetic shielding performance of this composites is comparable to comparable to some of the examples considered state-of-the-art found in Table 3. 1.

#### *3.7.1.2 MWs-Containing Samples.*

In the next step of the experiment, the MW-containing samples with increasing filling ratios (0.8 to 2.0 wt.%), i.e., samples #06 -#09, are studied.

For the MW-containing samples, the RL spectra presented in Figure 3. 10 show a maximum absorbance at around 8.7 and 9.3 GHz, with a weak positive relation between the frequency of the minimum contrary to the relation given by Equations 3.16 and 3.17. However, the closeness of these values added to the typical experimental error prevents stablishing a significant relation.

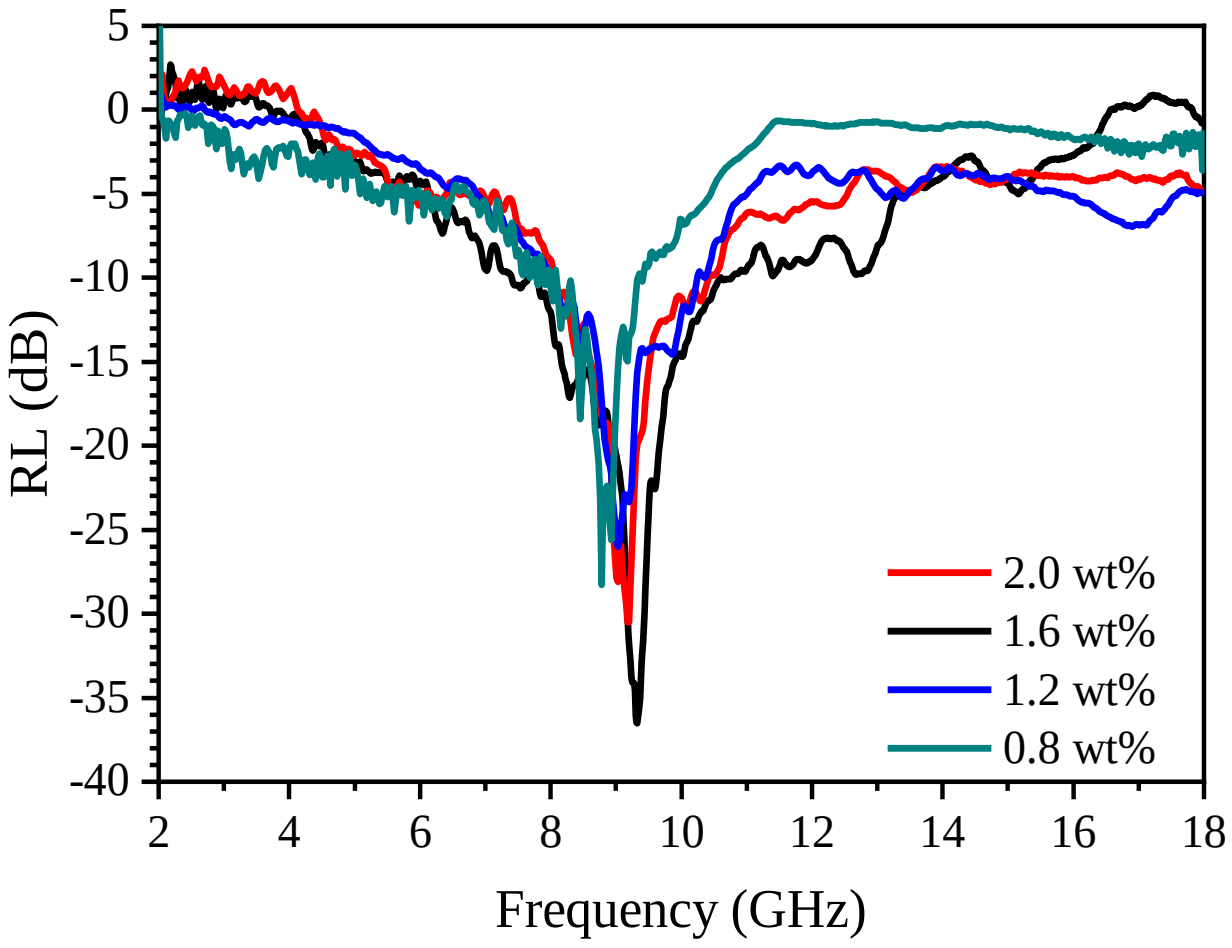

*Figure 3. 10 RL spectra for MW/ Paint ESMs, with different MWs filling ratios.*

As for the maximum absorbance, the minimum RL value ranges from -26 dB to -36 dB with increasing MWs concentration. These are significantly higher absorptions than those obtained for the FLMG-containing samples and are comparable to the performances reported previously for MWs (Table 3. 3).

*Table 3. 3 Comparative table for MWs reported in the scientific literature.*

| Material | Method | Thickness (mm) | RL (dB) | Frequency (GHz) | EAB (GHz) | Ref. |
|---|---|---|---|---|---|---|
| FeSiBCuNb MWs | FSM | 140 μm* | -45.9 | 10.0 | 1.8 | [4] |
| FeSiBCuNb MWs | FSM | 5 | -22.3 | 7.5 | 4.6 | [89] |
| FeSiBCuNb MWs | FSM | 220 μm | -36.5 | 9.3 | 3 | This |

*The exact RL, Frequency (of the maximum RL), and EAB were not provided in some references, thus they were dug out from the RL-f curves using a plot digitizer app (https://apps.automeris.io/wpd/). The focus has been paid to the best material combination in each report within the 2-18 GHz range. * Some works with MWs used a polymeric substrate as substrate for the ESM coating that affects the impedance matching conditions.*

#### *3.7.1.3 FLMG- And MW- Containing Samples.*

After confirming the electromagnetic shielding behaviour of both separately materials, both compositions were combined to obtain the FLMG + MWs containing samples, #10-#13, with a fixed filling ratio of 2.4 wt.%. The results are presented in Figure 3.12.

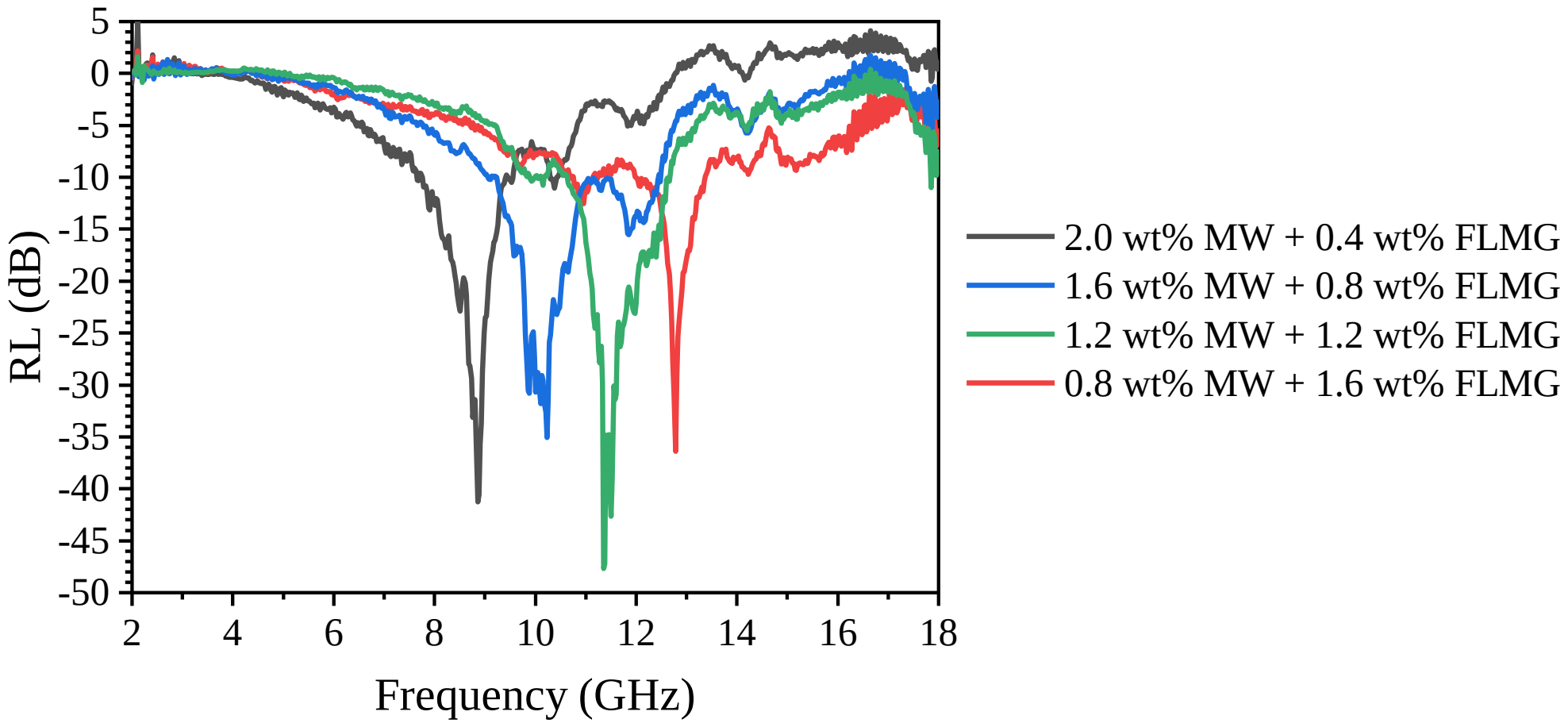

*Figure 3. 11 RL spectra for FLMG/MWs hybrids ESMs from Test 1 with a 2.4 wt. % total filling ratio and different compositions. With a coating thickness of 220 μm over a 1450 μm dielectric substrate.*

The performance of the combined FLMG and MWs samples revealed interesting and promising results. When increasing the FLMG filling ratio, substituting the MW, the frequency of maximum absorbance consistently shifted towards higher values. The results obtained for these samples are summarized in Table 3. 4.

*Table 3. 4 Results for the FLMG- and MW- containing samples.*

| Sample | FLMG content (%wt.) | MWs content (%wt.) | RL (dB) | Frequency (GHz) | EAB (GHz) |
|---|---|---|---|---|---|
| #10 | 0.4 | 2.0 | -41 | 8.8 | 1.4 |
| #11 | 0.8 | 1.6 | -35 | 10.2 | 2.1 |
| #12 | 1.2 | 1.2 | -47 | 11.4 | 3.3 |
| #13 | 1.6 | 0.8 | -36 | 12.7 | 1.8 |

These results indicate that the position of the minimum RL value is influenced by the composition of the combined FLMG and MWs samples, suggesting that the interaction between these materials plays a significant role in determining the frequency at which maximum microwave absorption occurs. Moreover, the tunability of the position of the minimum RL value by adjusting the FLMG and MWs concentrations offers the possibility of designing electromagnetic shields with targeted absorption frequencies, making them suitable for specific applications or frequency ranges.

Furthermore, the combined FLMG and MWs samples demonstrated higher absorbance values compared to the individual FLMG and MWs samples. This improvement in microwave absorption performance suggests that the combination of FLMG and MWs materials can lead to more effective electromagnetic shields, potentially outperforming single-material-based absorbers. The enhanced performance observed in the combined samples cannot be solely attributed to the combination of the individual performances of FLMG and MWs materials, indicating the presence of synergistic effects between the two materials that contribute to the improved microwave absorption properties. These synergistic effects may result from the unique structural and electrical properties of the FLMG and MWs materials, which can enhance the microwave absorption process through multiple reflection, scattering, and absorption mechanisms.

The RL spectra show sharp absorption peaks with EAB between 1.4 and 3.3 GHz. The maximum value is achieved for sample #12, with equal filling rations of FLMG and MW. For instance, sample #12 exhibits an absorption of 99% between 10.6 and 12.7 GHz, almost coincident with the frequency band used for direct broadcast satellite services (10.7 to 12.8 GHz) [90], signifying an interesting and potential application for the material.

In conclusion, the first experiment emphasizes two key findings. Firstly, a notable improvement in microwave absorption performance was observed when comparing graphite to FLMG, signifying the positive impact of the milling process. Secondly, the enhanced performance of combined FLMG and MWs samples underscores the potential of these materials when combined, resulting in superior microwave absorption properties. The observed synergistic effects, adjustable minimum RL values, and increased absorbance values collectively indicate that the combination of these materials can lead to highly effective electromagnetic shields across various applications.

### *3.7.2 Test 2: Addition of FLMG To MW.*

To further understand the role of FLMG on MW-based ESMs, a second test was designed so that the filling ratio of microwires was kept constant, at 2.4 wt.% and the filling ratio of FLMG was gradually increased from 0.1 wt.% to 1.6 wt.%.

The same setup and materials were used. A total of eight compositions were prepared, including one containing solely 2.4 wt.% of microwires. The larger set of samples, compared to the first test, was designed to gain statistical relevance and to allow a more gradual variation of the filling ratios. The coating thickness was 220 µm and three substrate thickness were tested, 1100, 1300 and 1500 µm (see Figure 3. 12).

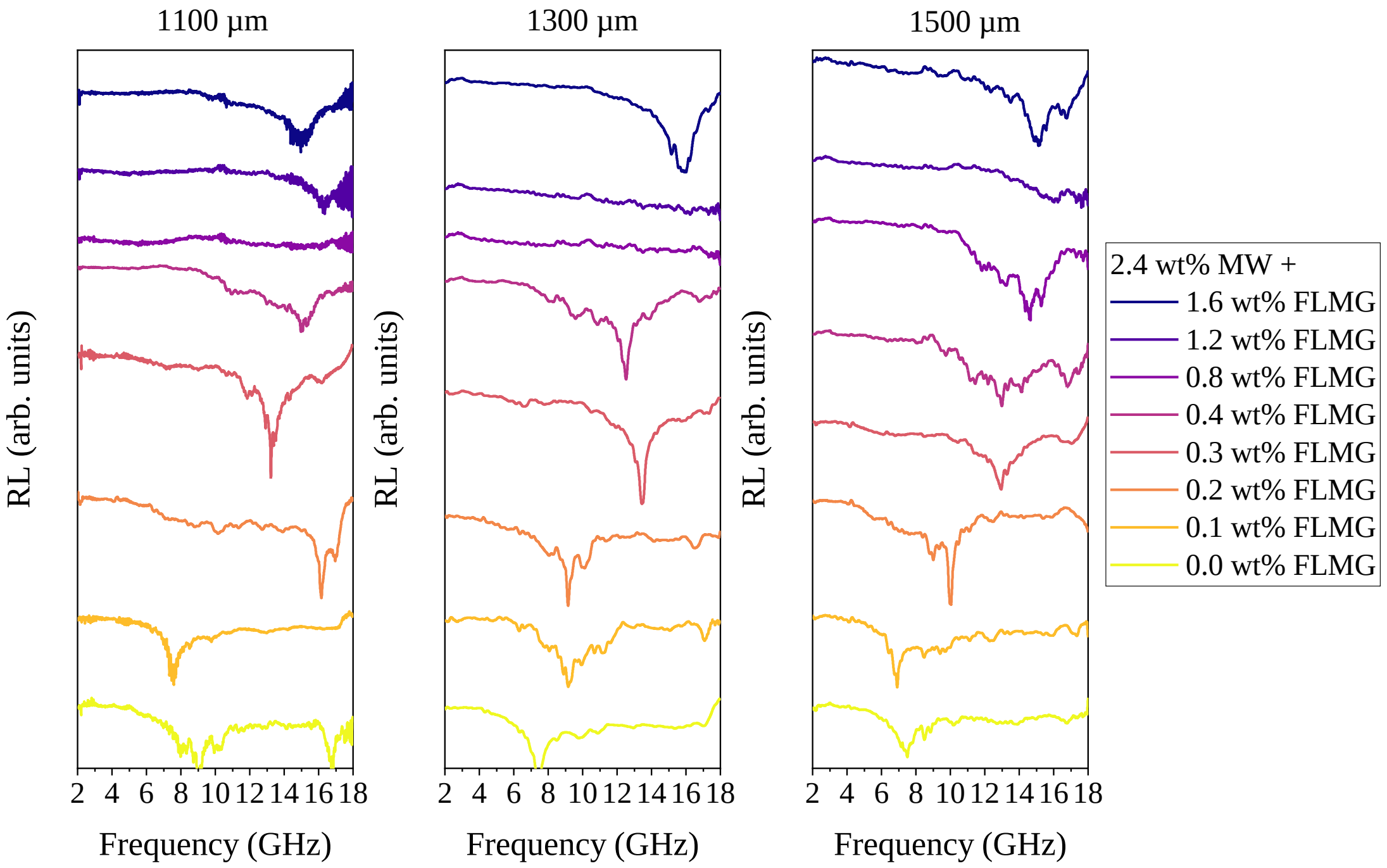

*Figure 3. 12 RL spectra for FLMG/MWs hybrid ESMs of Test 2, for different dielectric substrate thickness and different concentrations of FLMG.*

The thickness of the substrate has a strong effect on the behaviour of the ESM. The effect of the substrate thickness and it is best observed in the samples containing solely microwires. For these samples, the increase in the substrate thickness shifts the frequency toward lower values, the absorption peak becomes sharper, and the RL is decreased. Previous works at *IMA* have found that the thickness of the dielectric substrate plays a crucial role in optimizing the ensemble of measurement parameters and sample [64]. The sample containing only microwires achieves a maximum RL of -38.7 dB at 9.0 GHz and a QAB of 2.8 GHz, coherent with previous examples of amorphous microwire containing samples (see Table 3. 3). The dependence with the substrate thickness can be understood from Equation 3.17.

Based on the spectra shown in Figure 3. 12, there is a strong positive correlation between the addition of FLMG and a shift of the maximum absorbance towards higher frequency values. For instance, for the 1500 µm substrate set, the absorption peak frequency increased from approximately 6-10 GHz for 0-0.3 wt.% to 13-15 GHz for around 0.5 wt.% and 16-18 GHz for up to 1.6 wt.% of graphene. This first important observation highlights the impact of graphene concentration on the absorbance properties of the material.

Despite the difference with the samples from the first test, a general trend can be stablished of the role of FLMG in the frequency of the maximum absorbance. It is difficult to stablish and quantitative or theoretical expression of this behaviour considering the limitations about the models previously described and doing so falls out of the scope of this work. Further testing and the characterization of additional parameters, such as $\varepsilon_r$, would be required to unveil the absorption mechanisms in play for these materials.

This observation is interesting considering some examples in the literature where an increase in the weight fraction of GBM is expected to shift the maximum RL toward lower frequency values due to higher real and imaginary permittivity [76,77]. However, a possible explanation for the discrepancy is the synergistic effect between the amorphous MWs and the FLMG. In fact, a similar trend was reported for the increasing filler ratio of magnetic decorated graphene@CuS [78].

Furthermore, the maximum absorption seems to be independent of the amount of graphene and is typically between -20 and -40 dB, which indicates a high absorption performance for all samples that would be primarily determined by the amorphous microwire component of the composite. For example, for the 1500 µm substrate set the values range from -14.0 to -34.7 dB, with the minimum value corresponding to the 1.2 wt.% FLMG sample and the maximum corresponding to the 0.2 wt.% FLMG sample however the maximum absorption values do not consistently increase or decrease as the weight percentage of FLMG increases.

Therefore, based on the results from both tests, it can be concluded that using low filling ratios of FLMG to tune the minimum RL frequency does not negatively impact the maximum absorption values, which remain high (>96% absorption of the incoming wave) in all cases.

The effect of the FLMG addition to the EAB can be better observed in Figure 3. 13. Here, the RL spectra are represented as a heatmap, where the X axis represents frequency value, as in previous RL spectra, Y axis represents the filling ratio of FLMG in the FLMG/ MWs hybrids ESM. RL is given by the colour scale where red corresponds to 0 dB and blue to -35 dB. It is worth noting that the scale bar is limited to -35 dB while there are samples that achieve lower values. This is due to the adopted criteria that differences above 99.9% have little relevance for the study of ESMs.

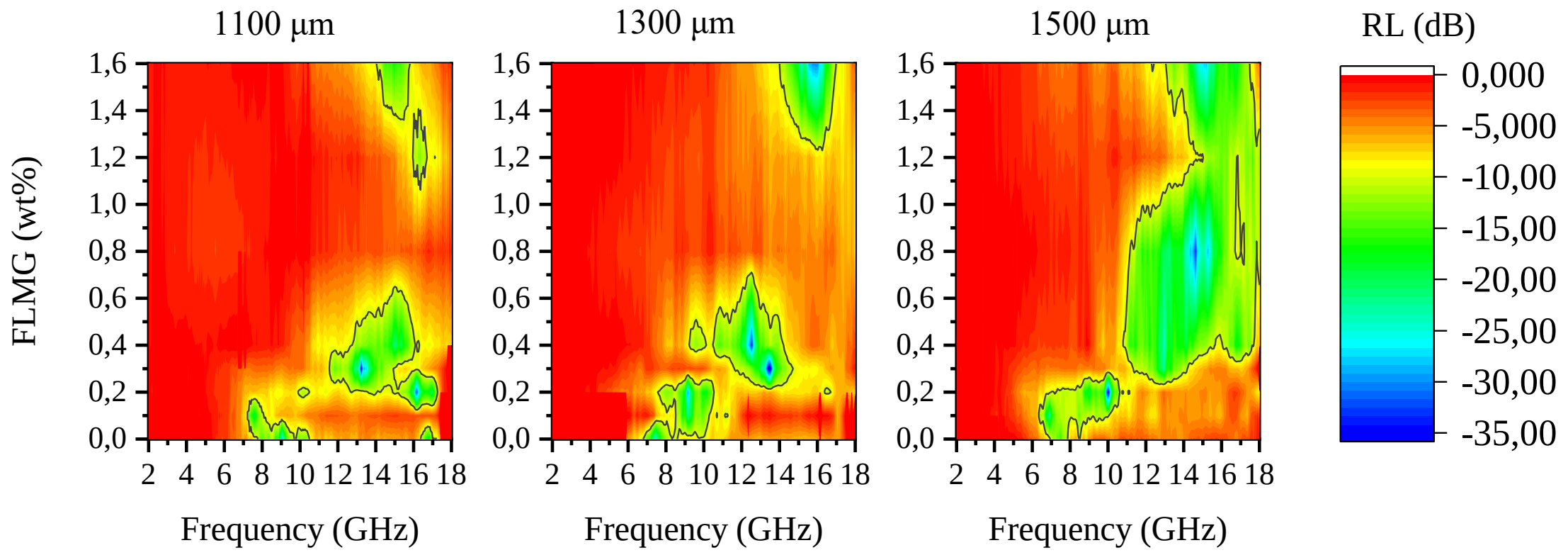

*Figure 3. 13 RL spectra in heatmap representation for FLMG/MWs hybrids ESMs of Test 2, for different dielectric substrate thickness and different concentrations of FLMG.*

As a general trend, the addition of FLMG broadens the EAB. For the 1500 µm substrate, the EAB was found to increase rapidly with concentrations from 0 to 0.8 wt.% up to >5 GHz and then lower back to 2 - 3 GHz for concentrations above 1 wt.%. The effects of high filling ratio are not as clear, but this suggests that the optimal concentration of graphene for achieving the widest absorption bandwidth lies within the range of 0.8 wt.% or lower. To put this into context, the ESM composites with a filling ratio of ~0.4 - 1 wt.% achieve an absorption of at least 99% across the whole $K_u$ band (12 - 18 GHz). The broadening of the absorption bandwidth could be attributed to the hierarchical structures within the mesoporosity of FLMG [91], although these phenomena should require further investigation.

The relevant figures of merit of the microwave absorption are summarized in Figure 3. 14. Similar effects have been previously reported. For instance, in $CoS_2$ nanocrystals embedded into reduced graphene oxide (RGO), increasing the RGO proportion in $CoS_2$/ RGO 1:2 composites led to an extreme enhancement of the MA performance. However, with a further increase in RGO proportion, the reflection loss (RL) becomes worse, attributed to an excessive increase in complex permittivity, leading to impedance mismatch of the composite [92]. In another study, the RL curves of three-dimensional reduced graphene oxide powder (3D-rGO) at a thickness of 5 mm (3D-rGO-5) were evaluated with different content from 1 wt.% to 15 wt.%. Results showed that with increasing content from 1 to 4 wt.%, the absorption performance in 2-4 GHz increased, but further increases in absorber content resulted in reduced absorption performance [93].

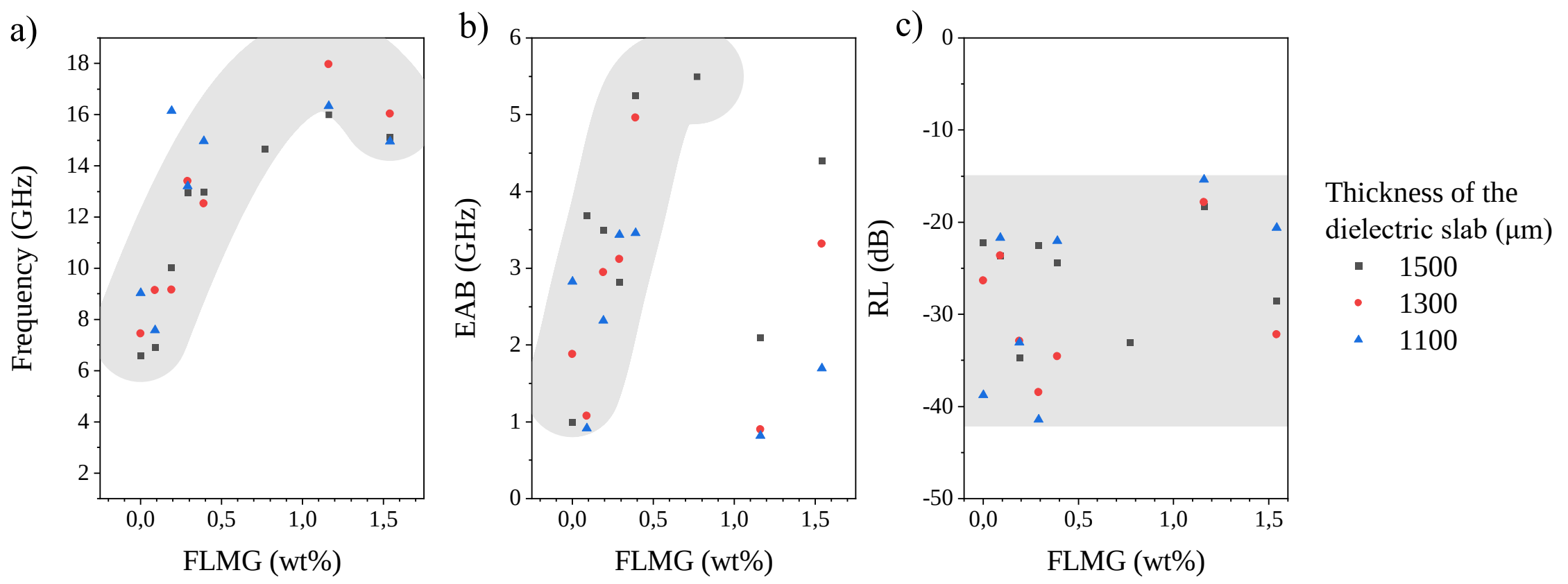

*Figure 3. 14 Extracted data from the second experiment. a) Frequency of the absorbance peak, b) EAB and c) maximum value of reflection loss, for samples with increasing content of FLMG. The differences with the dielectric substrate are expressed using three symbols with different colours (black square, red circle, and blue triangle).*

The relevance of this work can be attributed to two key factors. First, the use of direct RL characterization with the FSM setup, which offers accurate experimental analysis unhindered by theoretical models or presumptions. Second, this work explores the use of FLMG, a novel material developed from the works presented in Chapter 2, and amorphous microwires. a material in which *IMA* possess a comprehensive know-how and that is currently being used in commercial applications.

The observed frequency shift and the widening of the effective absorption bandwidth suggest that the composite material has potential for applications in electromagnetic interference shielding, radar stealth, and other microwave technologies. The manufacturing approach used in this study, being low-cost and environmentally friendly due to the lack of hazardous reactant, may improve the commercial feasibility of this ESM [16].

### *3.8 Conclusions.*

The use of electromagnetic shielding materials (ESMs) is crucial for protecting electronic devices from electromagnetic interference. The transmission line model, which is commonly used, has been shown to produce inaccurate results due to its underlying assumptions. As an alternative, the free-space measurement (FSM) method offers a non-contact approach that allows for direct measurement of the reflection loss without relying on such assumptions. In this thesis, we have used FSM as an empirical approach for two electromagnetic shielding experiments.

The first experiment used a paint-based composite containing FLMG, the graphene-based material obtained in chapter 1, and amorphous magnetic microwires. The results showed that the modification of the ESM composition, despite maintaining the filling ratio constant, allows for the effective fine-tuning of the microwave absorption properties of the material.

The second experiment explored the effect of FLMG to a paint-based composite material containing a fixed amount of amorphous magnetic microwires. The observed frequency shift and widening of the effective absorption bandwidth suggest that the composite material has potential for applications in electromagnetic interference shielding, radar stealth, and other microwave technologies.

The experiments presented in this chapter demonstrate that the use of FLMG can lead to the development of ultrathin composite ESMs with extraordinary performance and tuneable electromagnetic shielding characteristics. Thus, FLMG and amorphous magnetic microwires hold significant promise for future applications in electromagnetic shielding technologies. Overall, this work offers new insights into the design and development of ESMs, and the potential for their application in a range of fields.

### *3.9 References.*

# Chapter 4 : Graphene-based chemiresistive gas sensors.

*Chapter Introduction; Use of Graphene-Based Materials for Chemiresistive Sensors; Gas Sensors Characterization; Use of Few-Layered Mesoporous Graphene as Chemiresistive Sensors; Effect of UV On Multi-Layer Graphene Chemiresistive Sensors; Study of The Influence of Defects; Conclusions; References.*

### *4.1 Chapter Introduction.*

Gas pollution refers to the presence of harmful gases in the air, which can have negative impacts on human health and the environment. Some common examples of pollutants include nitrogen dioxide ($NO_2$), carbon monoxide (CO), sulphur dioxide ($SO_2$), ozone ($O_3$), volatile organic compounds (VOCs), ammonia ($NH_3$) or particulate matter (PM).

$NO_2$ is a reddish-brown gas commonly produced by combustion processes such as cars, power plants, and industrial facilities. It is a major air pollutant due to its negative impacts on human health and the environment. $NO_2$ is a respiratory irritant that can cause respiratory problems, particularly in individuals with asthma or other pre-existing respiratory conditions [1]. It also contributes to the formation of ground-level ozone. $NO_2$ can also have negative effects on vegetation and crops, reducing their growth and yield.

Regarding $NO_2$, Latza et al. found moderate evidence that exposures of 0.1 ppm for 24h or exposures to an annual mean of 0.026 ppm were related to adverse health effects, including increased hospital admissions and mortality, being children, adolescents, elderly, and asthmatics susceptible population of these effects [2]. A similar annual mean exposure limit (0.021 ppm) was indicated by the World Health Organization [3]. More recently, the National Institute for Occupational Safety and Health (USA) has established a recommended limit of 1 ppm for exposures up to 10 hours [4]. In contrast, the American Conference of Governmental Industrial Hygienists recommended a limit of 0.2 ppm for exposures up to 8 hours [5]. To put these numbers into context, the maximum hourly concentrations measured in urban areas in the UK, or the USA are around 0.4-0.5 ppm [6]. Of course, the term recommended limit refers to a maximum exposure in a given time, i.e., a level that should not be exceeded. As for lethality, the median lethal concentration (LC50) for one hour has been estimated at 174 ppm [7]. Conversely, the odour threshold is 0.1-0.4 ppm [8]. Hence, despite the several discrepancies found throughout the literature regarding the exact value of the exposure limits, monitoring systems capable of warning us against the sub-ppm presence of $NO_2$ below the human detection limits are an obvious necessity (see Table 4. 1). Under normal conditions, 1 ppm of $NO_2$ equals 1.914 mg/m$^3$.

*Table 4. 1 Recollection of relevant $NO_2$ limits.*

| Recommended exposure limits | | Estimated lethal concentrations | Concentrations in urban areas |
|---|---|---|---|
| Daily | Yearly | (1 hour) | UK-USA |
| 0.1-1 ppm | 0.02 ppm | 174 ppm | 0.1-1 ppm |

CO is a colourless, odourless, and tasteless gas. It is produced by incomplete combustion or some metallurgic processes, and it can be violently oxidized to the much more stable carbon dioxide. It is considered a dangerous air pollutant due to its toxic effects on human health.

When inhaled, CO interferes with the blood's ability to carry oxygen, potentially leading to headaches, nausea, dizziness/vertigo, and drowsiness/lethargy. These symptoms limit the ability of a subject to react to gas exposure and can ultimately be fatal. Approximately 400 people die from carbon monoxide (CO) poisoning in the United States each year, and an estimated 50,000 visit the emergency room [9]. Worldwide, the number of deaths due to carbon monoxide inhalation is estimated to be in the tens of thousands annually. Deaths and hospitalization related to CO poisoning are most often reported during the colder seasons, occurring mainly in winter, followed by spring and autumn [10]. The WHO recommends a 24 h exposure limit of 3.4 ppm and a 1 h exposure limit of 30.0 ppm [11]. Under normal conditions, 1 ppm of CO equals 1.165 mg/m$^3$.

$SO_2$ is a colourless gas with a strong, pungent odour. It is formed during the combustion of sulphur-containing fossil fuels, such as coal and oil, as well as from industrial processes such as the smelting of sulphide ores. It is considered a major pollutant that can harm human health and the environment, harming crops, forests, and bodies of water. $SO_2$ is soluble in water and can react to produce sulfuric acid, being a major contributor to acid rain.

Exposure to high levels of sulphur dioxide can cause irritation and respiratory problems, including bronchoconstriction and increased asthma symptoms. Symptoms of $SO_2$ poisoning include coughing, wheezing, shortness of breath, and chest tightness. In severe cases, $SO_2$ poisoning can cause respiratory distress DNA damage and even death [12]. Under normal conditions, 1 ppm of $SO_2$ equals 0.266 mg/m$^3$.

Recently it has been found that $SO_2$ can be generated endogenously in mammals, and it may play a role in physiological and pathophysiological regulation [13]. However, endogenous $SO_2$ should be distinguished from its gas-pollutant counterpart.

$O_3$ is a highly reactive gas that is beneficial and harmful to the environment, depending on its location. $O_3$ acts as a protective layer in the upper atmosphere, filtering out the sun's harmful ultraviolet (UV) radiation. However, when ozone is present at ground level, it can have harmful effects on human health and the environment [14].

Ground-level $O_3$ is formed through chemical reactions between oxides of nitrogen (NOx, including $NO_2$) and volatile organic compounds (VOCs) in sunlight. Exposure to ground-level $O_3$ can cause a range of health effects, including respiratory problems, such as coughing, wheezing, and shortness of breath, as well as irritation of the eyes and throat [15]. $O_3$ concentrations are usually higher during the warm months, and the recommended limit of exposure (daily) is 0.03 ppm, or 30 parts per billion (ppb) [11]. Under normal conditions, 1 ppm of CO equals 1.996 mg/m$^3$.

$NH_3$ is a colourless gas with a characteristic pungent odour that is commonly used in fertilizer production and refrigeration systems. It is the air pollutant we are most familiar with, as ammonia-based cleaning products are widely available. $NH_3$ can cause respiratory irritation, and exposure to high concentrations can be harmful to human health.

When released into the atmosphere, ammonia can contribute to air pollution, particularly in urban areas where high concentrations of ammonia can form particulate matter and secondary pollutants like nitrogen oxides and sulphates. $NH_3$ can also be produced endogenously during physiological and pathological processes [16]. Monitoring its presence can be useful for diagnosis during breath analysis, as will be further developed in the next chapter. Recommended exposure limits to $NH_3$ are 25-50 ppm (8h) and 1.47 ppm (24h) [17]. Under normal conditions, 1 ppm of $NH_3$ equals 17 mg/m$^3$.

VOCs are a large group of chemicals that can vaporize and enter the atmosphere. They are present in many common household and industrial products, including paints, adhesives, cleaning agents, and fuels. VOCs can contribute to poor air quality and have a range of negative impacts on both human health and the environment. The list includes benzene, formaldehyde, toluene, xylene, ethylene glycol or acetone. Their effects and exposure limits vary with the component. Generally, immediate effects include headaches, dizziness, and eye, nose, and throat irritation. As for exposure, the rule of thumb would be the lesser the better. Benzene and acetone will be further explored in the next chapter.

Finally, PM, is a term used to describe tiny particles suspended in the air. These particles vary in terms of composition and physical and chemical characteristics. Thus, PM is not a gas, but it is considered an atmospheric pollutant similar to the previously presented. PM is classified by its size, for example, particles with an aerodynamic diameter less than 2.5 µm (PM2.5) and less than 10 µm (PM10). PM2.5 and PM10 are classification often used due to being matter small enough to be inhaled (see Figure 4. 1).

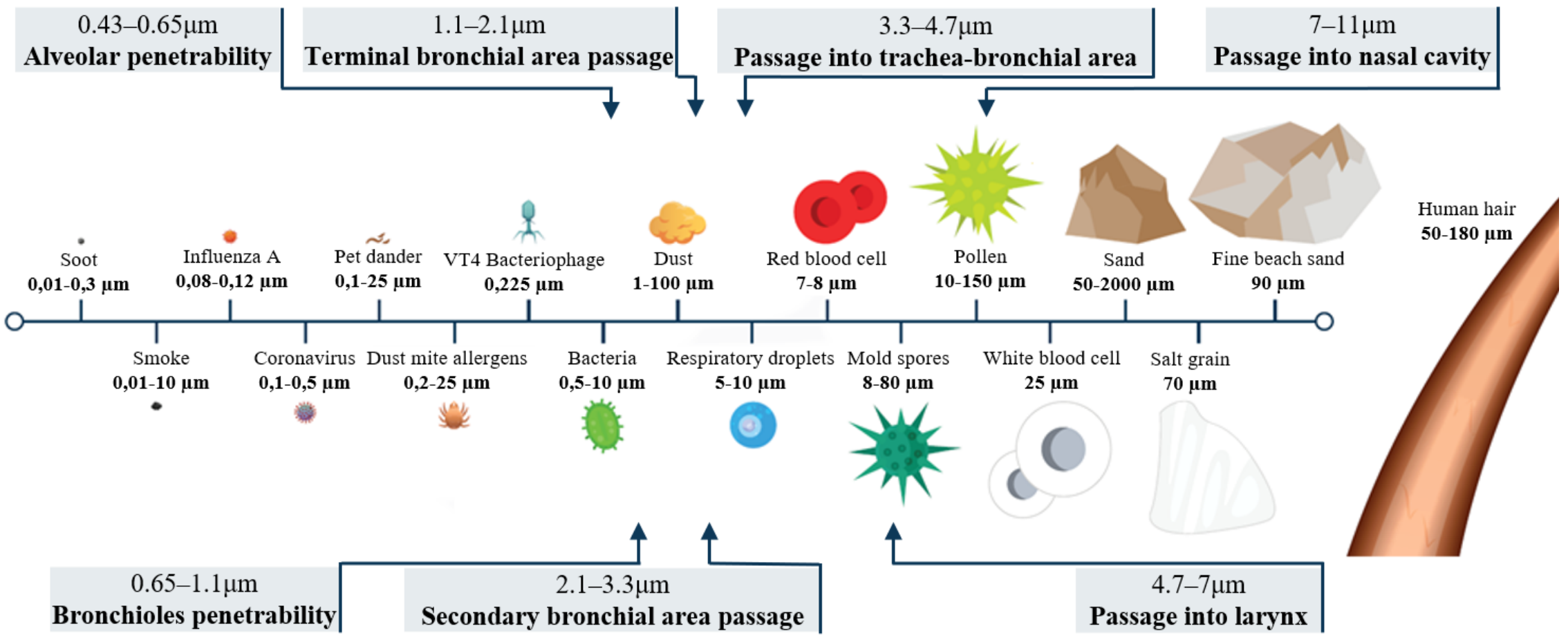

*Figure 4. 1 Illustrative comparison of different micrometric and sub-micrometric particles and their penetrability through the respiratory system. Adapted from* [15,18].

Exposure to particulate matter has been linked to a range of respiratory and cardiovascular problems, including asthma, bronchitis, and heart attacks. Particulate matter can also contribute to the formation of smog, acid rain, and reduced visibility. Exposures should be limited to 5 $\mu g/m^3$ (annual) and 15 $\mu g/m^3$ (daily) for PM2.5, and 15 $\mu g/m^3$ (annual) and 45 $\mu g/m^3$ (daily) for PM10 [11]. Unfortunately, several countries greatly exceed these limits, like areas from Bangladesh (15.4 times above the PM2.5 limit), Pakistan (13.4 times above the PM2.5 limit) or India (11.6 times above the PM2.5 limit) [18].

Several measures can be taken to reduce the risks associated with air pollutants, for example:

- Ventilation of enclosed spaces when the pollution source is indoors or avoid ventilation when the source is outdoors [17,18].
- In densely populated cities, the reduction and limitation of internal combustion vehicles traffic [19,20]. Cities like Madrid, Barcelona, London, Paris, or Amsterdam have introduced reduced speed limits, limitations related to vehicle emissions or substitution of automobile roads for pedestrian streets and green zones [21,22].
- Under heavy pollution conditions, face masks can be recommended or even mandatory, as face masks act as a filter for some air pollutants [23].
- Use of air purifiers with HEPA filters can significantly reduce the risks associated with air pollution in indoor spaces [24]. In addition, air purifiers equipped with charcoal (activated carbon) filters can also neutralize VOCs and unpleasant odours [25]

We can all relate through personal experience to some of these examples, especially since the CoViD-19 pandemic. Table 4. 2 presents a collection of air pollutants and sources for different environments.

*Table 4. 2 Typical air pollutants and their source in common environments. Extracted from* [26].

| Microenvironment | Sources | Pollutants |
|---|---|---|
| Home | Cooking, space heating, parked vehicles, hobbies, smoking, household products, pets, rodents, insects | PM, CO, $NO_x$, VOCs, allergens |
| Transportation environments | Vehicle and industrial emissions, road dust, background pollution, smoking | PM, including ultrafine PM, CO, $NO_x$, $O_3$, VOCs, aeroallergens, carcinogens |
| Streets | Vehicle emissions, road dust, background pollution | PM, including ultrafine PM, CO, $NO_x$, $O_3$, VOCs, carcinogens, lead |
| Work environments | Industrial processes, smoking, background pollution | PM, CO, VOCs, $NO_x$, carcinogens |
| Entertainment environments | Cooking and space heating, background pollution, smoking | PM, VOCs, carcinogens |

Monitoring devices can measure and track levels of these gases (PM is excluded from the category of pollutants for the rest of this chapter) and are crucial in preventing and mitigating their negative effects. There are various monitoring devices, including handheld devices and fixed monitoring stations. The use of these devices helps to identify sources of pollution and track changes in gas concentrations over time, allowing for timely intervention and effective mitigation strategies to be put in place. Additionally, gas monitoring devices play a critical role in supporting environmental regulations and policies aimed at reducing and controlling air pollution.

The most popular technologies for gas sensors are electrochemical, optic, and chemiresistive [27,28]. Novel approaches may also be developed for sensors with special requirements, such as surface acoustic wave (SAW), quartz crystal microbalance or wireless sensors (which will be discussed in the next chapter).

Electrochemical sensors are a type of gas sensor that use an electrolyte and electrodes to create an electrochemical reaction that detects the presence of a specific gas. The reaction produces a current that is proportional to the gas concentration, allowing for accurate measurement. Various cell designs have been created to target specific toxic gases such as carbon monoxide and hydrogen sulphide, as well as variations in oxygen levels. In addition to gas monitoring, electrochemical sensors can be used for biochemical analysis or food safety control. One common application of electrochemical sensors is in the automotive industry, where they are used as lambda sensors for oxygen monitoring [29–31].

Optical methods for gas sensing include both infrared (IR) sensors and laser-based sensors. IR sensors use IR radiation to detect the presence of specific gases by analysing the absorption or transmission spectra of the gas. Each gas component has a unique fingerprint in the IR spectrum, which allows for the identification and quantification of the gas. Laser-based sensors, on the other hand, use a highly collimated beam of light directed towards the gas sample, and the absorption or scattering of the light is measured. Laser-based sensors can use various wavelengths, including IR, depending on the analyte and measurement conditions. Both types of optical sensors can be used for similar purposes, but they differ in terms of sensitivity, selectivity, and cost [32,33].

Finally, in chemiresistive sensors, the interaction between the sensitive material and the target gas causes a change in the material's electrical resistance (more on this in the next section). Electrochemical and chemiresistive may appear similar but the key difference is that in an electrochemical sensor the reaction with the gas produces a flow of electrons that are measured as a change in current, whereas in chemiresistive the reaction with the gas changes the conductivity or resistance [34].

Each type of gas sensor has its own advantages and limitations, and the best choice depends on the specific application and the target gas. In this chapter, I will focus on chemiresistive gas sensors, and particularly the ones using graphene-based materials (GBMs).

## *4.2 Use of graphene-based materials for chemiresistive sensors.*

Chemiresistors based on metal oxides have gained widespread popularity as a gas detection technology for both domestic and industrial applications. These sensors work by measuring the change in electrical resistance of a metal oxide film when it is exposed to a specific gas. However, one of the major drawbacks of metal oxide-based sensors is that they require high operating temperatures for optimal performance. This can result in high power consumption, often ranging from tens to hundreds of milliwatts, and can also lead to decreased lifetime of the sensor due to degradation of the active material. Additionally, the implementation of microheaters to reach these high temperatures can be complex and costly. Despite these drawbacks, metal oxide-based sensors continue to be popular due to their high sensitivity, as well as their ability to detect a wide range of gases. Ongoing research is focused on developing new materials and optimizing sensor design to improve their performance and reduce their power consumption [35–39].

Recent research on metal oxide-based gas sensors has explored alternative methods to improve their performance. One such method involves the use of heterojunctions, which can enable these devices to operate at room temperature. For example, $MoS_2/SnO_2$ for the detection of $NO_2$ [40] or Ag-ZnO/$MoS_2$ for the detection of CO [41]. Another approach is to activate the sensing operation on metal oxide-based sensors using light. Some of the mechanisms involved in this approach are similar to the UV irradiation effect, which will be discussed in more detail later in this chapter [36,42–46].

Graphene and other GBMs are a recent alternative among the different active materials for chemiresistive gas sensors. Since graphene consists of an atom-thick layer, every carbon atom is a surface atom. In addition, its large specific surface (2630 $m^2/g$), high conductivity, and low noise-to-signal ratio make graphene and graphene-based materials ideal candidates for gas sensing applications [47–49]. Currently, the application of GBMs on chemiresistive gas sensors is mostly confined to scientific research.

Over a decade has passed since the first report of a graphene-based gas sensor, with a sensitivity down to a single molecule under highly controlled conditions [50]. During the following years, the number of similar devices reported for detecting different gaseous molecular species under conditions closer to real-life applications has increased dramatically [51–54]. These graphene-based devices have demonstrated their potential for room-temperature sensing $NO_2$, carbon monoxide (CO), sulphur dioxide ($SO_2$) or ammonia ($NH_3$), among other analytes, and significantly, the research interest in the material has increased [7,55–57].

Most GBMs, when used as active material in gas sensors, behave as p-type semiconductors. This means that their conductivity is associated to positive charge carriers, known as holes. However, pristine graphene exhibits neutral conductivity, it has a similar concentration of positive or negative charge carriers (electrons). When GBMs are exposed to oxidizer molecules under normal atmospheric conditions, such as oxygen or water, these molecules undergo a charge transfer process in which they capture electrons from the graphene lattice. This results in an imbalance in the concentration of charge carriers, with a higher concentration of holes than electrons. This imbalance in charge carrier concentration biases the conductivity of the GBMs to p-type [52,54,58–60].

The same principle is used for their gas-sensing operation. When oxidant analytes ($NO_x$, $SO_2$, $O_3$, formaldehyde…) attach to the p-type semiconductor they will further draw electrons increasing the number of positive charge carriers. The extra concentrations of holes will then lead to an increase the conductivity of the material, or conversely, decrease its resistivity. According to the Drude-Sommerfeld model:

$$\sigma = nq\mu \qquad \text{(Eq. 4. 2)}$$

Where σ is the conductivity, n is the charge carrier concentration, q is the charge of the carrier, and μ is the charge carrier mobility.

Conversely, when reducing analytes ($H_2$, $NH_3$, CO, methane…) are present, these molecules will draw holes, or analogously donate electrons. The reduced hole concentration will lead to increased resistance of the p-type semiconductor.

For the work carried out in this, we used graphene-based gas sensors to detect oxidizing, $NO_2$, and reducing gases, CO and $NH_3$. $NO_2$ was selected as a representative oxidizer gas due to its prevalence as an air pollutant and potentially harmful effects on human health. CO and $NH_3$ were chosen as representative reducing gases due to their common use in industrial applications and their potential effects on air quality.

Overall, graphene-based gas sensors have shown tremendous potential in detecting a wide range of gases under a variety of conditions. One of the key advantages of these sensors is their ability to operate at room temperature, unlike many conventional gas sensing technologies that require high operating temperatures. This not only simplifies the device design and reduces power consumption, but also allows them to detect combustible and explosive gases that can't be detected by high-temperature sensors. The low operating temperature also minimizes the risk of thermal drift and enhances the long-term stability of the device, making them suitable for real-world applications. In addition to these advantages, graphene-based gas sensors offer high sensitivity, fast response time, chemical and thermal stability, and miniaturization, making them an attractive option for gas detection in various settings. However, it is important to note that graphene-based gas sensors have their limitations.

One of these limitations is the issue of partial recovery, which occurs when physically adsorbed molecules stick to the material's surface [61]. Partial recovery can negatively affect the performance of graphene-based gas sensors by reducing their sensitivity and response time after each exposure to the analyte. Essentially, as more adsorption sites become blocked with physically adsorbed molecules, there are fewer active sites available for incoming molecules to bind to, leading to a decrease in the overall sensitivity of the sensor. Additionally, the presence of physically adsorbed molecules can decrease the response and increase the recovery times of the sensor, making it less efficient at detecting changes in gas concentrations over time. These effects can be especially pronounced in applications where fast and accurate gas detection is critical, such as in industrial or medical settings.

This limitation can be addressed through forced-desorption methods like thermal annealing [52,62–64] or UV irradiation [50,65–67]. In this research, we will explore the use of UV irradiation in a greater detail as a means of improving the recovery time and sensitivity of graphene-based gas sensors.

Another limitation of graphene-based gas sensors is their limited selectivity, as the change in electrical resistance of graphene is not always unique to a specific gas. This can result in cross-sensitivity, where the sensor responds similarly to multiple gases, making it difficult to distinguish between them. Researchers are exploring various approaches to improve the selectivity of graphene-based sensors, such as functionalizing the graphene surface with specific chemical groups [68], incorporating other materials such as metal nanoparticles [69–71], and introducing defects into the graphene lattice to modify its electronic properties [68,72].

In addition, the reproducibility and comparability of research on graphene-based gas sensors have been identified as significant challenges in the field. There are differences in the methodology used in different studies, such as the definition of sensitivity and the type of carrier gas used. This can make it difficult to compare results and draw meaningful conclusions from the data. Inconsistencies in reported results can also make it challenging to apply the findings in practical applications. Improving the reproducibility and comparability of research in this field will be essential to further advancing the development and use of graphene-based gas sensors. Standardized testing procedures and reporting guidelines may be one solution to this challenge. Additionally, collaborations and knowledge sharing among researchers could help address this issue and promote progress in the field.

To illustrate the impact of one of these variables, a test was conducted using different carrier gases (air and nitrogen) in the detection of an analyte. One of the key functions of the carrier gas is to generate a controlled and consistent atmosphere. The carrier gas acts as a buffer, helping to stabilize the simulated ambient conditions before the introduction of the analyte. In addition, the carrier gas is used to dilute the analyte to a desired concentration.

However, the carrier gas may interfere with the response of the device towards the analyte. To illustrate this effect, a test was done using synthetic air or nitrogen as carriers during an $NO_2$ calibration experiment.

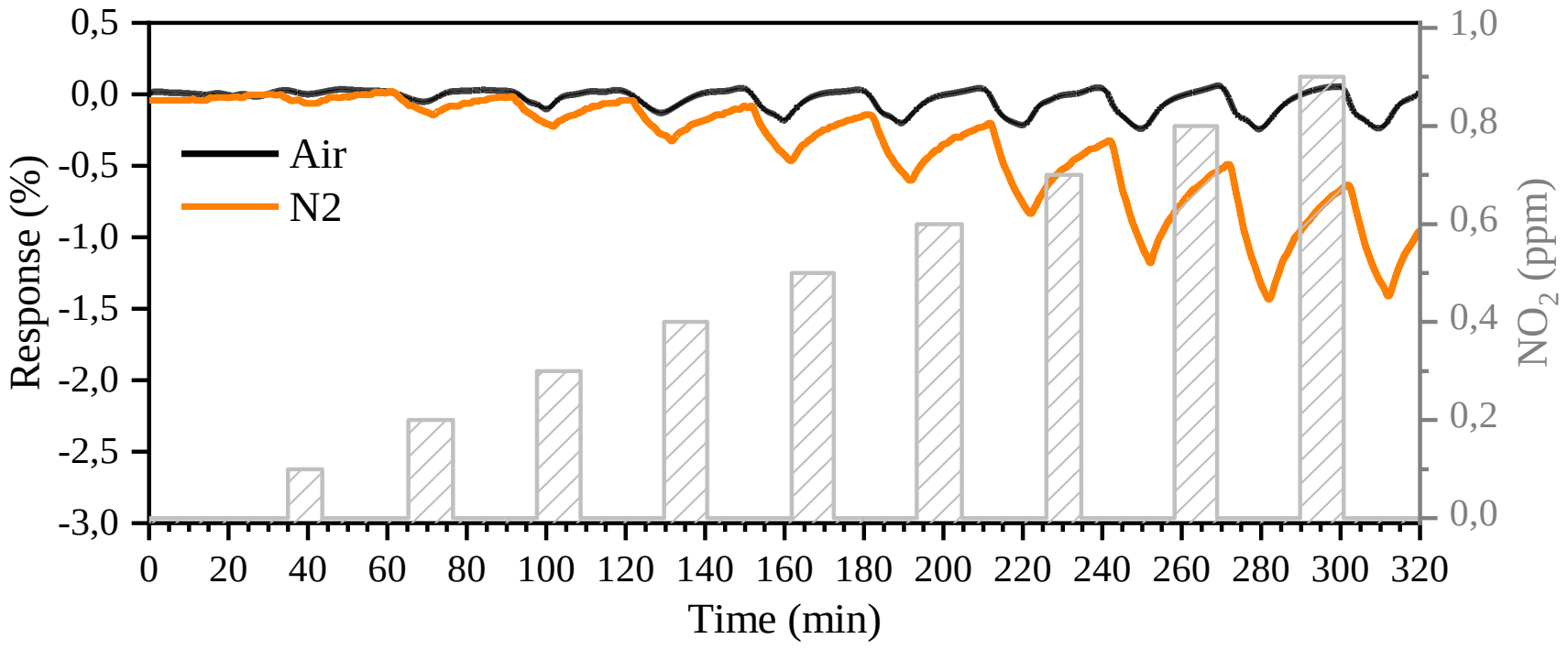

*Figure 4. 2 Comparison of a calibration test for $NO_2$ using air or $N_2$ as a carrier gas. This measure was taken with device MLG890 under UV@100 (see section Multi-Layer Graphene and UV of this chapter for more details).*

The results presented in Figure 4. 2 clearly demonstrate the significant impact that the choice of carrier gas can have on the response of graphene-based gas sensors. As previously noted, gases affect the resistance of graphene-based materials through charge transfer, and the presence of oxygen in the air can produce a similar effect to that of the target analyte (such as $NO_2$), leading to interference and decreased sensitivity of the sensor. The use of inert atmospheres, such as nitrogen or argon, can be valuable in understanding the underlying mechanisms involved in the operation of graphene-based sensors. However, it is essential to conduct research under conditions that closely resemble real-world scenarios to bridge the gap towards practical applications.

The definition of 'response' in graphene-based gas sensors can be particularly tricky, as it depends on exposure time and may increase until the steady state is reached. The steady state is a point at which the resistance of the sensor reaches an equilibrium state with the new environmental composition, i.e., the presence of the analyte, and will no longer change, except for noise or temperature drift [73,74]. This means that the response, when defined as the maximum change for a given exposure time, can vary depending on the duration of exposure. However, most researchers do not measure response at a steady state, as it can take a long time to achieve. This can lead to inconsistencies in reported results and make it difficult to compare different studies. In Table 4. 3 the response value is extracted from the maximum variation on the sensor's resistance caused by the exposure to an analyte. For that matter, it is important to clearly define the terminology, in this research, "response (R%)" is defined as the real-time change of the measured magnitude of the sensor, in the case of the chemiresistive sensors this magnitude. The maximum change of the response during the analyte exposure is defined as maximum response ($R_{Max.}$).

In sum, approaches currently being investigated to improve the performance of graphene-based gas sensors include using different types of GBMs, different degrees of functionalization or defects, and the use of UV irradiation or heating, as presented in Table 4. 3.

*Table 4. 3 Collection of previously reported graphene-based gas sensors with different approaches (material, defect engineering and UV irradiation use). Noteworthy, for the sake of comparison, the term "graphene" in the table is used in its common acceptance (see Chapter 2).*

| Material | Main approach | Analyte | $R_{Max.}$ | Relevant conclusions | Ref. |
|---|---|---|---|---|---|
| CVD-graphene | ~20 nm nanomesh by lithography | $NO_2$/ Air 1 ppm | 6% | | [75] |
| | | $NH_3$/ Air 5 ppm | 5.3% | | |
| CVD-graphene | Pristine | $NO_2$/ Air 200 ppm | 10% | The functionalization increases the response time (8x). Excessive treatment is harmful | [76] |
| | Ozone treatment (Oxygenated functional groups) | $NO_2$/ Air 200 ppm | 20% | | |
| Functionalised rGO | Magnetically aligned; Recovery under UV: 80 mW/cm$^2$ | $NO_2$/ Air 1 ppm | 55% | Sensor achieved full recovery under UV with neglectable side effects. | [77] |
| rGO | | $NO_2$/ $N_2$ 5 ppm | 13% | | [78] |
| | | $NH_3$/ $N_2$ 5 ppm | 2.5% | | |
| CNT | Continuous UV: 253.7 nm, 1.7 mW/cm$^2$ | NO/ Air 200 ppm | 36% | 5-fold increase but reduced lifetime under UV and material removal. | [79] |
| Mechanical exfoliation-Graphene | Pristine | $NO_2$/ $N_2$ 100 ppm | 11% | Sensor achieved full recovery under UV with neglectable side effects. | [80] |
| | Defective graphene (I(D)/I(G)= 0.24) | $NO_2$/ $N_2$ 100 ppm | 32% | | |
| | Defective graphene (I(D)/I(G)= 0.59) | $NO_2$/ $N_2$ 100 ppm | 18% | | |
| CVD-graphene | | CO/ Air 100 ppm | 3% | | [54] |
| | | $NO_2$/ Air 100 ppm | 18% | | |
| CVD | 1, 2, 3, and 4 layers | $NO_2$/ Air 25 ppm | 37.5% | Best performance obtained for bilayer | [81] |
| | | $NH_3$/ Air 1 ppm | 1% | | |
| CVD-graphene | High temperature operation (150°C) | $NH_3$/ Air 65 ppm | 2.1% | | [82] |
| CVD-graphene | Heat (200°C) and vacuum to force desorption | $NO_2$/ Air 10 ppm | 15% | | [83] |
| | | $NH_3$/ Air 10 ppm | 17.5% | | |
| CVD-graphene | Pristine | $NO_2$/ Air 100 ppm | 9.6% | | [68] |
| | Defective graphene (I(D)/I(G)= 0.04) | | 15% | | |
| | Defective graphene (I(D)/I(G)= 0.12) | | 20% | | |
| | Defective graphene (I(D)/I(G)= 0.38) | | 13.3% | | |
| | Defective graphene (I(D)/I(G)= 1) | | 7.8% | | |
| CVD-graphene | Pristine | $NO_2$/ He 200 ppm | 40% | | [72] |
| | | $NH_3$/ He 200 ppm | 3.5% | | |
| | Defective graphene (I(D)/I(G)= 0.1-1.1) | $NO_2$/ He 200 ppm | 53% | | |
| | | $NH_3$/ He 200 ppm | 25% | | |
| Commercial graphene substrate | Pristine | $NO_2$/ Air 5 ppm | 9.75% | Continuous high-temperature operation 150℃ was used to improve recovery | [84] |
| | Defective graphene (I(D)/I(G)= 0.065) | $NO_2$/ Air 5 ppm | 9.1% | | |
| CVD-graphene | Pristine | $NO_2$/ Air 100 ppm | 20% | | [85] |
| | Defective graphene (I(D)/I(G)= 0.459) | | 50% | | |
| | Defective graphene (I(D)/I(G)= 1.428) | | 80% | | |
| rGO on a flexible substrate | Printed on PET; Recovery under UV: 254 nm | $NO_2$/ Air 100 ppm | 6% | Sensor achieved full recovery under UV with neglectable side effects to rGO or PET. | [86] |
| | | $NH_3$/ Air 100 ppm | 21% | | |
| CVD-graphene on a flexible substrate | Recovery under UV: 254 nm, 2.5 mW/cm$^2$ | $NO_2$/ Air 2.5 ppm | 65% | Sensor achieved full recovery under UV with neglectable side effects. | [67] |
| CVD-graphene | Continuous UV: 370 nm | $NH_3$/ $N_2$ 0.6 ppm | 1.89% | 10-fold increase | [65] |
| CVD-graphene | Continuous UV: 253.7 nm, 1.7 mW/cm$^2$ | $NO_2$/ $N_2$ 0.4 ppb | 3.6% | UV significantly lowered the LoD with neglectable side effects. | [87] |
| | | $NH_3$/ Ar 2 ppb | 1.4% | | |
| CVD-graphene | Continuous UV: 265 nm, 1.68 mW/cm$^2$ | $NO_2$/ Air 100 ppm | 26% | 7-fold increase | [88] |
| | | $NH_3$/ Air 100 ppm | 1.6% | 3-fold increase | |
| | | CO/ Air 100 ppm | 1.2% | 6-fold increase | |

### *4.3 Gas Sensors Characterization.*

The performance of gas sensors is measured using equipment and techniques that have not been presented in Chapter 1. In particular, the experimental setup has been custom made, although some of its parts are commercially available, and the methodology for obtaining the figures of merit that define the sensor's performance need to be carefully described to avoid the issues previously discussed.

#### *4.3.1 Experimental Setup.*

The gas line is a crucial part of the experimental setup used to characterize gas sensors. It consists of several basic components, including gas cylinders, mass flow controllers, a gas cell, and a system of valves and tubing. In addition, instrumentation and data acquisition systems form part of the experimental setup.

The gas cylinders contain the gases to be analysed, i.e., analytes. The cylinder often contains a balance with a gas carrier with the required concentration of analyte, typically in the range of 1-100 ppm. In this work, gas cylinders containing $NO_2$ (1 ppm), CO (10 ppm), and $NH_3$ (50 ppm) were used, all provided by Nippon Gases, Spain.

The gas carrier is also fed from a gas cylinder. As previously stated, this gas is used to emulate the atmospheric conditions and dilute the analyte to concentrations lower than the ones given by their cylinders. Aside from the test previously presented which used $N_2$ as a carrier gas, the rest of the measurements were taken using synthetic air as a carrier. This air is obtained artificially and has a composition similar to natural air, mainly nitrogen and oxygen. Synthetic air was provided by Nippon Gases, Spain.

Importantly, the air used in this research did not contain humidity. Humidity used in some experiments was added by passing the carrier gas through a water-containing deposit known as a bubbler. As air passes de bubbler, it collects water molecules which are then carried out to the sensor. The relative humidity levels were calibrated by controlling the flow of air through the bubbler and using a thermohydrometer (RS 1364, RS Components, London, UK).

The gas cylinders are connected to mass flow meters (Bronkhorst High-Tech, Ruurlo, The Netherlands), devices containing electronic valves that control the flow of gas passing through. By combining the flow meters with a tubing system with T-connectors, it is thus possible to obtain a gas mixture or dilutions of the analyte gas to concentrations lower than those of the cylinder. For example, the electronic valves used in this work have a linear relationship between the gas flow and their voltage that reach a maximum of 100 ml/min at 2.5V, as represented in Figure 4. 3.

The software used in this setup has been custom-made using LabVIEW. Briefly, LabVIEW allows integration of the operation and monitoring of scientific instruments through intuitive and visual programming. The software has a user interface in which commands, in the form of buttons and drop-down menus, allow configuring the connection to the instruments or setting the experiment conditions. It may also include information windows containing the status of the instrument, graphs and plots, or possible errors.

For the experiments presented in this research, the LabVIEW programs simultaneously controlled and monitored the mass flow meters and the instruments measuring the resistance of the device (a Keithley 2400 Multimeter, Ohio, USA). The software then generates a .csv file containing all relevant information of the experiment along with chronometer data running since the beginning of the experiment (see Figure 4. 3). All this data is used to create the graphics that illustrate the change of the sensor's electrical resistance during the exposure to different atmospheres.

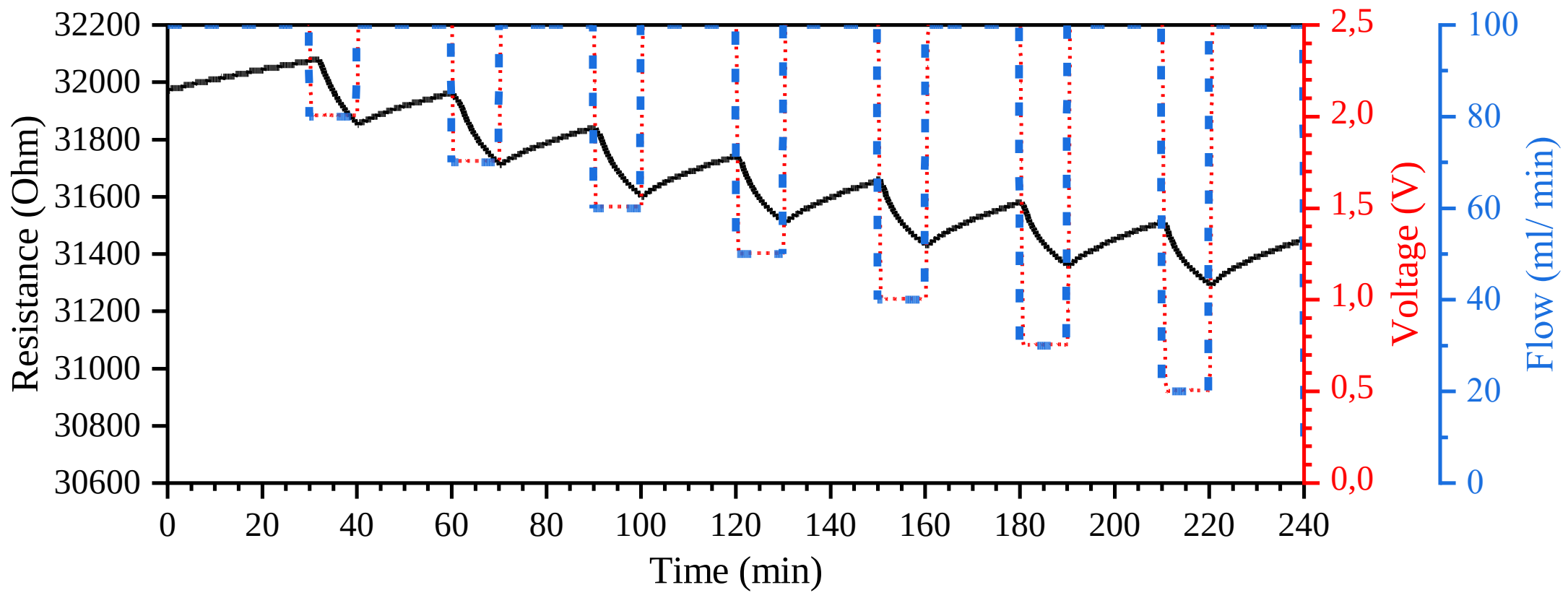

*Figure 4. 3 Raw data representation of a chemiresistive sensor response. The black line represents the real-time monitoring of the device's resistance. The red line indicated the voltage at which the valve operates, from 0 V to 2.5 V, and the blue line represents the flow of gas, in this case, air. This data belongs to a calibration test in which the analyte (not represented) is mixed to different concentrations.*

The gas mixture is then introduced into the gas cell, where the sensor is placed. These cells have been custom-made using 3D printing techniques for each of the sensors presented throughout this work. Depending on the sensor, the shape, size, and required connections for its operation will determine the design of the cell.

The cell may also contain additional components such as a UV light-emitting diode (LED). A UV LED (275 nm, CA3535 - CUD7GF1B, SeoulViosys, Korea) was used during some experiments that tested the sensor under UV irradiation.

For improved performance, the gas cell needs to have as little volume and as unrestricted flow as possible so that the changes in its atmosphere occurred without any significant delay. Whereas, for improved convenience, the cell is designed so that the sensor is easily replaced. Finally, it is crucial that the cell is air-thigh and made from a non-reactive and non-absorbing or desorbing material.

Due to the modular configuration of this setup, it can be easily adapted for different sensors. For example, both the chemiresistive and magnetoelastic resonance-based sensors may share most of the valve and tubing system, the gas carrier, the flow meters, and the portion of software that controls the mixing. The differences are found in the gas cell, the instrumentation for monitoring the sensors' response and the software portion responsible for the data acquisition (see Figure 4. 4). All the setup and gas-sensing characterization was done at the SENSAVAN's lab at the *Instituto de Tecnologías Físicas y de la Información* (*ITEFI*) from *Consejo Superior de Investigaciones Científicas* (*CSIC*), Spain [89–91].

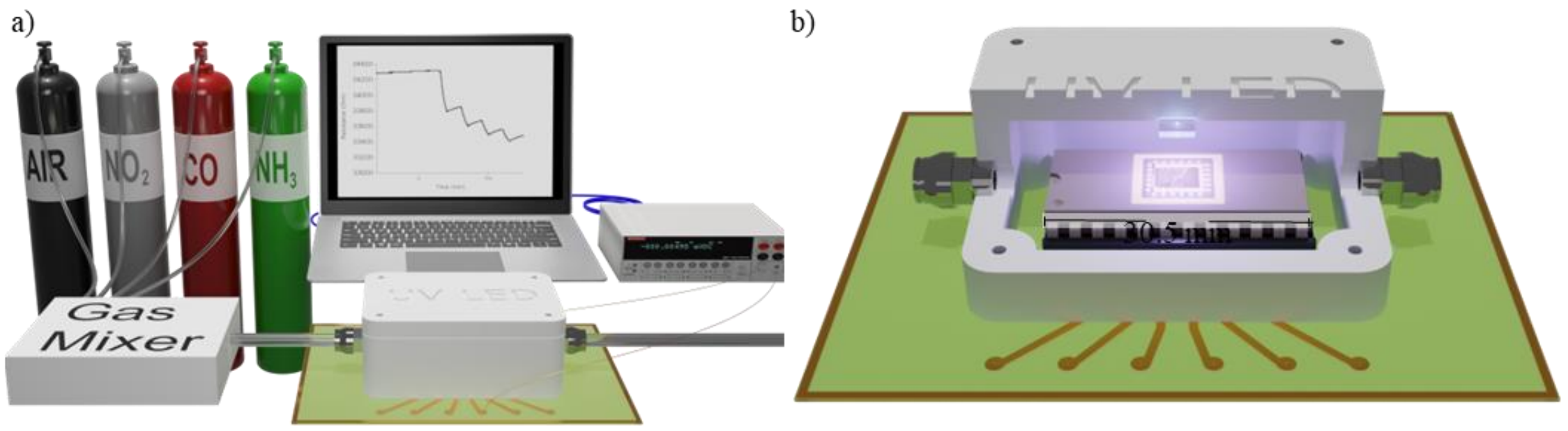

*Figure 4. 4 Illustration of the gas sensor characterization setup for the chemiresistive sensor (left) and gas cell with UV LED as used in the multi-layer graphene sensors experiments (right). Although this image illustrates the multilayer graphene sensor, it is mainly the same setup used for the few-layered mesoporous graphene sensors.*

### *4.3.2 Sequences And Methodology.*

The characterization of chemiresistive gas sensing devices involves the continuous measurement of resistance while exposing the device to different atmospheres. These experiments typically consist of sequence containing three basic steps: baseline, exposure, and recovery. Each sequence is carefully designed according to the sensor's behaviour and the intended purpose of the experiment.

First, the devices are left under constant pass of the carrier gas until a time-stable electrical resistance is achieved. This step is called baseline and mimics operational conditions prior to the analyte exposure. During the baseline, the device adsorbs and desorbs species until an equilibrium is reached. The stable value of electrical resistance is named $R_0$ and it can be extracted from a blank measurement, a measuring sequence without exposure cycles. Conversely, it can be defined as the resistance of the device immediately prior to the exposure to the gas.

During the baseline step, the sensor's signal is stabilized under the carrier gas. The exposure step involves exposing the sensor to a gas mixture, usually an analyte that is diluted in the carrier gas to a designed concentration. Finally, during the recovery step, the sensor is exposed again to the carrier gas to return to its initial state. Exposure and recovery may constitute a cycle, with a sequence having one or more cycles.

During the exposure step, the analyte flows through the gas cell in a determined concentration. During this time, the analyte molecules may be adsorbed by the sensor leading to a change in its electrical resistance. If the active sites of the sensors become fully occupied by adsorbed analyte, or if the device reaches an equilibrium between the adsorbed species and the atmospheric composition, no further change in its electrical will be observed. This state is often called steady-state or saturated and will be observed for some devices evaluated in this research. The response, previously defined as the real-time change of the measured magnitude of the sensor, is monitored via the variations in the sensor's resistances over time, and expressed as a percentage according to:

$$R\% = \frac{R - R_0}{R_0} * 100 \qquad \text{(Eq. 4. 2)}$$

Where, R is the measured electrical resistance of the device at each moment.

During the recovery step, the analyte is removed from the gas cell by passing the carrier gas. This causes the analyte, previously adsorbed on the sensor, to be released from the material's surface and the device tends to achieve a new equilibrium with the new atmospheric composition. Thus, the electrical resistance gradually recovers its value $R_0$. This process is often not immediate and if the sensors does not fully recover during this step, it may have a negative effect to its response during a subsequent exposure. This phenomenon will be further discussed during the presentation of various experiments.

The response time can be evaluated using the $\tau_{90}$ parameter, defined as the time required to achieve the 90% of the maximum response [89,92]. Strictly, $\tau_{90}$ should be calculated from steady-state conditions, however, it may occur that the device is unable to achieve such state during the exposure step, like the example depicted in Figure 4. 5.

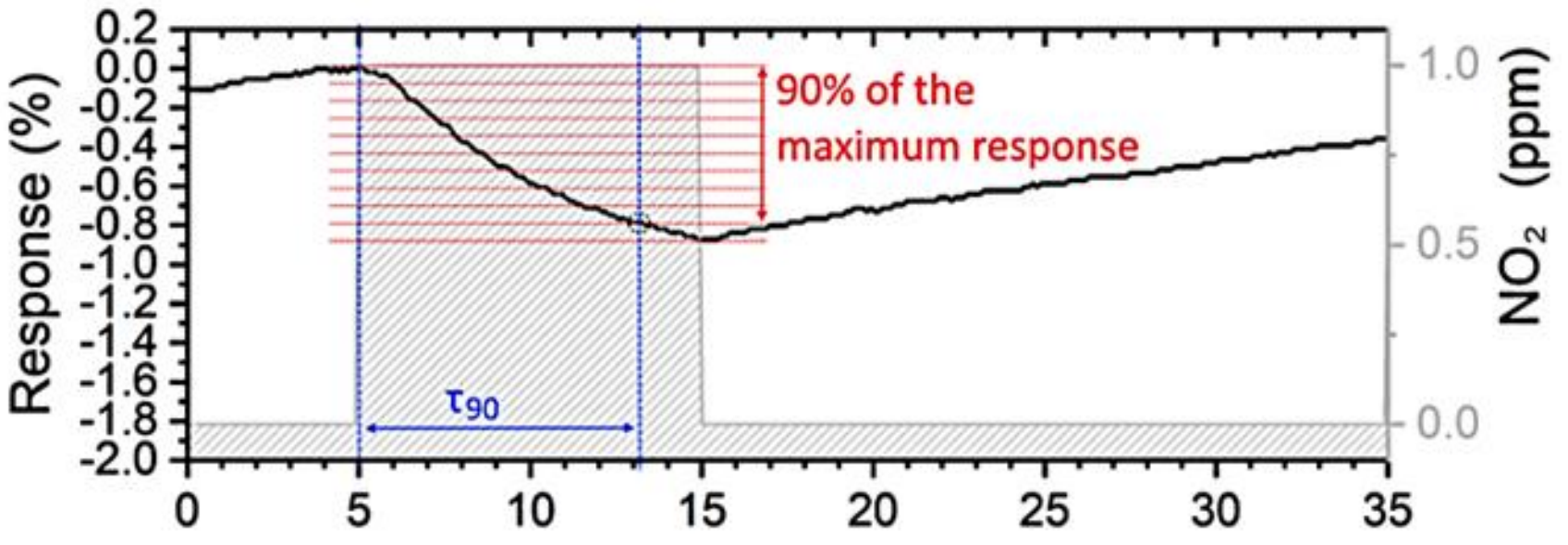

*Figure 4. 5 Schematic representation of $\tau_{90}$ parameter and its calculation. Rep. from* [91] *supporting information.*

The sensors response is calibrated using the sensitivity parameter. It is obtained by measuring $R_{Max.}$ for different analyte concentration. If $R_{Max.}$ has a linear dependency with the analyte concentration, the value of this slope provides the sensitivity of the sensor, typically expressed as the percentage of response per parts per million (%/ppm).

Finally, the limit of detection (LoD) is the lowest concentration of gas that can be reliably detected by the sensor. It is typically defined as the concentration of gas that produces an unequivocal signal:

$$LoD_{(RMSnoise,\text{Sensitivity})} = \frac{3 * RMSnoise}{Sensitivity} \qquad \text{(Eq. 4. 3)}$$

*Where RMS noise is the root mean squared noise of the device signal during a baseline measurement. The sequence to characterize the sensitivity and LoD are often called calibration sequences.*

## *4.4 Use Of Few-Layered Mesoporous Graphene as Chemiresistive Sensors.*

Since the performance of the sensors may be strongly influenced by defects and active surface, as explored in the examples from Table 4. 3, the question on how few-layered mesoporous graphene (FLMG) will perform in this application arises. FLMG is a GBM previously described in Chapter 2, briefly, it is an agglomeration of few-layered graphene domains with a certain degree of defectiveness with a mesoporous structure.

### *4.4.1 Device Preparation.*

In the first series of experiments, the goal is to evaluate the potential of FLMG as a material for use in gas sensors. Specifically, to investigate the gas sensing properties of FLMG and compare them to those of other graphene-based materials.

FLMG was prepared following the instructions presented in Chapter 2 for the oscillatory ball-mill. Briefly, using graphite as a precursor, 25 Hz milling frequency, a weight ratio of 1:25 [93]. Three milling times were tested, 100, 210 and 300 minutes (named FLMG-100, -210, and -300). Graphite was also used in this experiment as a reference.

The samples were first dispersed in n-methyl-2-pyrrolidone (NMP) with a concentration of 1 mg/mL using a sonication bath for 1 h, resulting in stable colloidal suspensions Figure 4. 6a). The sensitive material was then deposited on top of the interdigital electrodes by means of drop casting (Figure 4. 6b). Interdigitated electrodes (IDEs) are commonly used as transducers for gas sensors because of their high sensitivity and selectivity. The IDE substrate consists of two sets of metallic interdigitated electrodes that are perpendicular to each other and separated by a gap. Specifically, the substrate was fabricated on an FR4 glass epoxy board using gold-plated copper interdigital electrodes that consist of five fingers. Each finger measured 1.2 mm in length, and the width of both the fingers and the gap was 150 μm. The substrate contains four sets of IDEs allowing four device characterization simultaneously.

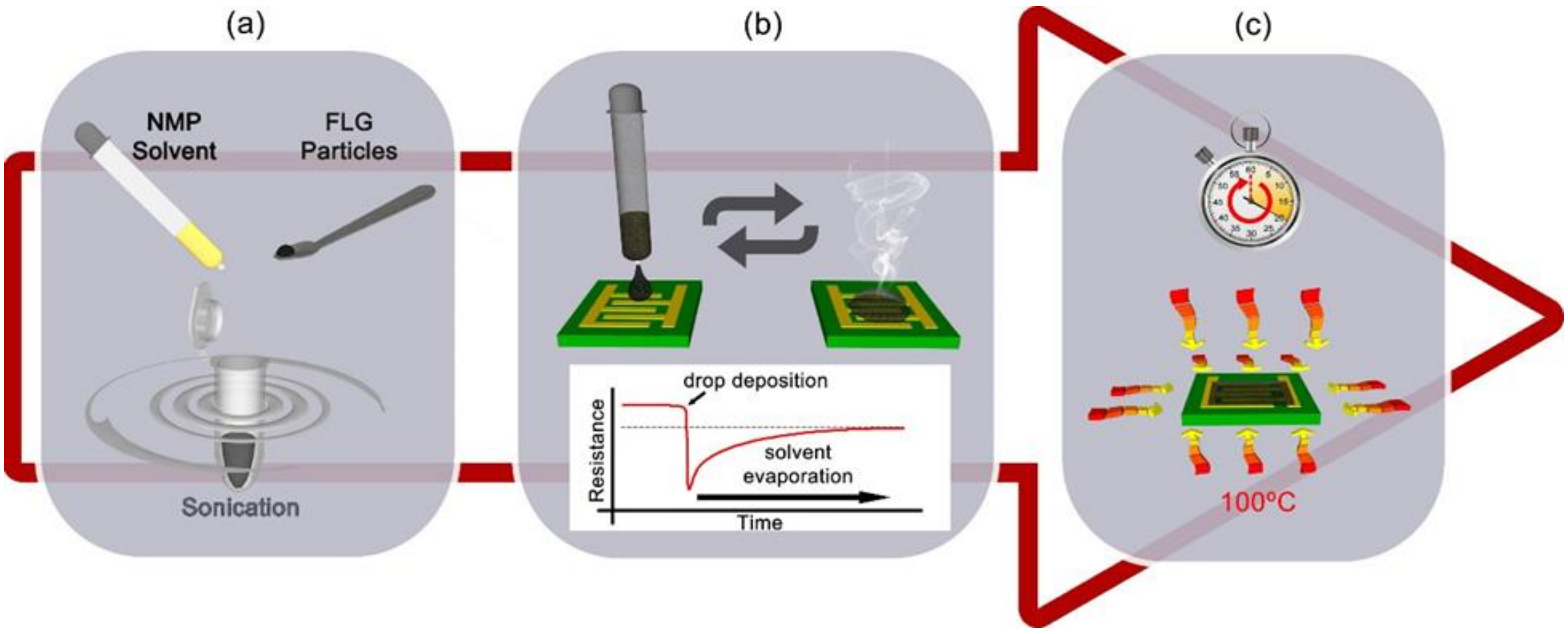

*Figure 4. 6 FLMG-based device fabrication. Rep. from* [89].

The deposition process was monitored and controlled by measuring the electrical resistance between the electrodes. The drop-casting process was continued until a pre-established electrical resistance was attained. The device was then placed in an oven at 100 Cº to evaporate the solvent fully (Figure 4. 6c) [94].

### *4.4.2 Results And Discussion.*

The first experiment aimed to study the response of the devices to $NO_2$ and the differences between the four samples. The sequence consisted of a single cycle with an exposure phase of 30 minutes followed by a recovery phase of 60 minutes. The device was exposed to 0.5 ppm of $NO_2$.

As shown in Figure 4. 7, all four graphene-based gas sensors responded to the presence of $NO_2$ by exhibiting a change in device resistance, as described previously. Specifically, the graphite sensor showed a shallow response of approximately 2.9% with a low response and recovery times. In contrast, the milled samples showed a much higher response of 14.4%.

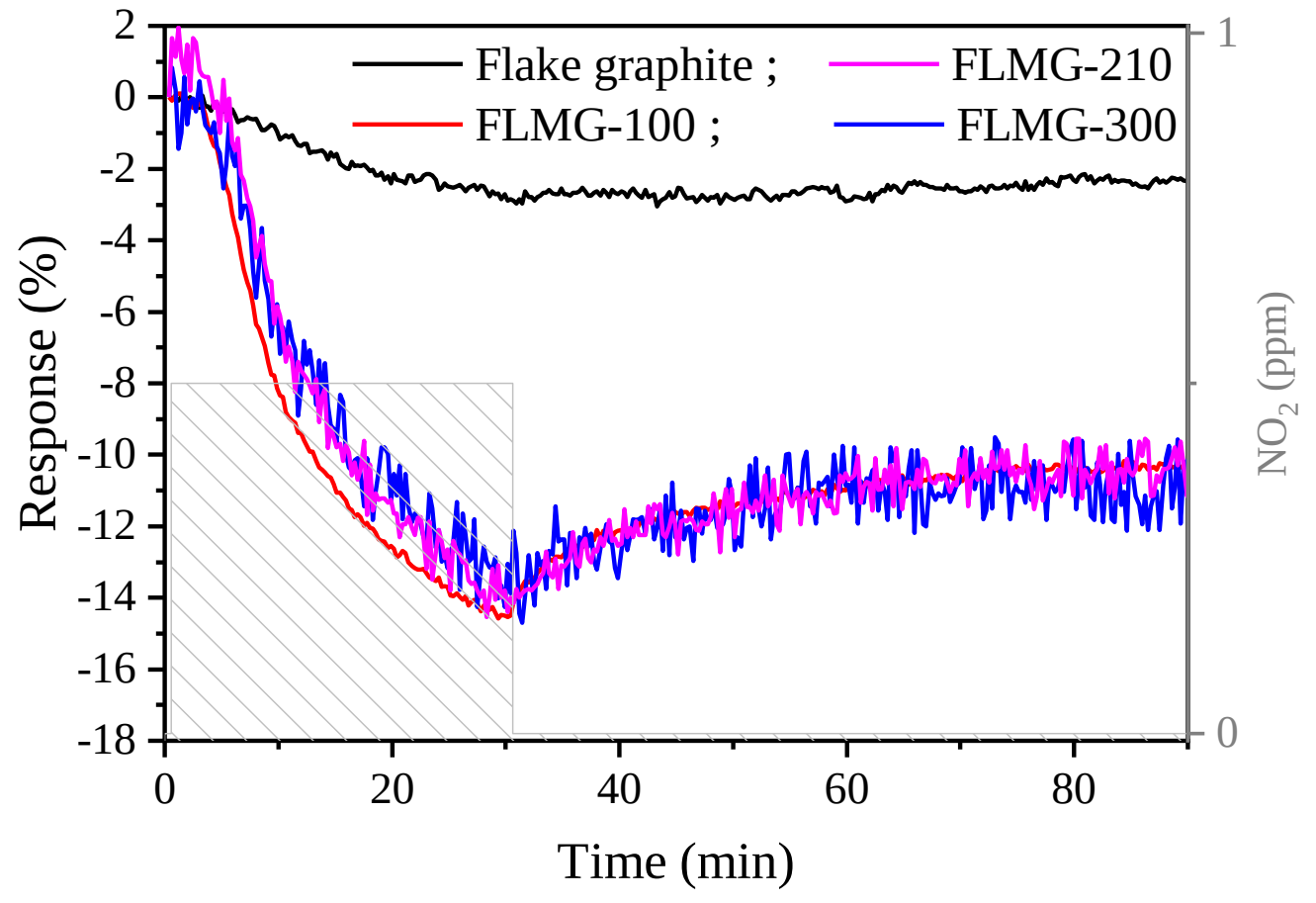

*Figure 4. 7 Response of Graphite, FLMG-100, -210, and -300 to 0.5 ppm of $NO_2$.*

In terms of recovery, all milled samples showed a slow recovery that did not reach the device's initial state, consistent with the partial recovery phenomenon previously described. However, the graphite sample showed no significant recovery after exposure to $NO_2$.

These results indicated that graphite's milling process improved the sensors' overall performance to $NO_2$. The improvements in sensitivity can be attributed to changes induced by the milling process on the graphite material, which resulted in reduced particle size and the formation of an agglomeration of few-layered graphene with a mesostructure that significantly increased the material's surface area, leading to improved interaction with the gas. Additionally, the milling process altered the crystallinity of the graphite, generating defects that serve as binding sites for the gaseous molecules and enhancing the material's reactivity.

However, differences in electrical noise were observed between the milled samples, depending on the milling time. The FLMG-100 sample had less noticeable noise compared to the FLMG-210 and FLMG-300, possibly due to less structural defects from a shorter milling process. Consequently, the remainder of the study primarily focuses on the 100-minute milled sample.

These results demonstrate the potential of FLMG-based gas sensors for detecting $NO_2$ and the importance of the milling process time to achieve the desired sensitivity and noise performance.

To improve the recovery time of the milled graphite sensors, a 275 nm UV diode with an optical output power of 11.5 mW was installed to facilitate the desorption of the $NO_2$ molecules from the active material, as previously reported in the literature [67,80]. The wavelength was chosen according to the absorbance peak observed for FLMG during the UV-Vis absorption characterization.

In Figure 4.8, UV irradiation's positive impact on the device’s performance is evident. Specifically, it markedly enhanced the recovery phase, while unexpectedly also improving the exposure phase response by 10% (up to 16%). Therefore, UV irradiation was retained throughout the FLMG-100 study.

It is worth noting that even under UV irradiation, the device suffered from partial recovery, i.e., 60 minutes after the exposure ended, the device’s resistance did not achieve its initial value.

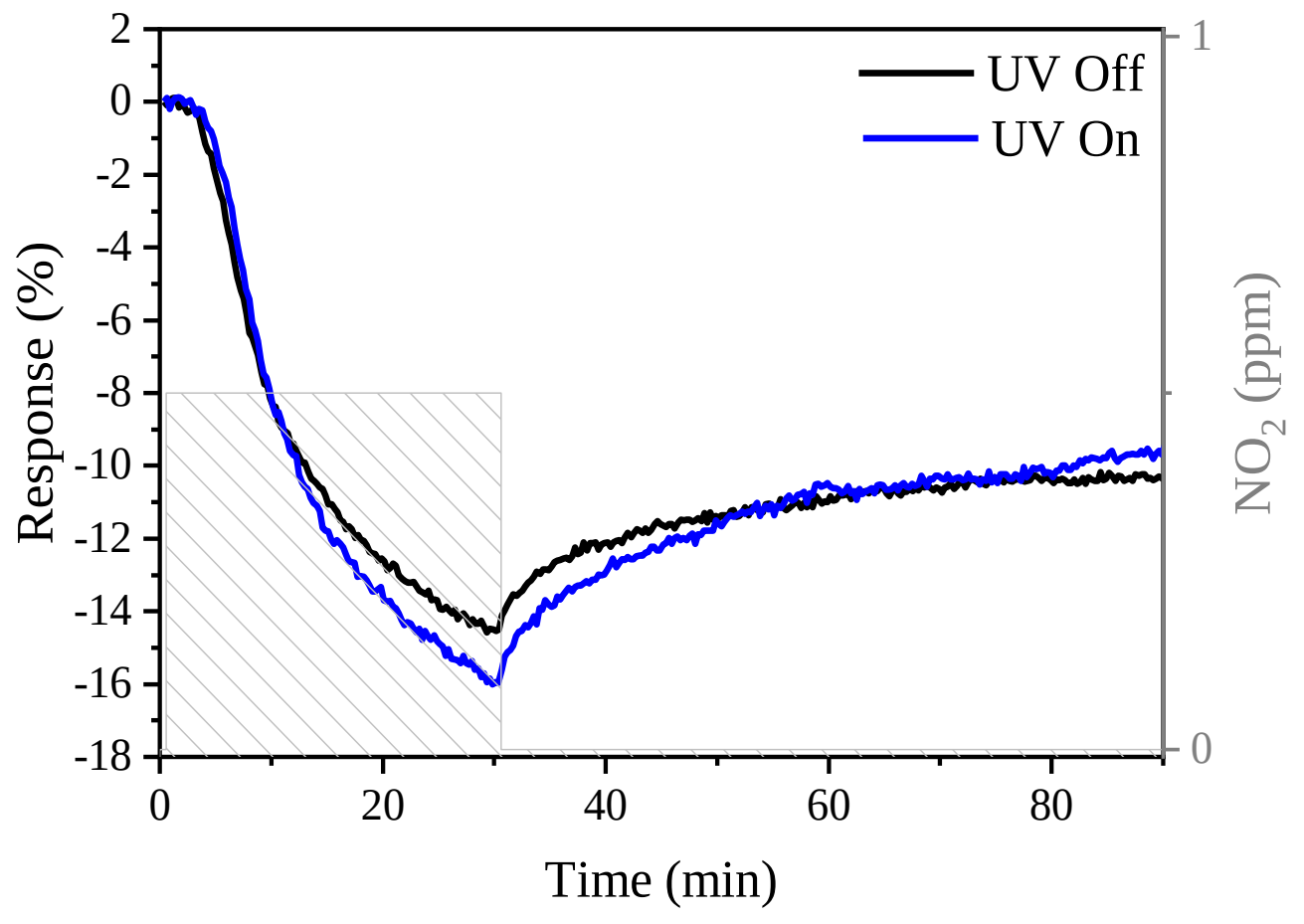

*Figure 4. 8 Response of FLMG-100 to 0.5 ppm of NO2 under UV vs no UV.*

The following test aimed to study the sensitivity and LoD of FLMG-100. For this test, the sequence consisted of eight cycles with an exposure phase of 10 minutes followed by a recovery phase of 50 minutes. The sensor was exposed to cycles with a concentration ranging from 0.2 to 0.5 ppm of $NO_2$. Each concentration was repeated twice.

As shown in Figure 4. 9, the calibration test for the gas sensor was conducted by exposing the device to increasing concentrations of the target gas. However, during the calibration test, the partial recovery phenomenon caused a non-linear response of the sensor to the gas concentration. This is due to the device not fully recovering during the recovery phase, resulting in a hysteresis effect that led to a downward drift in the measurement. Despite the reduced sensitivity caused by partial recovery, the LoD was determined to be 25 ppb, which is within the range of $NO_2$ exposure limits described in the introductory section of this chapter [94].

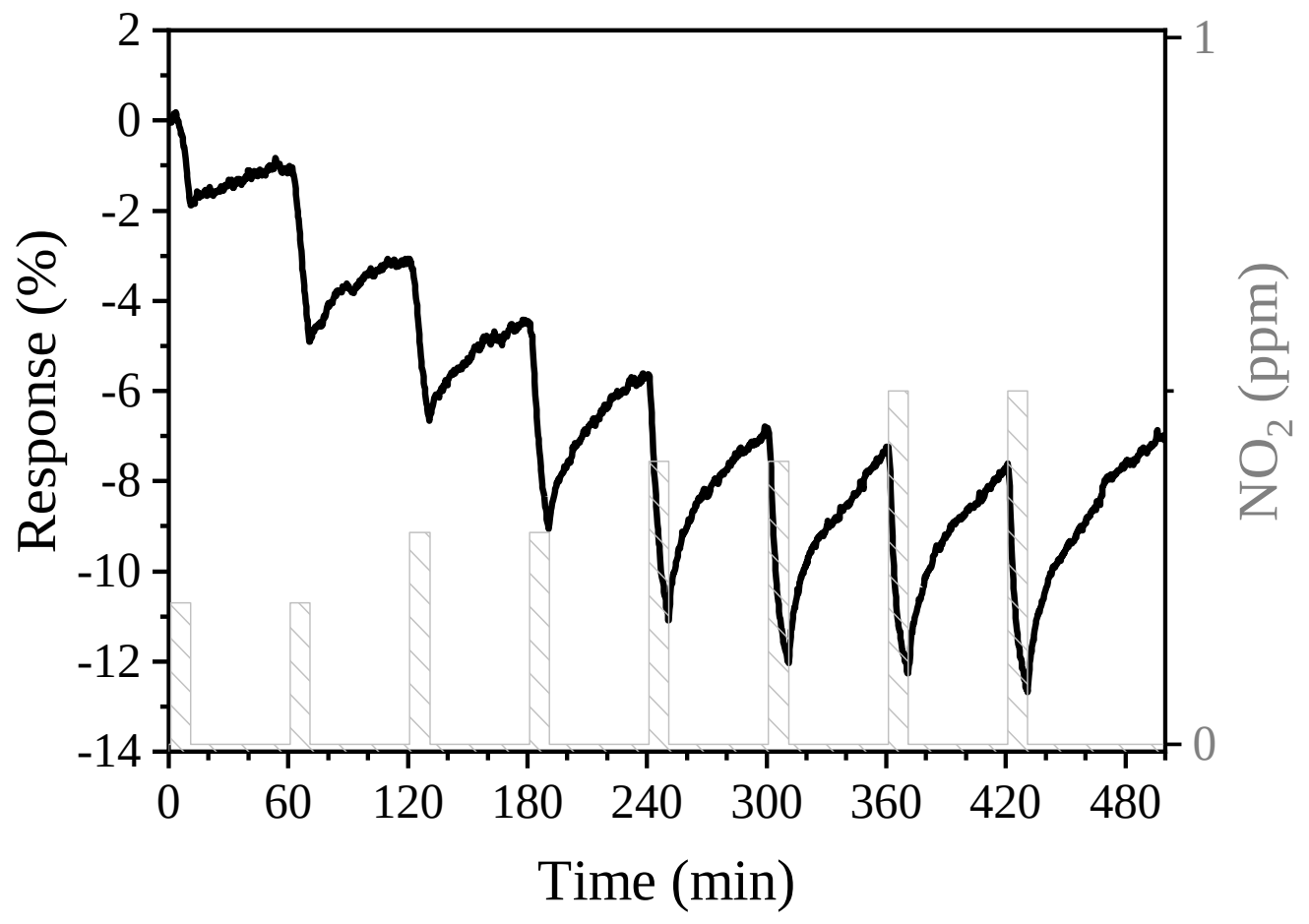

*Figure 4. 9 Response of FLMG-100 during the calibration test to $NO_2$ under UV.*

The ability to operate at room temperature, combined with its sub-ppm capabilities and limit of detection well below the recommended exposure levels for $NO_2$ demonstrates that FLMG, particularly FLMG-100, is a promising material for $NO_2$ sensing applications.

To ensure the successful development of the presented device, it is important to evaluate its selectivity, or ability to operate without interference from common molecules or other gas pollutants. While the device has demonstrated its capabilities under oxygen-containing atmospheres, the presence of reducing molecules such as $NH_3$ may interfere with the device's response to $NO_2$, potentially decreasing the maximum response or neutralizing the response altogether. This is because reducing molecules modify the device's resistance in the opposite way of oxidant molecules. A combination of both $NO_2$ and $NH_3$ could result in a null response for a given sensor with low selectivity, leading to the exposure of two dangerous gases without an alarm.

Therefore, the FLG-100 sensor was subsequently exposed to 50 ppm of $NH_3$ (maximum odor threshold). The sequence was designed similarly to that used to evaluate the response of the device to $NO_2$, i.e., a baseline plus a 30-min exposure phase and 60-min recovery phase.

Figure 4. 10 shows the exposure of FLMG-100 to 50 ppm resulted in a low response of around 1.5%. Therefore, it can be concluded that even high concentrations of $NH_3$ will not interfere with the sub-ppm detection of $NO_2$. Thus, this extremely high selectivity of the FLG-100 sensor towards $NO_2$ is highly advantageous for gas-sensing applications like air quality monitoring.

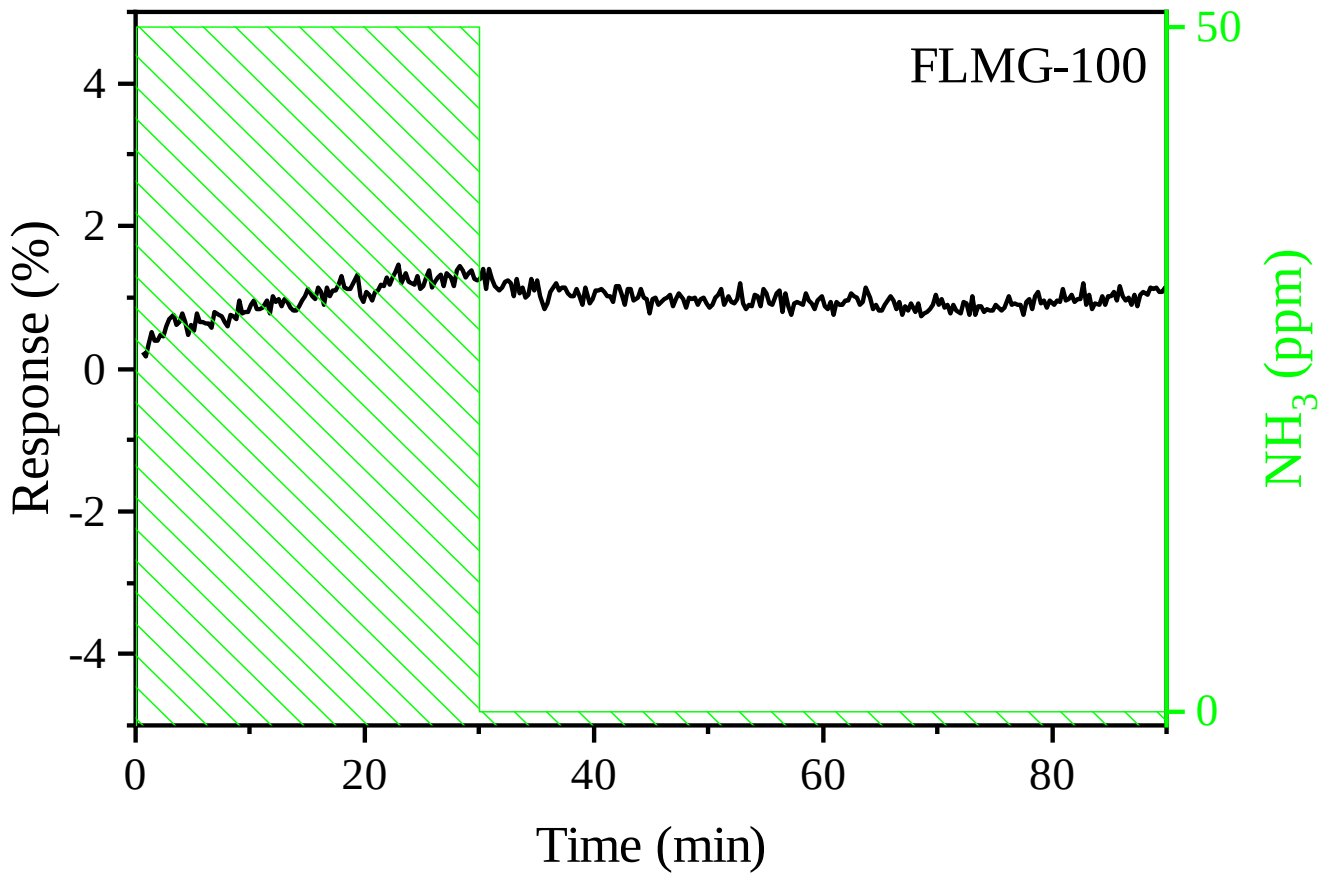

*Figure 4. 10 Real time response of FLMG-100 to 50ppm of $NH_3$ under UV.*

The impact of humidity on the gas sensing performance of the FLG-100 sensor was also assessed. As shown in Figure 4. 11, the sensor exhibited a similar sensitivity to 0.5 ppm $NO_2$ in environments with relative humidity levels of 0%, 10%, and 33% at 25°C. Moreover, the recovery process was slightly improved at higher humidity levels.

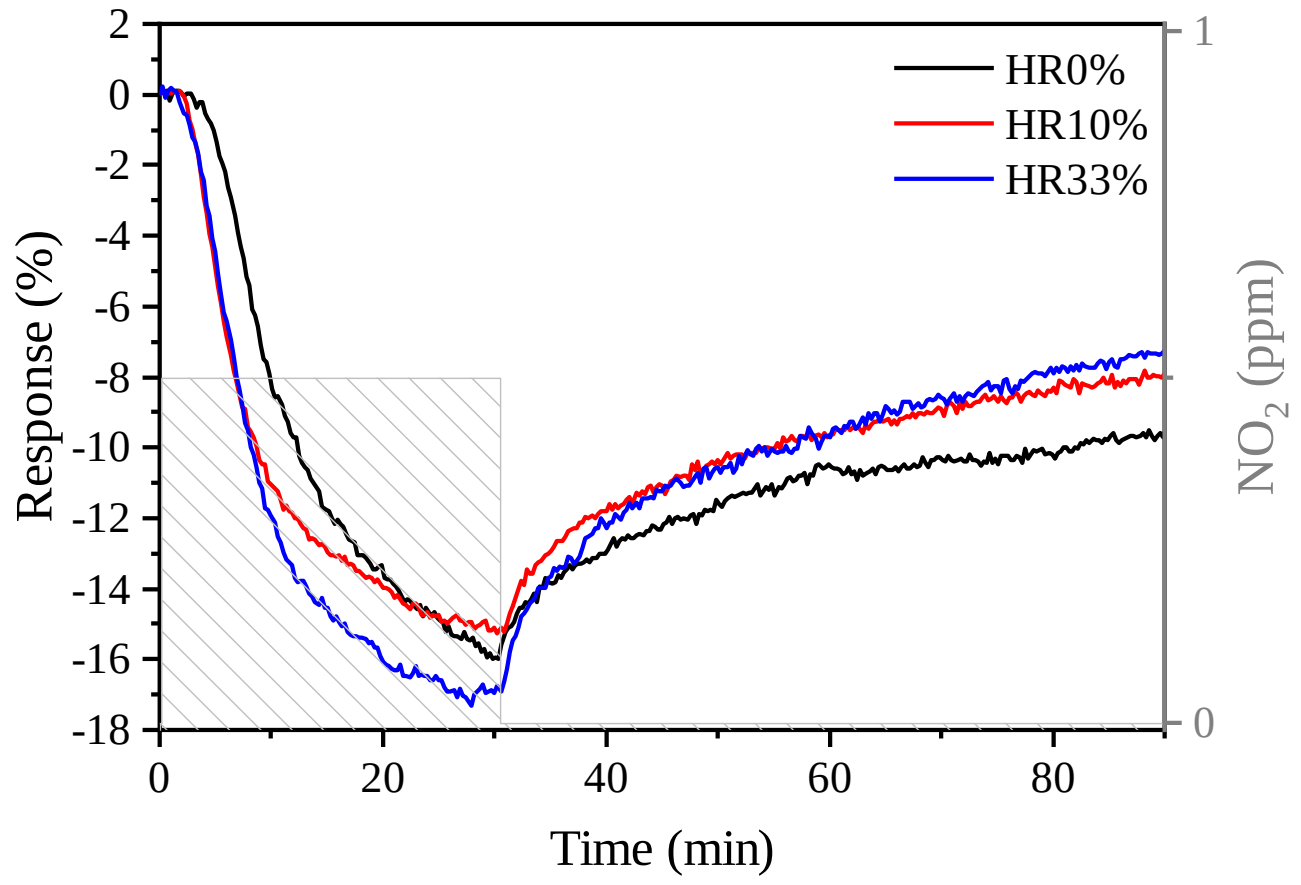

*Figure 4. 11 Response of FLMG-100 to 0.5 ppm of $NO_2$ under UV for different levels of relative humidity: 0%, 10% and 33%.*

These results highlight the sensor's versatility and robustness, as it can operate effectively across a wide range of environments with regular humidity levels without sacrificing sensitivity or recovery times. This makes the FLMG-100 sensor an attractive option for various multidisciplinary applications.

The behaviour of this device was explained in ref [89]. The sensitivity towards $NO_2$ was related to the vacancy-like defects observed from the Raman study for the 100-min milled FLMG sample, presented in Chapter 2. The affinity towards $NH_3$ and $H_2O$ was related to the presence of functional groups (hydroxyl, carboxyl, among other radicals) adhered to the surface of the material as observed from the infrared analysis, also presented in Chapter 2. The low affinity for water adhesion was also verified from the superhydrophobic behaviour of the FLMG samples.

Overall, the FLMG-100 sensor shows a high response to sub-ppm $NO_2$ concentrations due to the combination of mesoporosity and the structural defects generated in the active material by the milling synthesis. Furthermore, some major benefits of the proposed manufacturing method to develop $NO_2$ sensors are its reduced costs and its potential to be produced at a large scale.

## *4.5 Effect Of UV On Multi-Layer Graphene Chemiresistive Sensors.*

The effect of UV over the FLMG-100 performance, despite effectively increasing both the response and recovery behaviour. A 10% increase in the maximum response to 0.5 ppm of $NO_2$ (from 14.5 – 16%) could be attributed to the use of the 275 nm irradiation. However, it can be argued that the effect of UV is limited to the irradiated surface and that FLMG, being a mesoporous structure, has a sensitive amount of active surface that did not suffer the effect of UV irradiation during the device's operation.

Therefore, a second series of studies was carried out using a different GBM, multi-layer graphene (MLG). This work was part of a collaboration with Sten Vollebregt, Filiberto Ricciardella, and Leandro Sacco from the Department of Microelectronics at Delft University Technology (TU Delft).

This study was focused on two hypotheses, that UV would have a positive effect in boosting the operation of MLG-based devices and that this effect would be more pronounced than the observed for the FLMG-based devices.

MLG refers to a graphene material consisting of several layers of graphene stacked on top of each other. As opposed to FLMG the stacking has a larger degree of order and exposes a less rough surface. The second hypothesis is thus based on the idea that due to the high level of stacking order on MLG, most of the surface exposed to the gas will also be exposed to UV irradiation.

The MLG samples used in this work were grown by Vollebregt et al. using a chemical vapour deposition (CVD) technique.

### *4.5.1 Device Preparation.*

The effects of temperature control in CVD growth of graphene-based materials were presented in Chapter 2 but a brief recap is in order. In CVD, a substrate is placed in a furnace and heated to high temperatures, typically between 800-1000°C, and a hydrocarbon gas is introduced into the furnace. The hydrocarbon gas decomposes on the surface of the substrate, and the carbon atoms recombine to form graphene.

At higher temperatures, the decomposition rate of the hydrocarbon gas increases, leading to a higher concentration of carbon atoms on the substrate surface. This higher concentration promotes the nucleation of graphene, resulting in smaller and more densely packed graphene islands. As the temperature decreases, the carbon concentration on the substrate surface decreases, resulting in fewer and larger graphene islands.

However, it is vital to note that temperature is not the only variable affecting the growth process. Factors such as the catalyst's characteristics, including its thickness and solubility, as well as the growth duration, play equally crucial roles. Although these aspects extend beyond the purview of this thesis, acknowledging their impact is important for a comprehensive understanding of the process.

Furthermore, the growth rate of graphene is also affected by the temperature. At higher temperatures, the carbon atoms have higher mobility and can diffuse more easily, resulting in faster growth rates. At lower temperatures, the carbon atoms have lower mobility and diffusion rates, resulting in slower growth rates.

Therefore, temperature will affect both nucleation and growth rates, which in turn will affect the characteristics of the resulting MLG. In particular, the temperature will affect the defect ratio, which, as previously discussed, will affect the sensing performance [60]. With this context, two different MLG materials were tested.

MLG was synthesized by CVD on a pre-patterned Mo catalyst in an AIXTRON BlackMagic Pro reactor [95]. 20 sccm of methane ($CH_4$) was used as a carbon feedstock for 20 minutes in $Ar/H_2$ atmosphere at 25 mbar, and a growth temperature of 935°C [58,74,95,96].

The material was characterized with Raman spectroscopy, the results, presented in Chapter 2, revealed an I(D)/I(G) ratio of 0.266, indicating a relatively low defect contribution and an I(2D)/I(G) ratio of 0.366. In addition, AFM revealed an average roughness ($R_a$) of 17.4 ± 0.6 nm [91].

The sensing devices were fabricated by adopting the transfer-free process, further detailed in ref [95]. Through a few lithographic steps, a sputtered and patterned Mo layer (50 nm) was wet-etched after the growth of MLG. That way, MLG dropped on the $SiO_2/Si$ substrate at the pre-defined positions. Next, the MLG was contacted using 10/100 nm Cr/Au deposited using e-beam evaporation and patterned using a lift-off process. The device was named MLG935.

The results of the fabrication process are displayed in Figure 4. 12 through SEM images. The MLG strip is clearly distinguishable from the gold electrodes due to its darker colour. The contact between the two materials appears seamless, as evidenced by the image on the right.

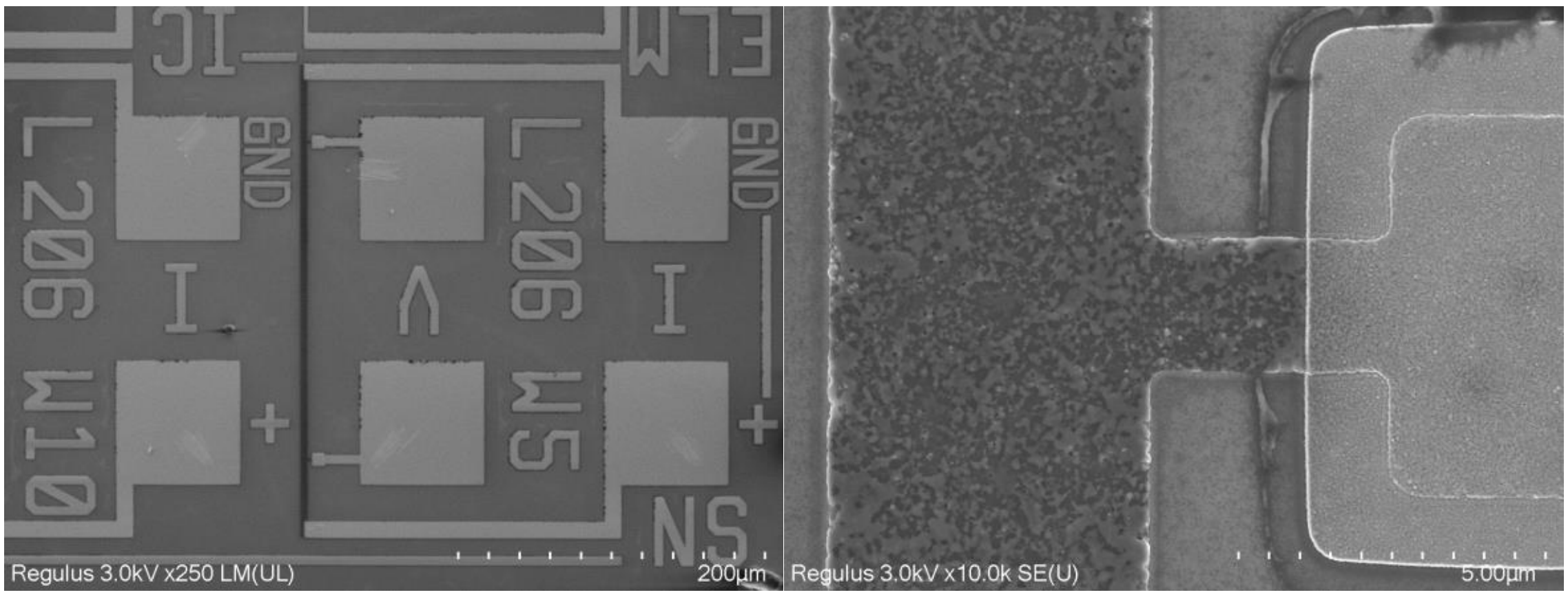

*Figure 4. 12 SEM images of the MLG935 device. The left image shows the complete MLG strip that acts as active part of the sensor. The right image shows a magnification where the strip is contacted to the electrode.*

*4.5.2 Results And Discussion.*

To evaluate the initial hypotheses, three gas measurement sequences were designed. Each sequence consists of a baseline step, where only carrier gas is flushed for 60 minutes (not fully shown in the figures); and cycles with a 10-minute exposure phase and a 20-minute recovery phase. All tests were performed under standard ambient temperature and pressure conditions. Each test specification is described as follows:

Test1 was designed to evaluate the sensing capability of MLG935 toward $NO_2$. It consisted of four consecutive cycles of 1 ppm of $NO_2$ under no UV irradiation, defined as UV@OFF.

Test2 was designed to study the effect of UV irradiation on the device. It consisted of four consecutive cycles of 1 ppm of $NO_2$ under UV irradiation at a power density of 68 $W/m^2$, defined as UV@100.

Test3 was designed to calibrate the device. It consisted of consecutive cycles with increasing $NO_2$ concentrations, from 0.2 – 1 ppm in 0.1 ppm steps, under UV@OFF and UV@100.

Thus, Test1 and Test2 are analogous to the response evaluation of FLMG presented in Figure 4. 7 and Figure 4. 8. It should be noted however that there are differences in the cycle's definition, i.e., exposure and recovery times. As previously discussed, sequences are designed according to the behaviour of the device and the purpose of the experiment. Test3 is analogous to the calibration test performed in Figure 4. 9.

The results for Test1 are presented in Figure 4. 13a. Under exposure to $NO_2$, the device displayed a decrease in resistance, indicating a p-type behaviour of the MLG [52,54,58–60]. Essentially, when the material is exposed to an oxidizing or electron acceptor molecule, a decrease in the material resistance is observed, similar to the behaviour observed for the FLMG devices. The response systematically decreased after each cycle. It was also noticed that, during the recovery phase, the device did not recover its initial conditions.

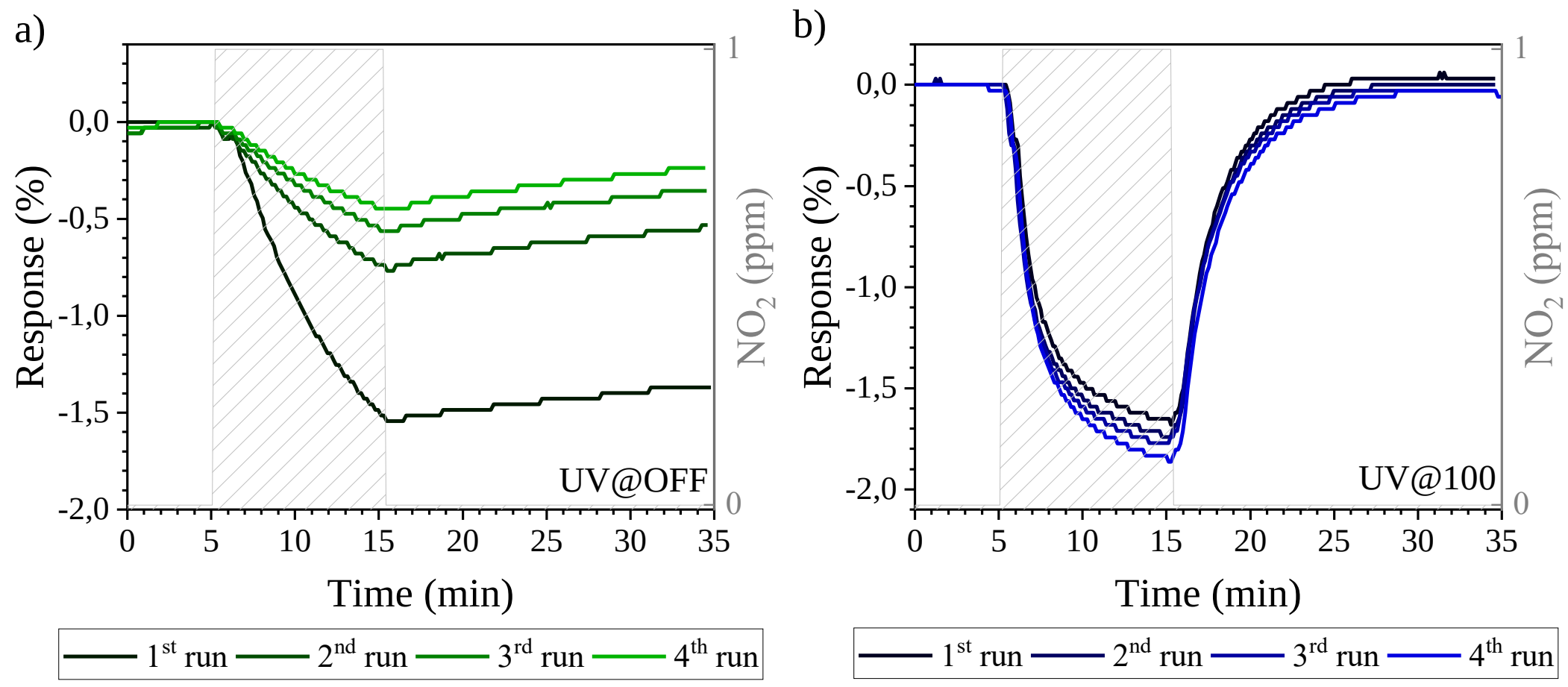

*Figure 4. 13 Response towards 1 ppm of $NO_2$ of (a) MLG935 under UV@OFF, and (b) under UV@100.*

To study the effect of UV irradiation on the device's operation, Test2 repeated the conditions of Test1 under UV@100. Results are presented in Figure 4. 13b. It was observed that, under UV irradiation, the response of the device remarkably tended to a steady state during the exposure phase differently from Test1. More importantly, the device showed a larger $R_{Max.}$ than during Test1. A 290% increase in the mean $R_{Max.}$ was observed, from 0.58 ± 0.15% to 1.70 ± 0.05% under UV irradiation. No significant decrease in the response was observed after each cycle.

The most relevant effect of UV irradiation was the full recovery during the recovery phase. Under UV@100, MLG935 could recover to their initial conditions approximately during the first 10 minutes of the recovery phase. Furthermore, no hysteresis is observed after each cycle of exposure at 1 ppm of $NO_2$, indicating that UV irradiation drastically improves the reproducibility of each exposure step and the reliability of both devices. The effects of the UV irradiation in terms of response and reproducibility are visible in Figure 4. 14, where the responses of the tested devices after each cycle with and without UV irradiation are reported.

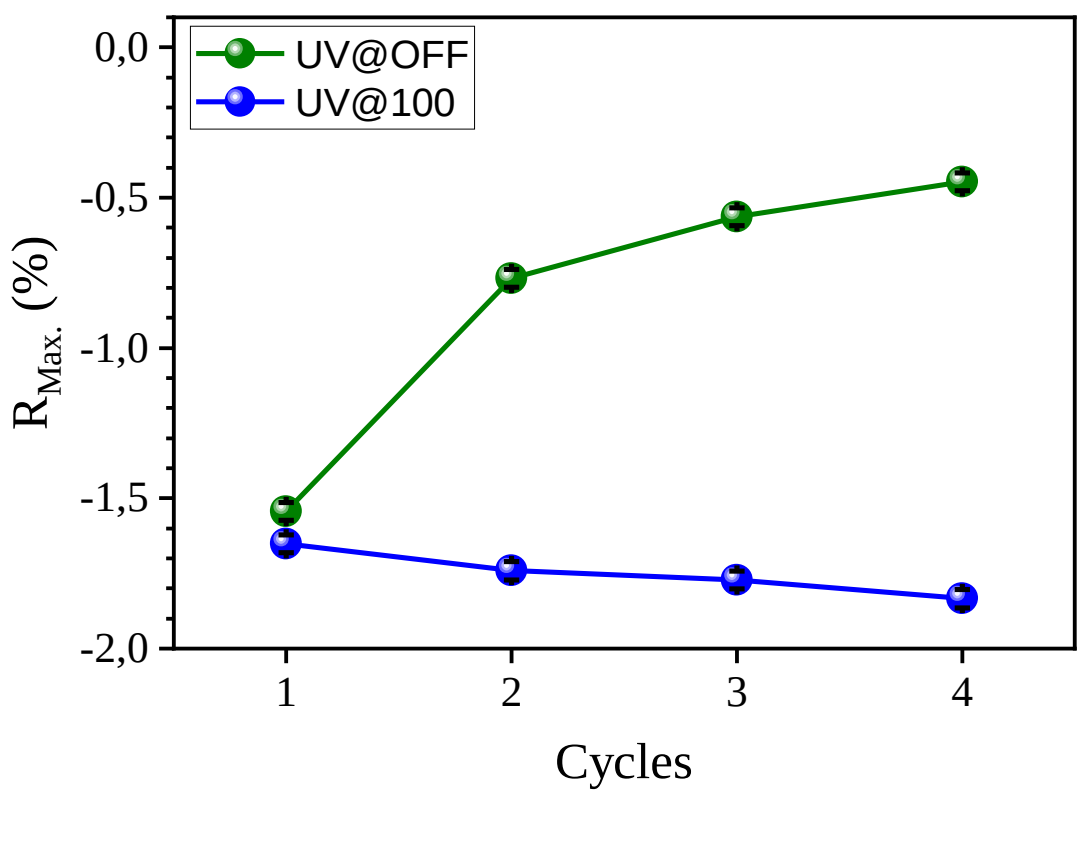

*Figure 4. 14 A comparison of the effect of UV irradiation in the response after each cycle.*

The response time was evaluated using the $\tau_{90}$ parameter. The $\tau_{90}$ values were 8.2 ± 1.0 for UV@OFF (Test1) and 5.6 ± 0.3 UV@100 (Test2). Meaning that $\tau_{90}$ was approximately 70% lower under UV irradiation. The results therefore remarkably show that the sensor achieves its maximum response at a faster rate when exposed to UV irradiation. In the case of UV@OFF, 90% of the $R_{Max.}$ value is achieved roughly after 8 min, indicating a near-to-linear response under the experiment's conditions. For UV@100, instead, during the first minutes of exposure, the response varies with a steeper slope followed by a more gradual slope, typically related to the saturation process where the device's response will no longer vary significantly. The 90% of the $R_{Max.}$ is then reached after about 6 min.

The effect of UV irradiation previously observed was further confirmed in Test3 (Figure 4. 15). Under dark conditions, i.e., UV@OFF, a downward drift with a systematic decrease of the response displayed that MLG935 suffered partial recovery, starting from the first exposure at 200 ppb. Under UV irradiation, no drift was observed, and the devices started each subsequent exposure phase from the initial conditions, meaning that no hysteresis was observed from two subsequent cycles.

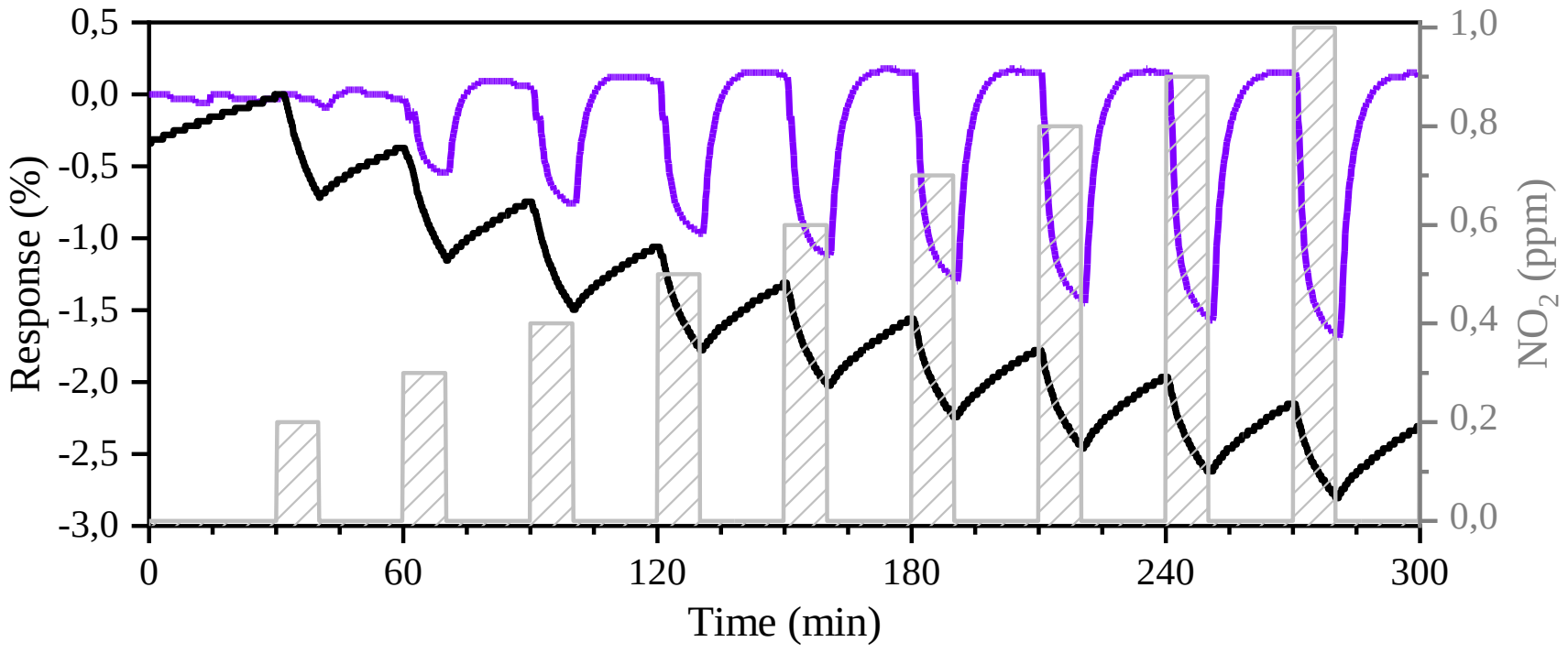

*Figure 4. 15 Real time response of MLG935 towards different concentrations of $NO_2$ under UV@OFF (black line) and UV@100 (purple line).*

To calibrate the device and calculate the LoD, the $R_{Max.}$ toward each concentration was extracted from Test3. A linear fit of these values provided the sensitivity which was then used to calculate LoD according to Equation 4. 3. The LoD error ($\Delta LoD$) was calculated using uncertainty propagation.

Figure 4. 16 presents the data for LoD calculation. LoD could only be calculated under UV@100 conditions, with a sensitivity of 1.9 %/ppm (absolute value) and an extrapolated LoD value of 31 ± 1 ppb. For UV@OFF, due to the partial recovery, the response did not increase with the concentration, and the linear fit does not converge to a coherent sensitivity so that the LoD can be calculated.

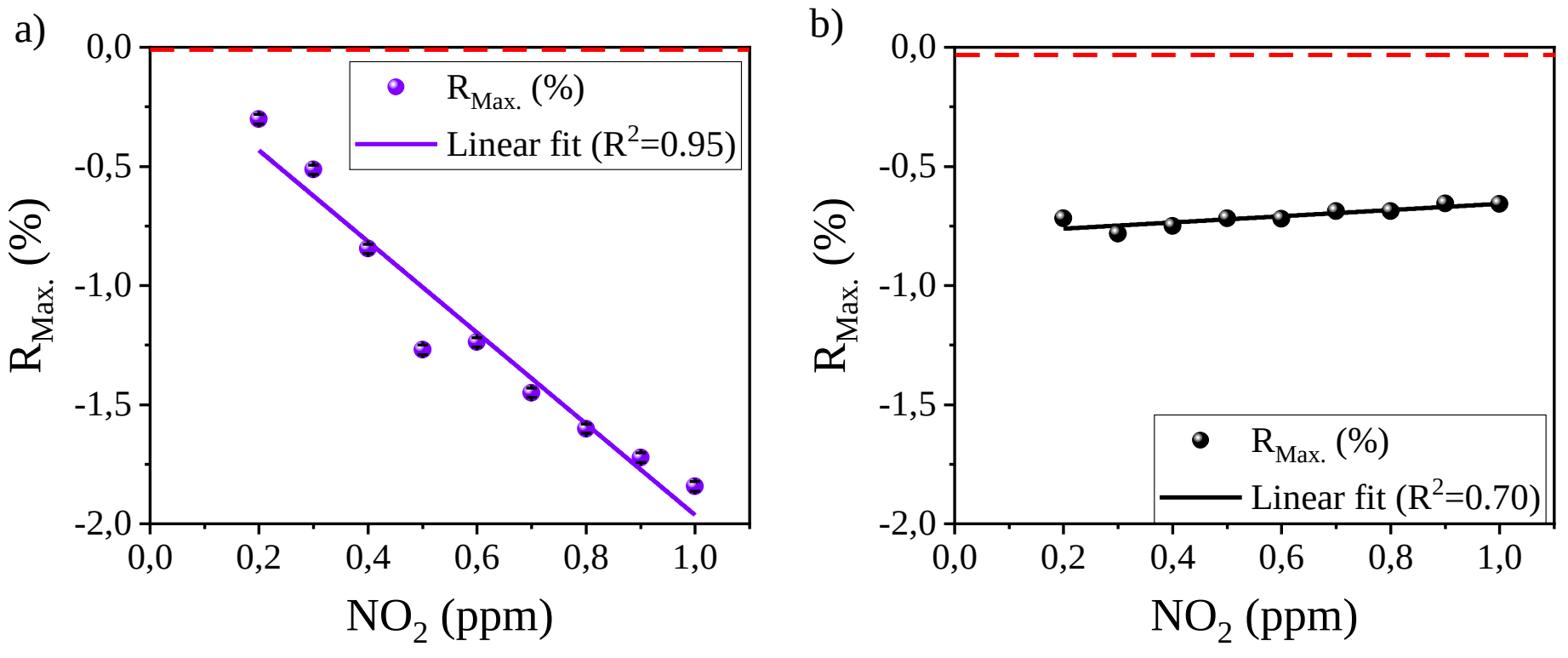

*Figure 4. 16 Calibration curve for MLG935, towards different concentrations of $NO_2$ under (a) UV@100 (purple line) and (b) UV@OFF (black line). The dotted red line represents the RMSnoise value used to calculate the LoD.*

Based on the results, it can be concluded that MLG-based devices are capable of detecting low concentrations of $NO_2$ in air. This is particularly noteworthy as previous records of this device only demonstrated detection in nitrogen atmospheres [60,74]. The influence of the gas carrier on the device's response has been previously demonstrated, with $NO_2$ detection being more effective in a nitrogen atmosphere due to the interference of oxygen in air.

While the MLG device exhibited a p-type response similar to that of FLMG-based devices, $NO_2$ acted as an electron acceptor, trapping electrons from the conduction band, and decreasing the MLG's resistivity [52,54,58,60], there were notable differences observed between the two. Specifically, FLMG-based devices demonstrated a significantly higher $R_{Max.}$ to $NO_2$. This could be due to the extra active surface area provided by the mesoporous structure, or a higher reactivity related to defects or chemical composition. Despite these differences, the results suggest that MLG-based devices hold great potential for commercial gas-sensing applications.

Furthermore, the results of the study confirmed the first hypothesis, UV irradiation had a positive effect on, $R_{Max.}$, the response rate, and recovery time of the sensors. Moreover, UV irradiation allowed for full recovery of the devices during the recovery phase. The improved recovery due to UV irradiation allowed to calibrate the response of the sensors under practical conditions, leading to the determination of a limit of detection below the recommended limits of exposure.

As for the second hypothesis, the effect of UV irradiation was significantly higher on MLG-based devices than on FLMG-based devices. As mentioned earlier, the mesoporous structure of the material allows for a considerable surface area that can interact with gases. However, it should be noted that not all this surface area is exposed to UV irradiation during the sensing process. Especially, FLMG-based devices did not achieve full recovery even under UV irradiation.

It has been clearly demonstrated that the use of UV irradiation during the testing of GBM-based devices is crucial for developing commercial applications since it helps to overcome their previously discussed limitations. However, concerns about increased power consumption and a potential decrease in device lifetime prompted a second series of tests on MLG935.

Test4 was designed as four consecutive cycles with 1 ppm of $NO_2$, analogously to Test2, with the difference that four different UV configurations were used, i.e., irradiation at half of the maximum power density of the diode (UV@50), and irradiation used solely during the recovery phase (UV@DES). The results are presented along with the results of configurations at UV@OFF and UV@100.

Figure 4. 17 shows the results of Test4. For UV@OFF, MLG935 exhibited the lowest $R_{Max.}$, 0.73%. In addition, the device exhibited poor recovery and suffered a downward drift with a decrease in its sensing performance.

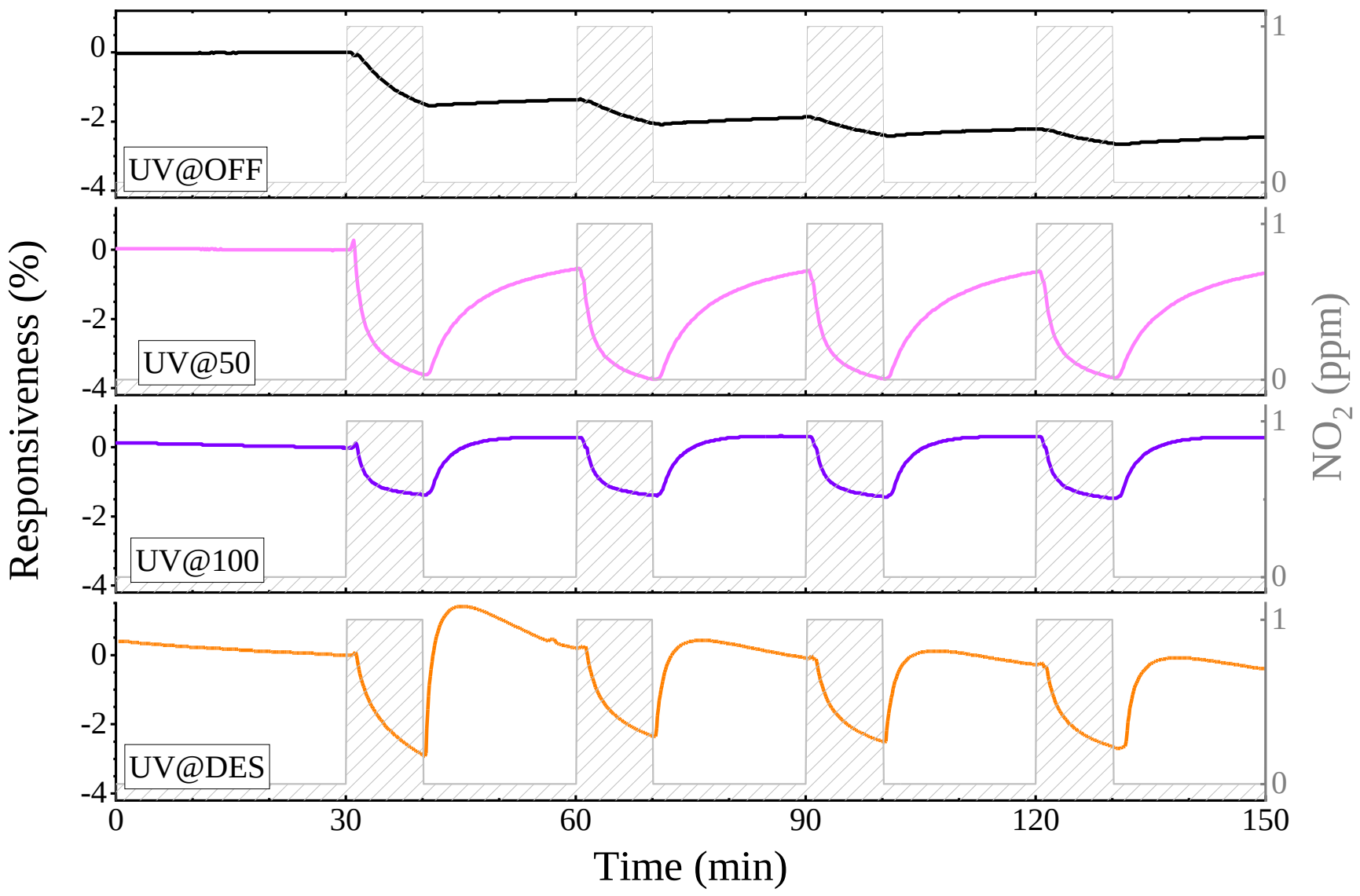

*Figure 4. 17 Test4, real time response of MLG935 towards 1 ppm of $NO_2$ under different UV irradiation settings.*

For UV@50, MLG935 exhibited extraordinary response. For all the cycles, the $R_{Max.}$ remains above 3%, which remarkably results in the highest $R_{Max.}$ observed in this work (see Figure 4. 18). During the recovery phase, the sensor fully recovered its initial conditions.

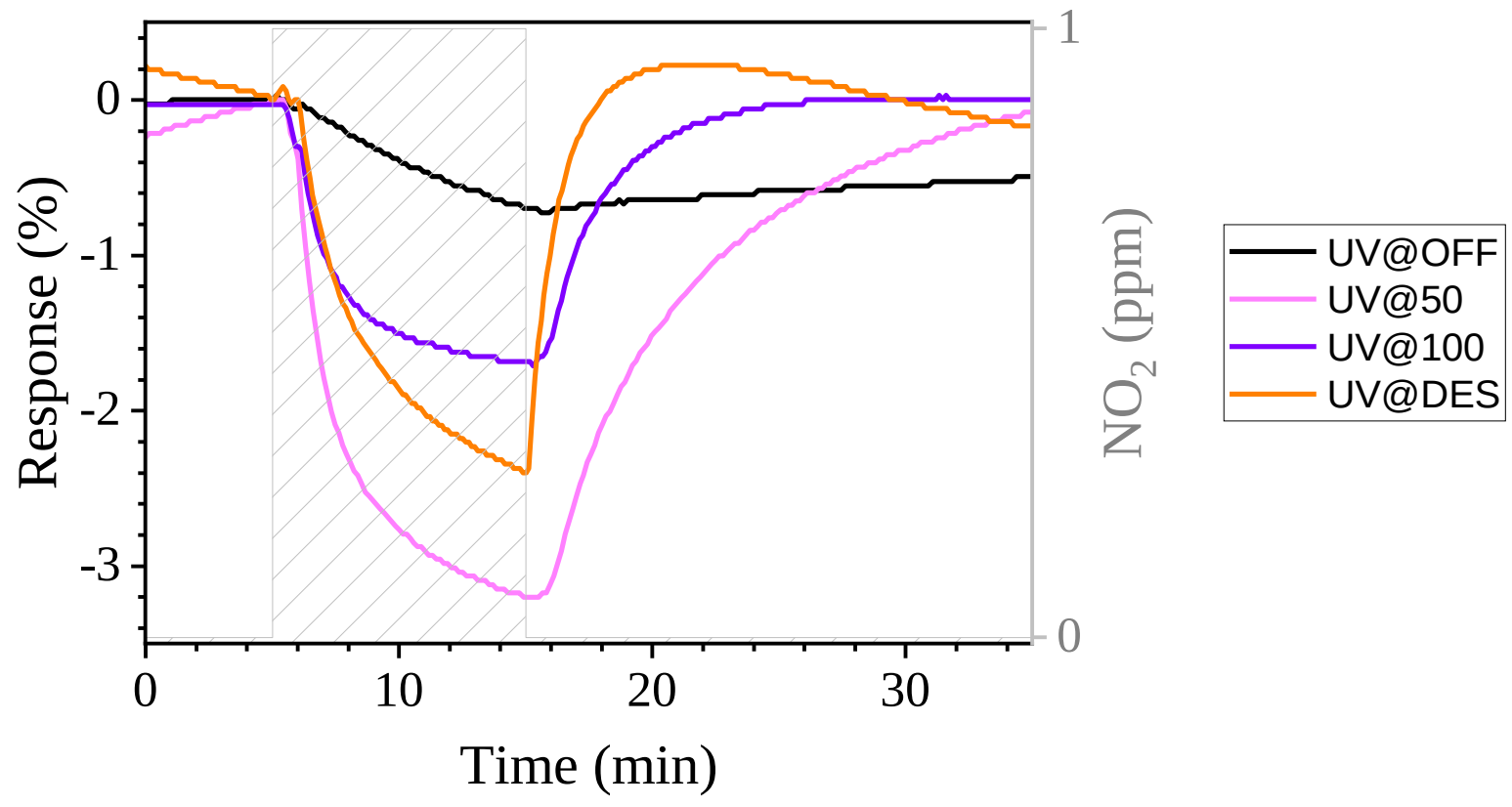

*Figure 4. 18 Second cycle from Test4 extracted and normalised.*

For UV@100, MLG935 exhibited a response of around 1.70 % for all cycles. Unlike under other UV conditions, for UV@100 the response decreased rapidly after the first few minutes of exposure, approaching a plateau around the maximum value. The sensor fully recovered its initial conditions, and excellent reproducibility was observed across the different cycles.

Finally, For UV@DES, MLG935 exhibited a large $R_{Max.}$, around 2.5 %. The response is more pronounced than under UV@OFF, achieving a behaviour between UV@100 and UV@50, despite being irradiated only during the exposure phase (see Table 4. 4). During the recovery phase, the recovery reached ~90% of the initial conditions.

*Table 4. 4 $R_{Max.}$ and $\tau_{90}$ of MLG935 towards 1 ppm of $NO_2$ under different UV irradiation configurations with graphical representation.*

| $NO_2$ | UV setting | $2^{nd}$ cycle | $3^{rd}$ cycle | $4^{th}$ cycle | Mean ± SD |
|---|---|---|---|---|---|
| $R_{Max.}$ (%) | UV@OFF | 0.73 | 0.56 | 0.44 | 0.58±0.15 |
| | UV@50 | 3.21 | 3.12 | 3.04 | 3.12±0.08 |
| | UV@100 | 1.69 | 1.66 | 1.75 | 1.70±0.05 |
| | UV@DES | 2.57 | 2.43 | 2.41 | 2.47±0.09 |
| $\tau_{90}$ (min) | UV@OFF | 8.75 | 9.03 | 8.44 | 8.74±0.30 |
| | UV@50 | 5.65 | 5.92 | 6.38 | 5.98±0.37 |
| | UV@100 | 5.62 | 5.28 | 5.44 | 5.45±0.17 |
| | UV@DES | 7.63 | 7.67 | 7.62 | 7.64±0.02 |

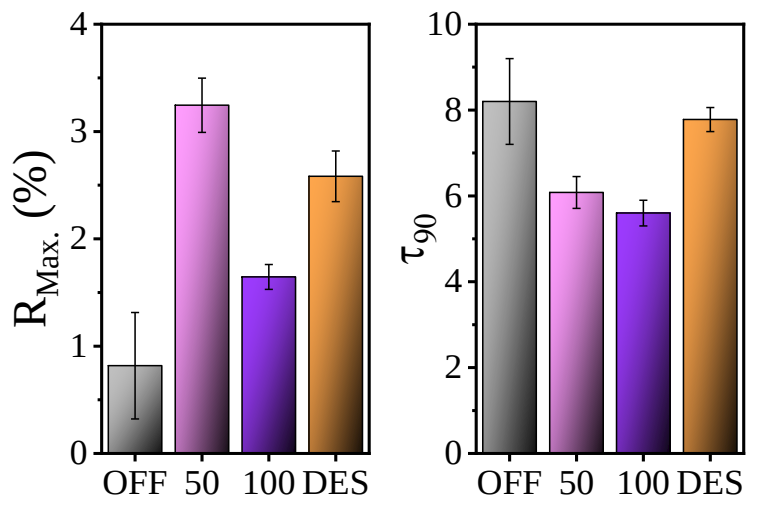

Regarding the response time, UV@OFF and UV@DES exhibited a similar $\tau_{90}$, around 8 minutes. The fastest time was achieved for UV@100, at around 5.5 minutes, followed by UV@50, at around 6 minutes.

Results from Test4 revealed that under UV irradiation, regardless of the specific configuration, MLG935's sensing performance was enhanced regarding its response to $NO_2$ and the recovery of the initial conditions during the recovery phase. Since the different UV irradiation conditions led to different results, Test4 showed that determining the UV irradiation required to achieve the best sensing performance for the sensor is not trivial. Establishing the optimal UV configuration might lead to a significant breakthrough in the scenario of the gas sensors working in environmental conditions.

A final test, Test5, was designed to evaluate the response of MLG935 towards reducing gases. The first sequence consisted of consecutive cycles with 1 ppm of CO under UV@OFF (Figure 4. 19a) and UV@50 (Figure 4. 19b). The UV irradiation was set at UV@50 configuration due to the promising results of Test4.

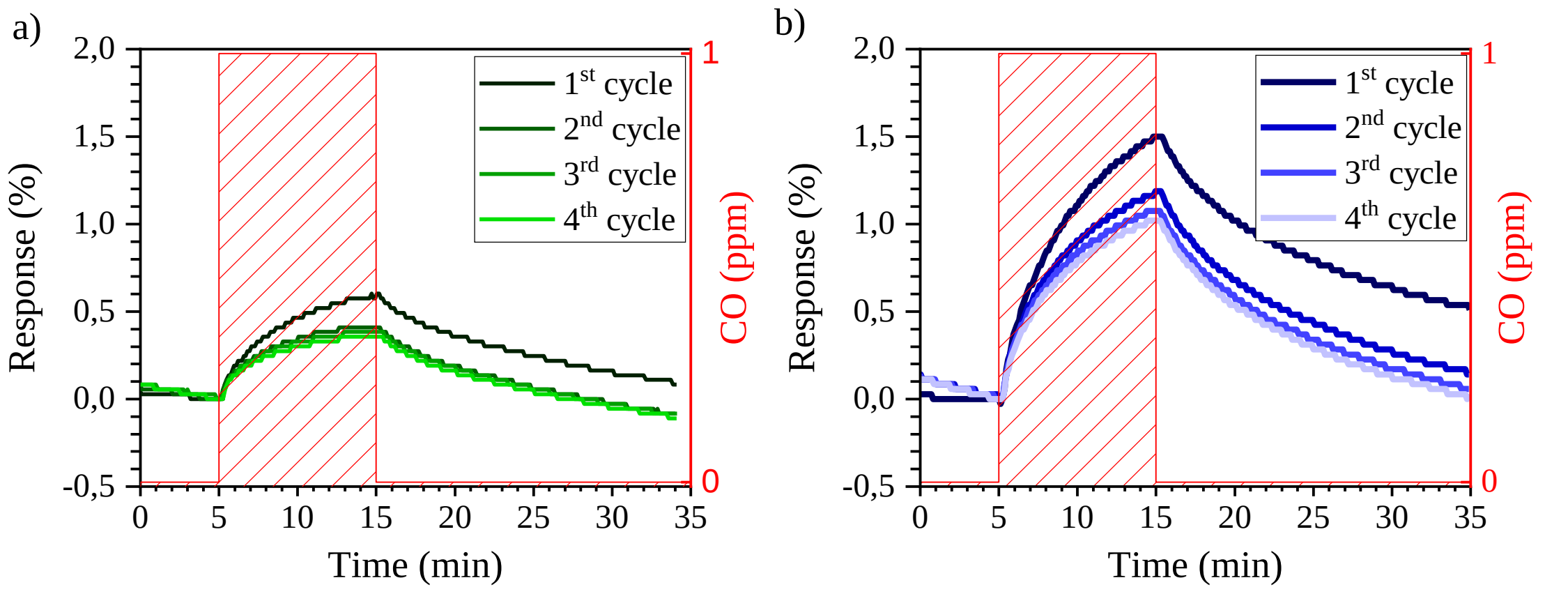

*Figure 4. 19 Real time response of MLG935 towards 1 ppm of CO performed under a) UV@OFF and b) UV@50.*

A similar experiment was performed for 10 ppm of $NH_3$ under UV@OFF (Figure 4. 20a) and UV@50 (Figure 4. 20b).

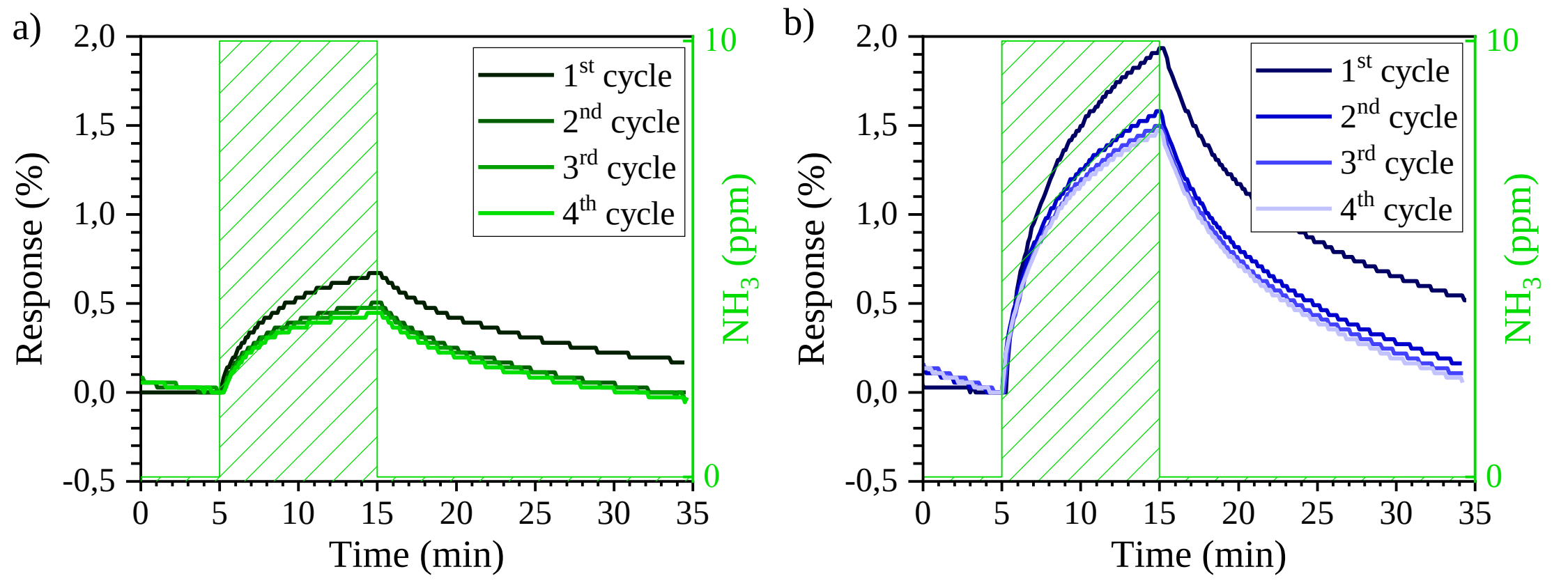

*Figure 4. 20 Real time response of MLG935 towards 10 ppm of $NH_3$ performed under a) UV@OFF and b) UV@50.*

MLG935 showed increased resistance when exposed to the reducing gases, further confirming the p-type behaviour, along with previous results towards $NO_2$. Essentially, when exposed to a reducing or electron donor molecule, such as CO or $NH_3$, electronic transfer decreases the number of holes, increasing the material's resistance [54,82,83,89,96].

For both gases, UV irradiation at UV@50 led to a three-fold RMax. enhancement (Table 4. 5). For the second and subsequent cycles, MLG935 exhibited slight variation in the response and fully recovered the initial conditions during the recovery phase, regardless of the UV irradiation, in contrast to the results from Test1 and Test2 for $NO_2$.

*Table 4. 5 $R_{Max.}$ of MLG935, towards 1 ppm of CO and 10 ppm of $NH_3$ under UV@OFF and UV@50.*

| $NO_2$ | UV setting | 1st cycle | 2nd cycle | 3rd cycle | 4th cycle | Mean ± SD |
|---|---|---|---|---|---|---|
| CO | UV@OFF | 0.60 | 0.41 | 0.39 | 0.35 | 0.44±0.11 |
| | UV@50 | 1.50 | 1.19 | 1.08 | 1.02 | 1.20±0.21 |
| NH3 | UV@OFF | 0.67 | 0.50 | 0.48 | 0.45 | 0.53±0.10 |
| | UV@50 | 1.93 | 1.58 | 1.50 | 1.47 | 1.62±0.21 |

In sum, under UV irradiation, different gas sensing dynamics were observed. A steeper slope during the first few minutes of exposure was observed for the sensor under UV@100. The response then approached a steady state with a slower variation. This stabilization of the response under gas exposure is typically related to a saturation process (Test2). The different slopes and the enabling of a steady state clearly indicated that different adsorption mechanisms took place under UV irradiation [58,97].

This change in the sensing dynamics under UV irradiation had two significant effects that are crucial to the potential application of the graphene-based devices. Mainly a substantial increase in the response during the exposure to $NO_2$ and better recovery of the initial conditions afterwards.

The response of the devices was increased by comparing the mean values for several cycles. For instance, results from Test4 demonstrated an increment of 290% (UV@100) - 550% (UV@50) compared to the response toward $NO_2$ without UV. A similar increase was observed by exposing the sensors to reducing molecules, 270% and 300% for CO and $NH_3$, respectively. Such increases were consistent with the few examples of graphene-based gas sensors operating under continuous UV exposure previously reported (Table 4.3).

Both initial hypotheses were confirmed after this study: (1) UV irradiation has a positive effect on the performance of MLG-based gas sensors, made with a GBM different than FLMG and (2) the effect of UV was more notable on an MLG-based sensor than that observed on a FLMG-based sensor (see Table 4. 6). Interestingly, the $R_{Max.}$ parameter for the FLMG-based sensor was higher than the obtained for the MLG one. Further confirming that the differences within the GBMs family lead to different performances, in this case, when used in gas-sensing applications.

*Table 4. 6 Summarized results of the different presented experiments.*

| Material | Main approach | Analyte | $R_{Max.}$ | Relevant conclusions | Ref. |
|---|---|---|---|---|---|
| Graphite | | $NO_2$/ Air 0.5 ppm | 3.32% | | This work |
| Ball-milled graphene (FLMG) | | $NO_2$/ Air 0.5 ppm | 14.52% | | |
| | Continuous UV: 275 nm | $NO_2$/ Air 0.5 ppm | 15.97% | Slight improve in response and recovery under UV | |
| | | $NH_3$/ Air 50 ppm | 1.5% | | |
| CVD-graphene | Continuous UV: 275 nm, 3.4 mW/cm$^2$ | $NO_2$/ Air 1 ppm | 3.1% | 5-fold increase | |
| | Continuous UV: 275 nm, 6.8 mW/cm$^2$ | | 1.7% | 3-fold increase | |
| | Continuous UV: 275 nm, 3.4 mW/cm$^2$ | CO/ Air 10 ppm | 1.2% | 3-fold increase | |
| | | $NH_3$/ Air 10 ppm | 1.7% | 3-fold increase | |

The $R_{Max.}$ enhancement is directly related to the increase in the device's sensitivity and ultimately lowers the LoD. UV irradiation led to a limit of $NO_2$ as low as 30 ppb, enabling to monitor $NO_2$ below the human toxicity threshold in practical scenarios.

The effect of UV irradiation on the performance of graphene-based gas sensors can be explained through two main mechanisms that occur simultaneously: The photogeneration of electron-hole pairs and the photodesorption of contaminants, often referred to as in-situ cleaning [98,99].

The incident photons create photogenerated electron-hole pairs. For graphene, the photogeneration is caused by absorption in the 275 nm wavelength, related to π-π* electron promotion, according to the following relation [100,101]:

$$\mathrm{hv} \rightarrow \mathrm{e}^{-}_{(\mathrm{hv})} + \mathrm{h}^{+}_{(\mathrm{hv})} \quad \text{(Eq. 4. 4)}$$

The photogenerated charge carriers can then interact with adsorbed molecules, such as water and oxygen species, both electron donors, and cause photodesorption [65,87,102]:

$$h^{+}_{(hv)} + O^{-}_{2(ads)} \rightarrow O_{2(gas)} \qquad \text{(Eq. 4. 5)}$$

Since the experiments used air as a carrier, the photogenerated electron can promote the additional adsorption of oxygen, leading to highly reactive photoinduced oxygen ions [103]:

$$O_{2(gas)} + e^{-}_{(hv)} \rightarrow O^{-}_{2(hv)} \qquad \text{(Eq. 4. 6)}$$

These reactive ions can act as binding sites for the analytes with higher affinity to graphene, like $NO_2$ [88,104]:

$$2NO_{2(gas)} + O^{-}_{2(hv)} + e^{-}_{(hv)} \rightarrow 2NO^{-}_{2(ads)} + O_{2(gas)} \qquad \text{(Eq. 4. 7)}$$

Meanwhile, the adsorption mechanisms that operate no irradiation conditions, i.e., the charge transfer from analytes, still occurs under UV irradiation but are further promoted by the excess of charge carriers from the photogeneration process [88,104].

The reversibility of the adsorption mechanisms was observed during the gas sensing tests. UV irradiation significantly promoted the full recovery of the initial conditions for both devices after the exposition to $NO_2$, improving reproducibility under serial cycles (Test2-Test4). However, for reducing analytes, it was observed that partial recovery was neglectable both with and without UV irradiation (Test5).

Furthermore, no permanent damage from the UV irradiation was observed in the sensors [91]. Thus, this confirms that after the photogenerated carriers recombine, the graphene layer returns to its pre-irradiated state with neglectable permanent damages [66,67,80,87,89,105–107].

Finally, Test4 demonstrated that tuning the power and timing of the UV irradiation configuration is not so trivial, as the best performance in terms of the $R_{Max.}$ of MLG935 towards $NO_2$ was obtained for UV@50, followed by UV@DES. Given these results, it is very likely that better sensing performance and even lower LoD can be achieved while simultaneously reducing the power requirements. However, the proper mechanisms relating the irradiation intensity with the performance variation are not fully understood yet.

### *4.6 Study Of the Influence of Defects.*

The studies previously presented have shown that gas sensors based on graphene-based materials (GBMs) exhibit promising results for detecting $NO_2$ and other gases, particularly when operating under continuous UV irradiation. However, the FLMG-based device and the MLG-based device strongly differed in terms of maximum response and response and recovery rates. It is important to note that the performance of GBM-based devices can strongly vary depending on their physicochemical characteristics, so far, the differences have been related to the presence of defects, differences in the active surface area or different functional groups. While the I(D)/I(G) ratio has been used to quantify defectivity in previous studies [68,72,80,85,108], little is known about the nature of these defects and how they affect the sensing performance.

In this section, the aim is to address this gap by comparing the gas-sensing properties of pristine graphene and multilayer graphene grown via CVD at different temperatures. Using the insights gained on the use of UV irradiation, continuous irradiation at UV@50 conditions are used to measure the devices at optimum conditions.

The MLG samples were synthesized by CVD in similar conditions to those used for MLG935, i.e., on a pre-patterned Mo catalyst in an AIXTRON BlackMagic Pro reactor [95]. 20 sccm of methane ($CH_4$) was used as a carbon feedstock for 20 minutes in Ar/$H_2$ atmosphere at 25 mbar [58,74,95,96]. Three growth temperatures were used, 850, 890 and 935 Cº. The characterization of the material has been presented in Chapter 2. The device was fabricated by the transfer-free method further detailed in ref [95].

The pristine graphene sample was obtained by micromechanical exfoliation from a graphite block [109] and then transferred to a Si wafer covered by 90 nm thick thermally grown $SiO_2$ [58].

The four devices, consequently named MechEx (for the mechanically exfoliated) and MLG850, MLG890 and MLG935' (for the CVD-grown samples at different temperatures). The device MLG935' used in this series of tests is different to the previously used (MLG935).

The sequences were prepared as a calibration test with 25-min exposure and 15-min recovery times. The devices were exposed to 0.1, 0.15, 0.25, 0.4, 0.65, and 1 ppm of $NO_2$. The logarithmic increment series was chosen instead of the linear one to optimize the use of resources, both time and gas flow.

These preliminary findings shown in Figure 4. 21 are currently being investigated. Nevertheless, some findings are worth mentioning such as the notably high response of MLG850 compared to MLG890 and MLG935', the differences in noise-to-signal ratio, and the p-type response of the MechEx device. Although further discussion is in too early stage to be presented here.

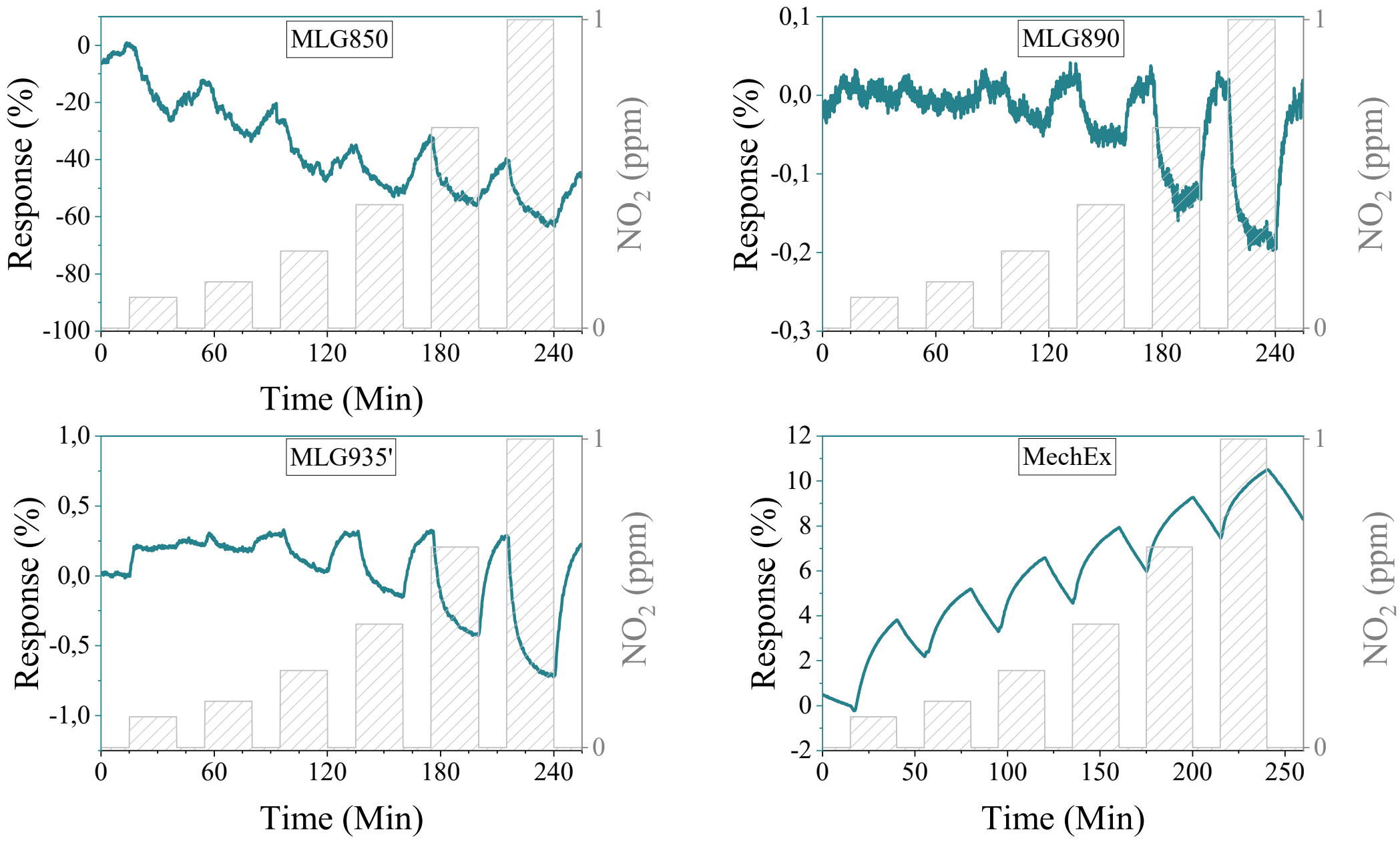

*Figure 4. 21 Calibration test for the MLG-based gas sensors related to the study of the effect of defects. Note that the response scales differ for each device.*

While initial data suggests a potential correlation between defect density and sensor performance, a comprehensive quantification and deeper understanding of this correlation will necessitate further investigation and is a promising direction for future work.

Remarkably, these findings support the concept explored in Chapter 2, which discusses how variations in the same synthesis technique can result in materials with distinct properties across the GBM spectrum. In terms of market viability, this has the potential to decrease production expenses and further advance the adoption of graphene-based gas sensors.

## *4.7 Conclusions.*

In this chapter, the development of graphene-based materials for gas-sensing applications was investigated. The research was motivated by the significant impact of gas pollutants on human health, particularly $NO_2$. Graphene-based chemiresistive gas sensors were identified as a promising technology due to their high sensitivity at room temperature.

An experimental study was conducted to test the sensing capabilities of few-layered mesoporous graphene (FLMG). A device was developed using a 100-min milled sample (FLMG100), which demonstrated an excellent sensing material for detecting sub-ppm $NO_2$ concentrations at room temperature in humid environments with low interference of reducing gases. The device showed a $NO_2$ limit of detection of 25 ppb, below the human exposure recommended limits. The FLMG100 performance was enhanced using UV irradiation, which reduced the response and recovery times. Continuous UV irradiation significantly improved the operation of FLMG-based gas sensors, surpassing the drawbacks of conventional graphene-based sensors, such as partial recovery.

The performance boost of UV irradiation was further studied on a multilayer graphene-based device obtained by CVD synthesis (MLG935), which showed that the device's performance could be increased while simultaneously lowering the power consumption using careful tuning of UV irradiation. The use of UV allowed for a low limit of $NO_2$ detection of just 31 ppb.

Further research on the effect of defects on graphene-based gas sensors was proposed, which has the potential to provide additional insights into the sensing mechanism and improve the sensors' performance.

Overall, the research presented in this chapter shows the great potential of graphene-based materials for gas sensing applications, and it contributes to a better understanding of the mechanisms behind their sensing performance. The results have significant implications for the development of advanced gas sensors that can be used to monitor air quality and protect public health. These findings demonstrate that the use of UV irradiation is an effective way to improve the performance of graphene-based gas sensors and pave the way towards successful commercial applications.

## *4.8 References.*

# Chapter 5: Magnetoelastic Resonance-Based Gas Sensors

*Chapter Introduction; Magnetoelastic materials; Resonance; Characterization; Gas sensor for breath analysis applications; Optimization of gas sensing devices; Conclusions; References.*

## *5.1 Chapter Introduction.*

The ability to remotely monitor the changes in the active material without direct contact with the measuring device is crucial for certain applications where physical access is limited. This is particularly important in the fields of biomedical applications and structural health monitoring [1,2]. While graphene-based chemiresistive sensors have shown to be excellent candidates for gas pollution monitoring, their intrinsic operation principle of physical contact with the measuring device may limit their use. Therefore, the use and optimization of magnetoelastic materials for sensing applications have gained significant attention due to their ability to operate remotely.

This chapter discusses the magnetoelastic effect and how it can be exploited to develop remotely operated sensors, a design of a non-standard measuring setup for real-time magnetoelastic resonance monitoring, the development of a functional magnetoelastic gas sensor, and the optimization of these sensors for practical applications.

## *5.2 Magnetoelastic materials.*

Magnetoelastic materials are a class of materials that exhibit a strong coupling between their magnetic and elastic properties. This unique characteristic makes them ideal for use in various applications, particularly for gas sensing.

### *5.2.1 Magnetoelastic effect.*

Magnetoelastic materials display a coupling between their magnetic and mechanical properties. When these materials undergo magnetization, they experience a mechanical stress response, resulting in physical strain. This phenomenon is referred to as the magnetoelastic, inverse magnetostrictive, or Joule effect, named after its discoverer [3].

The fundamental idea revolves around magnetization rotation in response to an applied magnetic field ($H_{apl}$) [4]. This rotation leads to changes in the arrangement of magnetic domains, resulting in either elongation or contraction ($\Delta L$), depending on the magnetostriction sign, of the magnetostrictive material [5] (Figure 5. 1).

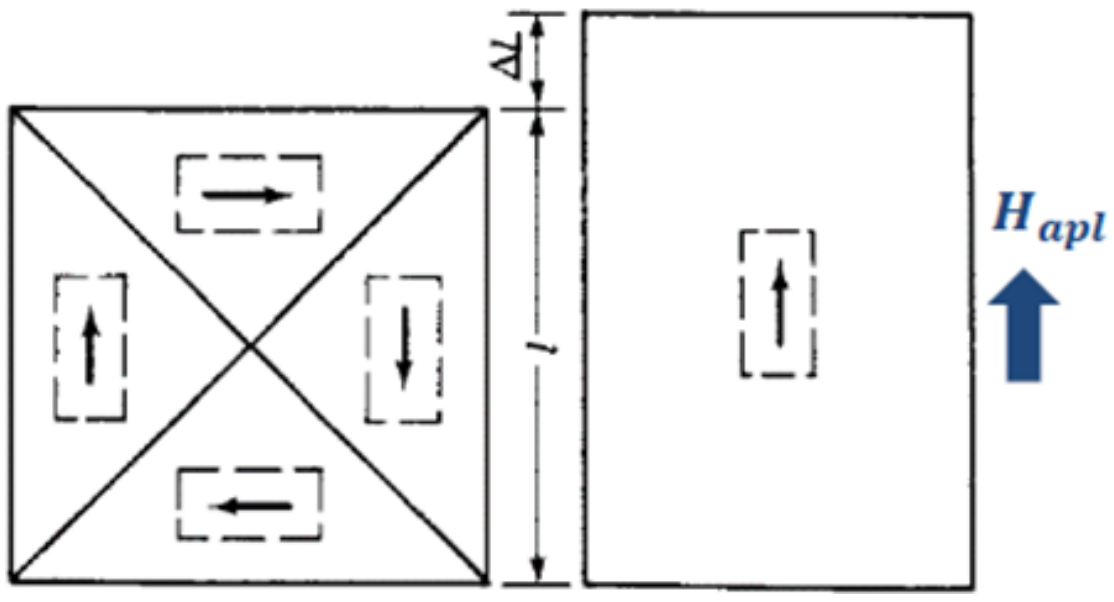

*Figure 5. 1 Traditional representation of the Joule effect.*

The magnitude of the inverse magnetostriction effect depends on several factors, including the strength of the applied magnetic field, the magnetic anisotropy of the material, and the orientation of the magnetic field relative to the crystal lattice of the material [6]. The strain is typically measured in parts per million, and as presented in Table 5. 1, it can reach values up to thousands of ppm [7–9].

*Table 5. 1 Comparison between different magnetostrictive materials* [4,10–13].

| Material | Description | Magnetostriction at saturation (ppm) | Saturation magnetic field (Oe) |
|---|---|---|---|
| Co | | -93 | 17900 |
| Ni | | -33- -50 | 6100 |
| Fe | | -7- -15 | 21500 |
| $Co_{50}Fe_{50}$ | | 87 | 24500 |
| $Ni_{50}Fe_{50}$ | | 19 | 16000 |
| Permalloy 45 | $Ni_{45}Fe_{55}$ | 27 | |
| Permalloy 80 | $Ni_{80}Fe_{15}Mo_5$ | <1.2 | |
| Permalloy 82 | $Ni_{82}Fe_{18}$ | 0 | 8000-12000 |
| $Fe_{73-x}Ni_xCr_5Si_{10}B_{12}$ | FeSiB-based amorphous alloys | 2.5-14 | 5900-11200 |
| Metglass 2826MB | FeSiB-based amorphous alloys | 11-12 | 8800 |
| Metglass 2605SC | FeSiB-based amorphous alloys | 30-60 | 16500 |
| $TbFe_2$ | | 2630 | 11000 |
| Galfenol | FeGa alloy | 400-500 | 300 |
| Terfenol-D | TbDyFe alloy | 1500-2000 | 1000-3000 |

Conversely, when a mechanical strain or stress is applied, this will affect the magnetic properties of the material. This is known as magnetostrictive or Villari (after its discoverer) effect [14]. Magnetostriction occurs because the crystal lattice of the material is distorted when it is subjected to mechanical strain or stress. This distortion changes the spacing and orientation of the atoms, which in turn affects the magnetic properties of the material. Specifically, the magnetic anisotropy of the material can change.

Considering an amorphous micro ribbon annealed under a magnetic field transversal to the ribbon's length -such is the case for the commercial Metglass 2826MB ribbons used in this thesis- we can picture an elongated parallelepiped with magnetic domains oriented perpendicularly to the long direction [15].

Under a magnetic field applied (H) the long direction, these domains will rotate and angle θ towards the direction of the field. Conversely, the anisotropy field $H_A$ represents the effective magnetic field required to overcome the anisotropy energy and rotate the magnetization direction away from its preferred alignment [16]. For an applied magnetic field below the anisotropy field ($H<H_A$) the magnetization along the longitudinal axis is given by:

$$M = M_s \cos\theta = M_s \frac{H}{H_A} \qquad \textit{(Eq. 5. 1)}$$

Where $M_s$ is the domain magnetization, equivalent to the saturation magnetization within the domain, and where $H_A=2K_u/M_s$ being $K_u$ is uniaxial anisotropy constant, related to the energy required to magnetize in the long direction.

Magnetic anisotropy indicates that the material has a preferred direction of magnetization, i.e., the material will have a larger susceptibility, or be more easily magnetized, in a specific direction than in the rest [6]. Magnetic anisotropy may have different origins, including magnetocrystalline anisotropy (given by the crystalline lattice), shape anisotropy (given by the demagnetization field), or, in this case, magnetoelastic anisotropy. In this context, a magnetoelastic material's ease of magnetization will be affected by the stress/ strain state in that direction. The susceptibility of the material may be described by:

$$\chi = \frac{dM}{dH} = \frac{M_s}{H_A} = \frac{M_s^2}{2K_u} \qquad \textit{(Eq. 5. 2)}$$

Under no magnetic field, each domain is naturally elongated $\varepsilon=\lambda_s$ in the transversal direction, and $\varepsilon=-\lambda_s/2$ in the longitudinal, i.e., each magnetic domain has a strain associated to its magnetization [15]. When the ribbon is under H, the domains will rotate and consequently the deformation will go from $-\lambda_s/2$ up to $\lambda_s$ by a total of $3\lambda_s/2$. From Equation 5.1, ε can be related to θ by:

$$\varepsilon = \frac{3}{2}\lambda_s \left(cos_\theta^2 - \frac{1}{3}\right) = \frac{3}{2}\lambda_s \left(\frac{H^2}{H_A^2} - \frac{1}{3}\right) \qquad \textit{(Eq. 5. 3)}$$

The magnetostrictive coefficient (d) quantifies the change in size or shape, $\varepsilon= \Delta L/L_0$, of a material in response to an applied magnetic field [15]. It is typically expressed in units of length per unit magnetic field.

The sign of d determines whether the material will experience compressive or tensile stress when subjected to a magnetic field. If the magnetoelastic coefficient is positive, as used in the example for Equation 5.3, the material will experience tensile stress when subjected to a magnetic field, while a negative magnetoelastic coefficient will result in compressive stress.

In general, materials such as nickel, gallium, and iron-containing alloys (including Terfenol-D and Metglass) exhibit a positive magnetostrictive coefficient, while cobalt-containing alloys and Terfenol-S have a negative magnetostrictive coefficient. However, it is important to note that the sign of the magnetostrictive coefficient can be significantly influenced by the material's composition. Some materials have a magnetostrictive coefficient of zero, such as Permalloy (Table 5. 1).

Regarding the magnitude of d, a higher magnetostrictive coefficient means the material will experience a larger elongation or contraction for a given magnetic field. However, this relation is not constant, as depicted in Figure 5. 2.

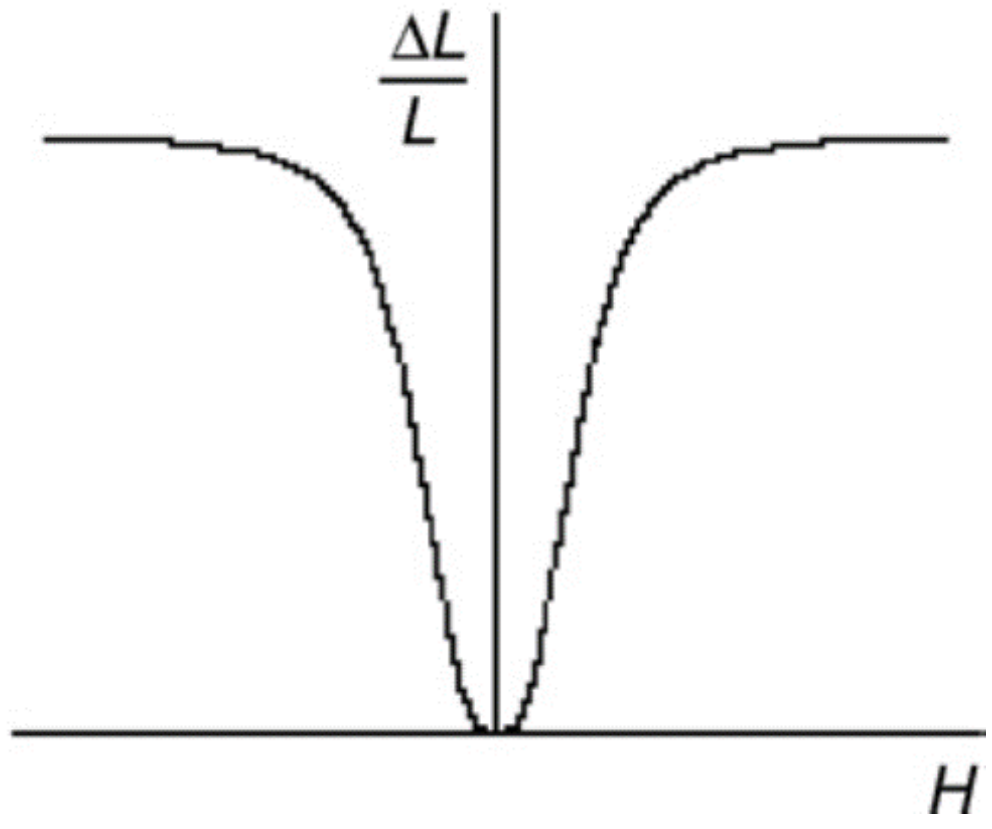

*Figure 5. 2 Typical magnetostriction curve (d) representation. Rep. from* [17].

Generally, d is considered lineal H<$H_A$, and maximum H=$H_A$, then d can be defined by:

$$d = \frac{d\varepsilon}{dH} = \frac{3\lambda_s H}{H_A^2}$$
$$d_{max} = \frac{3\lambda_s}{H_A} = \frac{3\lambda_s M_s}{2K_u} \quad \text{(Eq. 5. 4)}$$

The ease of magnetization is also affected by the magnetoelastic coefficient. When a magnetostrictive material is subjected to mechanical stress or strain, its magnetic properties can change, resulting in a change in the ease of magnetization. Specifically, if the magnetoelastic coefficient is positive, the ease of magnetization will decrease under tension and increase under compression. Conversely, if the magnetoelastic coefficient is negative, the ease of magnetization will decrease under compression and increase under tension.

If the ribbon from the previous example -with positive magnetostriction coefficient- is subjected a traction stress, σ, along the longitudinal direction, it will contribute to the magnetization by reducing the magnetization energy by:

$$K_u \rightarrow K_u - \frac{3}{2}\lambda_s \sigma \quad \text{(Eq. 5. 5)}$$

Thus, a high σ will switch the easy magnetization axis from the transversal direction to the longitudinal direction. In this case, the anisotropy field ($H_A$) is reduced by effect of strain to:

$$H_A = 2\frac{K_u}{M_s} \rightarrow H_{A\sigma} = \frac{2K_u - 3\lambda_s \sigma}{M_s} \quad \text{(Eq. 5. 6)}$$

If the term $H_A$ is replaced by $H_{A\sigma}$ in Equation 5.1, the magnetization M will achieve a higher value for a given magnetic field H when stress σ is present compared to the case with no stress. Note that the maximum magnetization value, $M_s$, is not altered by the presence of stress. Figure 5. 3. helps to illustrate this effect. In the figure, a schematic magnetization curve for a magnetoelastic material with a positive magnetostrictive coefficient is shown. The standard magnetization curve is represented by a black line.

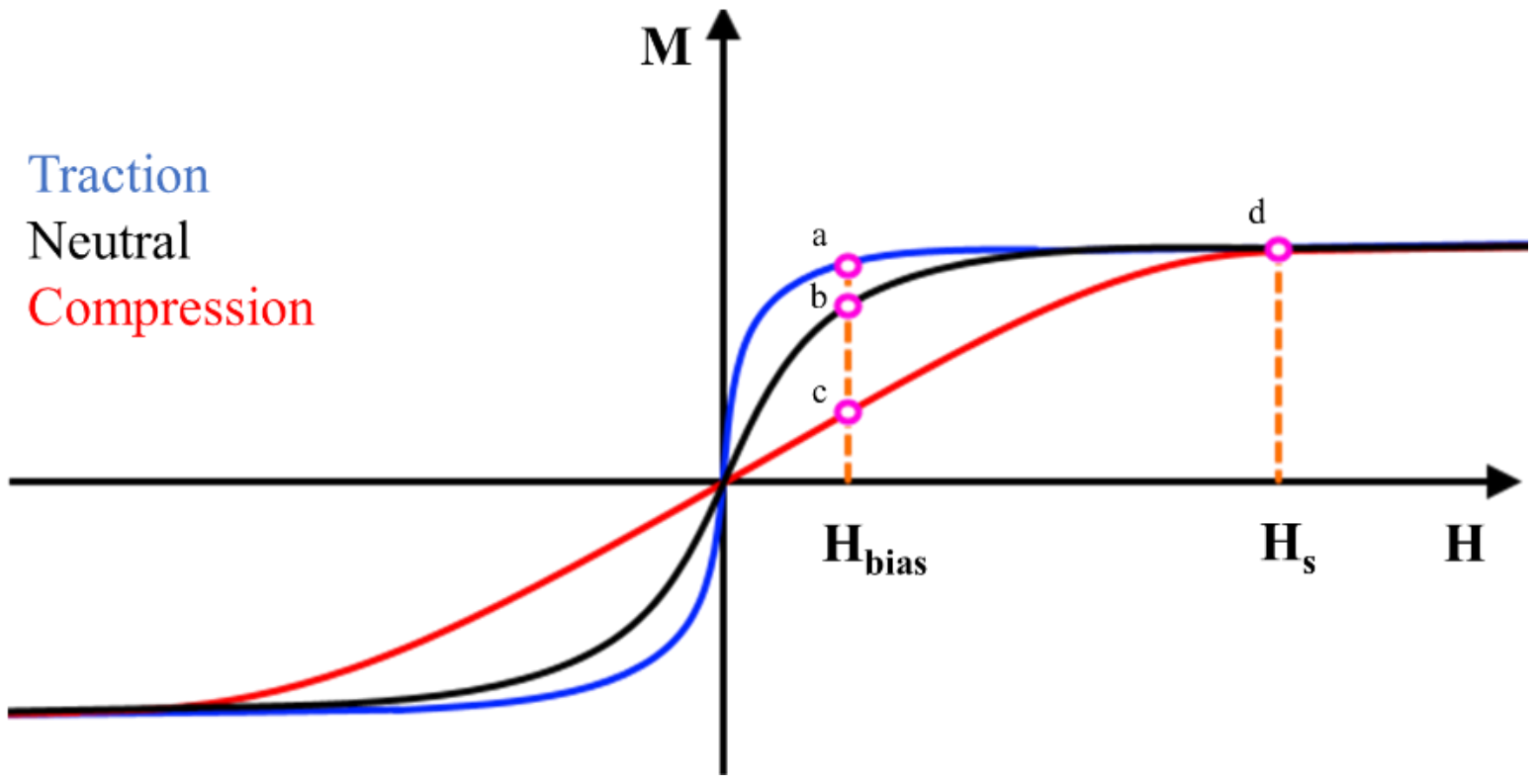

*Figure 5. 3 Schematic representation of the magnetization curve of a magnetoelastic material with a positive magnetostrictive coefficient under different stresses.*

If the material is subjected to positive stress, such as traction (represented by the blue line in the Figure 5. 3), then the magnetization in this direction will be favoured. This means that the same applied magnetic field will produce a higher magnetization in the direction of the applied stress compared to the standard magnetization curve. Conversely, if the material is under negative stress, such as compression (represented by the red line in the Figure 5. 3), then the magnetization in this direction will be unfavoured. This means that the same applied magnetic field will produce a lower magnetization in the direction of the applied stress compared to the standard magnetization curve.

Note that if under a constant magnetic field ($H_{bias}$), below the saturation field value, the material suffers a change in its stress state, this will directly affect its magnetization value. Under these circumstances, the tensioned $\sigma>0$ material, if relaxed, will have a magnetization change from point (a) to point (b), if then compressed, it will further vary from point (b) to point (c). This change in magnetization is easily using a conductive coil (Faraday's law of electromagnetic induction), laying the foundations for the development of the sensor that will be later discussed. At high applied magnetic field values, the magnetization is insensitive to the stress state (point (d)).

Since most materials have a positive Poisson coefficient, if they are elongated in one direction, they are compressed in the other to keep a constant volume. This means that for the material in the previous example, as the magnetization in one direction is favoured, it is necessarily and simultaneously unfavoured in the rest.

While the magnetostrictive phenomenon is an important factor in the operation of magnetoelastic sensors, this chapter will focus on the applied research and development of these sensors, rather than a fundamental study of magnetostriction. Therefore, this chapter will discuss the formal and practical aspects of the magnetostriction but will not go into extensive theoretical details.

Magnetoelastic materials have gained importance in various technological applications owing to their exceptional combination of magnetic and mechanical properties. One particularly promising application is the use of magnetoelastic materials for energy harvesting [18,19]. The magnetoelastic effect enables these materials to convert mechanical energy into magnetic energy, which can then be converted into electrical energy using techniques such as piezoelectric or electromagnetic conversion. While energy harvesters based on magnetoelastic materials may be less efficient than other types of harvesters like photovoltaic, thermoelectric, or piezoelectric devices [20], they offer several advantages, such as being more suitable for harvesting energy from low-frequency vibrations and being smaller and more mechanically durable [21,22].

Conversely, since these materials can also convert magnetic energy into mechanical energy, making them suitable for use in a wide range of actuation applications [23]. For instance, they have been explored for use in vibration control applications, where they can be used to reduce unwanted vibrations in mechanical systems [24–26]. In addition, magnetoelastic materials have been used in motors and other actuation devices, such as magnetic valves in additive manufacturing [27].

### *5.2.2 Fe-based amorphous magnetoelastic materials*

As previously described, magnetoelastic materials exhibit the ability to convert magnetic energy into mechanical energy. Terfenol-D is a well-known reference magnetostrictive material with a high degree of magnetostriction and the largest energy conversion efficiency due to its high magneto-mechanical coupling coefficient (k). However, the mechanical properties of Terfenol-D are relatively poor, the material is highly reactive and costly to produce [7,28].

In contrast, Fe-based amorphous alloys are synthesized by melt spinning. This rapid solidification technique involves casting a thin ribbon or wire of molten alloy onto a rapidly rotating copper wheel, which cools the material, resulting in an amorphous (non-crystalline) structure, known as the Taylor-Ulitovsky technique [29,30]. This technique allows the production of Fe-based amorphous alloys with a high yield and at low costs.

Despite exhibiting a lower magnetoelastic performance compared to Terfenol-D, amorphous alloys based on iron possess remarkable magneto-elastic properties and exceptional mechanical resistance, making them highly desirable for commercial applications [4]. One such example is Metglass 2826MB (Metglass, Conway, SC, USA), which features excellent magnetoelastic properties and has been extensively used in security tags [31]. These tags are commonly employed in retail stores to prevent theft by attaching them to merchandise. Equipped with a magnetic strip, the tags can be detected at store entrances and exits. When an item carrying the tag passes through the sensor, the magnetic strip triggers an alarm, alerting store personnel to potential theft. The magnet generating the magnetic field on the tag can be deactivated to disable the tag.

Fe-based amorphous alloys are usually made in the shape of ribbons [32] or wires [33], they possess a combination of high mechanical strength (~1000–1700 MPa), a high magnetoelastic coupling coefficient (k), up to 0.98, and magnetostriction on the order of $10^{-5}$ (dimensionless) [34,35].

The amorphous structure allows for a high degree of magnetic softness, which means that the magnetic domains in the material can be easily reoriented in response to an external magnetic field without the influence of an ordered atomic structure. In other words, there is no magneto crystalline anisotropy contribution. Then, the ease of magnetization in these materials is given only by the shape anisotropy or, more interestingly, by the magnetoelastic anisotropy.

The study on microwire use in sensing applications has focused mainly on the use of giant magnetostriction effect (GMI) [36–38] or other parameters [39–41] to make mainly stress or pressure sensors. However, the use of magnetoelastic microribbons for magnetoelastic resonance-based sensors is much more extended. For micro ribbons, it has been found that the highest k values have been obtained by thermal annealing under a magnetic field perpendicular to the ribbon's axis that induces a transverse homogeneous easy axis [42]. The thermal annealing of Fe-based amorphous materials is of particular interest as it leads to changes in their magnetic and mechanical properties, depending on the annealing temperature and composition. Thermal annealing will be later discussed in this chapter.

In the work presented in this chapter, two Fe-based amorphous alloys were used as magnetoelastic materials. Particularly, glass covered FeSiBNb ($Fe_{73}Si_{11}B_{13}Nb_3$) microwires and Metglass 2826MB ($Fe_{40}Ni_{38}Mo_4B_{18}$) microribbons.

The microwires are covered in a borosilicate glass shell (similar to those described in Chapter 3) which offers additional resistance to corrosion. The microribbons do not have such a cover. Instead, the presence of Ni in their composition provides such corrosion resistance [13,43].

### *5.2.3 On ΔE effect.*

The ΔE effect is a phenomenon observed in magnetoelastic materials where the elastic modulus of the material experiences a change under the influence of an applied magnetic field. To the best of my knowledge, was first described by H. Savage [44] and it is of particular interest when designed applications using magnetoelastic materials.

The stiffness of a material is given by the elastic modulus -or Young modulus- (E). Due to the unique properties of magnetoelastic materials, the stresses induced by a magnetic field modify their stiffness, causing that E, a typical constant value, is modified by H, and will achieve a maximum at $H_{A\sigma}$ [15,45]:

$$\frac{\Delta E}{E_H} = \frac{E_M - E_H}{E_H} = \frac{9\lambda_s^2 E_M H^2}{M_s H_{A\sigma}^3}$$
$$\left(\frac{\Delta E}{E_H}\right)_{Max} = \frac{9\lambda_s^2 E_M}{2K_u - 3\lambda_s \sigma} \qquad \textit{(Eq. 5. 7)}$$

Since E is a key mechanical parameter, the ΔE effect is often used to describe the dependence of k with H:

$$k_H = d\left(\frac{E_H}{\chi_\sigma}\right)^{1/2} = \left(1 + \frac{M_s H_{A\sigma}^3}{9\lambda_s^2 E_M H^2}\right)^{-1/2}$$
$$k_{Max} = \left(1 + \frac{2K_u - 3\lambda_s \sigma}{9\lambda_s^2 E_M}\right)^{-1/2} \qquad \textit{(Eq. 5. 8)}$$

Therefore, in sensor applications, it is essential to apply a magnetic field that results in the maximum k. Utilizing the optimal magnetic field ensures that the sensor achieves the highest sensitivity and performance, enabling more efficient and accurate detection and measurement of the targeted parameters.

This understanding of the ΔE effect is particularly relevant for the development of magnetoelastic-resonance based sensors, as these sensors rely on the interaction between the magnetic and elastic properties of the material to generate a resonance response [46].

### *5.3 Resonance.*

Resonance can be defined as a phenomenon that occurs when a system or object is subjected to an oscillating external force or input, causing the system to respond with an increased amplitude at specific frequencies, known as the resonant frequencies. Resonance is observed in a wide range of physical systems, including mechanical, electrical, optical systems, or acoustics.

These frequencies correspond to the natural frequencies of the system, where the energy transfer between the system's different modes of oscillation is most efficient, leading to a significant enhancement of the response or output.

The resonance frequency depends on the geometrical and mechanical parameters of the sample according to the equation that can be easily deduced assuming a free-standing solid parallelepiped resonator, where the mechanical waves propagate without losses [4]. The differential mass of the resonator is:

$$dm = \rho h e dx \qquad \textit{(Eq. 5. 9)}$$

Where ρ is the density of the material, h is the height, e is the depth, and x is the length (infinitesimal element). The wave induced into the material will propagate forward and backwards with forces that necessarily balance each other given by:

$$\overrightarrow{f_{(x)}} = \sigma_{(x)} S\vec{x};$$
$$\overrightarrow{f_{(x+dx)}} = -\sigma_{(x+dx)} S\vec{x} \qquad \textit{(Eq. 5. 10)}$$

Where σ is the stress and S is the cross section of the material (h*e). If treated like a stationary wave, this relation can be expressed in terms of the displacement (u) as:

$$-\sigma_{(x+dx)}S + \sigma_{(x)}S = \rho S dx \frac{\partial^2 u}{\partial t^2};$$
$$-\frac{\partial \sigma}{\partial x} = \frac{\partial^2 u}{\partial t^2} \quad \text{(Eq. 5. 11)}$$

Note that from since S has been eliminated from the equation, the rest of the calculation will not be dependant of the cross-sectional shape of the object. Rather, the only geometrical parameter left is its length. Hooke's Law relates σ with the elongation (ε) through E (ε=σ/E). The elongation can be described as:

$$\frac{\Delta L}{L} = \frac{u_{(x+dx)} - u_{(x)}}{dx} = \frac{\partial u}{\partial x};$$
$$\sigma = E\,\frac{\partial u}{\partial x} \quad \text{(Eq. 5. 12)}$$

By combining Equations 5.11 and 5.12:

$$\frac{\partial^2 u}{\partial x^2} - \frac{\rho}{E}\frac{\partial^2 u}{\partial t^2} = 0 \quad \text{(Eq. 5. 13)}$$

The ρ and E can be related to the velocity of propagation of the wave within the material ($c=(E/\rho)^{-1}$). Using the harmonic solutions:

$$u_{(t,x)} = u_{(x)}e^{i2\pi f t} = U_1 e^{i\frac{2\pi f}{c}x} + U_1 e^{-i\frac{2\pi f}{c}x} \quad \text{(Eq. 5. 14)}$$

Where i is the imaginary number and f is the frequency. The following boundary conditions are imposed: (1) At x=0, $u_{(0)}$=0, i.e., the displacement at the centre of the parallelepiped is null (see Figure 5. 4). And (2) $\sigma_{(L/2)}= \sigma_{(-L/2)}= 0$, i.e., at the extremes of the parallelepiped, located and half lengths (L/2) from its centre, there is no stress.

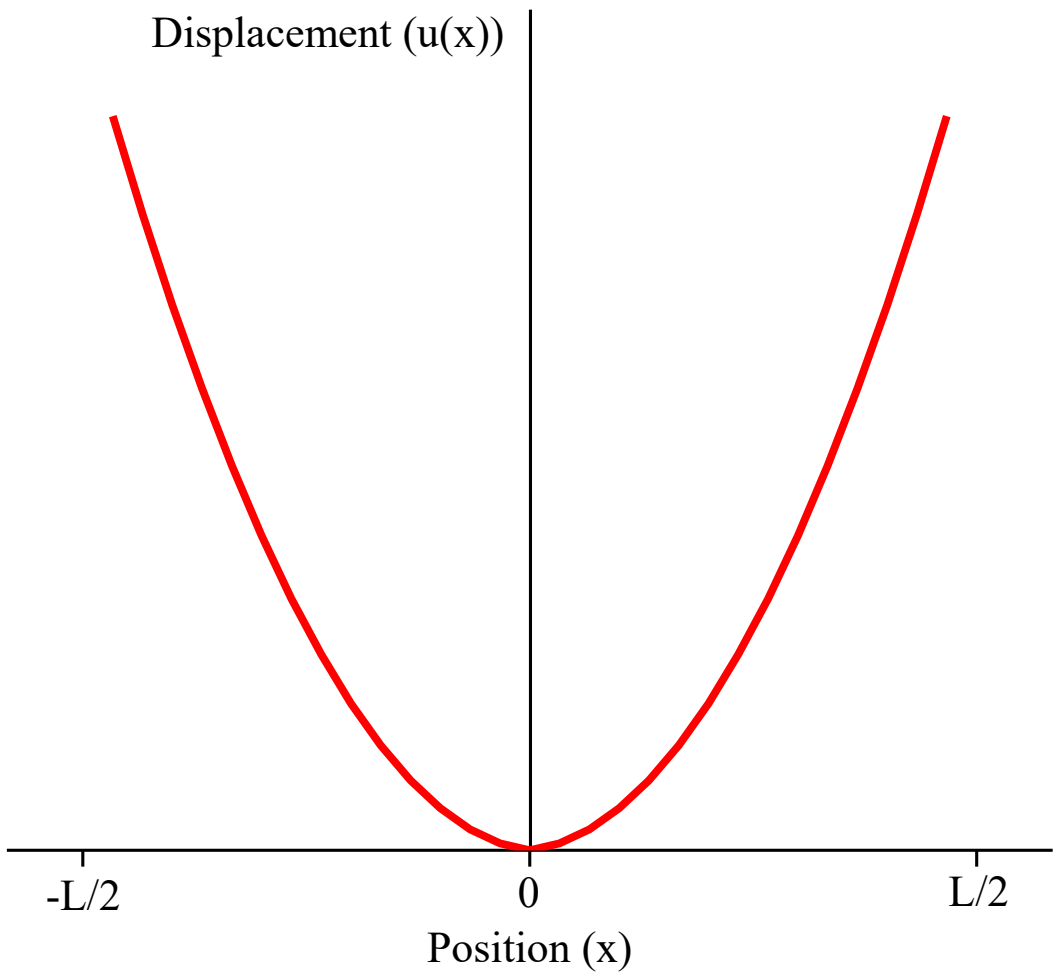

*Figure 5. 4 Schematic representation of the displacement in the fundamental mode of a one-dimensional free-standing element.*

These constraints come from the initial conditions of the system being one-dimensional and freestanding. Applying them along with the harmonic solutions of Equation 5.14 results in:

$$\frac{2\pi f}{c}L = \pi(2p + 1) \quad \text{(Eq. 5. 15)}$$

Where p is a positive integer number. Solving for f will provide the solution for the resonance frequencies stablished in this system. Particularly, solving for p=0 will provide the fundamental resonance frequency:

$$f_0 = \frac{1}{2L}\sqrt{\frac{E}{\rho}} \qquad \textit{(Eq. 5. 16)}$$

Equation 5.16 is a simplified solution; however, it highlights some important relations within the system. For instance, the resonance frequency only depends on the geometric parameter of length, and not from the thickness or height as the cross-sectional dependence was eliminated in a previous step. This can be easily demonstrated as the resonance frequency of a ribbon and a microwire are approximately the same, with the differences due to density and mechanical properties. The resonance frequency will also depend on mechanical parameters and composition through E and ρ. Since this system is one-dimensional, it neglects the effect of the deformation of the material in the perpendicular directions, which can be included using the Poisson ratio (ν) [46]:

$$f_0 = \frac{1}{2L}\sqrt{\frac{E}{\rho(1-\nu^2)}} \qquad \textit{(Eq. 5. 17)}$$

Equation 5.17 can be checked with the parameters of a Metglass 2826MB micro ribbon such as the ones used in commercial security tags, i.e., a length of 37 mm, elastic modulus of 152 GPa, ρ of 7900 $kg/m^3$ and ν of 0.33 [47]. The calculated resonance frequency is 62.8 kHz for the first harmonic and 125.6 kHz for the second harmonic. The experimentally measured resonance frequency was 58.6 kHz for the first harmonic and 118.8 kHz for the second harmonic [48].

The comparison between theoretical and experimental values reveal that Equation 5.17 is a reliable predictor of the resonance frequency value. However, there are certain factors that may account for the observed deviations between the theoretical and experimental values. One of these factors is the fact that the micro ribbon is not entirely free-standing in actual operating conditions, as it rests on a surface that introduces some degree of friction. Furthermore, due to the manufacturing process, there is a certain level of variability in the composition and properties of the micro ribbons, which may also contribute to the observed discrepancies.

### *5.3.1 Magnetoelastic resonance.*

As previously discussed, resonance can be understood from the propagation of a mechanical wave within a material. Since magnetoelastic materials possess the unique feature of coupled magnetic and elastic properties, a time-varying magnetic field will induce a strain in the form of a propagating mechanical wave. By subjecting a magnetoelastic material to an alternating magnetic field, the material will oscillate, and, eventually, resonate at frequencies determined by Equation 5.15.

Resonance is sensitive to a range of parameters that can influence the mechanical and magnetic properties of the magnetoelastic material. This means that the magnetoelastic resonator can act as a standalone sensor for monitoring temperature and magnetic fields [49,50], density and viscosity of a medium [51–53], or mechanical stresses, including tension, torsion, or pressure [54–56]. More interesting, if the resonator is coated with a functional material that changes its mass as a response to external parameters, the potential for magnetoelastic resonance-based sensor is vastly incremented as it will be further elaborated in this chapter. The operation for all these sensors is based on monitoring the changes in the magnetoelastic resonance.

For instance, when the magnetoelastic resonator is immersed in a viscous liquid, the resonant frequency decreases due to dissipative shear force created by the medium. For a low viscosity, the relation between the frequency shift and the mediums viscosity is described by [46]:

$$\Delta f = -\frac{\sqrt{\pi f_0 \eta \rho_l}}{2\pi\rho e} \qquad (Eq.\ 5.\ 18)$$

Where η is the viscosity of the liquid and $\rho_l$ its density. Note that Equation 5.18 is dependent on e (thickness), a geometrical parameter of the resonator. For instance, this effect was confirmed by Pilar Marín et al. after studying the magnetoelastic resonance of a microwire immersed in ethanol, petrol, and oil [52].

The magnetoelastic resonance can be detected by a variety of systems. The frequency of the mechanical oscillation can be detected with a microphone through the emission of acoustic waves, or it can be detected by optical means with a combination of a laser emitter and a photodetector [46]. However, the most popular method is by means of the electromagnetic conversion using a pick-up coil [43,57–62].

*5.3.2 Influence of mass loadings.*

The resonance frequency is extremely sensitive to a wide range of parameters. The most relevant for the field of gas sensor development is its sensitivity to mass loadings.

If the magnetoelastic resonator is uniformly coated with a mass (Δm), then ρ in Equation 5.18 can be substituted by ($m_0$+Δm)/ Sx, where $m_0$ is the original mass of the resonator. Solving the equation leads to a new resonance frequency ($f_{\Delta m}$) [46,63]:

$$f_{\Delta m} = f_0 \sqrt{\frac{1}{1 + \Delta m / m_0}} \qquad (Eq.\ 5.\ 19)$$

Equation 5.19 can be simplified assuming that Δm is relatively small compared to $m_0$ leading to the well-known expression [64,65]:

$$\Delta f = f_{\Delta m} - f_0 = -f_0 \frac{\Delta m}{2 m_0} \qquad (Eq.\ 5.\ 20)$$

Equation 5.20 describes an interesting relationship between the frequency shift of the resonance (Δf) and the mass variation of the coating. Specifically, when the coating mass is increased, the resonant frequency shifts towards lower values (redshift), while when the coating mass is decreased, the resonant frequency shifts towards higher values (blueshift).

This relationship has practical implications in various applications. For example, in gas sensing operations, when gaseous molecules attach to the active material coated on the resonator, the resonator's mass increases, causing its frequency to redshift. This effect can be used to detect the presence and concentration of gases [57,59]. Similarly, in corrosion or degradation monitoring applications, as the mass of the coating or resonator decreases due to the degradation process, its resonance will blueshift, indicating the degree of degradation [66].

*5.4 Characterization.*

The elastic waves propagate throughout the magnetoelastic materials causing changes in the magnetization which in turn leads to changes in the magnetic flux around the resonator that can be detected with a detecting coil that transform the magnetic flux into an electrical signal.

This setup typically consists of an exciting coil, that generates an alternating magnetic field in a frequency sweep, and a detecting coil. The differences in the intensity of the signal between each coil will provide a gain value, expressed in dB. The gain varies with frequency around the resonant frequency providing a spectrum similar to the one presented in Figure 5. 5.

The magnetoelastic resonance can be characterized by the following figures of merit: its $f_r$, antiresonance frequency ($f_a$), amplitude (A), magneto-mechanical coupling coefficient (k), and quality factor (Q). Here $f_r$ and $f_a$ are given by the maximum and minimum gain value in the spectrum, respectively (illustrated in the next section). Then, from the difference between gain values at $f_r$ and $f_a$, A can be calculated (A= $f_a$- $f_r$). Finally, k can be calculated from $f_r$ and $f_a$ as:

$$k = \sqrt{\frac{\pi^2}{8}\left[1 - \left(\frac{f_r}{f_a}\right)^2\right]} \qquad \textit{(Eq. 5. 21)}$$

Finally, Q is defined as the ratio of the energy stored in the device during one cycle of vibration to the energy dissipated per cycle [67,68]:

$$Q \stackrel{\text{def}}{=} 2\pi \times \frac{energy\ stored}{energy\ dissipated\ per\ cycle} \qquad \textit{(Eq. 5. 22)}$$

A high-Q device has a narrow resonance peak, indicating that it can store and release energy efficiently, while a low-Q device has a broad resonance peak, indicating that it is dissipating energy. A high-Q device is desirable because it can produce a sharp resonance peak with a high amplitude, which makes it easier to detect and measure small changes in the resonant frequency [68]. Q can be experimentally calculated from:

$$Q = \frac{f_c}{BW} \qquad \textit{(Eq. 5. 23)}$$

Where $f_c$ is the peak centre and BW is the 3 dB bandwidth (from the maximum amplitude).

Although Q is generally related to the magnetoelastic resonator, it can also be used to evaluate other resonating systems. For instance, Q was used to evaluate the oscillating system for magnetoelastic resonance characterization that will be presented in a later section of this chapter.

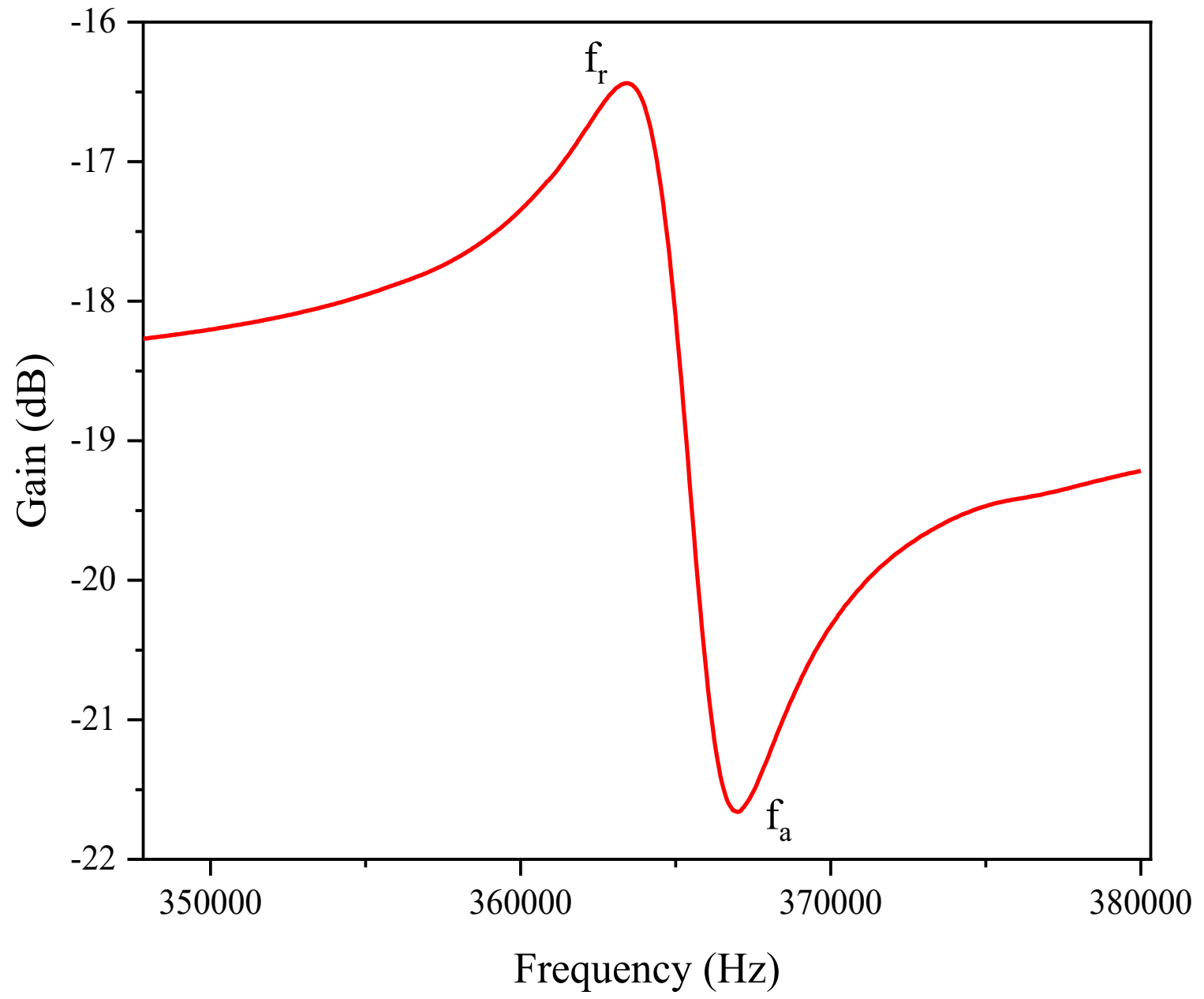

*Figure 5. 5 Typical spectrum of magnetoelastic resonance frequency sweep, in this case, from a 6.6 x 1 mm Metglass 2826MB micro ribbon.*

### *5.4.1 Single coil setup.*

The magnetoelastic characterization setup underwent several iterations. Initially, a custom-made system consisting of a single coil was used, in which the ribbon or microwire was placed. In this setup, the single coil acted as both an emitter and receiver, and the gain values were obtained through changes in the coil's impedance. A Red Pitaya system controlled with LabView-based data acquisition software was used to feed the coil with a time-varying signal.

The Red Pitaya is an open-source electronic instrumentation platform system that provides functionalities similar to an oscilloscope, spectrum analyser, signal generator, or logic analyser among others, and is a low-cost, versatile, and compact device. LabView, on the other hand, is a programming software that enables the control of instrumentation, measurement, and automation.

The accuracy of the measurement can be adjusted by indicating the desired frequency range and the total number of samples in the acquisition. In doing this, high accuracies in steps of 1 Hz can be obtained for a specific frequency range.

In addition, a pair of identical larger coils, or Helmholtz coils, were used to create a uniform steady magnetic field to submit the magnetoelastic material to a tuneable bias field. The system is depicted in Figure 5. 6.

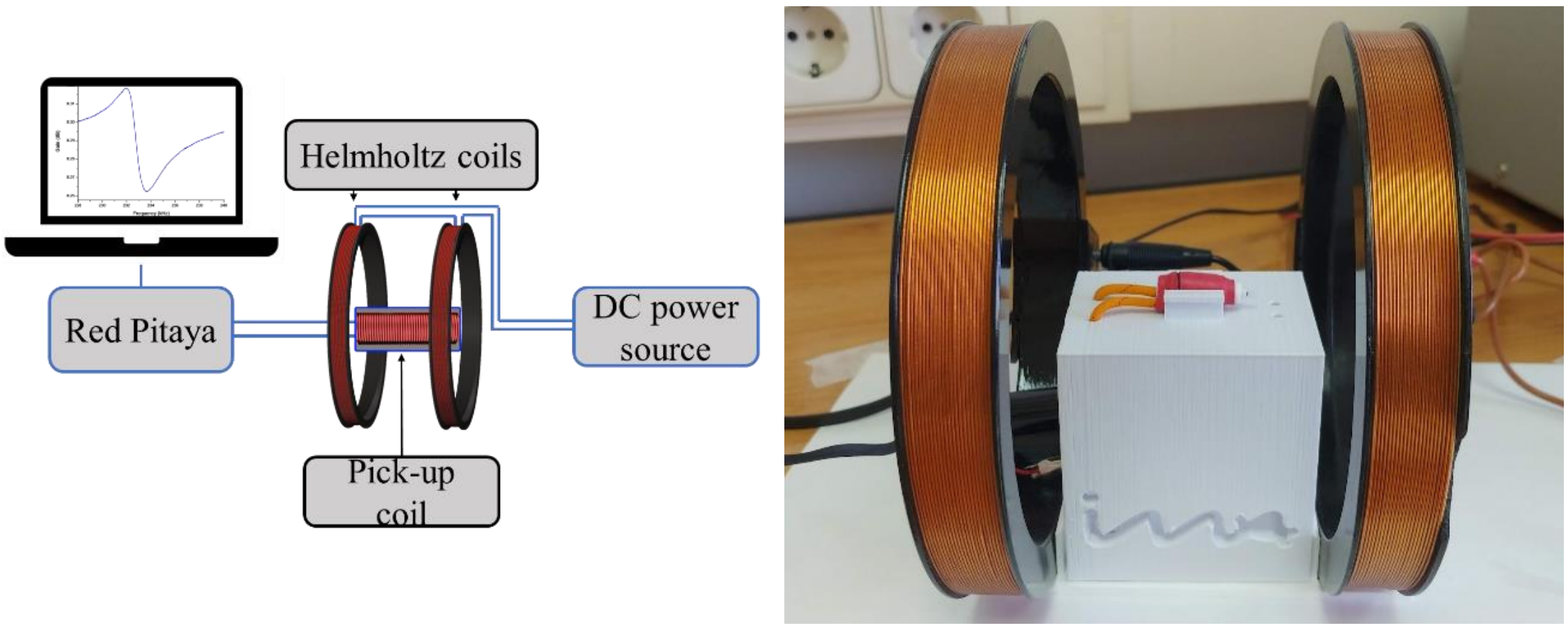

*Figure 5. 6 (left) Schematic representation of the magnetoelastic resonance characterization setup, Rep. from* [50]. *(right) Photography of the setup.*

The single coil, or pick-up coil as defined in Figure 5. 6, is carefully designed for each sample. As an example, for microwires, a 15 mm long, 2.4 mm in diameter coil with 1000 turns of 0.2 mm copper wire was used for 7 - 12 mm long microwires and a 55 mm long, 20 mm in diameter coil with 3000 turns of 0.5 mm copper wire was used for 23.5 - 37 mm long micro ribbons [50].

The geometry of the coil is designed so that the turns are as close as possible to the sample to maximize the picked flux variations. It is important to note the manufacturing limitations for small coils. For instance, the inner bobbin or spool was manufactured using additive manufacturing techniques by mean of fused deposition 3D printing. In addition, the bobbin needs to be manipulable enough to manually create hundreds or thousands of turns with the copper wire.

The number of turns and the geometry of the coil also affects the measurement through the reactance of the coil. Briefly, the reactance is similar to the resistance in AC circuits, meaning it can difficult the flow. To avoid these effects, the number of turns (N) is selected according to the coils geometry to equal the reactance of the coil to the input impedance of the Red Pitaya and other similar systems (50 Ohms) according to:

$$N = \sqrt{\frac{50\, l}{\omega \mu S}} \qquad \textit{(Eq 5. 26)}$$

Where l is the coil length, ω is a frequency value, in this case the expected resonance frequency value of the sample, μ is the free space magnetic permeability and S is the coil cross-section.

Since the gain is dependent on the coil configuration, the A parameter can only be directly compared in measurements taken with the same configuration. However, the parameters dependent on the frequency ($f_r$, $f_a$, and k) do not suffer such limitation.

This setup obtains one measure per second. Therefore, a typical frequency sweep consisting of 1000 points, e.g., a sweep from 50 to 60 kHz with 10 Hz steps, lasts <15 minutes. Other typical values range at around tens of seconds per step. Although this technique reveals a range of magnetoelastic, like A and k, parameters it is unable to monitor fast reactions in real time and, consequently, has hindered potential applications of magnetoelastic resonance-based sensors [48].

It is worth mentioning the approach to solve the slow measurement taken by B. Sisniega and her colleagues at the Basque Country University. Using numerical fitting they were able to extract $f_r$ from significantly narrower sweeps, reducing the measuring time [68]. Although this method may not be strictly considered real-time, as it relies in discrete measurement, it has allowed researcher to monitor fast-occurring precipitation reactions [62].

*5.4.2 Twin coil setup and real-time monitorization.*

The latest iteration consists of a twin coil setup. In this setup, two identical coils are used, the exciting coil and the detecting coil are connected to the output and input port, respectively, of a Red Pitaya system, which in this case acts as a spectrum analyzer (see Figure 5. 7).

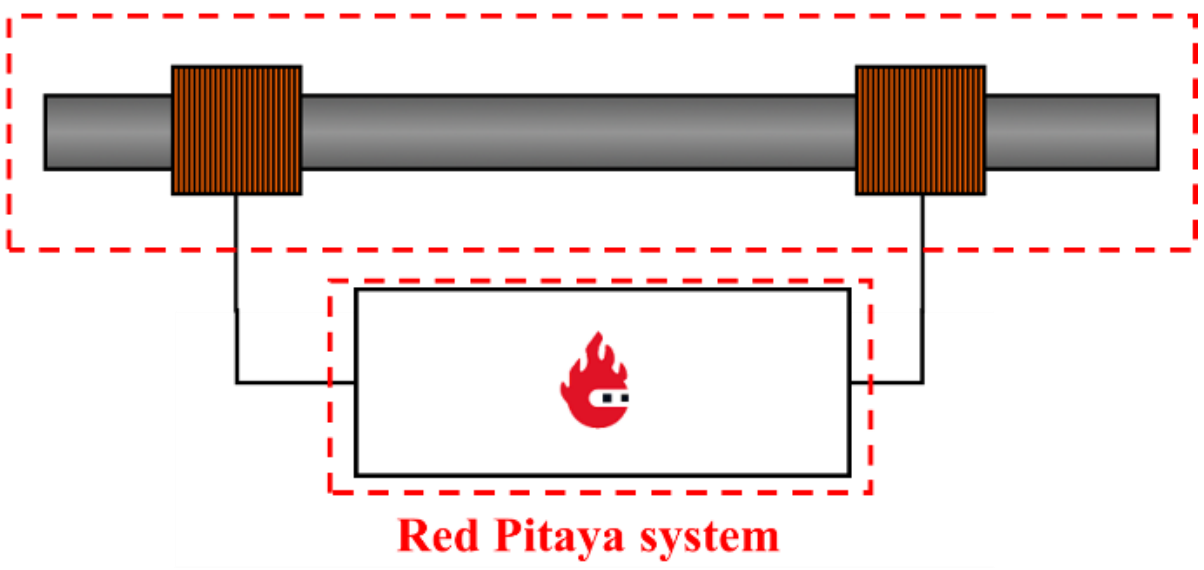

*Figure 5. 7 Twin coil magnetoelastic resonance characterization setup.*

Notably, the exciting and receiving coils are placed at a specific distance from each other to avoid any direct induction of the magnetic field generated by the exciting coil on the receiving coil. Rather, the induction on the receiving coil is due to the magnetic flux variation caused by the propagating elastic waves that originate at the opposite end of the resonator. Which is to say, the two coils are communicated only through the magnetoelastic sample.

Experimentally, this is evaluated by comparing the base gain levels with and without the resonator. For instance, Figure 5. 8 shows and example of this evaluation. These results reveal that due to the susceptibility of the resonator, the base gain level is approximately 10 dB higher when the resonator is placed between the coil and thus, the induction of one coil to the other can be neglected.

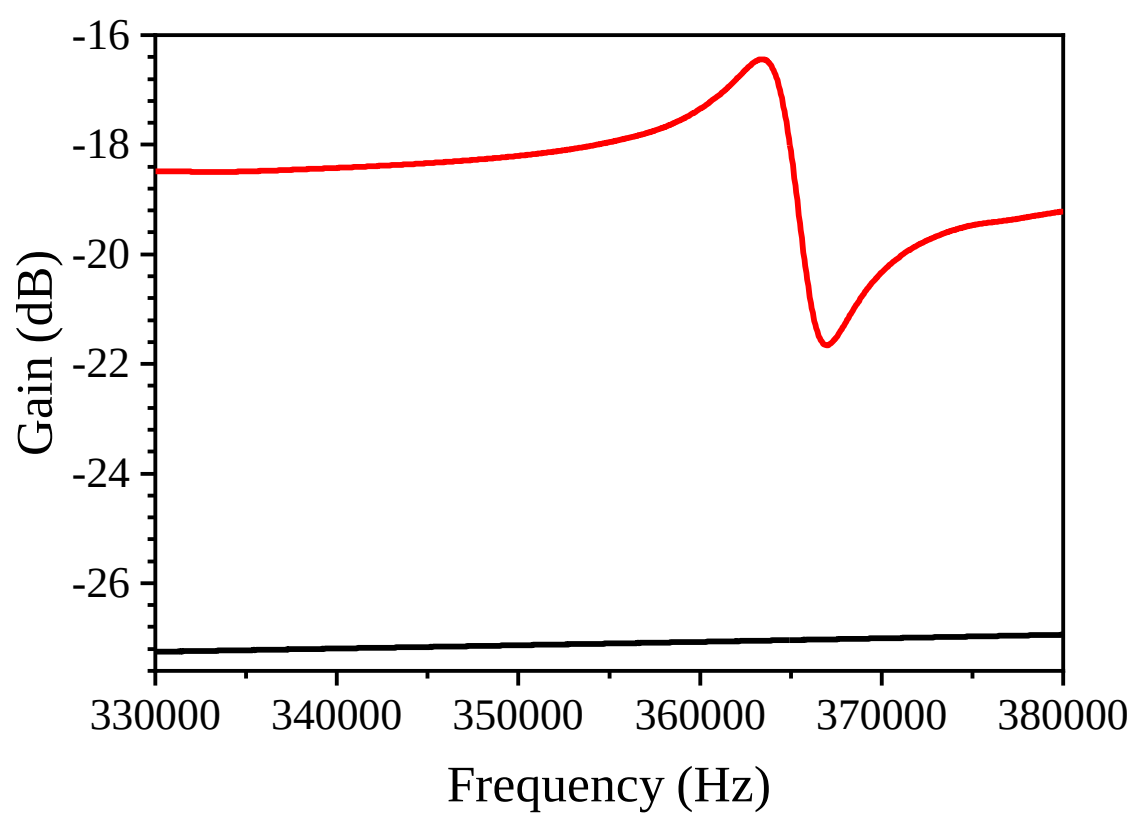

*Figure 5. 8 Gain spectrum of the twin-coil system with (red line) and without (black line) a resonator in place.*

Over this revision of the system, two significant upgrades were applied, a real-time monitorization capability and the integration of the bias field generation through and offset in the exciting signal.

The real-time monitorization capability was achieved by transforming the setup into an oscillating circuit as depicted in Figure 5. 9. Due to the characteristic change in the susceptibility of the magnetoelastic transducer, the circuit will oscillate, with a radiofrequency (RF) signal travelling the loop that matches the transducer's resonance frequency [69]. The oscillator system comprised of three basic circuit blocks:

- An amplifier.
- A feedback network (magnetostrictive resonator).
- A passive bandpass filter.

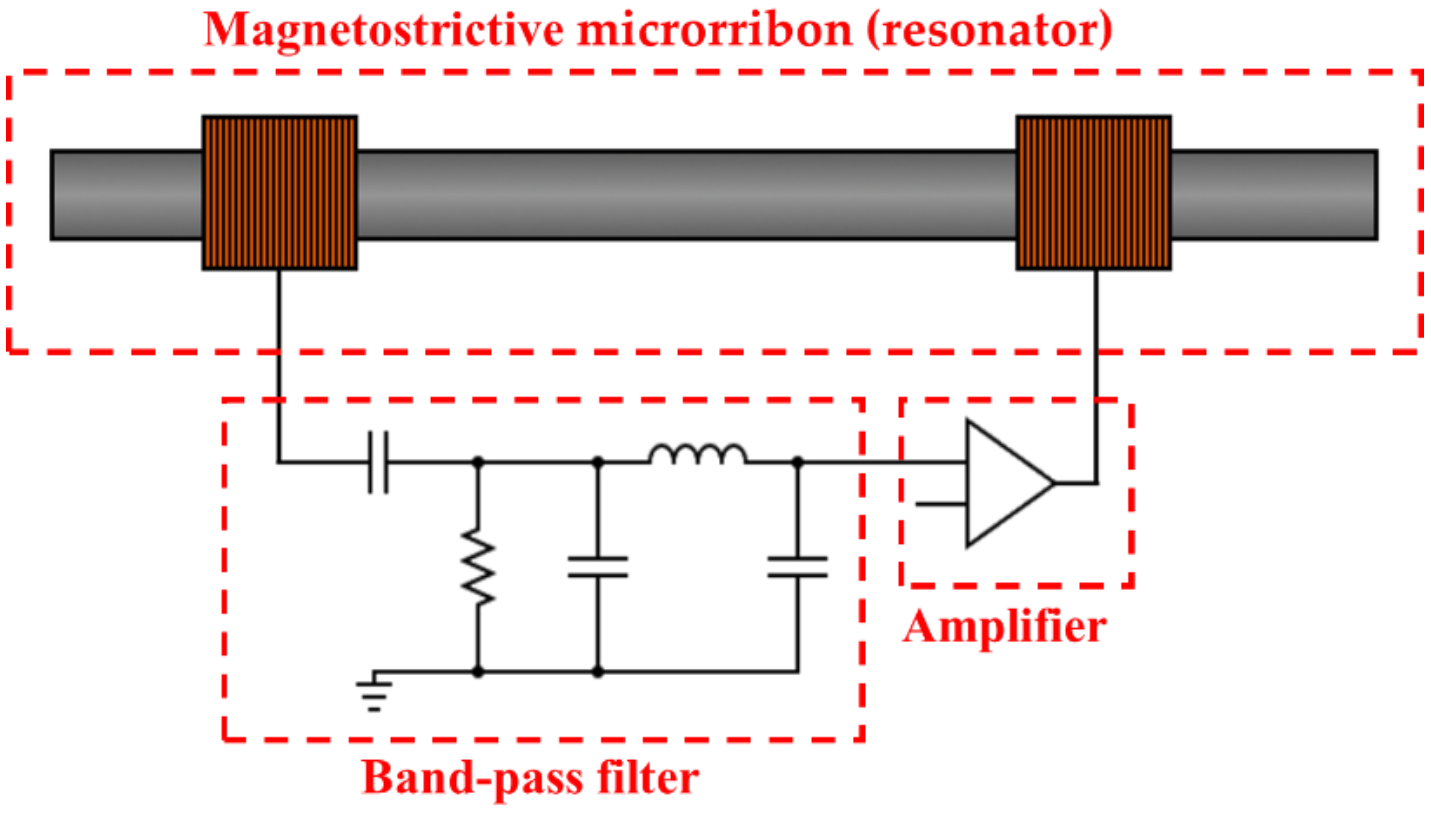

*Figure 5. 9 Schematics of the oscillator circuit.*

The amplifier was connected to a power source and was the only active element of the loop. The amplifier generates certain degree of noise that induces random multiple-frequency signals in the loop, from which the resonance frequency is the only one energetically favored to travel through the entire loop. The Barkhausen criteria should be satisfied, with a total phase shift of $2\pi n$ (with n being an integer other than zero) and a loop gain equal to the unity to ensure that the system works as an oscillator. When these criteria are fulfilled, the frequency of the oscillator circuit is autonomously and immediately synchronized upon any disturbance in the transducer.

The concept can be better understood by imagining the system as stationary. The frequencies contained in the noise signal are transformed into alternating magnetic fields that penetrate the magnetoelastic resonator. As per the definition of resonance, one frequency, $f_r$, or rather highly narrow range of frequencies, will travel through the resonator in the form of elastic waves without significant loss. These elastic waves will induce an electric signal into the second coil. The signal induced in the second coil are filtered using a passive bandpass filter that eliminates, for instance induction from the 50 or 60 Hz from the power grid or signal induced by other vibration modes or harmonics of the resonator. The filtered signal is then fed to the amplifier that produces an amplified $f_r$ and further noise. If the resonators $f_r$ is altered, then the process repeats again using the amplifiers noise as seed.

However, the system is not stationary, hence, the process previously can be described as autonomous and immediate.

The amplified $f_r$ can be easily detected. In a recent work $f_r$ was remotely detected using a probing coil connected to a frequency meter [48]. It is important to highlight that, as shown in Figure 5. 10, this probing system was a separated circuit independent of the main oscillating circuit. However, probing the circuit or using the acoustic or optical methods previously discussed are also possible options.

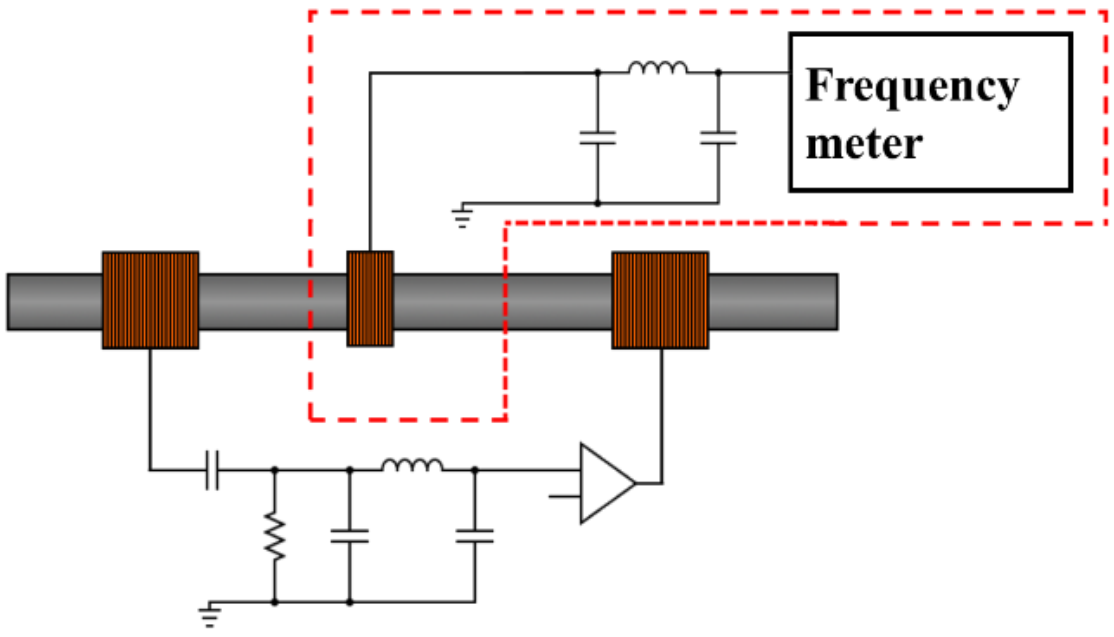

*Figure 5. 10 Probing system for real-time resonance frequency monitoring.*

The oscillating system needs to be tuned for each resonator, as was the case for the single coil system. Apart from the coil design that was previously discussed, the bandpass filter needs to be adjusted for the new resonator frequencies ($f_r$ and its harmonics), and the amplifier is adjusted for the gain values related to the sample and coils.

The first iteration of this system was rudimentary built over a protoboard system whereas the latest iteration was refined with more precise filters and soldered components (see Figure 5. 11). Interestingly, the system can be further integrated and miniaturized into a single component which constitutes the next intended iteration.

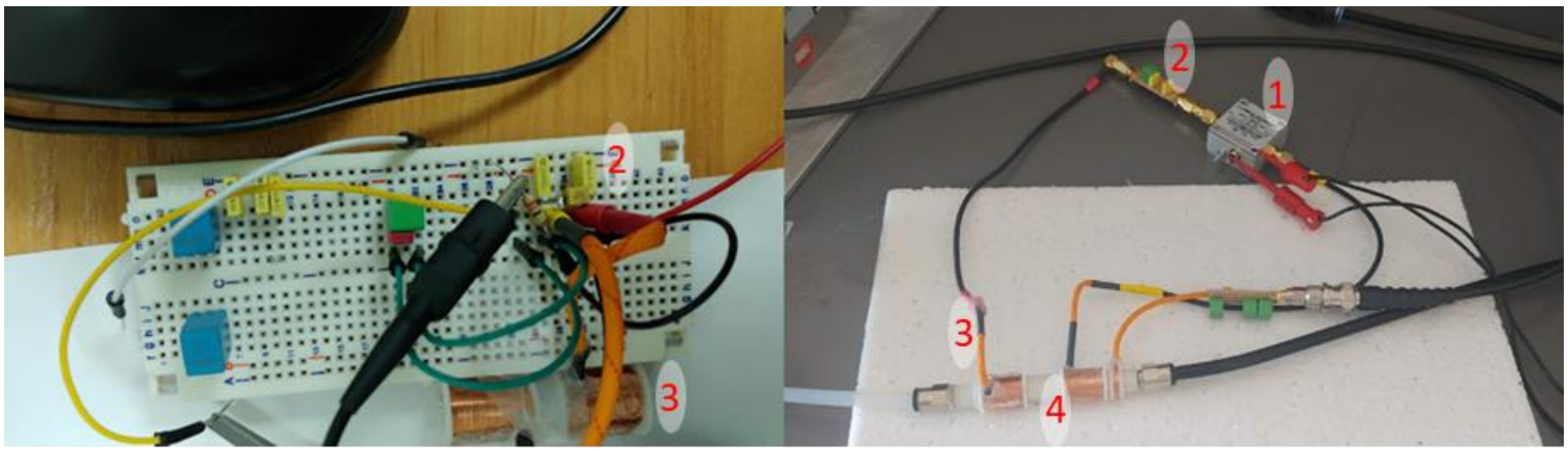

*Figure 5. 11 Evolution of the real time magnetoelastic resonance monitoring system (left to right). The right image also shows the gas cell chamber used for the gas sensing application reported in* [48]. *Some of the relevant components have been highlighted: (1) The amplifiers, (2) the bandpass filters, (3) the twin coils, and (4) the gas cell.*

The second significant upgrade is the integration of the bias field generation through and offsets in the exciting signal. This substitutes the use of permanent magnets or Helmholtz coils [50]. Instead, by using a bias tee the exciting coil produced a combination of a steady and alternating magnetic field.

Permanent magnets offer a convenient form of steady field generation with great integrability [48]. However, permanent magnets and particularly the magnetic strips found in security tags suffer from a great disadvantage related to the reproducibility of their characteristics. One the one hand, these magnets can be easily demagnetized, it was observed during their manipulation that meeting ferromagnetic tools (scissors, screwdrivers, or others) could significantly affect the strip's magnetization. On the other hand, these magnetic strips have poor corrosion resistance, which may have a negative effect on potential applications.

As for Helmholtz coils, they allow great control over the bias field value [50]. Although their use for research and lab is highly recommended on a potential application, they pose an obvious experimental complexity when designing a commercial sensing device.

A bias tee is an electronic component that is used to combine or separate high frequency signals and DC power in a single transmission line. It typically consists of three ports: an input port, an output port, and a DC port. The input and output ports are designed to pass high frequency signals, while the DC port is used to inject or extract DC power. It combines the tuneability and robustness of Helmholtz coils with the ease of integration of a permanent magnet.

However, a main drawback needs to be highlighted. The additional current flowing through the coil causes heating that affects the intensity of the alternating signal. Since the DC power source compensates the increment of resistance by fixing the output intensity, the bias field is not affected. The alternating signal on the other hand diminishes due to the increased resistance, which is translated as an overall loss of gain and a decreased exciting signal. This effect can be observed in Figure 5. 12, where the arrows along the spectra indicate different days of measurement or rather a break in the heating process during a bias magnetic field-dependent magnetoelastic resonance characterization.

Figure 5 12 illustrates a top view of the magnetoelastic resonance, the reason for the variation of $f_r$, the maximum seen red in the figure, with the bias field, will be discussed in a following section.

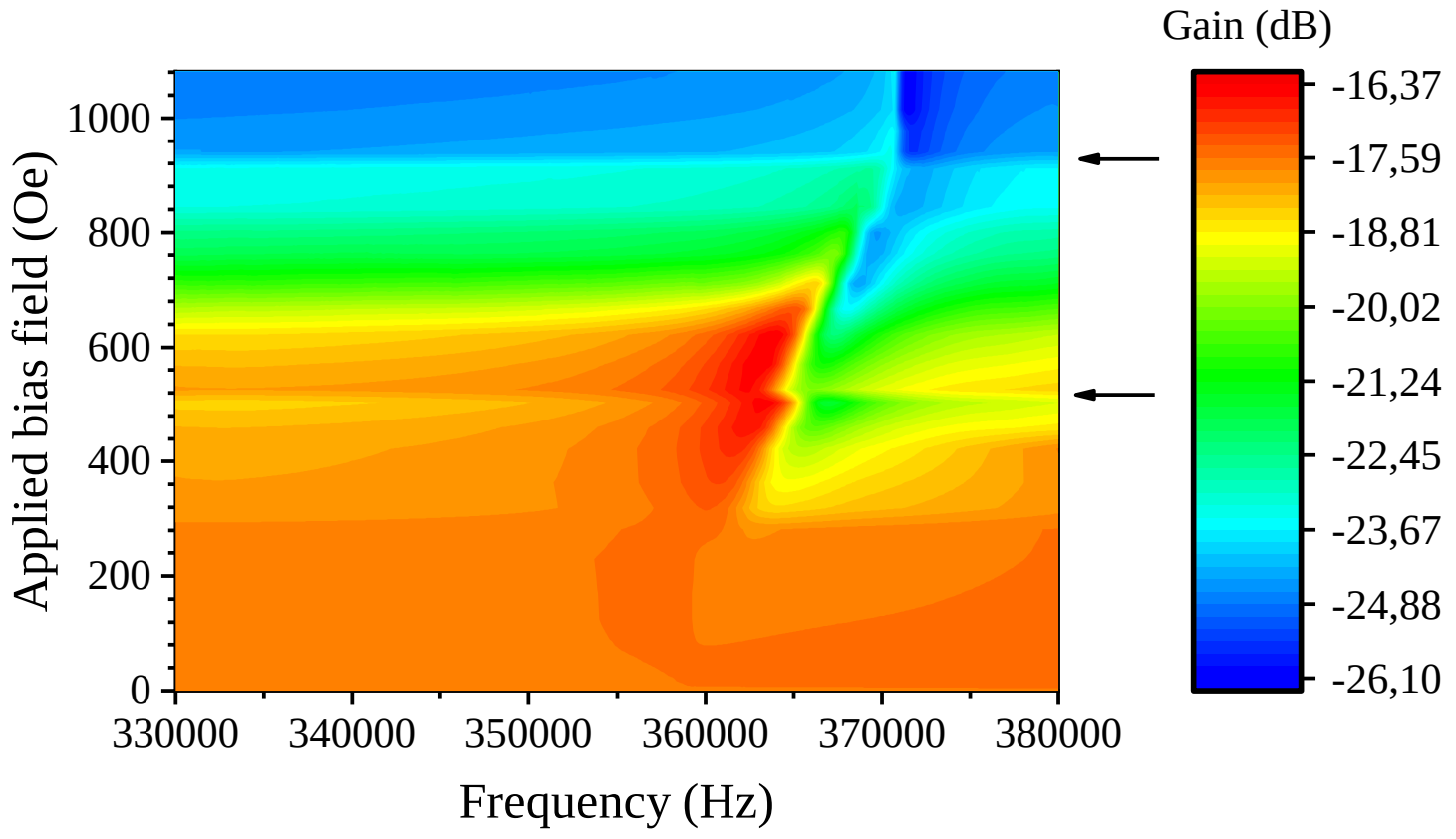

*Figure 5. 12 Bias magnetic field-dependent magnetoelastic resonance spectra for a 6.6 x 1 mm Metglass 2826MB microribbon. The bias field was produced with the bias tee configuration.*

Contrary to Helmholtz coil arrangement or a permanent magnet, the bias tee generates the bias field in a localized region of the resonator, i.e., in the extreme of the resonator by one of the twin coils. This however has not produced any negative noticeable effect on the measurement.

## *5.5 Magnetoelastic Behaviour of Amorphous Microribbons and Microwires.*

A series of test were performed to verify how different parameters affect the magnetoelastic resonance, such as the resonator geometry, the influence of applied bias field through the ΔE effect and the influence of mass loadings.

Two materials were used for this test Metglass 2826MB micro ribbons (MR) with an original size of 37 x 6.60 x 0.02 mm, and FeSiBNb microwires (MW) with 67 ± 1 μm metallic core diameter and a Pyrex shell up to 90 ± 3 [50]. In addition, the single coil setup was used for characterization of the magnetoelastic resonance and obtaining its figures of merit.

### *5.5.1 Influence of the length on magnetoelastic behaviour.*

From Equation 5.17, the length is a determinant factor of the frequency value for the magnetoelastic resonance. On that instance and having miniaturization as a potential goal for sensor developing, it is worth exploring the effect of the sample geometry, and particularly its length, on the magnetoelastic resonance.

For this test, the previously described MR and MW were cut to various sizes. MW were cut to lengths of 12.00, 10.00, 8.50, and 7.00 mm, whereas MR were cut to lengths of 32.50, 30.00, 26.50, and 23.50 mm.

Two different coils were used, one with 15 mm long, 2.4 mm in diameter and 1000 turns of 0.2 mm copper wire was used for MW and another with 55 mm long, 20 mm in diameter and 3000 turns of 0.5 mm copper wire for MR. All measurements were done with the same applied bias field, 5.15 Oe.

Results are shown in Figure 5.13. Notably, the resonance frequency shifts towards higher values with shorter lengths as predicted by Equation 5.17.

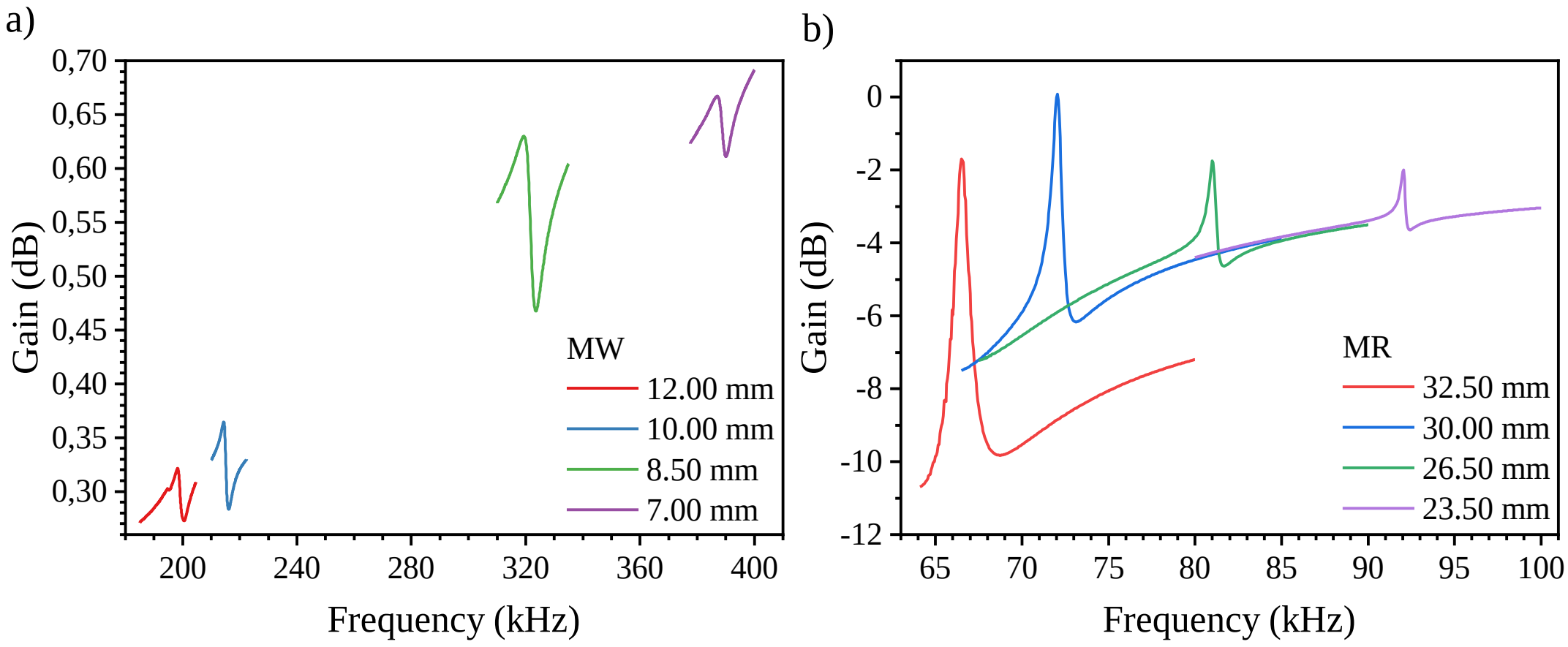

*Figure 5. 13 Magnetoelastic resonance spectra for a) MW and b) MR with different lengths.*

If the variations on the magnetoelastic resonance were limited to the change in physical dimensions, a decrease in A would be expected due to the decreased mass of magnetic material oscillating within the measuring coils. However, the resonators shape also affect the magnetic properties of the materials through its effect on the shape anisotropy previously discussed. To further study this effect it is necessary to also consider the magnetic aspect of the material.

### *5.5.2 Effect of the applied bias field.*

In this regard, the applied bias field was added as a variable. As previously discussed, the bias field affect the resonance through the ΔE effect. Particularly, the ΔE effect dictates that when the applied bias field matches the $H_A$, a measure of the resistance to magnetisation rotation, of the sample a critical point (maximum or minimum) can be observed in the figures of merit $f_r$, A and k [15,45].

Figure 5. 14 is presented as an example of the measuring protocol, i.e., the magnetoelastic resonance spectra is obtained for various bias fields. Figure 5.14 clearly demonstrates the effect of the applied bias field on the MR's magnetoelastic resonance, in particular, the resonance frequency achieves a minimum at ~8 Oe, a value that according to the ΔE effect would match the $H_A$ of the sample.

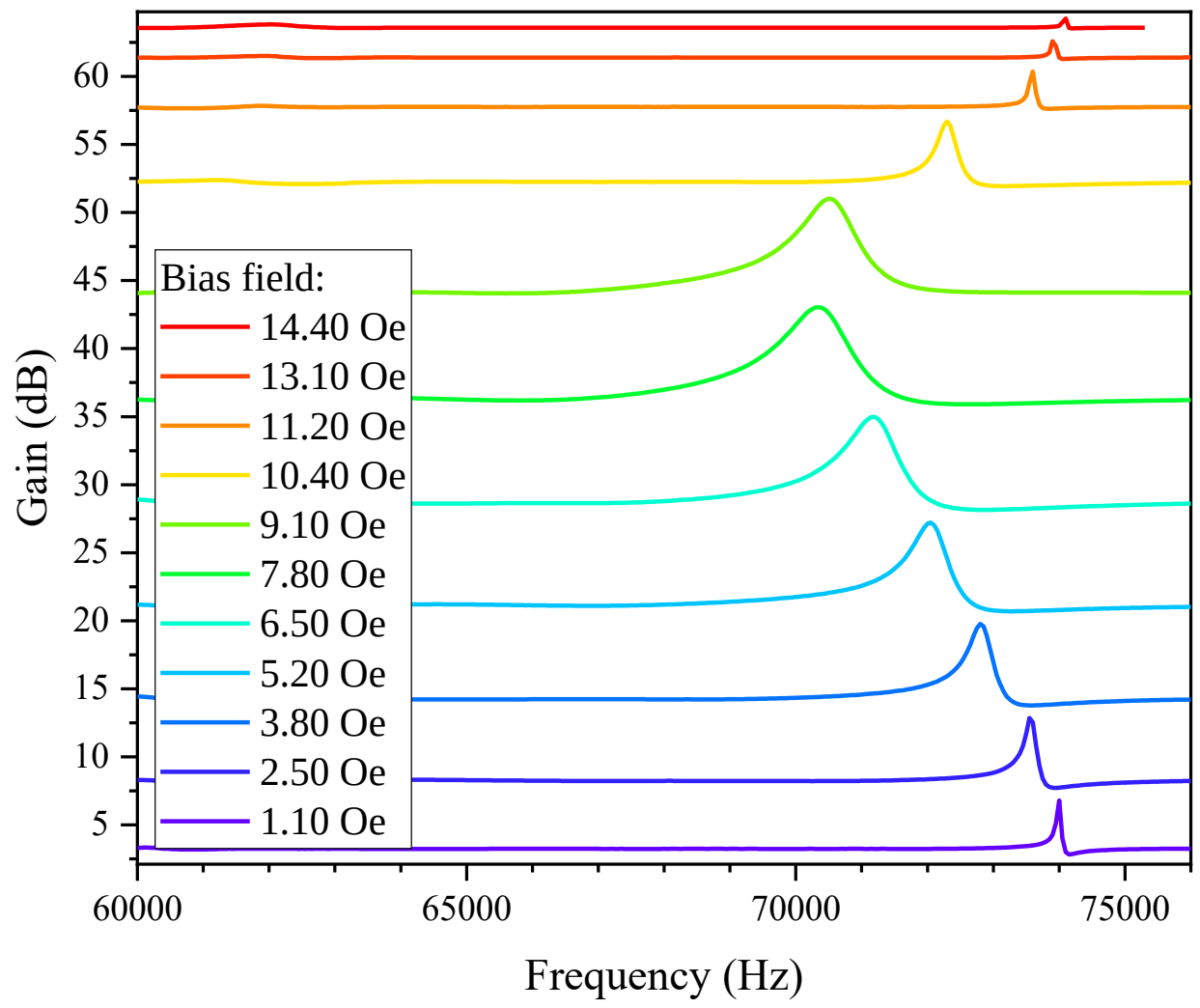

*Figure 5. 14 Magnetoelastic resonance spectra for the influence of the applied bias field on a 30 mm MR.*

Interestingly, for no applied bias field, no magnetoelastic resonance was detected. Conversely, for extremely high applied bias field, the magnetoelastic resonance decays until it is also no longer observed. This phenomenon can be understood by looking at Figure 5. 3, as previously discussed, the critical parameter that governates the magnetoelastic resonance is the ability of transversal magnetic domains to rotate and elongate the resonator.

The figures of merit are then extracted for the MR samples and presented in Figure 5. 15. In Figure 5. 15a, $f_r$ shift of the MR was studied with respect to its length and the applied bias field. Aside from the previously observed frequency shift with the resonator's length, a minimum in the frequency shift can be observed for all lengths as a function of the applied bias field.

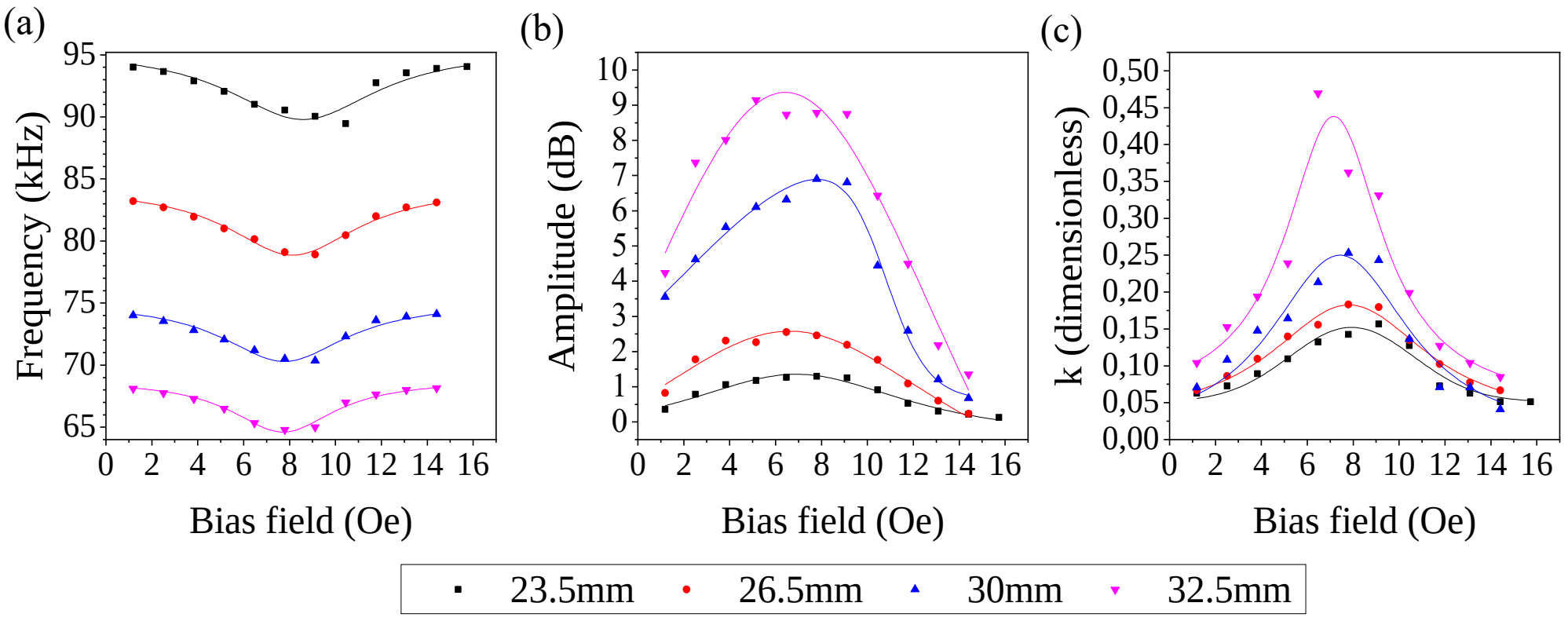

*Figure 5. 15 Relation of bias and a) $f_r$, b) A and c) k for the different MR with different lengths.*

In Figure 5. 15b, A of the MR is shown as a function of its length and the applied bias field. As the length decreases, A also decreases. This is due to the expected reduction of the magnetoelastic mass vibrating inside the coil, leading to a lesser variance in impedance. The optimal operating conditions defined by A can be identified by the maximum shown in Figure 5. 15b., higher A are easily detected and thus allow for easier measurement.

Figure 5. 15c shows k of the MR as a function of their length and the applied bias field. As a function of the length, k decreases for shorter lengths. This variation can be explained by the increase of the demagnetization field, given by the less favourable shape factor of shorter ribbons. As the strength of magneto-mechanical coupling depends upon the ease of rotation of magnetization, the highest values of k indicate better efficiency on the uptake of magnetic energy for the mechanical vibration. The maximum observed in Figure 5. 15c thus defines the optimal measuring conditions.

Similar to MR, $f_r$ of MW was also studied as a function of their length and the applied bias field, as shown in Figure 5. 16a. The $f_r$ increases linearly with decreasing MW length, consistent with the expected behaviour from Equation 5.17. A minimum in the $f_r$ shift can be observed for all MW lengths as a function of the applied bias field, indicating the optimal operating conditions (due to the scale, the minimum is better observed in Figure 5.17).

Figure 5. 16b shows A of the MW as a function of their length and the applied bias field. A clear dependence as a function of the length cannot be drawn from these plots; this is most probably due to imperfections on the wires derived from their production (heterogeneous composition and thickness) or their manipulation (shattering of the Pyrex shell). However, a maximum on A is clearly observed, which defines the optimal bias conditions.

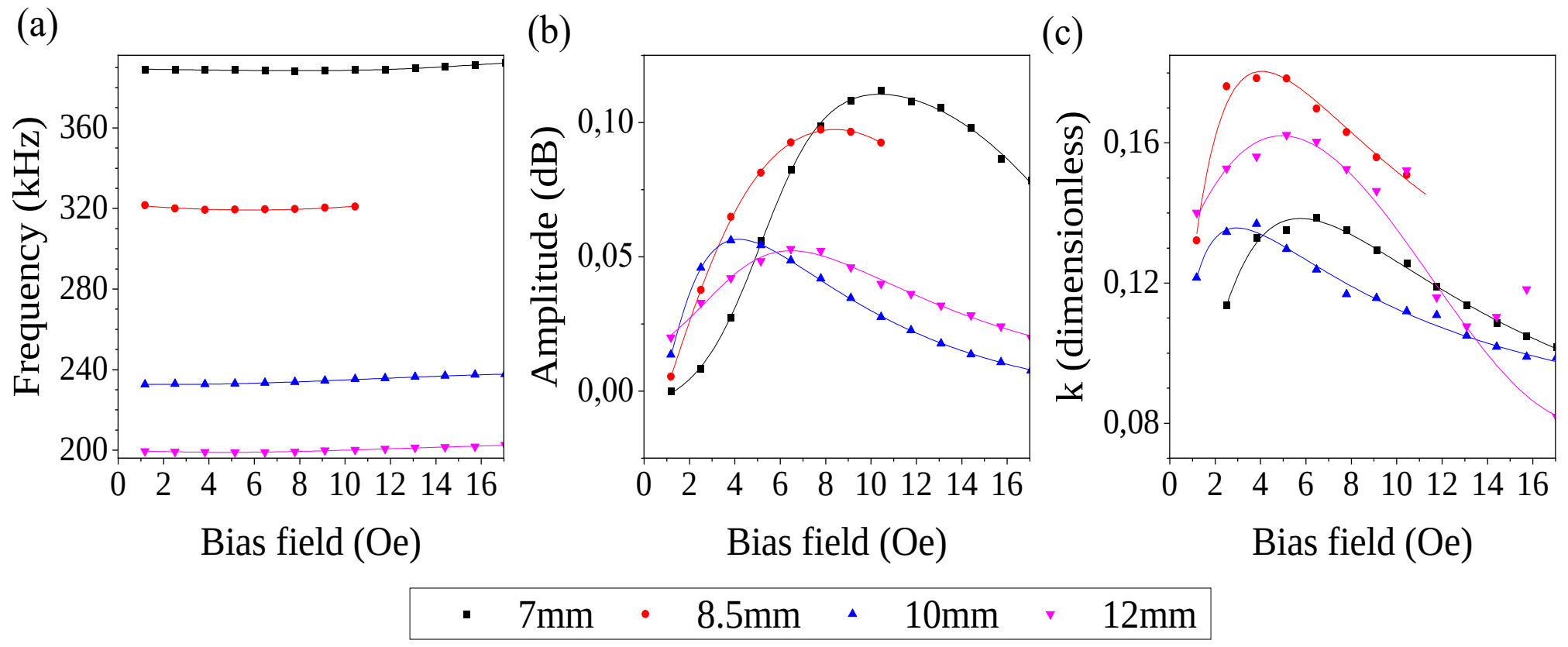

*Figure 5. 16 Relation of bias and a) $f_r$, b) A and c) k for the different MW with different lengths.*

Figure 5. 16c shows k of the MW as a function of their length and the applied bias field. A clear dependence on the length cannot be observed from these plots, similarly to the A results. However, an optimal bias field can be defined by the maximum in the coupling coefficient observed in Figure 5. 16c.

Figure 5. 17 summarizes the optimal conditions for MR and MW according to the ΔE effect. Interestingly, different bias field values are necessary to achieve optimal conditions for each parameter.

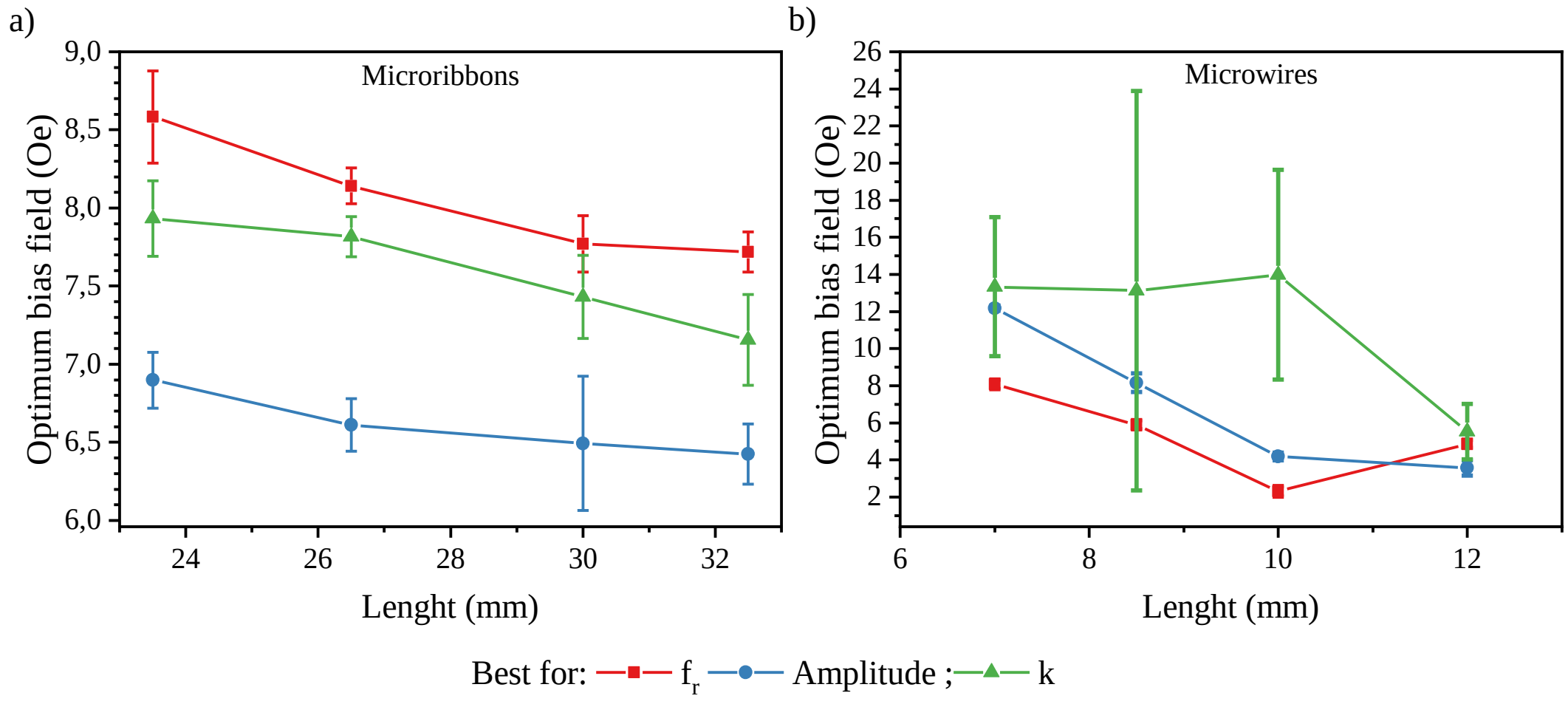

*Figure 5. 17 Magnetic field for minimum $f_r$ and maximum A and k for a) MR and b) MW with different lengths.*

For MR, a consistent decrease in the optimum bias field is observed for increasing lengths throughout all the figures of merit. Hysteresis loops for MR (see Figure 5. 18a) show a slight increase in $H_A$, defined as the magnetic field value where the magnetisation slope breaks and reaches $M_S$, is observed as the length of the resonator is reduced.

In this case, shorter MR are harder to magnetize in the long direction. From the definition of shape anisotropy given in the General Introduction (Equation 0. 2), shape anisotropy is mainly determined by the aspect ratio, which favours magnetization in the long direction for larger MR. For MR the change on aspect ratio (0.20-0.28), does not seem a determinant factor, as $H_A$ can be identified within the range of optimum bias field observed in Figure 5. 17a, i.e., 6 – 9 Oe. Then, the substantial variations of k with the resonator's length may have a different origin. For instance, it could be related to a decrease in the stability of transversal magnetic domains [70] when the MR lengths is decreased, or defects introduced during the cutting process. Regardless, length reduction had a negative effect on the MR magnetoelastic resonance.

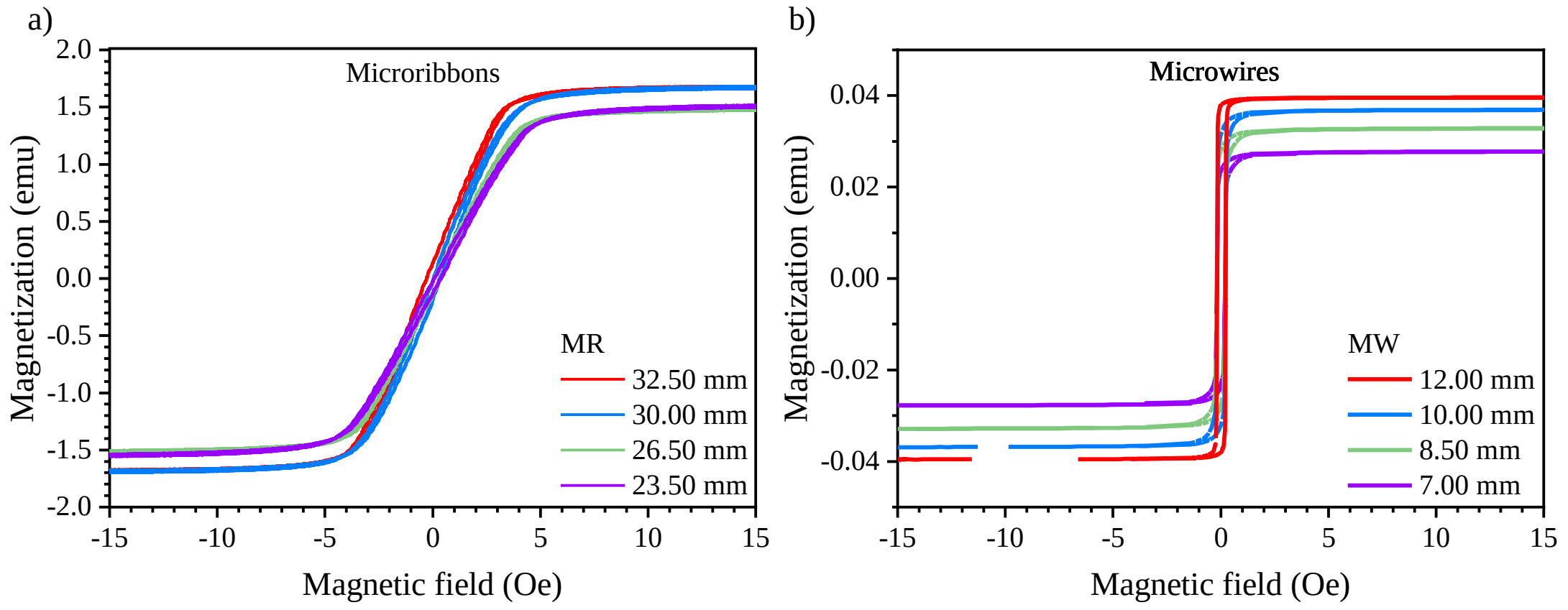

*Figure 5. 18 Hysteresis loops for a) MR and b) MW with different lengths.*

Conversely, for MWs, the results presented in both Figure 5. 16 and Figure 5. 17 indicate that length reduction had a limited effect on the samples magnetoelastic resonance aside from its effect in the resonance frequency. The shape factor of MWs varies from 0.006 to 0.01 when cut, meaning that even the shorter MW can be considered a relatively highly elongated object. Thus, the lack of influence on the shape anisotropy could explain how MWs are less affected by the length reduction, no significant differences are observed in the hysteresis loops either, besides of the reduced magnetization related to the reduced amount of material (see Figure 5. 18b). Nevertheless, MWs appear to be extremely sensitive to manipulation and experimental error and a conclusion for the behaviour requires further experimentation.

Overall, it was verified that there is an optimal bias field in terms of magnetoelastic resonance and that it may be influenced by the type of resonator. Careful characterization of the resonator in this regard is thus required for their potential use in sensing applications.

### *5.5.3 Effect of mass loadings.*

A first test was designed to evaluate Equation 5. 19 and Equation 5. 20. Briefly, the resonance frequency of a set of two MR (37 mm long) and two MW (10 mm long) was obtained before and after a mass loading.

These resonators were dip-coated in a solution of polyepichlorohydrin (PECH [71]) in acetone (12 mg/ml). Before and after coating, the resonators were weighed using a high-precision microbalance (± 0.001 mg; XP23, Mettler Toledo). Their resonances were characterized using a single coil resonance characterization system. The results are presented in Table 5. 2.

*Table 5. 2 Resonator characteristics for the mass loading effect on $f_r$ test.*

| Resonator | $m_0$ (mg) | $f_0$ (kHz) | Δf (kHz) | Measured Δm (mg) | Calculated Δm (mg) |
|---|---|---|---|---|---|
| MR 1 | 45.096 | 57.46 | -0.60 | 0.355 | 0.601 |
| MR 2 | 43.067 | 58.76 | -0.14 | 0.138 | 0.206 |
| MW 1 | 0.444 | 229.74 | -7.74 | 0.029 | 0.031 |
| MW 2 | 0.440 | 234.20 | -7.19 | 0.026 | 0.028 |

Interestingly, all the coated samples showed a redshift in resonance frequency as predicted by Equations 5.19 and 5.20. However, the micro ribbon sample values displayed a much worse correspondence with the calculated values (up to ±169.3%), compared to the microwire samples, which showed good agreement between the measured and calculated values (up to ±3.5%).

The discrepancies between the calculated and measured values can be easily understood by considering the assumptions of Equation 5. 19, which assumes a uniformly applied coating. This assumption holds true for the microwire samples, which have a low surface area and a slippery shape that prevents the deposition of large polymeric solution deposits. However, for the micro ribbon samples, which have a larger surface area and a concave shape, visually noticeable irregularities in the PECH deposit were observed.

In addition, the sensitivity, a sensing evaluation parameter that was introduced in Chapter 4, was calculated from the measured Δm, leading to values of 266.67- 277.00 Hz/ mg for the microwires and 1.01- 1.69 Hz/ mg for the microribbons. The difference in these sensitivities can be partially explained by the differences in the base resonance frequency. As stated in Equation 5.19, the higher the $f_0$, the higher the Δf for the same Δm.

Additionally, according to Equation 5. 17, $f_0$ can be adjusted by altering the length of the resonator. Shorter resonator elements will resonate at higher frequencies, theoretically leading to higher sensitivities.

To investigate this effect, a second series of tests were performed on the resonators with different lengths previously described. Specifically, MR with lengths of 32.5 mm, 30 mm, 26.5 mm, and 23.5 mm, and MW with lengths of 12 mm, 10 mm, 8.5 mm, and 7 mm were dip-coated twice in the same PECH/acetone solution used in the previous tests. The weight and resonance of each sample were measured after each dip coating.

Figure 5. 19 shows an example of the results for a 12 mm MW in which the effect of the mass loading over the magnetoelastic resonance is clearly observed.

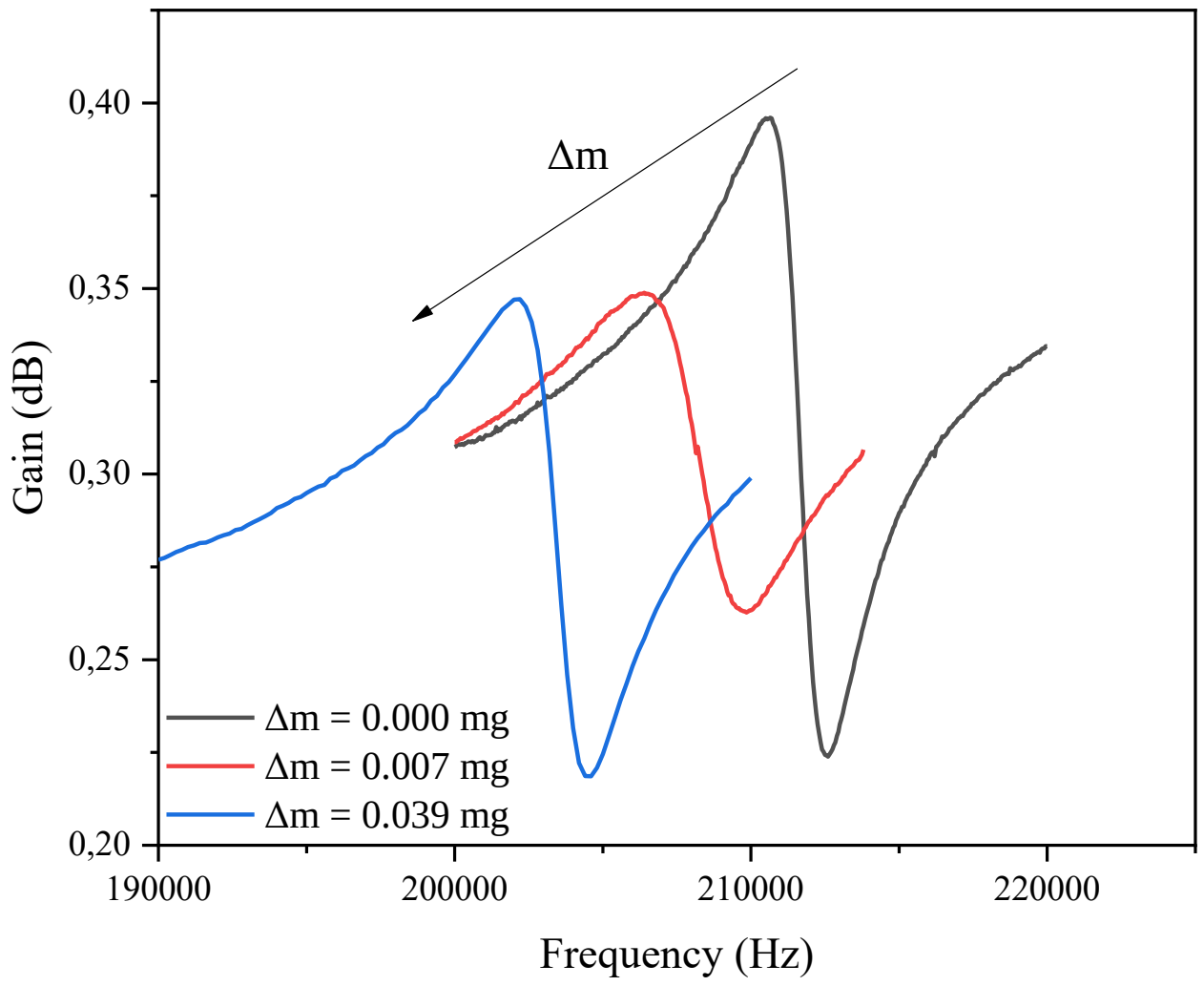

*Figure 5. 19 Magnetoelastic resonance spectra for the influence of mass loading on a 12 mm MW.*

From the resulting spectra, the position of the resonance frequency is extracted and presented in Figure 5. 20 where the blue line represents values calculated from Equation 5. 19 and the black lines represent the experimental values.

The sensitivities were calculated from a lineal fit of the experimental values. For micro ribbons, the sensitivity was 2.45 Hz/ mg ($R^2$=0.86), 1.68 Hz/ mg ($R^2$=0.99), 2.34 Hz/ mg ($R^2$=0.95) and 2.56 Hz/ mg ($R^2$=0.93) for lengths of 32 to 23.5 mm respectively. For microwires, the sensitivity was 216.45 Hz/ mg ($R^2$=0.99), 446.43 Hz/ mg ($R^2$=0.96), 271.00 Hz/ mg ($R^2$=0.98) and 649.35 Hz/ mg ($R^2$=0.73) for lengths of 12 to 7 mm respectively.

These results suggest that there is not a significant confirmation ($p$=0.05) of the hypothesis that a shorter resonator would lead to a more sensitive device based on the experimental values in micro ribbons or microwires. However, it is important to note that this study represents a preliminary investigation into the influence of mass loading in the magnetoelastic resonance, and a larger statistical sample size would be required to draw definitive conclusions.

While these findings do not necessarily invalidate Equation 5. 19, it is crucial to recognize that the experimental error, which is largely influenced by the quality of the coating and the measurement system, is a significant factor that cannot be ignored when calibrating these devices. Therefore, it may be necessary to calibrate each resonator experimentally to obtain accurate results, rather than relying solely on the general expression provided by Equation 5. 19. This underscores the importance of carefully considering experimental factors when designing and interpreting studies involving magnetoelastic resonators.

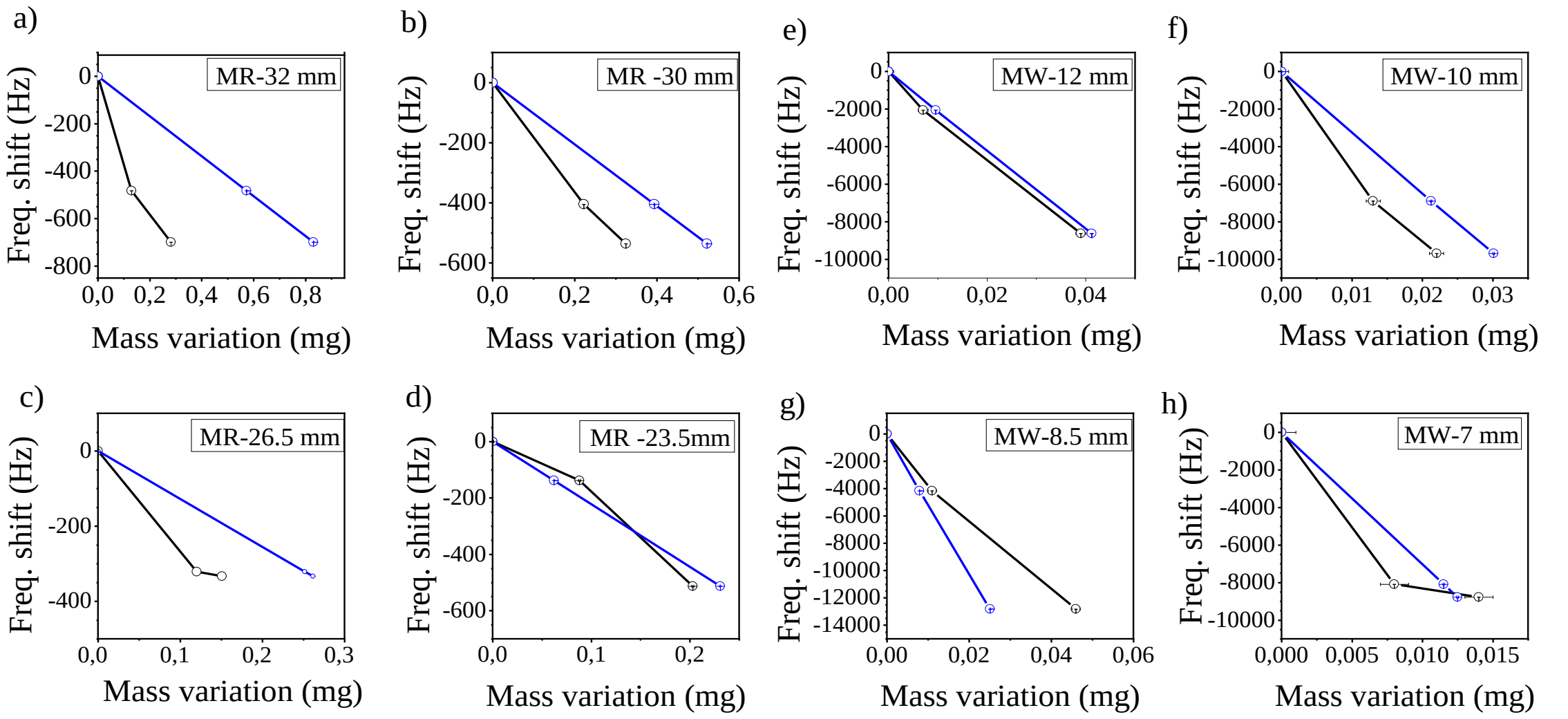

*Figure 5. 20 a)-d) micro ribbons samples, marked as MR and their length, and e)-h) the microwire samples, marked as MW and their length. The black line represents experimental data and the blue line data calculated with Eq. 5.19.*

In addition, the linearity of the Δf with Δm can be evaluated from the $R^2$ values of the lineal fit. Only two out of the eight samples showed excellent linearity ($R^2 \geq 0.99$). It is worth noting that for sensing applications, it is desirable to operate within a linear response range. However, magnetoelastic devices may deviate from linearity under higher loading. This issue has been previously addressed by A. Sagasti, B. Sisniega, and colleagues, who demonstrated that the sensitivity slope decreases with heavier loadings and proposed a second power expansion for Equation 5.21 [43,72]:

$$\frac{\Delta f}{f_0} = -a\frac{\Delta \mathrm{m}}{\mathrm{m}_0} + b\left(\frac{\Delta \mathrm{m}}{\mathrm{m}_0}\right)^2 \qquad \textit{(Eq 5. 21)}$$

Where a and b are calibration parameters unique for each resonator and extracted from experimental values.

Interestingly, the mass sensitivity of the magnetoelastic resonator can be optimized. For instance, the oscillation at the resonant frequency has points of maximum amplitude (see Figure 5. 21) known as hot spots. These hot spots are more sensitive to variations in the mass loading than other parts of the resonator [65]. If, for instance, a uniform coating would lead to significant dampening with negative effects on the potential sensor, the coating could be discretely applied only over these hot spots. Conversely, the nodes of each resonance have null oscillation and would thus show a neglectable to response to variations in the mass loading.

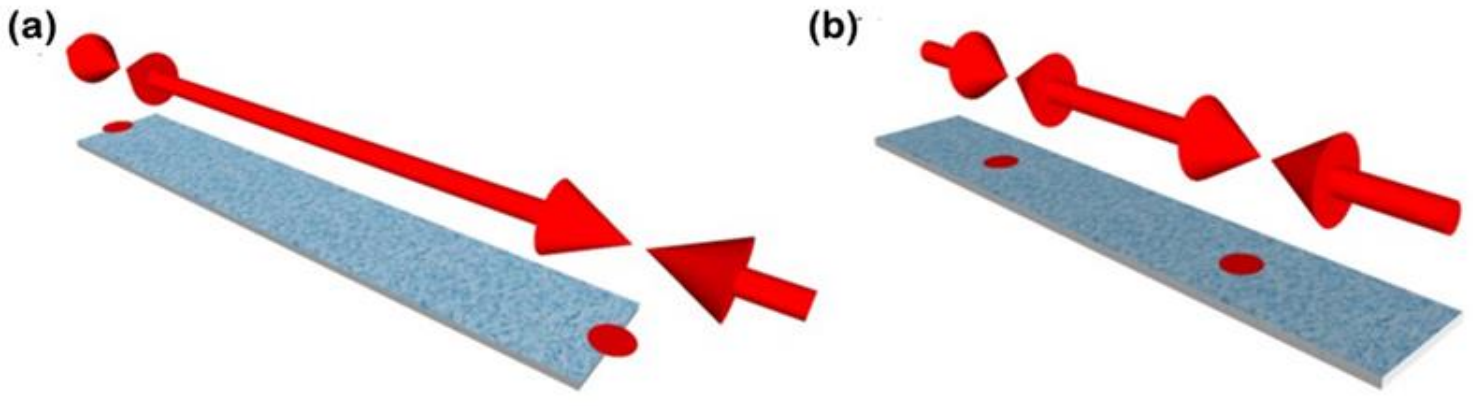

*Figure 5. 21 Hot-spots of a resonator at a) the first and b) the second harmonic of its resonance. Rep. from* [48].

The equation describing the effect of mass loading variation on the resonance frequency (Equation 5. 20) can be modified for discrete coating resulting in [73]:

$$\frac{\Delta f}{f_0} = -\frac{\Delta \mathrm{m}}{2\mathrm{m}_0}\left[1 - \cos\frac{\pi x}{L}\right] \qquad \textit{(Eq. 5. 24)}$$

Where x is the position of the discrete coating from the edge of a L long resonator.

Discrete coating can optimize the functionalization of the resonator and, more interestingly, can be used to create a multi-parameter sensor. By coating different sensitive materials on specific hot spots of different harmonics of the resonance, the frequency shift of each harmonic will be caused primarily by the response of that sensitive material, resulting in a sensor capable of detecting multiple parameters [74,75].

Traditionally, the reduction of the transducer size has been used as a technique to increase the mass sensitivity of the system. However, downsizing magnetoelastic platforms can result in the reduction of the resonance quality factor and signal intensity -as will be demonstrated for micro ribbons in the next section-, as well as the surface area available for active layers to detect specific chemical or biological compounds. As an alternative, recent research has focused on optimizing the geometry of the magnetoelastic resonator to improve its sensitivity without reducing its size [47,76,77].

By changing the resonator shape, it is possible to reduce the mass density in a hot sensing area, increase its resonant frequency, and optimize the tension to further enhance its sensitivity. These findings highlight the potential for novel geometries to be used in the development of highly sensitive and selective multi-parameter magnetoelastic resonance sensors.

The mass loading effect on magnetoelastic resonance has led to the development of highly sensitive and selective sensors for a variety of applications. By functionalizing the surface of the magnetoelastic resonator with specific recognition elements, sensitive polymers, antibodies, enzymes or DNA probes, the sensors can detect viruses [78], bacteria [79–81], biomolecules [82,83], biological processes [84,85] or gaseous molecules [48,57,59–61].

## *5.6 Gas sensor for breath analysis applications.*

In the preceding chapter, we thoroughly examined the role of gas sensors in pollution monitoring. To recap, gaseous pollutants, including nitrogen dioxide, carbon monoxide, and volatile organic compounds (VOCs), are known to induce or worsen respiratory and other health issues [86–89]. Consequently, creating devices that can swiftly detect these gases, even at low concentrations, is vital for reducing exposure and mitigating the risks to human health [90,91]. Solid-state chemical sensors, which operate on principles such as impedance, resistivity, and piezoelectricity, are among the most advanced technologies. These sensors offer high sensitivity, rapid response, low cost, and room temperature operation. However, they share a common drawback concerning instrumentation: the requirement for a physical connection, e.g., electrical wiring, which restricts application possibilities.

This constraint on physical connections is particularly significant in health-related applications, an area where gas-sensing applications are garnering increased interest [92–94].

The goal is to develop new analytical systems with sufficient capability and portability to serve as alternatives to traditional analytical systems [95]. Although conventional systems are accurate, they tend to be bulky and expensive. As a result, they are not commonly employed in scenarios that demand analytical investigation, such as air quality monitoring, safety, or medicine [96].

### *5.6.1 Breath analysis.*

Breath analysis, a medical application, is a particularly promising area for gas sensor usage. This technique offers several advantages over traditional diagnostic methods, such as its non-invasive nature, user-friendliness, and rapid results [97]. Breath analysis identifies biomarkers —molecules present in exhaled human breath— that are associated with abnormal biological processes or external factors [98,99]. The detection of these molecules can be influenced by factors like smoking, drug use, age, diet, or body mass index [100–102].

For example, ammonia has garnered significant attention as a biomarker [103]. Irregular ammonia levels in exhaled breath have been primarily associated with conditions such as type 2 diabetes [104] and bacterial infection [105], kidney malfunction [106], or asthma [107].

Ammonia is produced through several metabolic processes, including the breakdown of amino acids and urea in the liver. In healthy individuals, the liver converts ammonia into urea, which is then excreted in urine. However, in certain conditions such as liver disease, kidney disease, or metabolic disorders, the liver may not be able to process ammonia efficiently, resulting in elevated ammonia levels in the bloodstream and subsequently in the breath [108,109]. Additionally, certain bacteria in the oral and gut microbiome can produce ammonia through their metabolic processes [110].

Acetone is another key biomarker connected to blood glucose levels in diabetic patients [111,112]. As a result, detecting ammonia and acetone as gaseous biomarkers, in contrast to blood glucose meters that necessitate punctures, could offer a non-invasive, user-friendly, and risk-free alternative for daily monitoring of diabetic patients [104,113].

Acetone, a ketone body, is produced in the liver through the breakdown of fatty acids and is excreted through exhaled breath [114,115]. In individuals with uncontrolled diabetes, the production of ketone bodies, such as acetone, increases due to insulin deficiency, leading to an elevation in the concentration of acetone in exhaled breath [113]. While acetone is primarily used as a biomarker for diabetes, recent studies have also suggested its potential as a biomarker for other medical conditions, such as alcoholism, lung cancer and heart failure [116]

Some biomarkers may arise from external factors, like benzene. For instance, elevated benzene concentrations have been observed in breath due to smoking habits or air pollution. In such instances, this biomarker reflects the benzene accumulation within the body [101,102]. However, the specific interest in benzene arises from its potential to interfere with and disrupt the functioning of gas sensors targeting other biomarkers.

Water in the gas phase is another important analyte for gas sensors. As the body naturally produces it in high concentrations, humidity sensors have emerged as a valuable tool in biomedical applications. For instance, real-time humidity monitoring enables differentiation between exhaled and ambient air, thus helping to identify irregular breathing patterns caused by sleep apnoea, asthma, or cardiac arrest [97,117].

Apart from gas sensors, other techniques can be employed to analyse exhaled breath, including gas chromatography-mass spectrometry (GC-MS) [118,119]. In GC-MS a sample is first vaporized and separated into its individual components by passing it through a GC column. The separated components then enter the MS detector, where they are ionized and fragmented. The mass-to-charge ratio (m/z) of the resulting ions is measured, producing a mass spectrum unique to each compound. This technique provides high sensitivity, selectivity, and the ability to identify and quantify a wide range of compounds. However, this technique suffers from protracted diagnosing time, high cost and bulky volume caused by various pre-treatment processes [97].

However, there are still some challenges associated with breath analysis, such as the need for standardization and validation of analytical methods, as well as the identification of biomarkers specific to certain diseases. Despite these challenges, breath analysis remains a promising area of research for disease detection and monitoring.

Magnetoelastic resonance-based gas sensors are particularly appealing for breath analysis applications due to their low-cost, remote operation, and unneeded integrated power source, as they only operate under the externally applied magnetic field [63,65]. In addition, this enormous advantage means they can be easily miniaturized or made into portable sensors [120,121].

Effective diagnosis requires detecting a mixture of biomarkers, called exhaled breath profiles, which can contain more than 3000 compounds [122]. Building an array of sensors with different selectivity allows for the detection and identification of multiple gases or compounds in a single measurement. Each sensor in the array can be designed to selectively respond to a particular gas or compound, allowing for a diverse range of analytes to be detected. This approach offers several advantages over a single sensor, including higher selectivity and sensitivity, reduced false positives, and the ability to identify complex gas mixtures. The work presented in this thesis has focused on the development of a single sensor, either chemiresistive or magnetoelastic resonance based. However, in future work, the development of an array of sensors with different selectivity is envisioned to further enhance the sensing capabilities and broaden the range of detectable analytes.

*5.6.2 Device Preparation.*

The magnetoelastic resonator used for this work was a 37.00 mm × 6.60 mm × 0.02 mm Metglass 2826 MB magnetoelastic micro ribbon.

The sensitive layer consisted of electrospun PVP nanofibers directly deposited over the microribbon with an electrospinning method. First, PVP (Mw = 360000 g/mol, ref: 81440 Sigma-Aldrich, Burlington, MA, USA 81440) was dissolved in distilled water at a 1:4 weight ratio and stirred until homogeneous. The solution was then degasified under vacuum conditions until any visible air bubbles were re-moved. Next, the solution was loaded into a syringe with a metal needle connected to a high-voltage power supply and placed at a 14 cm distance from the microribbon that acted as the grounded collector. Finally, the solution was extruded at a flow rate of 5 µL/min for 30 min at a voltage of 14 kV. The electrospinning was done at the SENSAVAN's lab at the *Instituto de Tecnologías Físicas y de la Información* (*ITEFI*) from *Consejo Superior de Investigaciones Científicas* (*CSIC*), Spain.

The deposit was evaluated using scanning electron microscopy (SEM). Despite PVP being non-conductive, no static charge artifacts were observed. The sensitive layer was successfully deposited over the transducer. Due to the characteristics of the electrospinning technique, the water based PVP solution dried and formed polymeric nanofibers, with a thickness in the 100 nm range, on the target (Figure 5. 22).

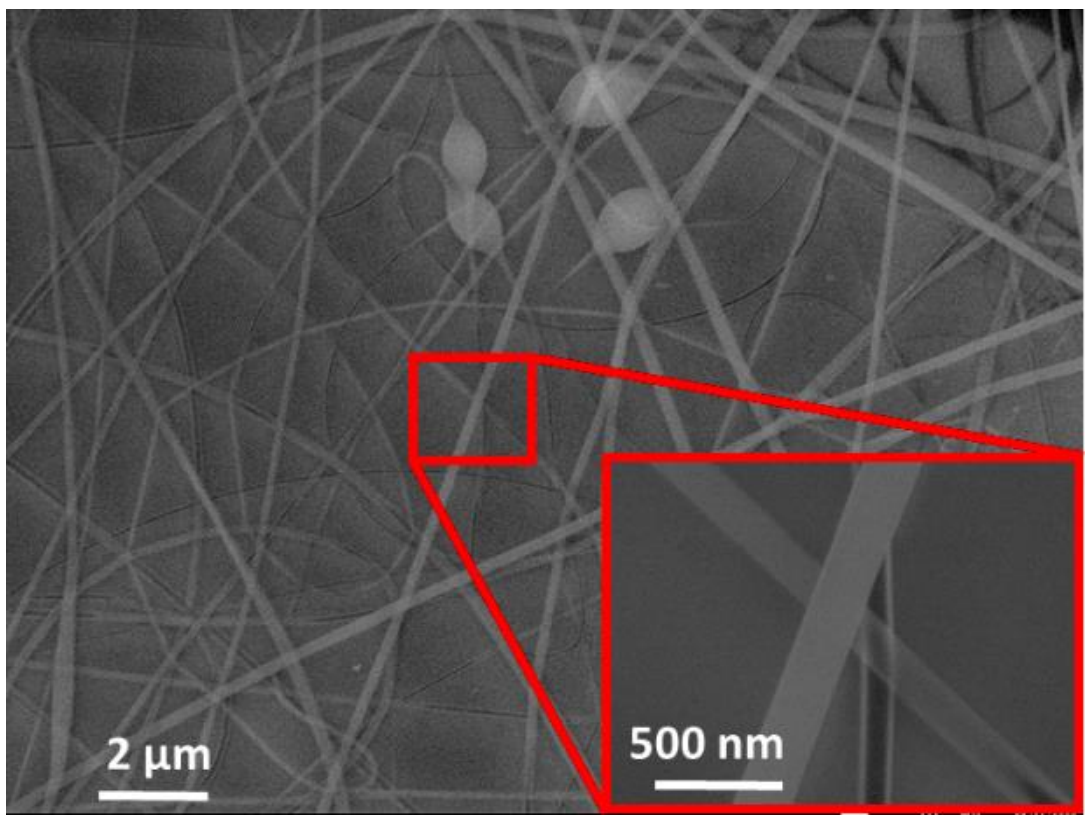

*Figure 5. 22 SEM images of the electrospun sensitive layer deposited over the resonator. Rep from* [48].

The nanofibers were evenly deposited on the transducer's surface mainly due to the good electrical conductivity of the magnetoelastic micro ribbon, which promoted a homogeneous electric field distribution during the electrospinning process.

The gas sensor, based on the functionalized magnetoelastic resonator (MEGS) was placed, with the deposited side up, in a custom-made 3D-printed PLA airtight cell (with a volume of ~15 mL) connected to an automatized gas sample generator (as anticipated in Figure 5. ). Inside the cell was a 30.85 × 6.00 × 0.05 mm permanent magnet (~6.50 Oe) (Figure 5. 23). The magnet creates a bias field near the ribbon's $H_A$, which induces magnetization in the transducer, typically between 2.50 and 11.30 Oe [70]. This magnetization maximizes the magnetoelastic coupling coefficient value and improves magnetoelastic effect detection [42,123].

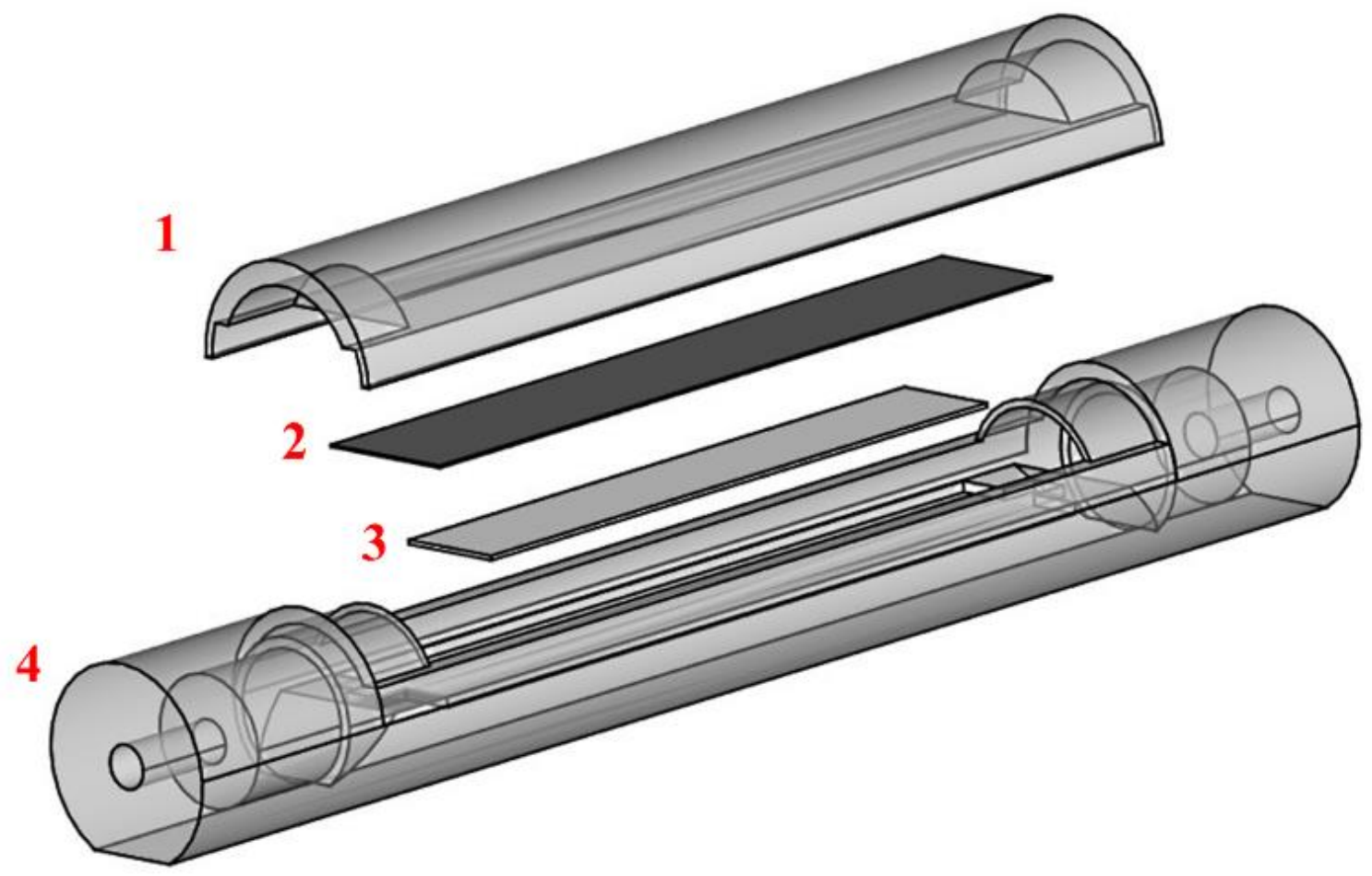

*Figure 5. 23 Schematics of the sensor cell, which includes cover (1), magnetoelastic microribbon (2), permanent magnet (3), and main body (4).*

The magnetoelastic resonance was characterized before and after the PVP deposit, the characterization setup used a twin coil configuration paired with the Red Pitaya system. The set of twin coils had a diameter of 12 mm, a length of 17 mm, and 300 turns of 0.15 mm copper wire. The frequency sweep was configured to include the first and second harmonic of the magnetoelastic resonance. Results are presented in Figure 5. 24.

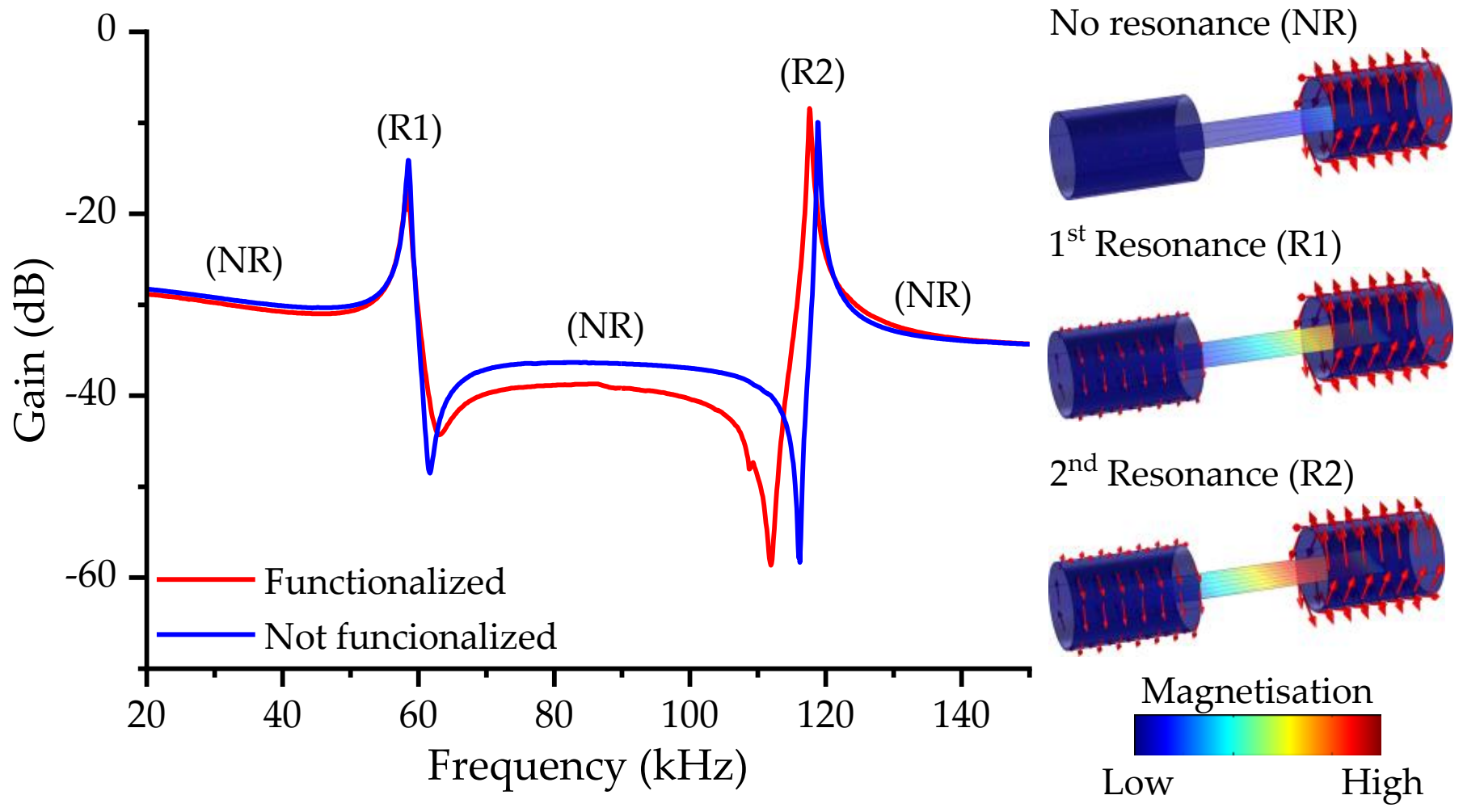

*Figure 5. 24 Frequency spectra of the transducer with and without the sensitive layer functionalization. COMSOL simulation results are used to illustrate the magnetoelastic resonance in the gain spectra for the no resonance state (NR) and the first (R1) and second (R2) harmonic of the magnetoelastic resonance. Rep. from* [48].

Although in general, the first harmonic is usually greater when centred coils are used, in this case, due to the placement of the coils on the magnetoelastic ribbon sides, the second harmonic was favoured against the first [48].

COMSOL Multiphysics simulations are also presented in Figure 5. 24. Susceptibility was parametrized with a relatively high value for the resonance case and a relatively low value for the non-resonance case, both values being arbitrary. These simulations confirm that during the second harmonic, a larger induction is produced in the receiving coil compared to the induction during the first harmonic [48].

In addition, a redshift is observed after the PVP functionalization as predicted by Equation 5. 20. At the same time, the deposit had a neglectable effect on the dampening of the resonance amplitude, which remained largely unaltered, indicating the robustness of the magnetoelastic strips as a transducer for MEGSs.

For the gas sensing measurements, the oscillator system previously described was used with the same set of twin coils. The remote probing was done with a coil, with a diameter of 12 mm, length of 2 mm, and 35 turns of 0.15 mm copper wire, connected to a frequency meter (Agilent 53131A, Agilent, Santa Clara, CA, USA).

The passive band filter was tuned to select the second resonance frequency harmonic. The second harmonic of $f_r$ was chosen for the MEGS operation because of its higher amplitude, which should also improve the sensing performance, and its intrinsic higher sensitivity (higher $f_0$, Equation 5. 20).

The quality factor of the oscillator, i.e., the whole circuit including the MEGS, was calculated using Equation 5. 23 and a high-resolution measure (Figure 5. 25). Under continuous-wave operation, the amplifier can compensate for the losses of the magnetoelastic resonator in the oscillator. When such gain is considered, the Q factor re-lates to the whole oscillator system instead of the magnetoelastic resonator alone. With the linewidth of 9 Hz for a frequency of ~119 kHz, a high-quality factor of $Q_{oscillator} > 13,000$ was obtained.

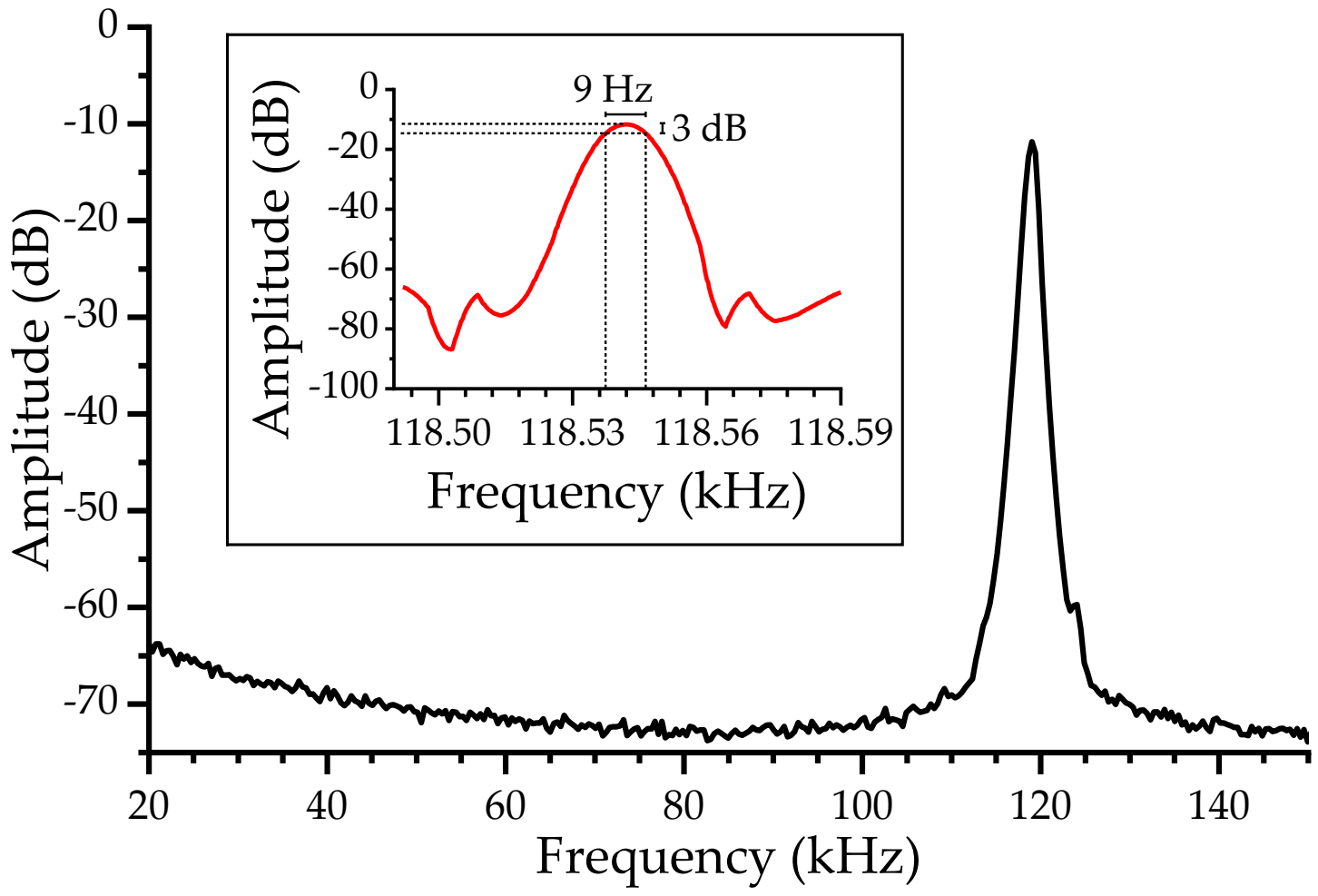

*Figure 5. 25 High-resolution frequency spectra for the quality factor determination with 13 Hz steps (black line) and 0.2 Hz steps (red line).*

### *5.6.3 Gas Sensor Characterization.*

The characterized system was then submitted to gas sensing experiments. Four analytes were tested, including ammonia, acetone, benzene, and humidity as per their importance in breath analysis previously discussed. The sensing response was evaluated in with a characterization setup similar to the one used on chemiresistive gas sensor that was comprehensively described in the previous chapter.

Few notable modifications were done over the gas line to adapt its operation to the magnetoelastic resonance-based sensors, the gas cell, as previously described, and the instrumentation (see Figure 5. 26). Adjustments to the data acquisition software were performed to adapt the instrumentation control and monitoring to a frequency meter to operate the real-time magnetoelastic resonance system. All gas measurements and setups were done at at the SENSAVAN's lab, *ITEFI.*

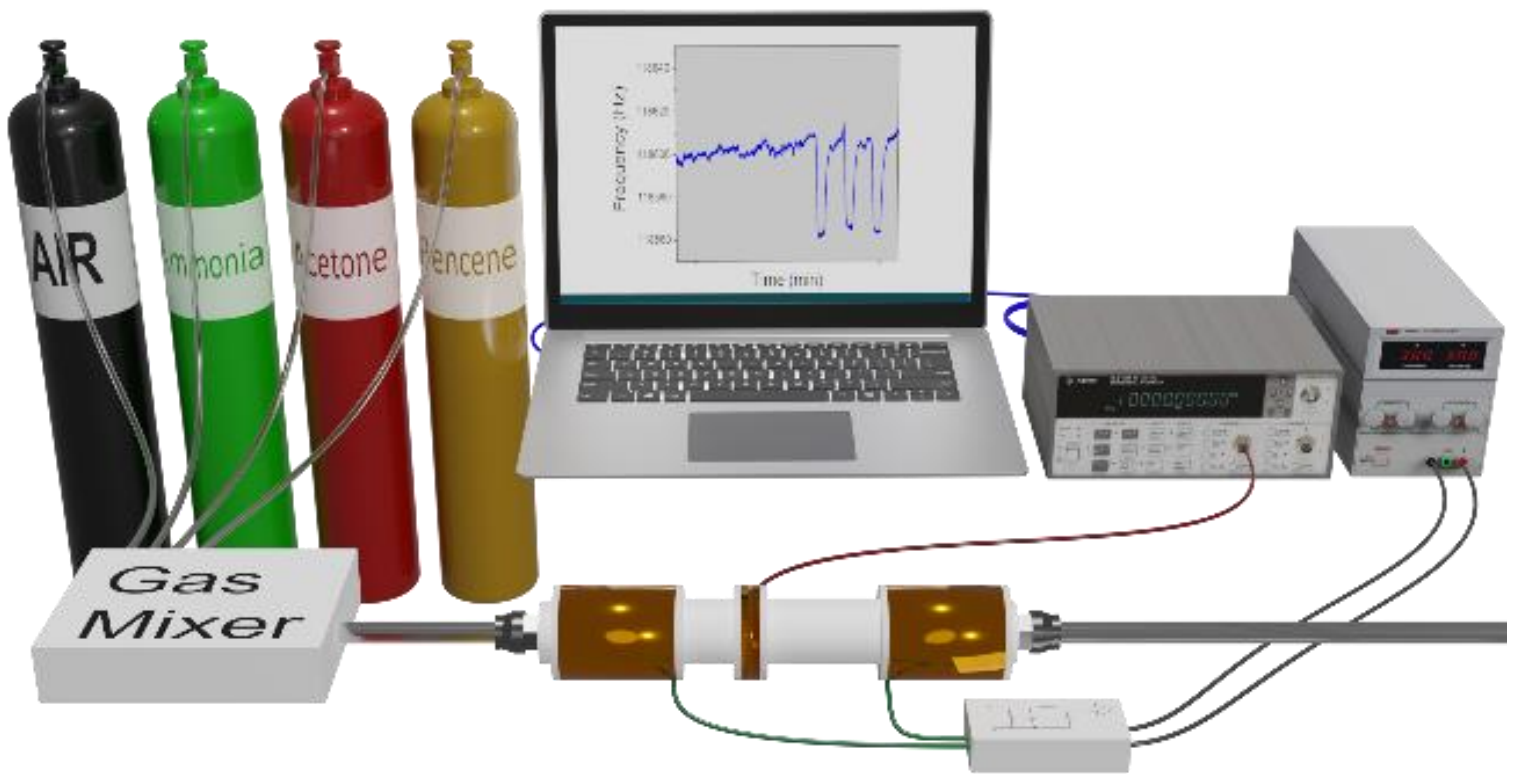

*Figure 5. 26 Magnetoelastic resonance-based gas sensor characterization setup.*

The sequences consisted of a baseline step, where only synthetic dry air was flushed in the sensor chamber, and cycles with an exposure phase, where the sensor was tested with the gas mixture containing the analyte, and a purge phase, where the device was again exposed only to the carrier gas.

For the humidity test, synthetic air was passed through a water bubbler and mixed with the carrier to create different humidity conditions to which to expose the device. The mix was calibrated with a humidity and temperature meter (RS 1364, RS Components, London, UK) to achieve the desired RH values. The sequence consisted of a 60-min baseline step followed by three cycles at each RH level (17%, 36%, 54%, 73%, and 95%). Each cycle had a 5-min exposure and a 1-min purge phase.

For the acetone, ammonia, and benzene tests, the analytes (from 50 ppm balance air cylinders) were mixed, using synthetic dry air as carrier at a concentration of 40 ppm. All gases were provided by Nippon Gases, Madrid, Spain. The sequences consisted of three different runs, one for each analyte, with a 60-min baseline step and three cycles, with a 2-min exposure phase and a 5-min purge phase.

The experiments were performed in a temperature-controlled room at 25° C. The air flow inside the cell was set to 100 mL·min$^{-1}$ in all cases.

Three parameters were used to evaluate the sensor's performance: response, limit of detection (LoD), and $\tau_{90}$. These parameters were extensively discussed in the previous chapter. Briefly, the response represents the amount of change in the measured physical magnitude of the sensor. In this case:

$$\Delta f = f_r - f_i \qquad \textit{(Eq. 5. 25)}$$

Where $f_i$ is the resonance frequency prior to the exposure. LoD, is the lowest concentration of gas that can be reliably detected by the sensor, and $\tau_{90}$ the time required to achieve the 90% of the response.

### *5.6.4 Results And Discussion*

First, the response of the device towards water in the form of humidity was evaluated. As water acts as a solvent of PVP, $H_2O$ molecules are trapped in the sensitive material's nanofibers, incrementing the deposited mass over the transducer. A negative resonant frequency shift, as predicted by Equation 5.20, was observed during the real-time monitoring under the exposure to different RH levels (Figure 5.27).

The sensor responded with high reproducibility. For instance, after three exposures to the same RH level, the relative standard deviation of the response was ≤ 4% (Figure 5. 28a). In general, the sensor device showed a fast recovery with a low baseline drift for tested RH.

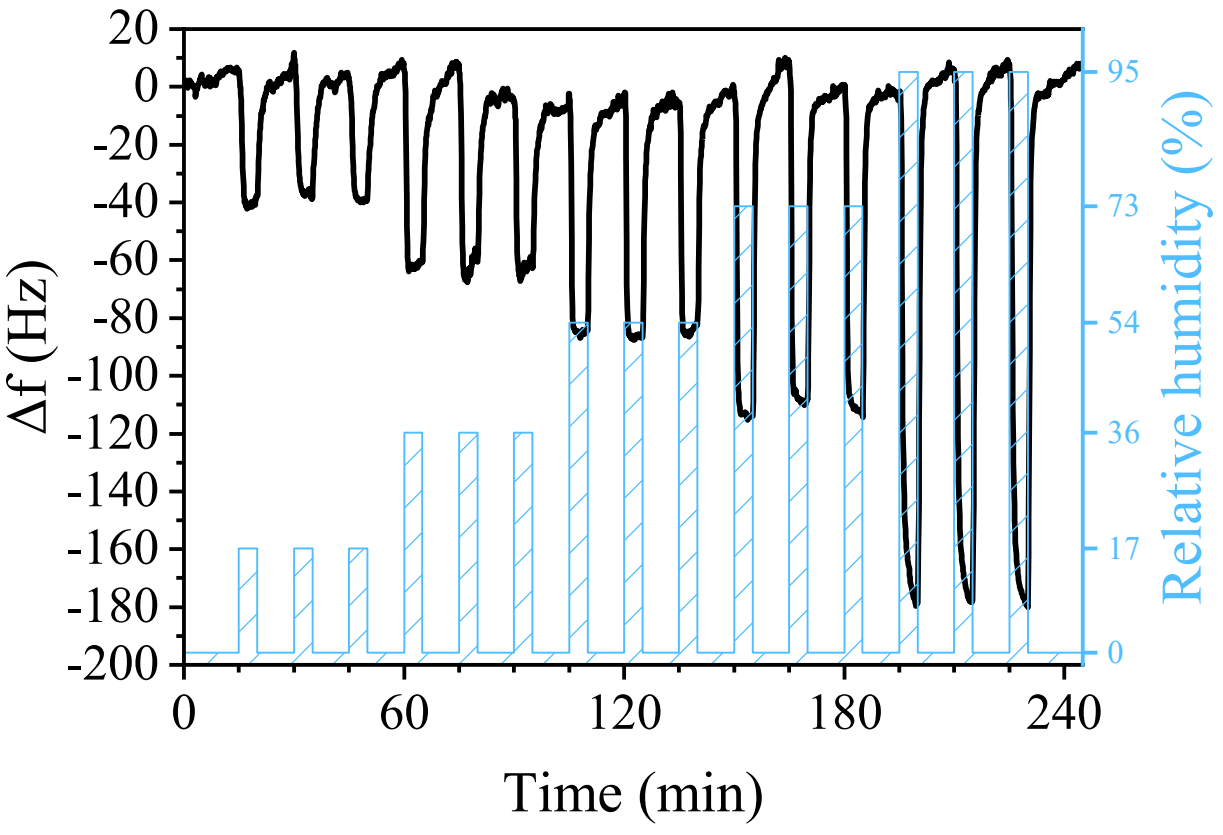

*Figure 5. 27 Real-time frequency shifts of the device exposed to different RH levels.*

The sensitivity of the device, defined as the response to a specific concentration of the analyte (expressed in Hz/%), was obtained by fitting the values provided in Figure 5. 28a. A linear response regime for levels up to RH = 73% was established. The linear fit provided a sensitivity of −1.17 ± 0.1 Hz/% ($R^2$= 0.97). However, for high RH, i.e., 95%, the sensors' behaviour deviated from the linearity. This effect may be explained by a large total mass of the active layer relative to the mass of the transducer [43] or, most probably, by condensation effects. The resulting LoD was a RH level of 4.7 ± 0.4%.

Figure 5. 28b shows that the sensor's response toward water is fast, as 90% of the maximum change was achieved during the first two minutes or less. The fast response of the device was indicated by the resulting $\tau_{90}$. However, the response of the device could have been delayed significantly due to the effect of the water bubbler, i.e., the bubbler acts like a virtual pre-gas cell increasing the chamber constant [124] of the experiment. This delay was also observed during the calibration with the hygrometer.

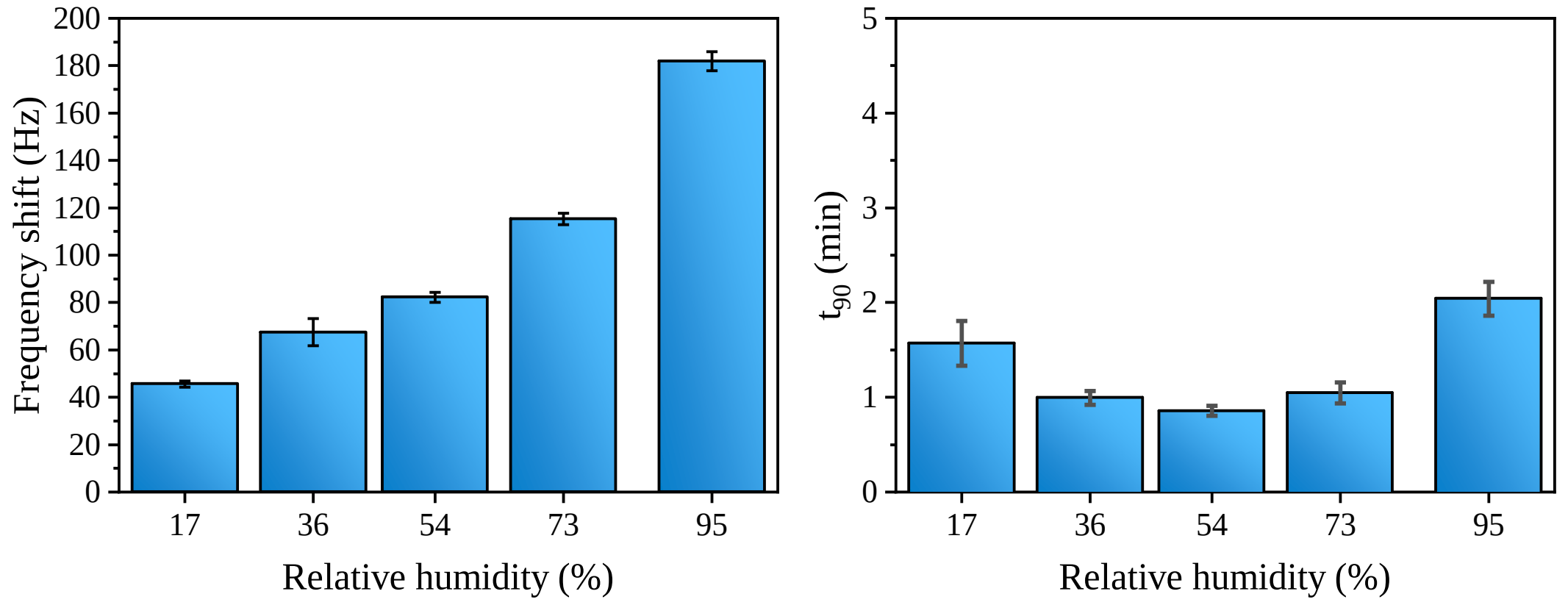

*Figure 5. 28 (a) Calibration curve and (b) $\tau_{90}$ values for the sensor humidity test. Rep. from* [48].

The sensor was then exposed to natural breath; during this experiment, a researcher breathed normally onto the sensor. As the exhaled breath reached the device, an immediate response was observed. The response, consisting of a negative frequency shift, was most probably caused by a broad set of analytes in the breath, but mainly by its high humidity concentration (Figure 5.29).

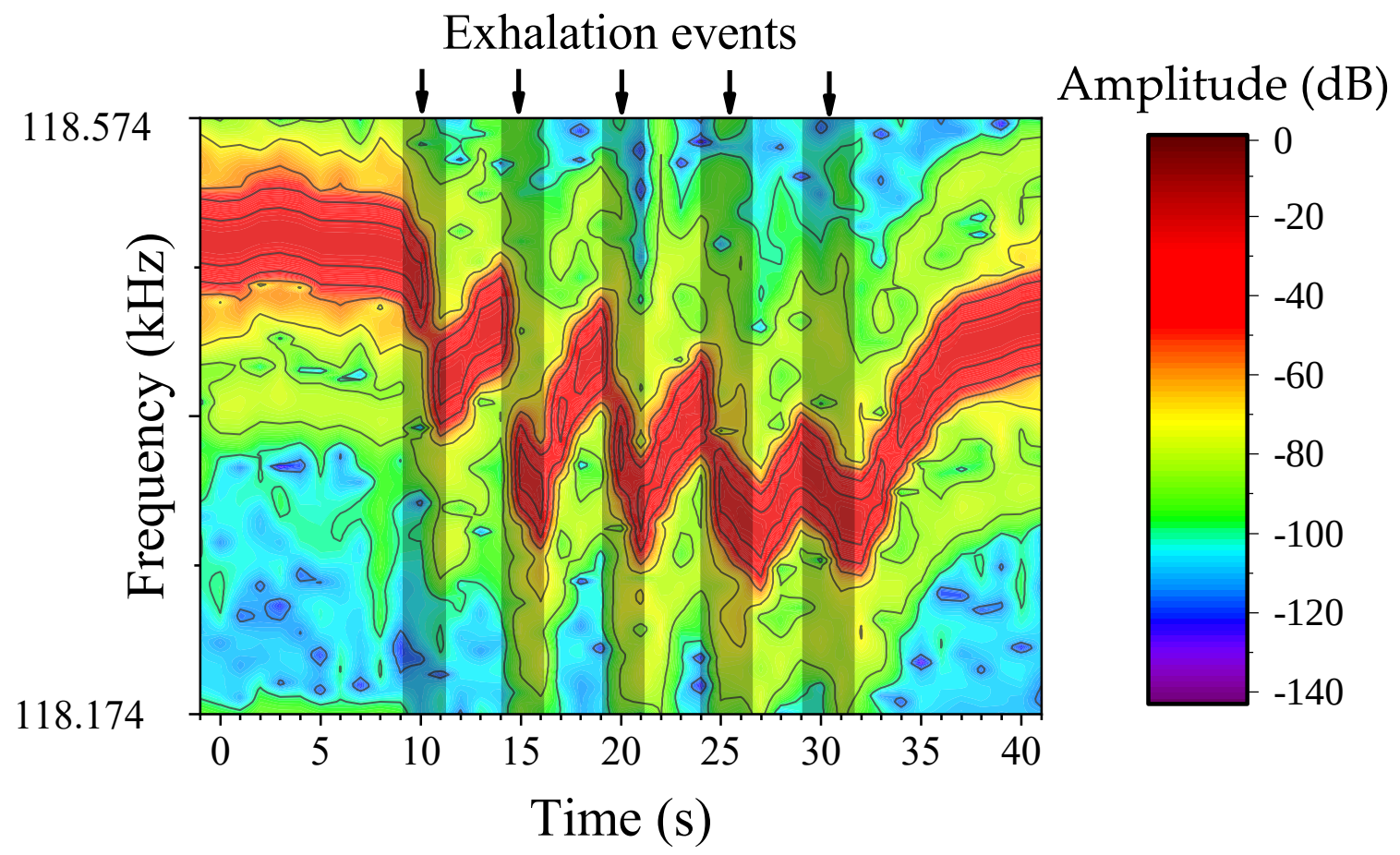

*Figure 5.29 Breath detection test, the resonance frequency (red area) shifts to lower values as a response to exhalation events.*

Although the exact compositional profile cannot be obtained with the current device, the excellent response to natural breath proved the utility of the sensor in breath analysis applications. Furthermore, the resonance amplitude remains unaltered, indicating that the transducer's resonance did not suffer any noticeable damping or distortion during the breath exposures.

The MEGS was then evaluated for acetone detection. The sensor responded to 40 ppm of gaseous acetone with a negative frequency shift, analogously to the behaviour observed during the humidity detection test related to the absorption of acetone molecules in the sensitive layer (Figure 5. 30a). Acetone detection exhibited high reproducibility with slight variation between exposure cycles. The mean response value was 44.1 ± 4.8 kHz.

The test was repeated for ammonia detection under the same conditions. The sensor's response also indicated that the ammonia molecules became attached to the sensitive layer, as observed during the humidity or acetone tests (Figure 5. 30b). The device exhibited a similar response among the three exposures, further confirming the excellent reproducibility of the MEGSs operation. The response was higher than that to the acetone exposure, with a mean value of 112.6 ± 11.0 kHz.

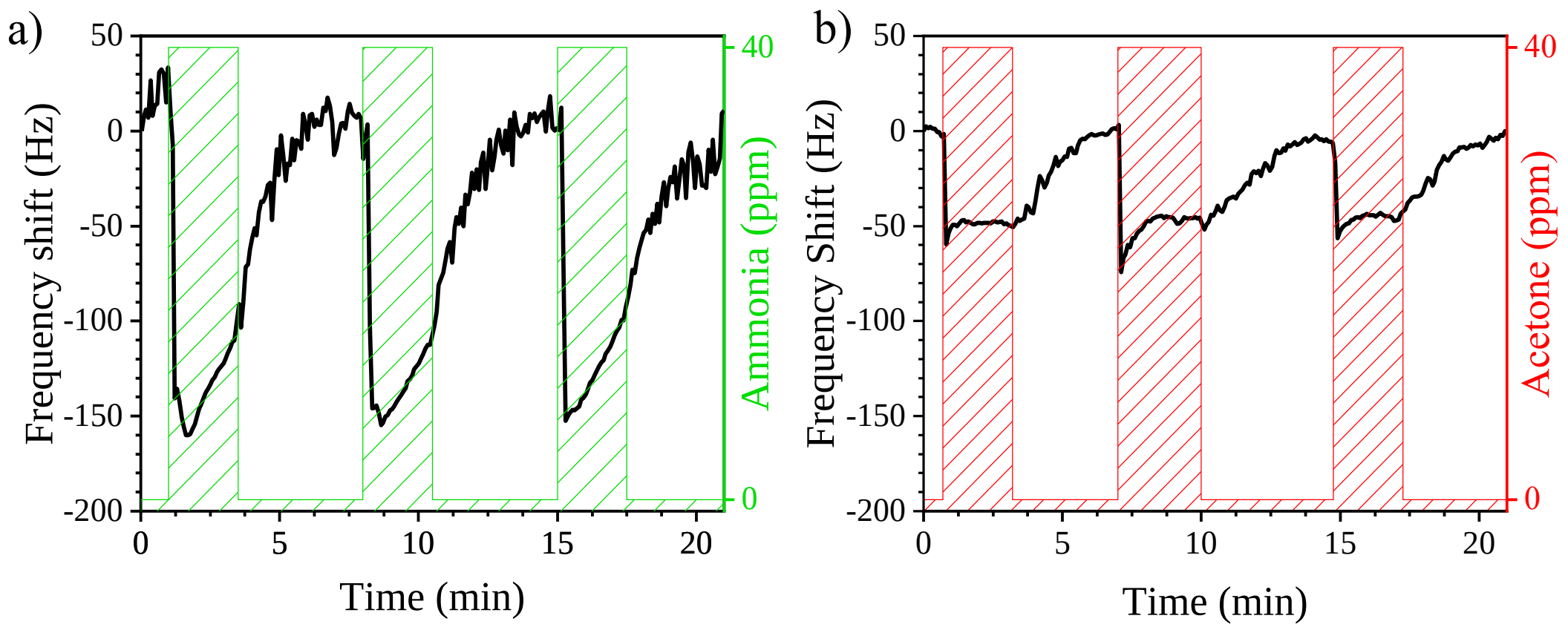

*Figure 5. 30 Real-time frequency shifts of the device exposed to 40 ppm of (a) acetone (gas) and (b) ammonia (gas).*

The sensor showed an immediate response toward acetone and ammonia. However, a higher time was required to recover the base frequency compared to the humidity test. This may be related to a slowing down of the diffusion of acetone and ammonia molecules within the PVP nanofibers with respect to water molecules.

Overall, the MEGS detected acetone and ammonia at a 40-ppm concentration. Furthermore, the sensor exhibited high reproducibility, owing to the full recovery during the purge phase, i.e., the frequency achieved its initial value (frequency shift equal to zero). A sought-after phenomenon in sensing technologies as discussed in Chapter 4.

The last test was performed to study the effect of a common interferent biomarker in breath analysis, benzene (Figure 5. 31a).

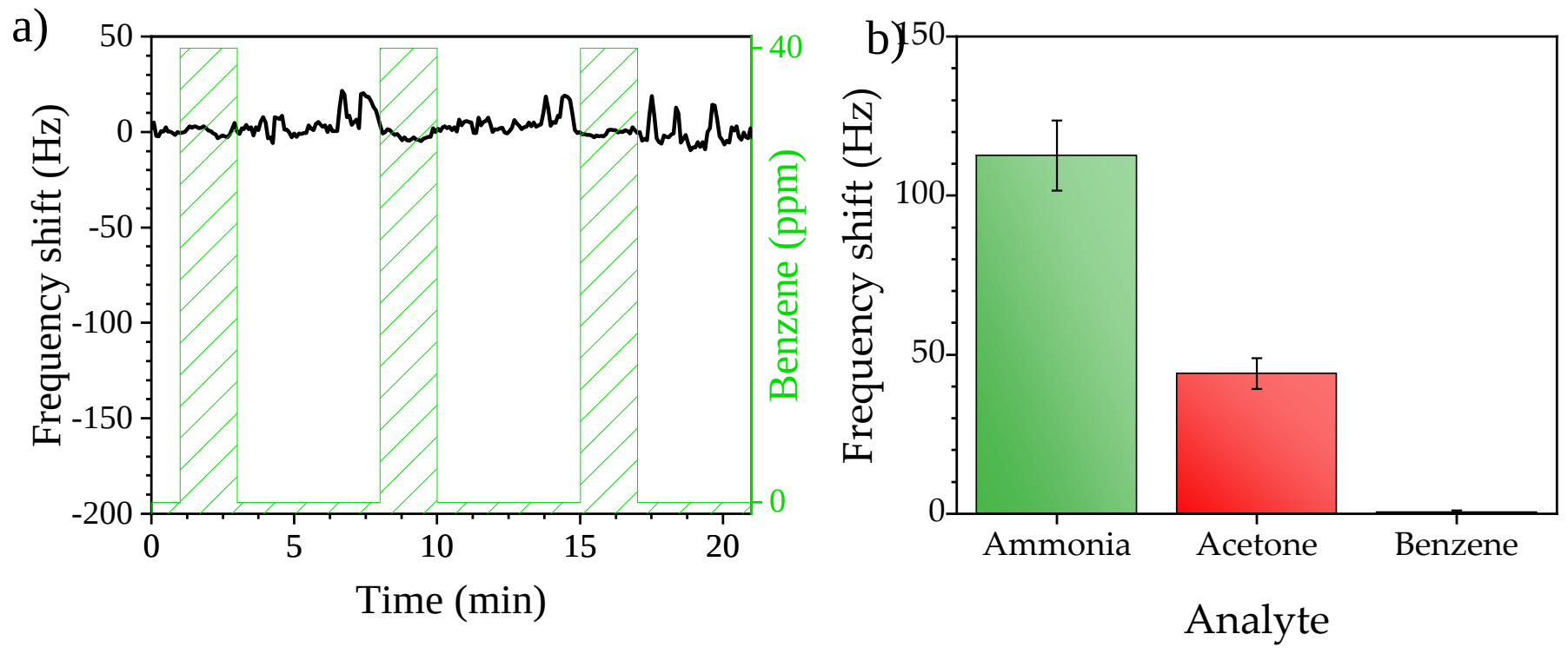

*Figure 5. 31 (a) Real-time frequency shift of the device exposed to 40 ppm of benzene (gas) and (b) comparison between the MEGS response to 40 ppm of ammonia, acetone, and benzene.*

The MEGSs exhibited a neglectable response to 40 ppm of benzene, with a mean value of 0.5 ± 0.5 kHz. Compared to water, acetone, and ammonia, the sensor demonstrated an excellent performance, as the device showed an insignificant response to benzene. According to the relationship between the experimental responses, PVP nanofibers can bind with polar molecules, e.g., water, acetone, or ammonia with insignificant interaction with non-polar molecules such as benzene. In addition, the final selectivity results from a wide range of other solvation parameters [125,126]. The comparison between the three tested biomarkers highlights the fact that the device can target acetone and ammonia biomarkers without interference from benzene presence in exhaled breath (Figure 5. 31b).

Overall, the viability and performance of a magnetoelastic resonance-based gas sensor and the real-time monitoring system were validated. The proof-of-concept study demonstrated the successful detection of biomarkers related to breath diagnosis using functionalized magnetoelastic resonators. The real-time monitoring system allowed for the detection of fast phenomena, which is crucial for developing responsive and reliable sensors, and facilitated the assessment of the sensor's response to various concentrations of gaseous biomarkers and even real exhalation events. While this study focused on using magnetoelastic resonators for detecting gaseous biomarkers in breath analysis, the techniques, and systems we developed have broader applications. They could be used with various kinds of magnetoelastic sensors, making it possible to monitor and sense a wide variety of other parameters or conditions.

### *5.7 Optimization of gas sensing devices.*

Throughout this chapter, several ways to optimize magnetoelastic resonance-based sensors have been explored or anticipated. To summarize:

- Shape
- Thermal annealing
- Functionalization
- Instrumentation

These four approaches provide a range of possibilities for improving the performance of magnetoelastic resonance-based gas sensing devices. By changing the shape of the resonator, we can reduce its mass density in certain areas, which increases its resonant frequency and sensitivity to changes in the environment. Additionally, thermal annealing can improve the magnetoelastic resonance figures of merit and thus enhance the sensor's performance. Functionalization can also enhance the sensitivity and selectivity of the sensor by coating the resonator with different sensitive materials. Finally, proper instrumentation can help reduce noise and improve the detection limit of the sensor. By combining these approaches, it is possible to optimize the performance of magnetoelastic resonance-based gas sensing devices for a wide range of applications.

### *5.7.1 Shape.*

In addition to changes in the length of magnetoelastic microwires that has already been presented, variations in diameter can also be used to optimize the performance of magnetoelastic resonance-based gas sensing devices.

In this aspect, microwires with a composition $Fe_{73}Si_{11}B_{13}Nb_3$, length of 10 mm, and four different diameters were tested (see Table 5. 3).

*Table 5. 3 Tested microwires with different diameters.*

| Group | Core diameter (µm) | Shell diameter (µm) |
|---|---|---|
| 1 | 79.3 | 100.6 |
| 2 | 70.7 | 84.2 |
| 3 | 61.7 | 78.2 |
| 4 | 36.7 | 47.9 |

In this experiment, microwires with four different diameters were used to investigate the figures of merit of the magnetoelastic resonance. Interestingly, the thinner wires from group 4 did not produce observable magnetoelastic resonance, while the other three diameters showed a dispersion in the values of resonance $f_r$, A, and k coefficient in a large sample of microwires. Results are shown in Figure 5. 32.

In terms of A, a trend of decreasing A with increasing wire diameter was expected, as the A is related to the amount of mass causing variations of flux during the magnetoelastic resonance. The collected data showed that, that the group 1 MWs had and A of 0.238 ± 0.065 dB (see Figure 5. 32a), the group 2 MWs, 0.215 ± 0.051 dB (see Figure 5. 32d) and the thinner MWs, group 3, 0.112 ± 0.063 dB (see Figure 5. 32g). Thus, A clearly decreases as the diameter of the microwires decreases ($p=0.05$).

As for $f_r$, the values were mainly included in the 220-250 kHz range (see Figure 5. 32b, e, and h). This parameter is largely dependent of the resonators length as indicated by Equation 5. 17 so no significant variations are expected due to the different diameters, and none is observed. However, changes in the quality of cut may explain the dispersion in the measured $f_r$, this quality includes the length of the microwire but also in the fracture and loss of glass shell fragments near the cut.

A significant dispersion in the values of k was observed among the different microwire thicknesses (see Figure 5. 32c, f, and i). Statistical analysis revealed a weak correlation between the thickness and k ($p = 0.048$).

As previously mentioned, k of magnetoelastic resonators discussed in this chapter is influenced by the ease of rotation of magnetic domains that are not aligned with the longitudinal direction.

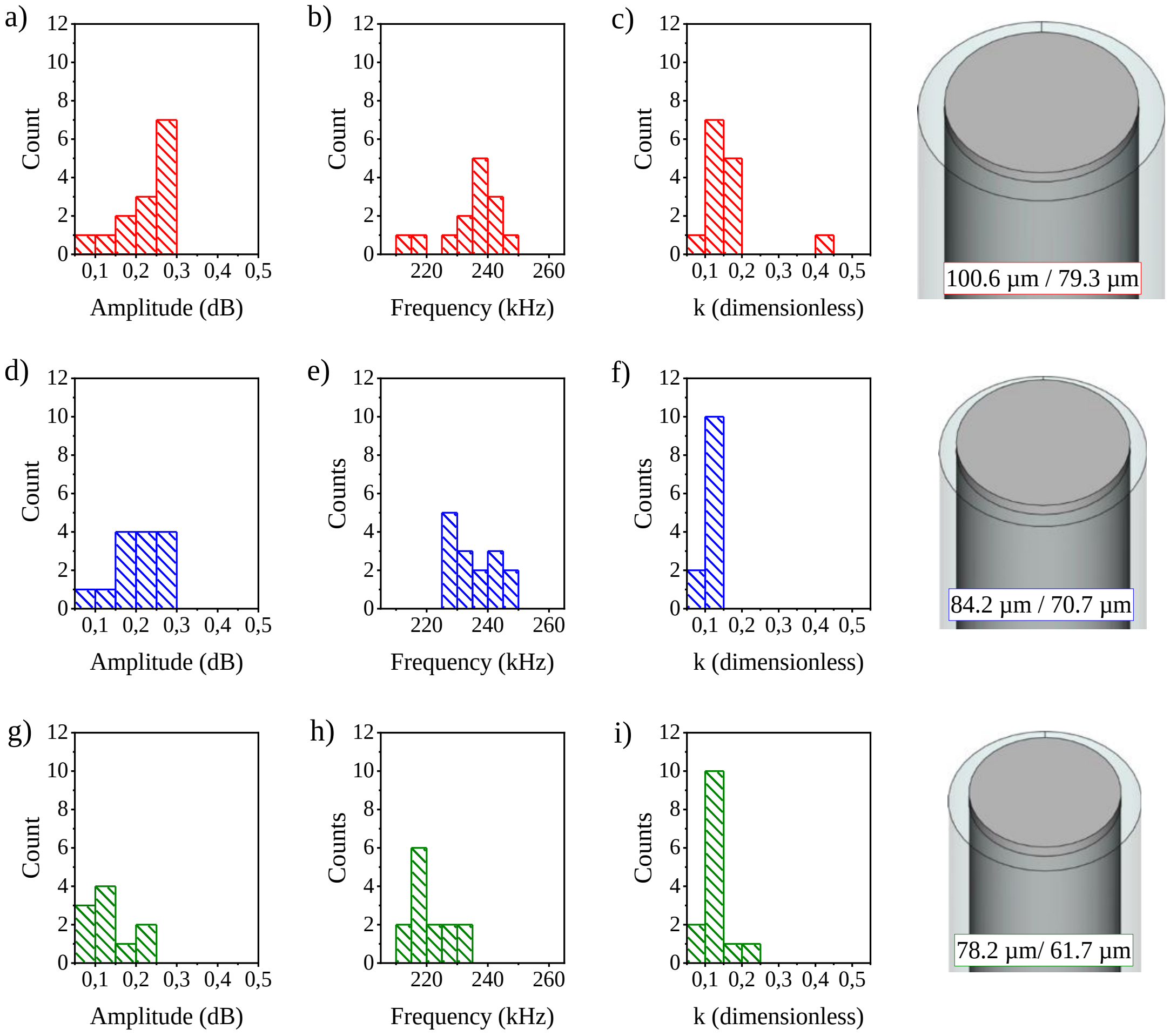

*Figure 5. 32 Histograms of the magnetoelastic resonance figures of merit (A, $f_r$, and k) for the tested microwires.*

The magnetic domain structure in microwire typically consist of a large longitudinal domain and a crown of radial domains as depicted in Figure 5. 33 [127]. When magnetized in the long direction, the rotation of the radial domains is responsible for the magnetostriction of microwires.

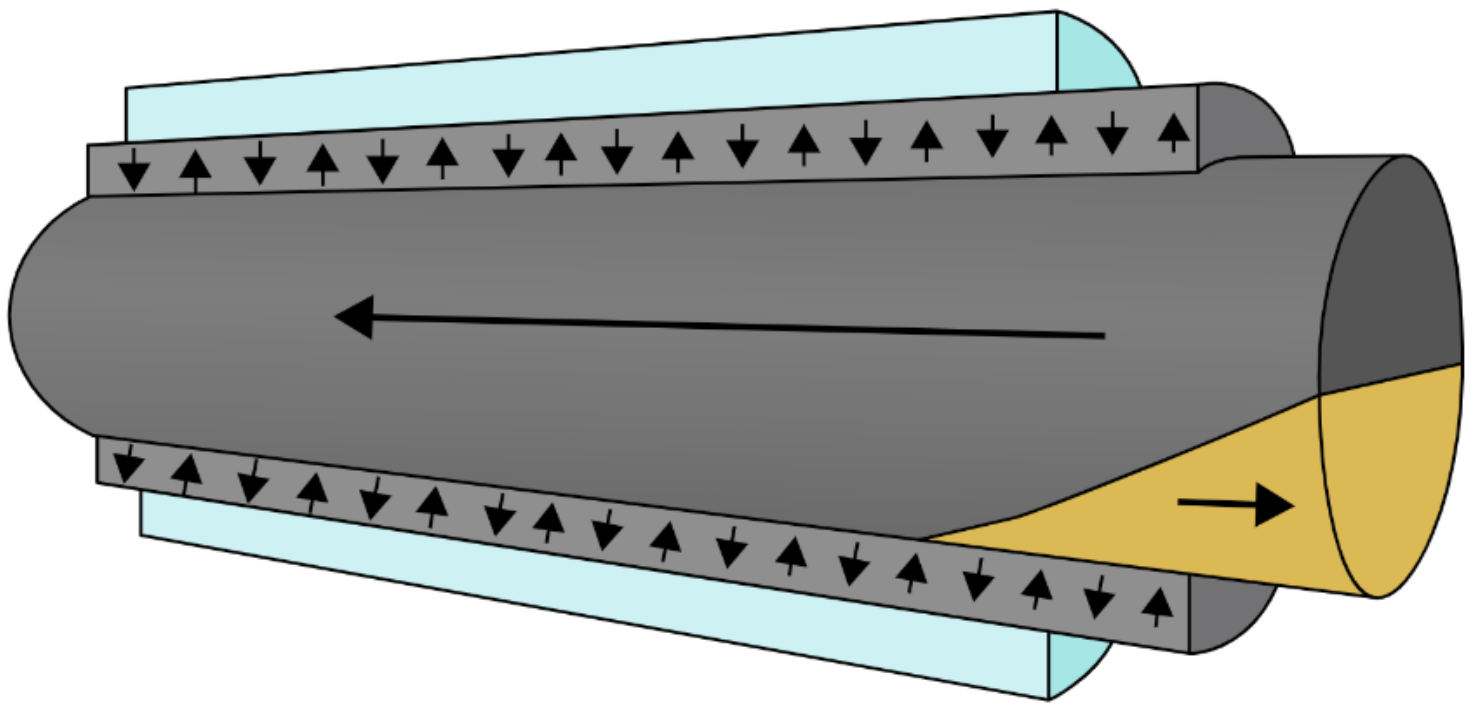

*Figure 5. 33 Typical magnetic domain structure in a Fe-based amorphous microwire. Rep. from* [127].

This domain structure is largely determined by the geometry of the microwires, as it results from the balance of the different magnetic energies. For instance, the existence of a single longitudinal domain will be favourable for a wire with infinite, or practically infinite, shape factor, and the wire will exhibit a bi-stable magnetic behaviour. As the thickness increases respect the length, the magnetostatic energy will promote the apparition of transversal magnetic domains and break the bistability. This can be observed in the hysteresis loops for longitudinal magnetization of the collection of microwires previously described.

The obtained magnetic hysteresis loops in Figure 5. 34 reveal interesting trends with respect to the thickness of the microwires. The thinner microwire shows a more squared shape loop. A perfect squared loop is typical of bi-stable behaviour where the magnetization process is given by a single and large Barkhausen jump through domain wall propagation. However, as the thickness is increase, the angles of the squared shape are smoothed, indicating the presence of magnetic domains that require stronger magnetic fields to be oriented in the longitudinal direction.

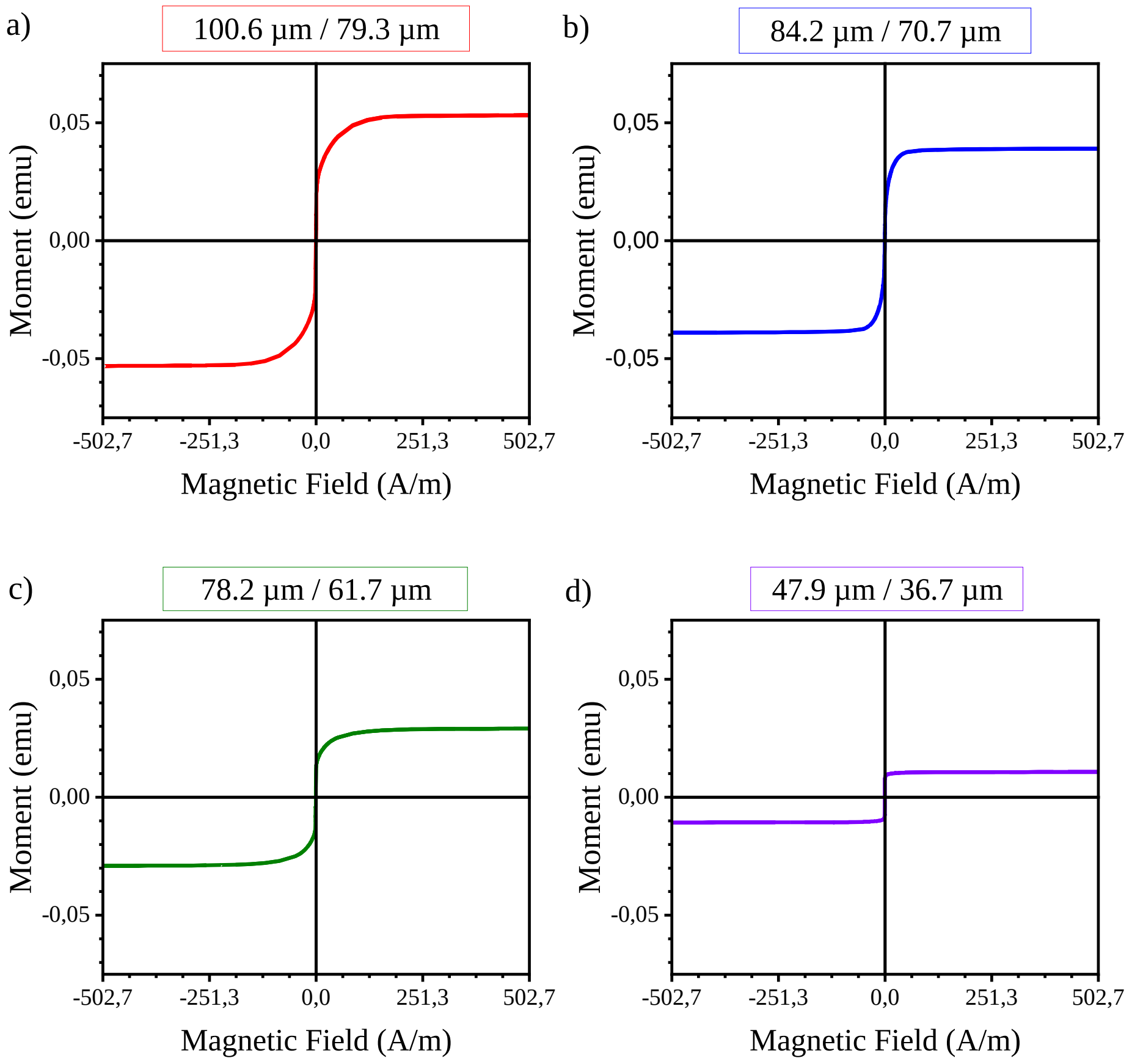

*Figure 5. 34 Hysteresis loops for microwires with four different shell and core diameters (indicated above each loop).*

These results predict that thicker microwires are better suited for magnetoelastic resonance-based applications, favored by the presence of magnetic domains transversal to the magnetization direction, despite the weak correlation previously observed between thickness and k.

### *5.7.2 Thermal annealing.*

The amorphous state is a metastable state, meaning that it is obtained under conditions that do not allow the material to reach equilibrium (due to ultra-fast cooling). As a result, when heat is applied to the system, relaxation, crystallization, or other phase transitions may occur. These processes are thermally activated and may be endothermic or exothermic, as observed in the differential scanning calorimetry (DSC) spectrum (see Figure 5. 35).

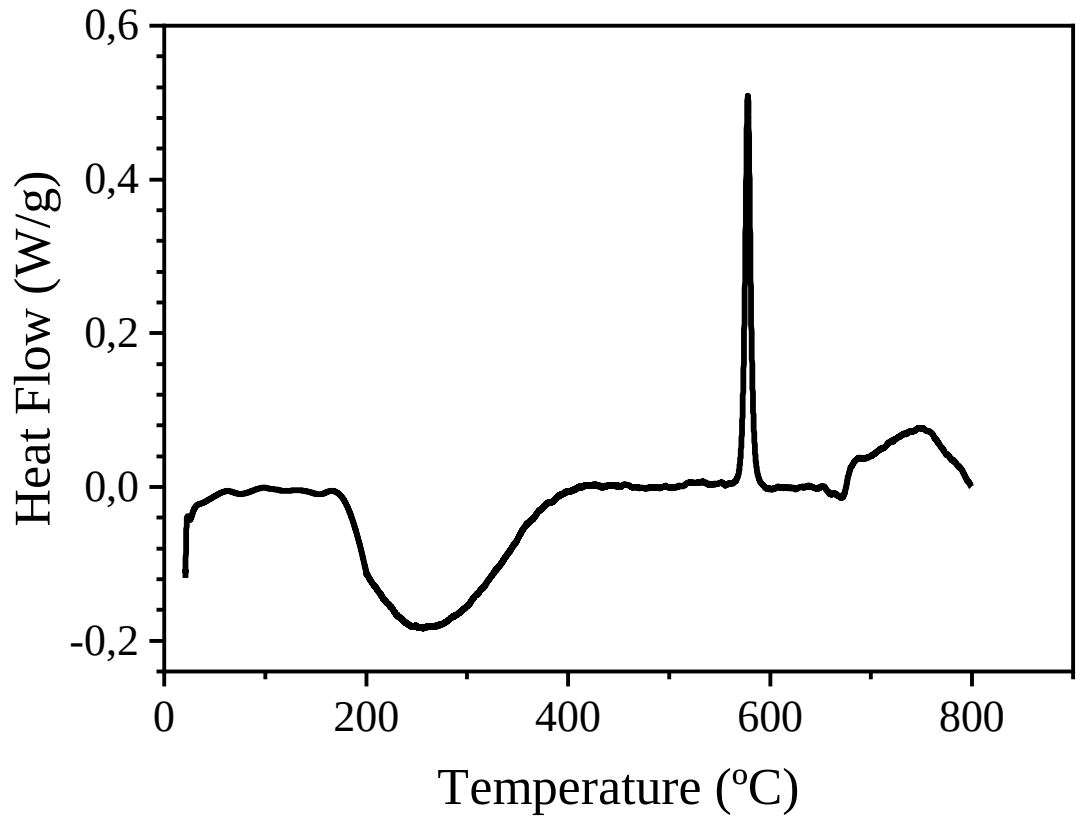

*Figure 5. 35 DSC spectrum for a typical Fe-based amorphous alloy ($Fe_{73}Si_{11}B_{13}Nb_3$).*

First, an endothermic process initiates around 200 Cº. In Figure 5. 35, this process is observed as a valley in the spectrum. This process can be related to structural relaxation from the stress originated during the ultra-fast cooling temperature. M. Vazquez et al. reported that magnetic modifications were detected even though no significant changes in the structure were observed, i.e., the amorphous state was preserved, while performing annealing in this range of temperature [128]. As for magnetic properties, the material can still be considered soft magnetic [129].

At temperatures between 530-580ºC, partial loss of the amorphous state occurs due to the nucleation of alpha-FeSi nanocrystals within the FeSiB amorphous matrix. This process is observed as a sharp peak in Figure 5. 35. The nanocrystals typically have a size of around 10 nm, which is significantly smaller than the exchange correlation length of 40-50 nm, resulting in further magnetic softening of the material. The crystalline phase increases the mechanical stiffness and exhibits a significantly lower magnetostriction effect than the amorphous phase, which negatively affects the magnetoelastic properties of the material during this process [128]. The crystalline phase increases up to approximately 70-80 % of the total volume.

It is worth noting that the presence of Cu and Nb in the FeSiB alloy plays a crucial role in the nano crystallization process. Cu acts as a promoter of crystallization, leading to a lower crystallization temperature and facilitating the nucleation of nanocrystals. In contrast, Nb acts as an inhibitor of crystallization, causing the crystallization temperature to rise and impeding the growth of nanocrystals. The combination of these two elements enables the production of a finely dispersed distribution of nanocrystals.

The remaining amorphous phase crystallizes into intermetallic FeB compounds, typically $Fe_3B$ [130]. This crystallization occurs at temperatures above 550ºC and may occur simultaneously with the nano crystallization process. The fully crystallized phase becomes brittle [131] and magnetically hard with strong crystalline anisotropy whereas the magnetostriction is further decreased as the amorphous state disappears [128,132].

Interestingly, in microwires have a unique distribution of residual stresses from the manufacturing process. Due to the different thermal expansion coefficients between the borosilicate glass and the metallic nucleus, microwires are compressed in the radial direction as depicted in Figure 5. 5. Conversely, this compression may be translated into traction stress in the longitudinal direction through the Poisson ratio previously described.

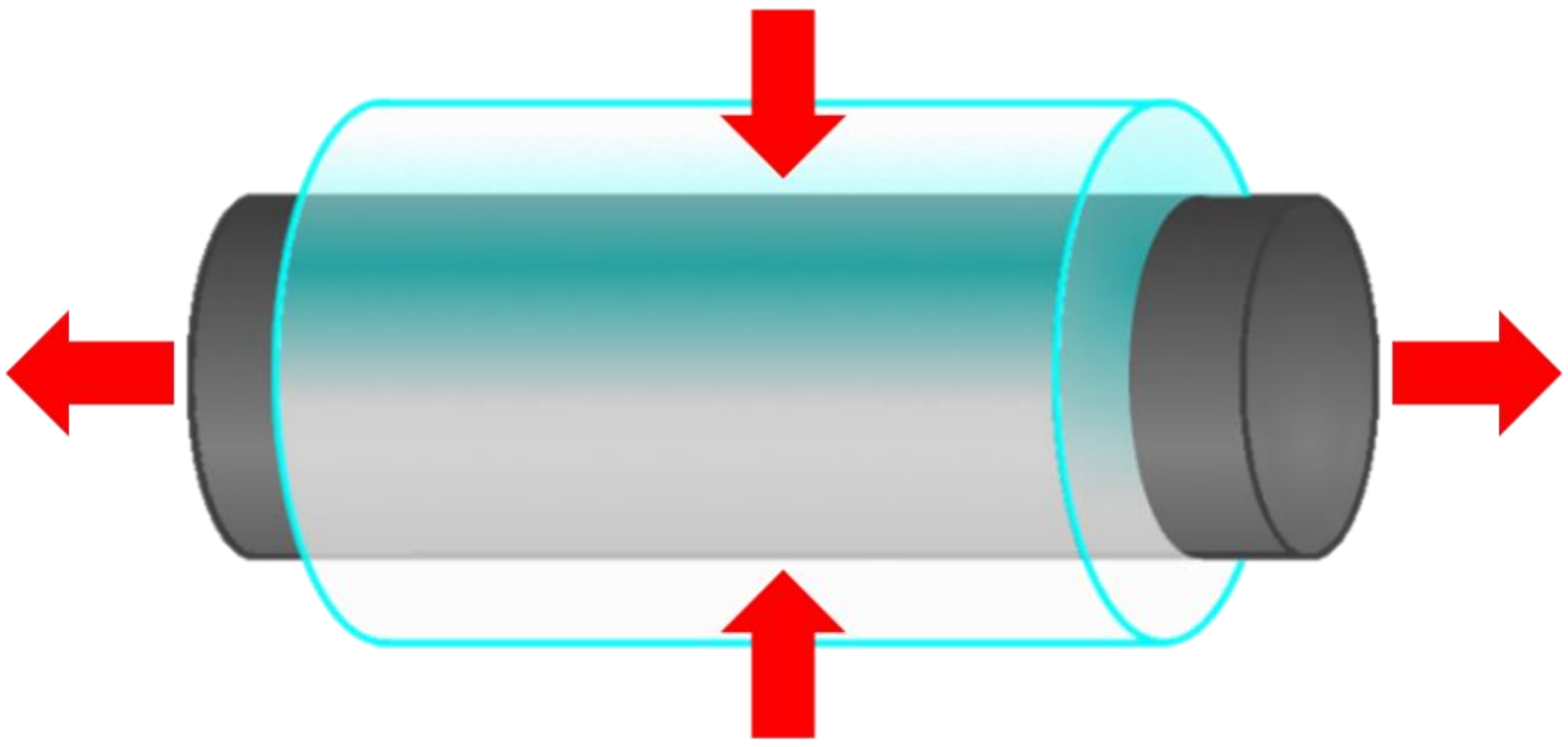

*Figure 5. 5 Schematics of the stress distribution on an as-cast microwire.*

Due to the positive magnetostriction of the Fe-based amorphous microwires, the stress distribution induces a strong magnetoelastic anisotropy in the longitudinal direction. The stress distribution is heavily influenced by the shell/ core diameter ratio. Chiriac et al. found that, on the one hand, as the metallic nucleus is thicker for a constant glass shell, the axial magnetic anisotropy constant is decreased. On the other hand, if the glass shell is thicker for a constant metallic nucleus, the axial magnetic anisotropy constant is increased [133].

The strong axial magnetic anisotropy leads to the magnetic bistable behaviour of microwires. P. Corte-Leon et al. reported how a Fe-based amorphous microwires losses its magnetic bi-stability when thermally annealed at temperatures between 300-350 ºC.

From the discussion above, the stress relaxation could be responsible for a decrease of the axial magnetic anisotropy and the presence of magnetic domains with the magnetization oriented not in the longitudinal direction, but on a radial or circular directions.

Thus, thermal annealing treatments to improve the magnetoelastic performance of Fe-based amorphous materials should be carried out at temperatures below the crystallization threshold, approximately <500ºC.

Given that the magnetoelastic effect, and in particular, the magnetoelastic resonance, relies on the rotation of magnetic domains to produce the deformation that couples magnetic and mechanical properties, the thermal annealing of Fe-based microwires is expected to enhance the figures of merit for magnetoelastic resonance.

To evaluate this hypothesis, microwires of three different diameters were subjected to thermal annealing at temperatures ranging from 200-500 ºC, and their magnetoelastic resonance figures of merit were subsequently compared.

Accounting for the variation in parameters previously observed in these microwires, the following methodology was used. A total of fifty-four microwires with three different diameters were used. Each diameter group (of eighteen microwires) was divided into five subgroups, with each subgroup consisting of three microwires that were thermally annealed at different temperatures ranging from 200°C to 450°C. The magnetoelastic resonance figures of merit of each microwire was characterized before and after the thermal annealing process. The changes in the figure of merit were then presented as variations (%) over their original values for each subgroup of microwires. Results are presented in Figure 5. 37.

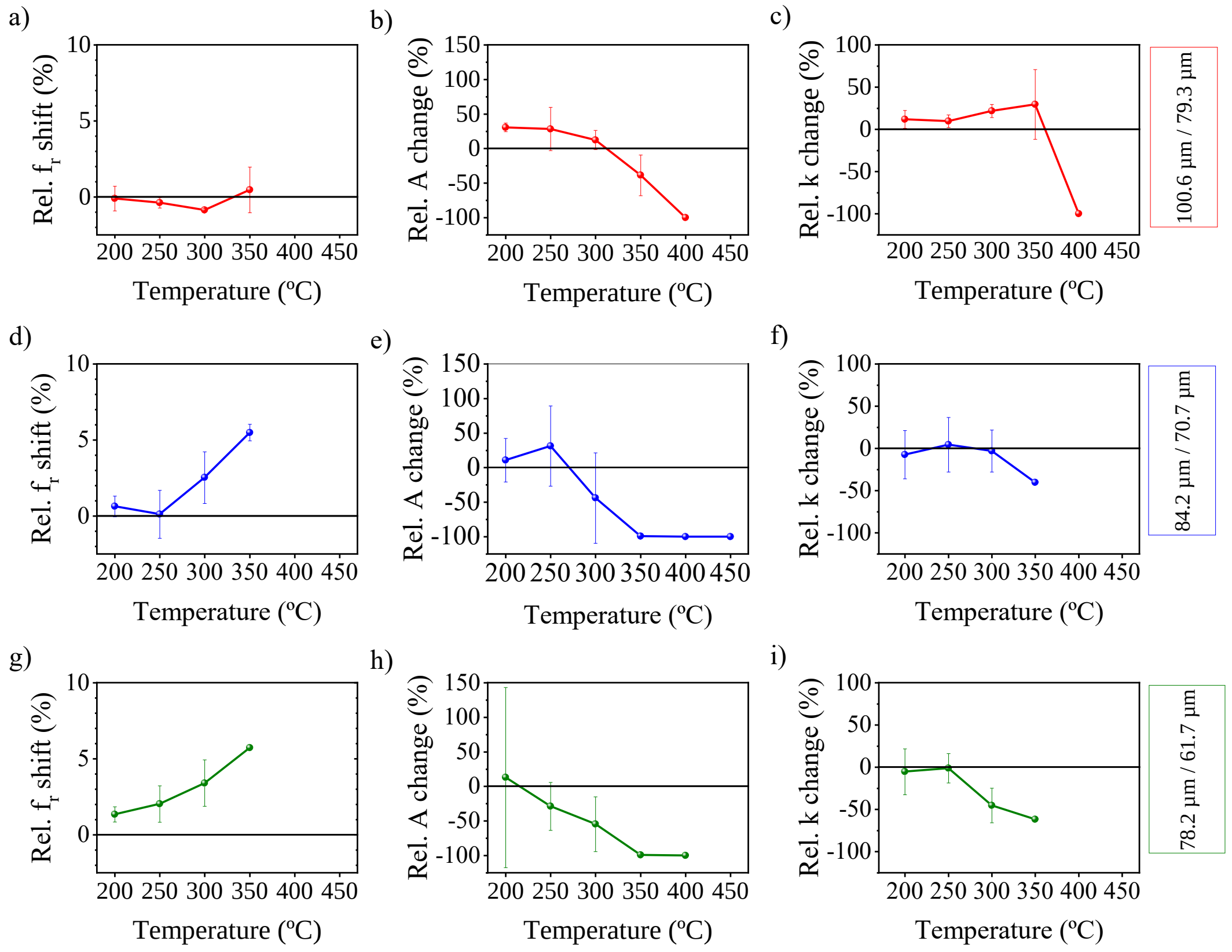

*Figure 5. 37 Relative variations for the figures of merit ($f_r$, A, and k) of the magnetoelastic resonance for the different microwires.*

For all microwires, $f_r$ is increased with the temperature of annealing. It was previously discussed that the elastic modulus affects $f_r$ as described in Equation 5. 9, i.e., thus the results indicate that the thermal annealing leads to higher E (Figure 5. 37a, d, and g).

The thickest microwires annealed at 200-300°C showed an increase in A, while those annealed at >350°C showed the largest decrease in A (Figure 5. 37b). For the medium-thickness microwires, annealed at 200-250°C an increase in medium value of A was observed within large error bars, for 350ºC a significant decrease was observed (Figure 5. 37e). For the thinner microwires a huge variation of the results was observed for annealing at 200°C with a huge increase (163%) in a single sample and moderated decreases in the other two (~60%), for higher temperatures a decrease was consistently observed (Figure 5. 37h). The A values for microwires annealed at 400°C and 450°C were not measured in all cases due to the complete loss of magnetoelastic resonance.

Finally, k was consistently increased in the thicker microwires at 200-350ºC. For the other two set of microwires, such an increase was not observed (Figure 5. 37c, f and i).

Overall, the microwires with 100.6 and 79.3 µm of shell and core diameters, respectively, have shown the best characteristics towards their use in magnetoelastic resonance-based sensors. On the one hand, their base figures of merit are significantly better than the ones of thinner microwires. On the other hand, the result of the thermal annealing indicates that they can be further improved using annealing at temperatures of 200-300ºC.

A future work will focus on the development of a gas sensors using these thick annealed microwires with the advanced instrumentation generated during this chapter. It is expected that the miniaturization of devices using microwires will bring a significant advance to the field of magnetoelastic-resonance based sensors.

### *5.7.3 Functional coatings.*

New functional coatings are currently being investigated to improve the response of the magnetoelastic resonance-based biosensor previously presented. Although the first experiment, carried out using a PVP functional coating, yielded interesting results, increase the sensitivity and selectivity of gas sensors is crucial for their potential application as has been thoroughly discussed during this chapter and the previous one. In that regard, different polymers used as functional coating may increase the range of detectable analytes as well as the overall performance of the device [126,134].

The functional coating has been electrospun over a magnetoelastic MR in a similar manner as for the breath analysis sensor. Three polymers have been chosen for this work, polyacrylonitrile (PAN) [135], polystyrene (PS) [136], and poly(styrene-alt-maleic anhydride) (PSMA) [136].

PAN (Mw ~ 150,000 g/mol, Sigma-Aldrich, Burlington, MA, USA 81440) was dissolved in dimethylformamide (DMF) at 10% wt. The solution was stirred until homogeneous and degasified. Next, the solution was loaded into a syringe with a metal needle connected to a high-voltage power supply and placed at a 14 cm distance from the microribbon that acted as the grounded collector. Finally, the solution was extruded at a flow rate of 5 μL/min for 30 min at a voltage of 14 kV.

As seen in Figure 5. , the PAN was successfully deposited in the form of nanofibers in an equivalent manner than the PVP previously tested. The nanofibers exhibit an overall homogeneous aspect with diameters of less than 500 nm.

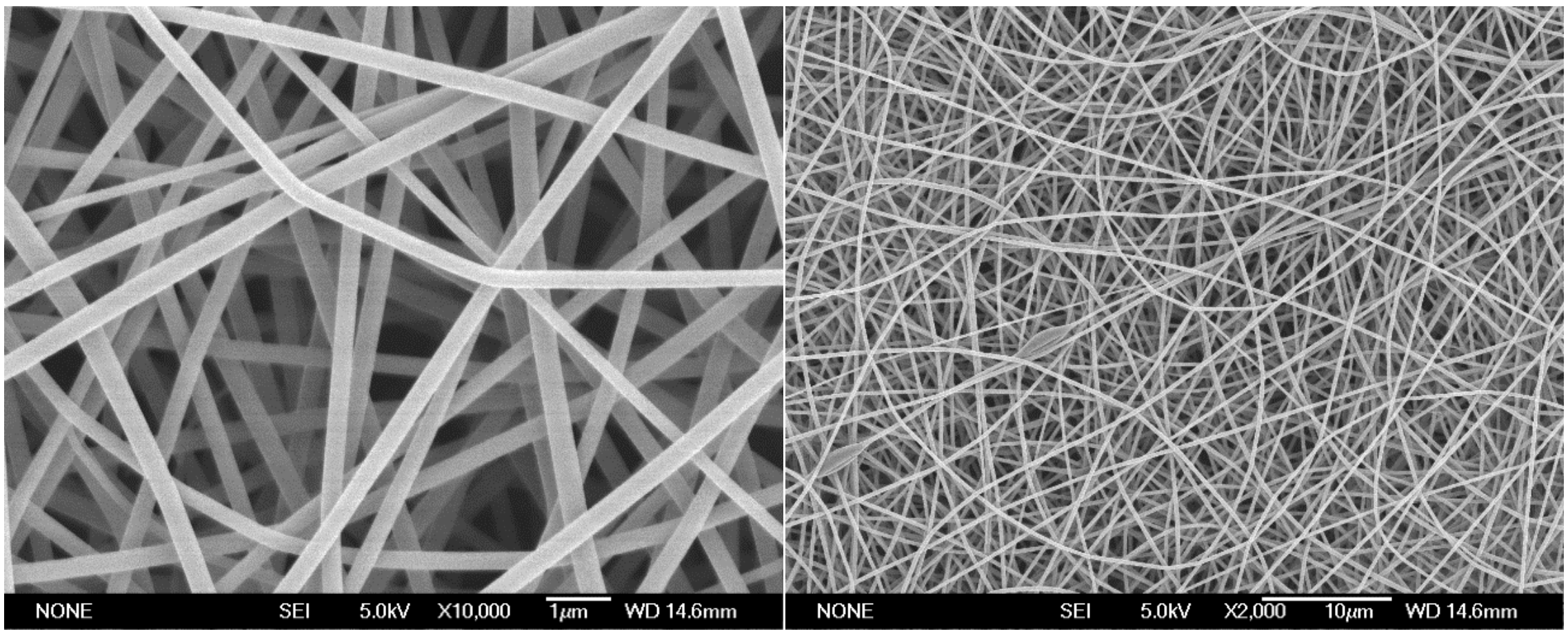

*Figure 5. 38 SEM image of PAN nanofibers deposit at different magnifications.*

PS (Mw ~ 192,000 g/mol, Sigma-Aldrich, Burlington, MA, USA 81440) was dissolved in DMF at 20% w/v. The solution and electrospinning process were the same as for the PAN.

The PS nanofiber shown in Figure 5.  were deposited in a less homogeneous manner than the PAN nanofibers. In particular, the PS nanofiber exhibit a beaded morphology probably caused by droplets of the polymeric solution. These beads present various sizes and a rough surface. The nanofibers present a heterogeneous diameter typically below 500 nm.

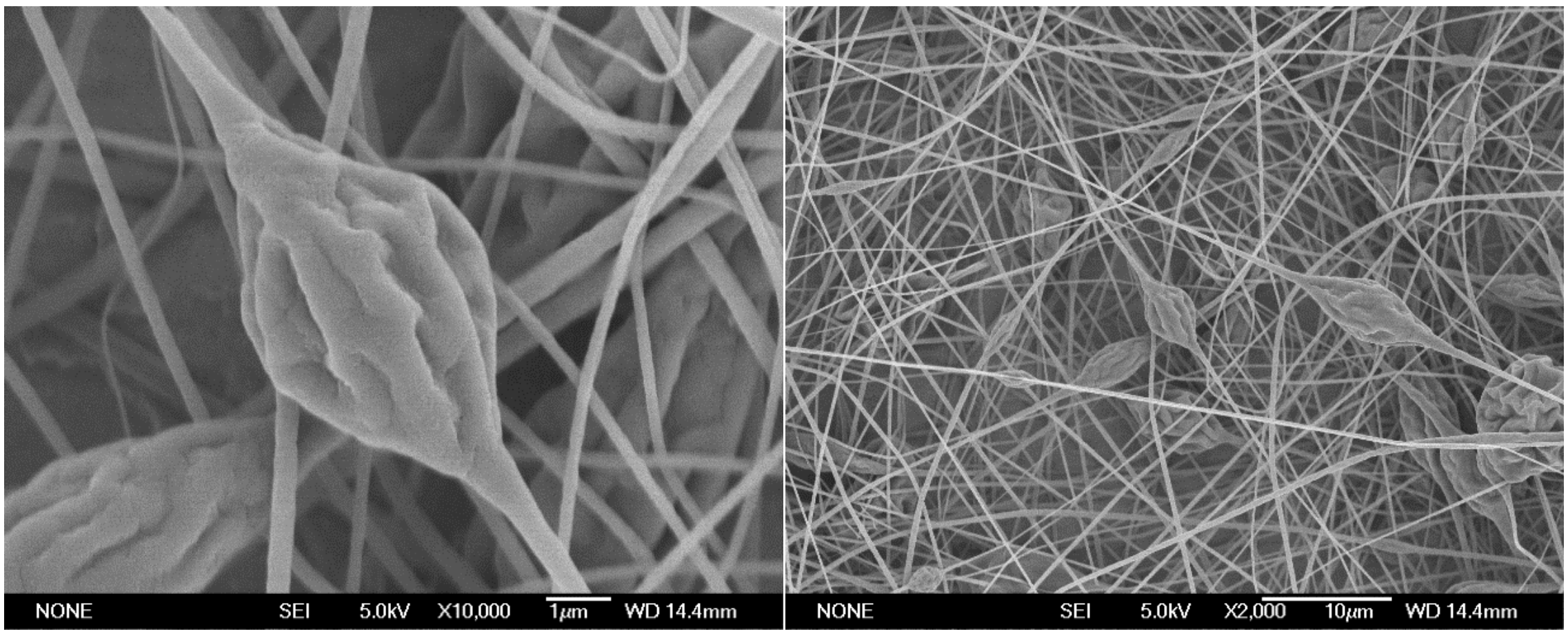

*Figure 5. 39 SEM image of PS nanofibers deposit at different magnifications.*

Finally, the PSMA (Mw ~ 350,000 g/mol, Sigma-Aldrich, Burlington, MA, USA 81440) was dissolved in DMF at 20% w/v. The solution and electrospinning process were the same as for the previous polymers.

Figure 5.  shows the PSMA nanofiber successfully deposited over the MR. In this case, a slight beaded morphology can be observed in a less notable manner that for the PS nanofibers. The PSMA nanofibers present an overall homogeneous aspect with diameters less than 600 nm.

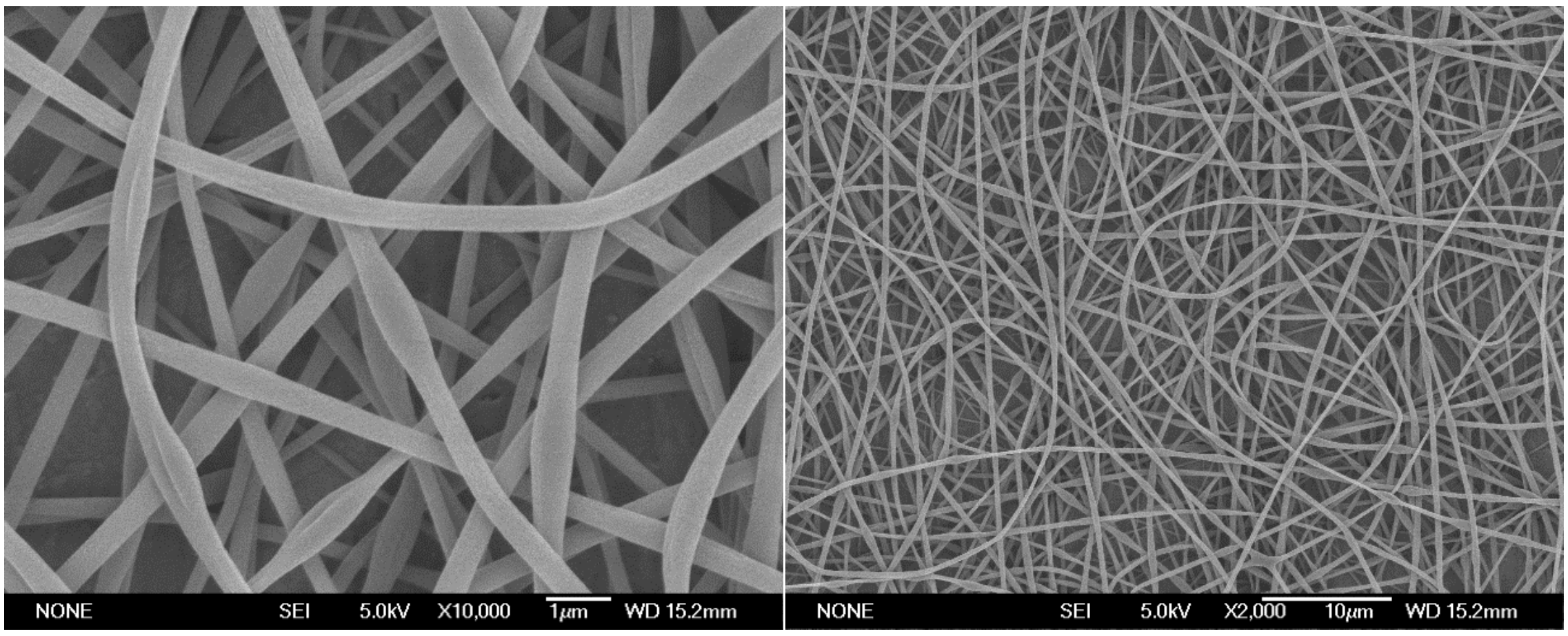

*Figure 5. 40 SEM image of PSMA nanofibers deposit at different magnifications.*

The effect of these functional coating on the sensor's response is currently being investigated using the same real-time monitorization setup previously presented. However, from previous reports [126,134,136] a significant improvement in the device's performance, in terms of sensitivity, selectivity, or both, is expected.

### 5.8 Conclusions.

In this chapter, the use of the magnetoelastic effect in gas sensing applications was explored, with a particular focus on amorphous Fe-based magnetoelastic materials. Various parameters, such as bias field and mass loading, were shown to influence the resonance frequency of the magnetoelastic sensor, which can be utilized to develop different types of sensors, particularly gas sensors.

The development of instrumentation to remotely monitor and characterize magnetoelastic resonance was explored, including the frequency sweep spectra approach and a novel real-time magnetoelastic resonance monitoring system. The advantages and limitations of each setup were discussed, and a real-time monitoring system was developed to tackle the limitations of the frequency sweep approach towards fast-occurring reactions. A significant advancement in the application of magnetoelastic resonance-based sensors in gas sensing was achieved through this real-time monitoring system.

Additionally, the development of a magnetoelastic resonance-based sensor was presented, featuring real-time monitoring, contactless operation, low cost, and easy adaptability for monitoring various gas biomarkers in exhaled breath. Relatively low concentrations (40 ppm) of acetone or ammonia and humidity levels as low as 5% RH were detected by the sensor with excellent recovery and reproducibility. This proof-of-concept represents a significant advancement for the application of magnetoelastic resonance-based sensors in gas sensing, particularly in biomedical applications such as breath analysis.

Finally, a comprehensive experimental study on the optimization of magnetostrictive materials was presented, which will further enhance the performance of this type of sensor. The work presented in this chapter is expected to bring a significant jump forward in the field of magnetostrictive sensors, paving the way for new advancements in gas sensing applications, especially in medical and environmental monitoring applications.

### *5.9 References.*

# **Chapter 6 : Closing remarks.**

*Summary Of Key Findings; Discussion of The Implications; Future Research Directions.*

## 6.1 Summary of Key Findings.

In this PhD thesis, four major research lines have been explored: graphene-based materials synthesis, electromagnetic shielding materials, graphene-based chemiresistive gas sensors, and magnetoelastic resonance-based gas sensors.

### 6.1.1 Graphene-Based Materials Synthesis

A comprehensive review of the family of graphene and graphene-based materials (GBMs) and their synthesis has been presented, with a focus on the ball-milling technique. Three distinct mills —planetary, high-energy, and oscillatory— were investigated, with the most emphasis on the latter. Various conditions and parameters were explored, yielding promising results for large-scale GBM production. The study revealed that different mill types and conditions resulted in the creation of GBMs with unique attributes, such as exfoliation degree, lateral size, defectiveness, and oxidation degree.

A notable outcome was the obtention of few-layered mesoporous graphene (FLMG), which represents a paradigm shift in GBM production, emphasizing quantity while accepting structural defects as desirable features for specific applications. The ball-milling technique was proven to be suitable for GBM synthesis, with the potential for tuning physicochemical characteristics by adjusting parameters. The use of natural graphitic powder as a precursor and the absence of additives or reactants highlighted the viability of large-scale FLMG production through ball-milling, with the oscillatory mill as the primary focus, and the planetary and high-energy mills warranting further exploration.

### 6.1.2 Electromagnetic Shielding Materials.

Microwave absorbance applications have been explored, specifically in shielding electromagnetic radiation within the 2-18 GHz frequency range. The study was built upon the long-standing expertise at the *Instituto de Magnetismo Aplicado* (*IMA*) and introduced GBMs for microwave absorption. An experimental approach employing free-space measurements (FSM) was utilized to characterize microwave-absorbing materials, as opposed to the popular theoretical approach based on transmission line theory. The research investigated the use of FLMG and amorphous magnetic microwires (MW) in electromagnetic shielding materials (ESMs) and evaluated their performance using the FSM method.

Key findings from this study include the improvement in microwave absorption performance when comparing graphite to FLMG, highlighting the positive impact of the milling process. Additionally, the enhanced performance of FLMG and MW samples demonstrates the synergistic potential of these materials, resulting in superior microwave absorption properties. By altering the concentration of FLMG, the microwave absorption properties of the composite material could be effectively tuned. The research also revealed the potential of the composite material for applications in electromagnetic interference shielding, radar stealth, and other microwave technologies. Overall, the use of FLMG can lead to the development of ultrathin composite ESMs with exceptional performance, such as > 99% shielding across the X-band, and tuneable microwave absorbance characteristics, offering significant promise for future applications in microwave-absorbing technologies.

### 6.1.3 Graphene-Based Chemiresistive Sensors

Building upon the foundational knowledge provided by the SENSAVAN group at *Instituto de Tecnologías Físicas y de la Información* (*ITEFI*), this research investigates the potential of GBMs, specifically FLMG and multi-layer graphene (MLG), in gas sensing applications, with a particular emphasis on $NO_2$ detection.

The experimental work demonstrates the potential of FLMG as a cost-effective and scalable technology for monitoring gas pollution, specifically $NO_2$ down to 25 ppb, with neglectable interference from ammonia or humidity. Moreover, the research using MLG highlights the use of UV irradiation to address the main drawbacks of graphene-based gas sensors compared to commercially accepted alternatives. Continuous UV irradiation during the device's operation was found to increase sensitivity, response (up to 500%), and, more importantly, recovery, greatly improving the overall performance of the sensor. The findings contribute to the understanding of sensing mechanisms and pave the way for the development of advanced gas sensors for air quality monitoring and public health protection.

#### *6.1.4 Magnetoelastic Resonance-Based Sensors*

This research line explored the use of magnetoelastic materials in sensing applications, particularly for gas sensing. Magnetoelastic materials exhibited a strong coupling between their magnetic and elastic properties, making them ideal for various sensing applications, including remote sensing where physical access was limited. The chapter discussed the design of a non-standard measuring setup for real-time magnetoelastic resonance monitoring, the development of a functional magnetoelastic gas sensor, and the optimization of these sensors for practical applications.

The research validated the viability and performance of a magnetoelastic resonance-based gas sensor and the real-time monitoring system, demonstrating the successful detection of low concentrations of ammonia and acetone, biomarkers related to breath diagnosis, using functionalized magnetoelastic resonators. The potential of magnetoelastic resonators and the measuring system could be applied to a range of sensing and monitoring applications. The findings in this section revealed the promising future of magnetoelastic resonance-based sensors in gas sensing applications, especially in breath analysis. The optimization of these sensors and the development of a real-time monitoring system offered new avenues for future research, with the potential to improve diagnostic techniques for various diseases and conditions and contribute to human health.

### *6.2 Discussion of the implications*

This research carries substantial implications for the scientific community, industries, and practical applications, addressing specific challenges and advancing current understanding. Primarily, it pioneers a scalable production method for graphene-based materials (GBMs) and an innovative real-time monitoring system for magnetoelastic gas sensors.

The development of a large-scale production method for GBMs holds transformational potential across numerous applications. In one hand, for electromagnetic shielding applications, GBMs can address the limitations of existing materials such as poor corrosion resistance, inadequate mechanical properties, excessive thickness or weight, and suboptimal absorbance performance. By enhancing these properties, GBMs can provide a robust and efficient solution for electromagnetic shielding. On the other hand, the production of small-sized and low cost graphene-based gas sensors able to detect NO2 below human toxicity thresholds can significantly impact the air quality in urban and industrial environments and reduce health associated risks.

While the full implications await practical implementation, the potential impact of the research is evident. The granting of two patents in relation to these advancements underscores their significance and promising future in the field.

The real-time monitoring system for magnetoelastic sensors promises further progress in sensor development, benefitting sectors like healthcare and environmental monitoring with remote, real-time tracking of fast occuring reactions.

### *6.3 Future Research Directions.*

Throughout the work presented in this research, numerous preliminary results and intriguing research ideas have emerged, indicating potential future research directions in the field.

#### *6.3.1 Graphene Oxide Production Using Ball Milling Techniques.*

While there is a promising potential for obtaining graphene oxide using ball milling techniques, the experimental research has only reached a preliminary stage, limited to the characterization step. Further research in this area could expand the applicability range of this technique and enhance our understanding of graphene oxide production.

### *6.3.2 Understanding Electromagnetic Shielding Mechanisms in Devices.*

There is currently a significant focus on developing dielectric and magnetic lossy materials. However, the microwave shielding mechanisms at play when these materials are integrated into devices, such as active coatings over substrates, are not yet fully understood. Future research should delve deeper into these mechanisms to optimize the performance of electromagnetic shielding materials in real-world applications.

### *6.3.3 Chemiresistive graphene-based sensor.*

Two potential future directions have been identified for chemiresistive graphene-based sensors. First, advancing the technology readiness levels and commercialization of the FLMG-based sensors. Second, exploring the role of defects in MLG-based sensors, with an emphasis on understanding the impact of defect type rather than defect quantity on sensitivity and selectivity.

### *6.3.4 Real-time magnetoelastic resonance-based sensors.*

The development of real-time magnetoelastic resonance-based sensors offers exciting opportunities for progress in remote sensing, not only for breath analysis but also for other applications. Future research should investigate various optimization approaches, including functionalization and miniaturization, to enhance the performance and versatility of these sensors.

# Annex 1: Academic and Scientific Contributions.

*Research articles; Patents; Scientific events; Bachelor's and Master's degree theses.*

*A.1 Research articles*

*A.1.1 2021*

| **Title:** | **Ultrasensitive $NO_2$ gas sensor with insignificant $NH_3$-interference based on a few-layered mesoporous graphene.** | | |
|---|---|---|---|
| **Authors:** | Daniel Matatagui; Jesús López-Sánchez; **Alvaro Peña**; Aída Serrano; Adolfo del Campo; Oscar Rodríguez de la Fuente; Noemí Carmona; Elena Navarro; Pilar Marín; María del Carmen Horrillo. | | |
| **Journal:** | Sensors and Actuators B: Chemical | Volume, issue or pages: | 335 |
| **Year:** | 2021 | DOI: | 10.1016/j.snb.2021.129657 |
| **Abstract:** | Few-layered mesoporous graphene (FLMG) is employed as a sensing material to develop an innovative and high-sensitivity room temperature $NO_2$ sensor through a simple manufacturing process. For this purpose, sensing material is optimized at 100 min by a high-energy milling process where natural graphite is used as a precursor: it is an inexpensive, sustainable and suitable active material. The large number of defects created and the enhanced degree of mesoporosity produced during the milling process determine the physical principles of operation of the designed device. $NO_2$ gas sensing tests reveal an improved and selective performance with a change in resistance of ~16 % at 0.5 ppm under ultraviolet photo-activation, establishing a detection limit around ~25 ppb. Interestingly, the response of the developed sensor to humidity is independent in the measured range (0–33 % relative humidity at 25 °C) and the dependency to the presence of $NH_3$ is rather poor as well (~1.5 % at 50 ppm). | | |

This publication focused on the synthesis of Few-layered Mesoporous Graphene (FLMG) and the investigation of its use in gas sensing applications. The study marked the first collaboration between researchers from Instituto de Magnetismo Aplicado and Instituto de Tecnologías Físicas y de la Información, with further collaboration of researchers from Universidad Complutense de Madrid and Instituto de Cerámica y Vidrio. The content of this publication has been extensively covered in Chapters 2 and 4.

The research is notable for its elegant simplicity, which is a testament to the high quality of the work. It showcases that a functional and high-performance sensor can be developed using cost-effective materials (like graphite), accessible techniques (ball-milling and drop-casting) and monitored with available technologies (a digital multimeter).

As for my contribution to this work, I was involved in the conceptualization, synthesis and ball-mill use, materials characterization, preparation of the sensing device and experimental setup, gas measurements, data curation, creation of figures, discussion, and writing of the original draft.

| **Title:** | **Study of magnetoelastic resonance for chemical sensors: Ribbons vs microwires** | | |
|---|---|---|---|
| **Authors:** | **Alvaro Peña**; Daniel Matatagui; Carlos Cruz; Patricia De La Presa; Pilar Marin; María del Carmen Horrillo. | | |
| **Journal:** | IEEE - 13th Spanish Conference on Electron Devices (CDE), Sevilla, Spain, 2021 | Volume, issue, or pages: | pp. 106-109 |
| **Year:** | 2021 | DOI: | 10.1109/CDE52135.2021.9455747 |
| **Abstract:** | In this work, the magnetoelastic resonance behaviour has been studied in amorphous metallic ribbons and microwires using a custom-made setup. First, optimal setup conditions were determined for both devices, then the frequency shift dependence on polymer mass deposition was studied. Both devices show a predictable response to the mass deposition of the polymer tested, making them suitable for contactless chemical sensors. | | |

This publication examines the impact of amorphous magnetic materials' characteristics on their magnetoelastic resonance figures of merit in sensing applications. The study marked the beginning of a new research line focused on magnetoelastic resonance-based gas sensors, which is extensively discussed in Chapter 5.

During this work, numerous challenges arose, primarily concerning the experimental setup. Nonetheless, the knowledge gained from this experience has been applied to subsequent projects involving magnetoelastic materials at the IMA and continues to be utilized.

My contributions to this work included conceptualization, sample preparation and characterization (the samples were commercially available), experimental setup design, data curation, creation of figures, discussion, and writing of the original draft. This research was published as a proceeding at the CDE2021 conference.

A.1.2 2022

| **Title:** | **Real-Time Monitoring of Breath Biomarkers with A Magnetoelastic Contactless Gas Sensor: A Proof of Concept** | | | |
|---|---|---|---|---|
| **Authors:** | **Alvaro Peña**; Juan Diego Aguilera; Daniel Matatagui; Patricia de la Presa; Carmen Horrillo; Antonio Hernando; Pilar Marín. | | | |
| **Journal:** | Biosensors | Volume, issue, or pages: | | |
| **Year:** | 2022 | DOI: | 10.3390/bios12100871 | |
| **Abstract:** | In the quest for effective gas sensors for breath analysis, magnetoelastic resonance-based gas sensors (MEGSs) are remarkable candidates. Thanks to their intrinsic contactless operation, they can be used as non-invasive and portable devices. However, traditional monitoring techniques are bound to slow detection, which hinders their application to fast bio-related reactions. Here we present a method for real-time monitoring of the resonance frequency, with a proof of concept for real-time monitoring of gaseous biomarkers based on resonance frequency. This method was validated with a MEGS based on a Metglass 2826 MB microribbon with a polyvinylpyrrolidone (PVP) nanofiber electrospun functionalization. The device provided a low-noise (RMS = 1.7 Hz), fast (<2 min), and highly reproducible response to humidity ($\Delta f$ = 46–182 Hz for 17–95% RH), ammonia ($\Delta f$ = 112 Hz for 40 ppm), and acetone ($\Delta f$ = 44 Hz for 40 ppm). These analytes are highly important in biomedical applications, particularly ammonia and acetone, which are biomarkers related to diseases such as diabetes. Furthermore, the capability of distinguishing between breath and regular air was demonstrated with real breath measurements. The sensor also exhibited strong resistance to benzene, a common gaseous interferent in breath analysis. | | | |

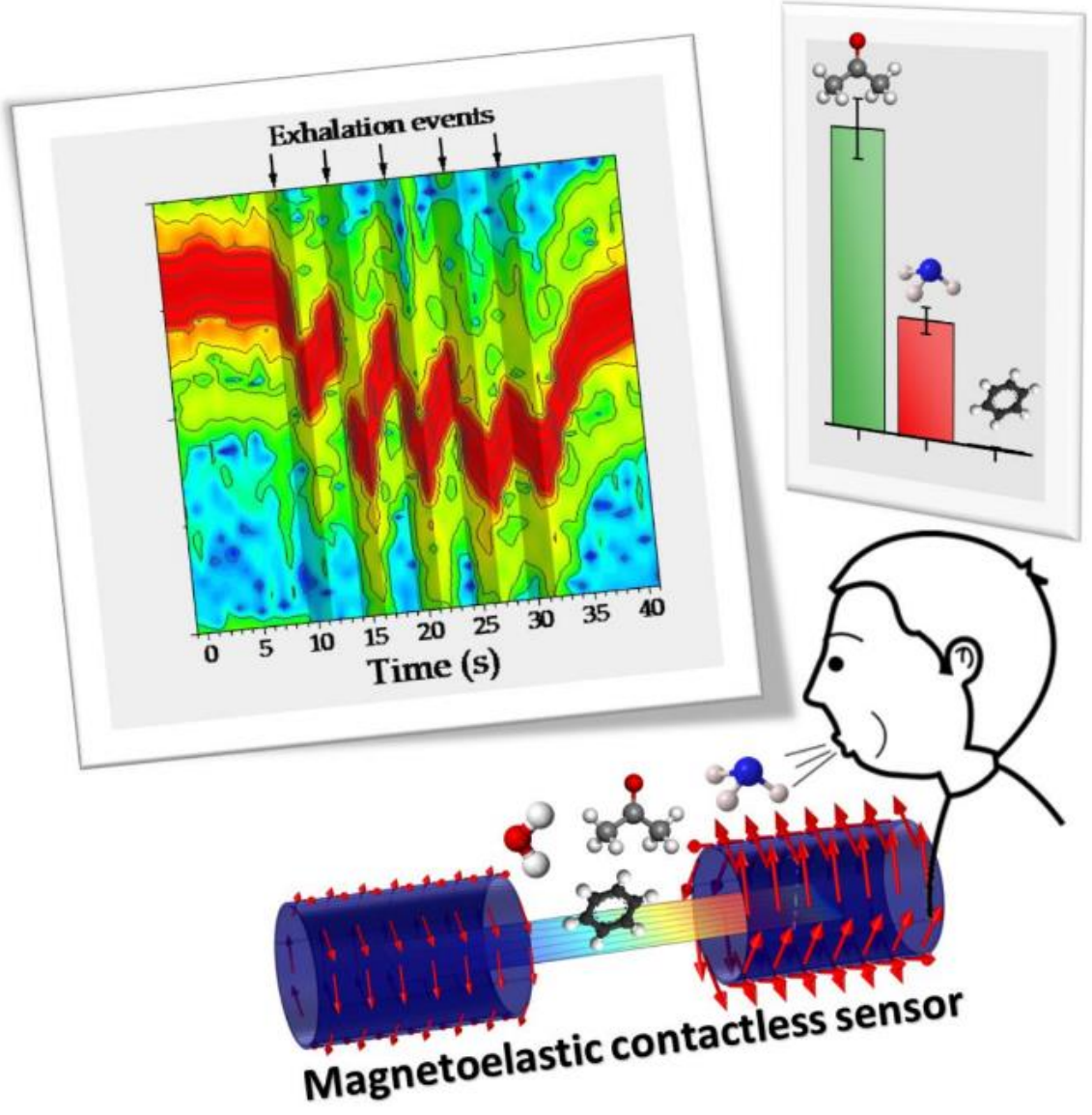

*Figure A. 1 Graphical abstract of the publication. The online version (mdpi.com/2079-6374/12/10/871) contains an animated video summary of the work.*

Building on the foundation of the previous publication, which marked the initial exploration into magnetoelastic resonance-based sensors, this study serves as a robust proof of concept. Notably, the paper presents a real-time monitoring system for magnetoelastic resonance and explores its potential for breath analysis applications. The content of this publication is thoroughly discussed in Chapter 5.

My contributions to this work included sample preparation and characterization (the micro ribbons were commercially available), the design and assembly of the sensing device and experimental setup, gas measurements, data curation, creation of figures and animations, discussion, and writing of the original draft. This research was published as part of a special issue in Biosensors and Bioelectronic Devices.

| Title: | **A feasible pathway to stabilize monoclinic and tetragonal phase coexistence in barium titanate-based ceramics** | | | |
|---|---|---|---|---|
| **Authors:** | Jallouli Necib; Jesús López-Sánchez; Fernando Rubio-Marcos; Aída Serrano; Elena Navarro; **Álvaro Peña**; Mnasri Taoufik; Mourad Smari; Rocío Estefanía Rojas-Hernández; Noemí Carmona; Pilar Marín. | | | |
| **Journal:** | Journal of Materials Chemistry C | Volume, issue, or pages: | | 46 |
| **Year:** | 2022 | DOI: | 10.1039/D2TC04265G | |
| **Abstract:** | Multiphase coexistence has attracted significant interest in recent years because its control has entailed a significant breakthrough for the piezoelectric activity enhancement of lead-free piezoelectric oxides. However, the comprehension of phase coexistence still has many controversies, including an adequate synthesis process and/or the role played by crystalline phases in functional properties. In this study, functional barium titanate [$BaTiO_3$, (BTO)]-based materials with tunable functional properties were obtained by compositional modification via Bismuth (Bi) doping. Towards this aim, we systematically synthesized BTO-based materials by a sol–gel method, focusing on the control of Bi substitution in the $BaTiO_3$ structure. In particular, we found that the substitution of $Bi^{+3}$ leads to the stabilization of a monoclinic–tetragonal (M–T) phase boundary close to room temperature, which facilities the polarization process of the system. As a surprising result, we believe that the simple and cost-effective strategy and design principles described in this work open up the possibility of obtaining BTO-based lead-free ceramics with enhanced properties induced by the stabilization of the phase coexistence, expanding their application range. | | | |

This article has not been referenced or discussed during the thesis as it does not belong to any of the presented research lines. However, it represents a side collaboration with Jallouli Necib, who spent his research stay at *Instituto de Magnetismo Aplicado* (*IMA*). His research focused on the synthesis and characterization of multi-ferro (ferromagnetic and ferroelectric) barium titanates.

My participation in this work was limited to magnetic characterization using the vibrating sample magnetometer, partial discussion, and final draft revision.

A.2.3 2023

| Title: | **Generation of Defective Few-Layered Graphene Mesostructures by High-Energy Ball Milling and Their Combination with FeSiCuNbB Microwires for Reinforcing Microwave Absorbing Properties** | | | |
|---|---|---|---|---|
| **Authors:** | Jesús López-Sánchez; **Álvaro Peña**; Aída Serrano; Adolfo del Campo; Óscar Rodríguez de la Fuente; Noemí Carmona; Daniel Matatagui; María del Carmen Horrillo; Juan Rubio-Zuazo; Elena Navarro; | | | |
| **Journal:** | ACS Applied Materials & Interfaces | Volume, issue, or pages: | | 15, 2, 3507–3521 |
| **Year:** | 2023 | DOI: | 10.1021/acsami.2c19886 | |
| **Abstract:** | Defective few-layered graphene mesostructures (DFLGMs) are produced from graphite flakes by high-energy milling processes. We obtain an accurate control of the generated mesostructures, as well as of the amount and classification of the structural defects formed, providing a functional material for microwave absorption purposes. Working under far-field conditions, competitive values of minimum reflection loss coefficient ($RL_{min}$) = −21.76 dB and EAB = 4.77 dB are achieved when DFLGMs are immersed in paints at a low volume fraction (1.95%). One step forward is developed by combining them with the excellent absorption behavior that offers amorphous $Fe_{73.5}Si_{13.5}B_9Cu_1Nb$ microwires (MWs), varying their filling contents, which are below 3%. We obtain a RLmin improvement of 47% (−53.08 dB) and an EAB enhancement of 137% (4 dB) compared to those obtained by MW-based paints. Furthermore, a $f_{min}$ tunability is demonstrated, | | | |

| | |
|---|---|
| | maintaining similar $RL_{min}$ and EAB values, irrespective of an ideal matching thickness. In this scenario, the Maxwell-Garnet standard model is valid, and dielectric losses mainly come from multiple reflections, interfacial and dielectric polarizations, which greatly boost the microwave attenuation of MWs. The present concept can remarkably enhance not only the MW attenuation but can also apply to other microwave absorption architectures of technological interest by adding low quantities of DFLGMs. |

This publication delves deeper into the synthesis of FLGMs than the first publication and investigates their use in microwave-absorbing applications when combined with amorphous magnetic microwires. The content of this publication is covered in Chapters 2 and 3.

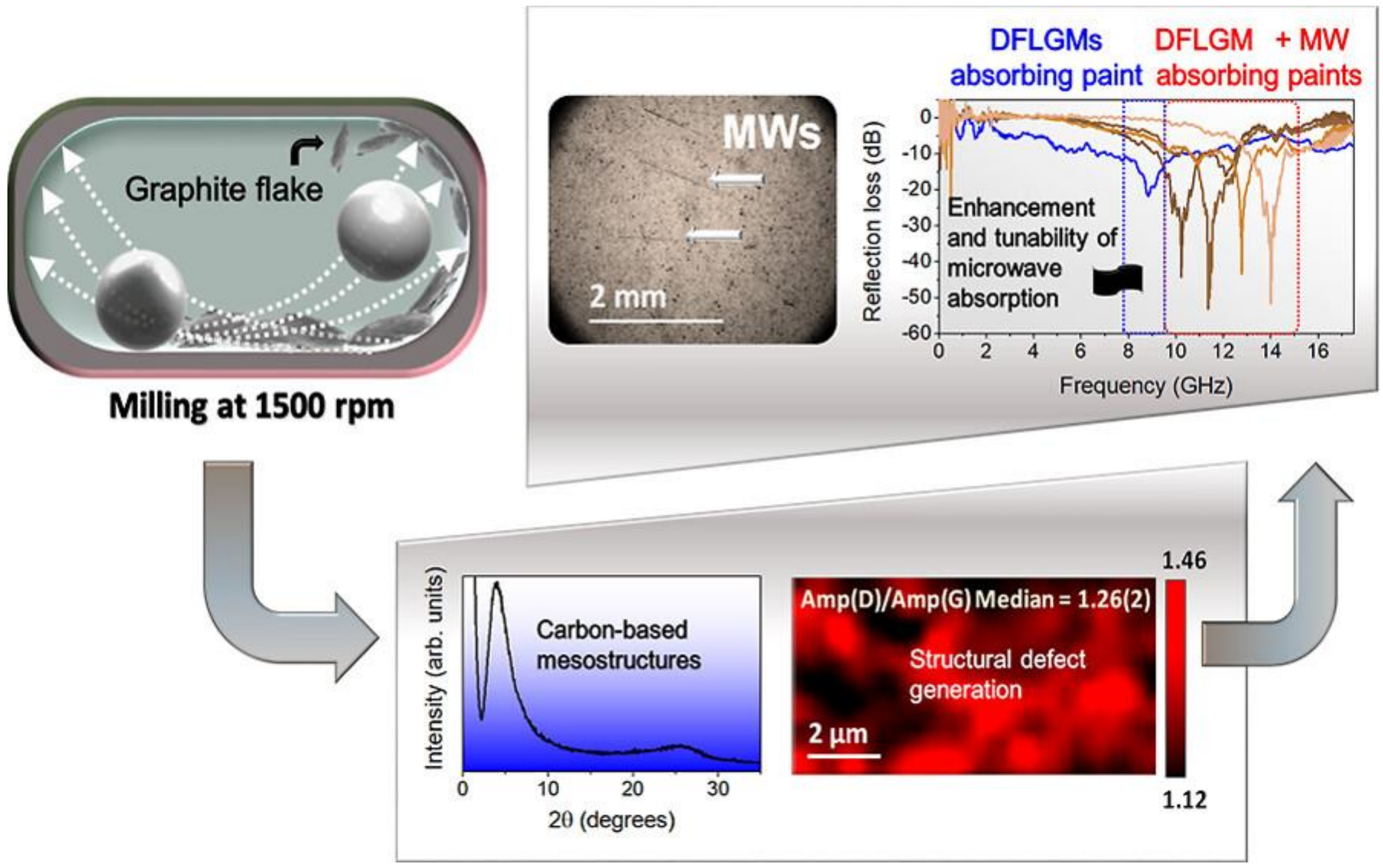

*Figure A. 2 Graphical abstract of the publication. The online version (pubs.acs.org/doi/10.1021/acsami.2c19886) contains an animated video summary of the work.*

My involvement in this work includes conceptualization, synthesis and ball-mill use, materials characterization, preparation of paint samples, microwave absorbance measurements, data curation, creation of figures and animations, discussion, and writing of the original draft.

| **Title:** | **Optimization of multilayer graphene-based gas sensors by ultraviolet photoactivation** | | |
|---|---|---|---|
| **Authors:** | **Alvaro Peña**; Daniel Matatagui; Filiberto Ricciardella; Leandro Sacco; Sten Vollebregt; Daniel Otero; Jesús López-Sánchez; Pilar Marín; María del Carmen Horrillo | | |
| **Journal:** | Applied Surface Science | Volume, issue, or pages: | 610 |
| **Year:** | 2023 | DOI: | 10.1016/j.apsusc.2022.155393 |
| **Abstract:** | Nitrogen dioxide ($NO_2$) is a potential hazard to human health at low concentrations, below one part per million (ppm). $NO_2$ can be monitored using gas sensors based on multi-layered graphene operating at ambient temperature. However, reliable detection of concentrations on the order of parts per million and lower is hindered by partial recovery and lack of reproducibility of the sensors after exposure. We show how to overcome these longstanding problems using ultraviolet (UV) light. When exposed to $NO_2$, the sensor response is enhanced by 290 % − 550 % under a 275 nm wavelength light emitting diode irradiation. Furthermore, the sensor's initial state is completely restored after exposure to the target gas. UV irradiation at 68 W/m2 reduces the $NO_2$ detection limit to 30 parts per billion (ppb) at room temperature. We investigated sensor performance optimization for UV irradiation with different power densities and target gases, such as carbon oxide and ammonia. Improved sensitivity, recovery, and reproducibility of UV-assisted graphene-based gas sensors make them suitable for widespread environmental applications. | | |

This publication centred on the utilization and optimization of UV irradiation for graphene-based gas sensors. The study represented the inaugural collaboration with researchers Filiberto Ricciardella, Leandro Sacco, and Sten Vollebregt from TU Delft. A significant portion of Chapter 4 is dedicated to the findings of this publication.

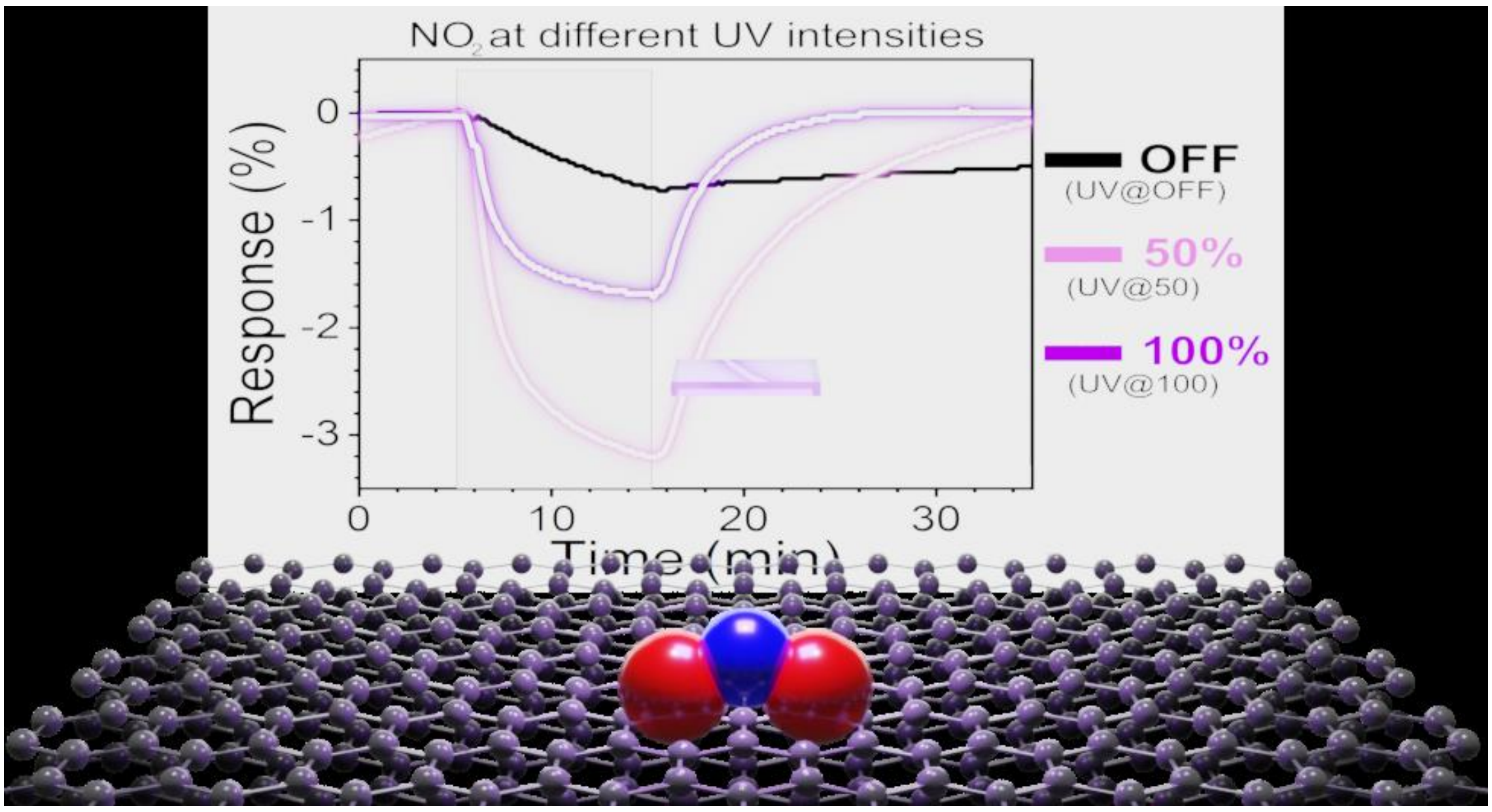

*Figure A. 3 Graphical abstract of the publication. The online version (sciencedirect.com/science/article/pii/S016943322202921X) contains an animated video summary of the work.*

My contributions to this work included conceptualization, materials characterization, preparation of the sensing device and experimental setup, gas measurements, data curation, creation of figures and animations, discussion, and writing of the original draft. The MLG samples were prepared by the researchers at TU Delft.

*A.2 Patents*

<table>
<tr><td>Title:</td><td colspan="3">Obtención a gran escala en un solo paso y a temperatura ambiente de material compuesto por pocas láminas de grafeno con un alto grado de defectos mediante molienda mecánica seca oscilatoria de alta energía.</td></tr>
<tr><td>Authors:</td><td colspan="3">Pilar Marín; Elena Navarro; Jesús López Sánchez; Álvaro Peña; María del Carmen Horrillo; Daniel Matatagui</td></tr>
<tr><td>Office:</td><td colspan="3">Spanish office; PCT</td></tr>
<tr><td>Year:</td><td>2020</td><td>Number:</td><td>ES2779151 (B2)</td></tr>
<tr><td>Claims:</td><td colspan="3">1. Método de producción a gran escala de grafeno de pocas capas (FLG) con alto grado de defectos a partir de grafito por molienda de bolas de alta energía por movimiento oscilante en seco caracterizado porque:<br>a. El precursor es grafito en forma de escamas con una longitud comprendida entre 2 y 50 µm y un espesor menor de 100 nm, y la molienda se realiza en un recipiente metálico recubierto de un material que posee una dureza de Knoop superior a 1000 kg/mm2 en presencia de bolas del mismo material, con una frecuencia de oscilación entre 1000 y 1500 rpm y una relación de masas entre la bola o bolas y el grafito empleado de entre 1:20 y 1:30.<br>2. Método de producción a gran escala de grafeno de pocas capas (FLG), según la reivindicación q, porque el proceso de molienda se realiza a temperatura y presión ambiente, en atmósfera inerte y sin aditivos ni tratamientos de procesado posteriores<br>3. Método de producción a gran escala de grafeno de pocas capas (FLG), según reivindicaciones 1, donde las escamas de grafito son monocristalinas.</td></tr>
</table>

| | |
|---|---|
| | 4. Método de producción a gran escala de grafeno de pocas capas (FLG), según reivindicación 1, donde la relación de masas entre el grafito y la bola o bolas empleados es preferiblemente 1:25.<br>5. Método de producción a gran escala de grafeno de pocas capas (FLG), según reivindicaciones anteriores, donde el material que recubre el recipiente metálico y forma las bolas es carburo de tungsteno.<br>6. Método de producción a gran escala de grafeno de pocas capas (FLG), según reivindicación 5, donde la molienda se realiza con una única bola de carburo de tungsteno.<br>7. Método de producción a gran escala de grafeno de pocas capas (FLG), según reivindicación 6, donde la bola posee una relación de volumen de entre 0,8:1 y 1,2:1, preferiblemente 1:1 con respecto al grafito empleado y de volumen entre 1:48 y 1:52, preferiblemente 1:50 con respecto al recipiente contenedor.<br>8. Método de producción a gran escala de grafeno de pocas capas (FLG), según reivindicaciones anteriores, donde el tiempo de molienda varía entre 20 y 300 minutos para obtener estructuras mesoporosas.<br>9. Método de producción a gran escala de grafeno de pocas capas (FLG), según reivindicación 8, donde el tiempo de molienda varía entre 20 y 100 minutos para obtener distribuciones de tamaño distintas correspondientes al grafito de partida y a FLG.<br>10. Método de producción a gran escala de grafeno de pocas capas (FLG), según reivindicación 8, donde el tiempo de molienda varía entre 120 y 300 minutos para obtener distribuciones de tamaño homogéneas correspondientes a FLG.<br>11. Método de producción a gran escala de grafeno de pocas capas (FLG), según reivindicación 8, donde se obtiene FLG con un espesor comprendido entre 3 y 10 capas de grafeno según el tiempo de molienda de forma que el número de capas disminuye según aumenta el tiempo de molienda. |

This patent refers to the FLMG synthesis method, which was discussed in Chapter 2.

Since the patent application was presented at the Spanish office, the original text is in Spanish, as is presented here to avoid the possible loss of legal nuances. Briefly, the claims state the following:

The method for large-scale production of few-layer graphene (FLG) with a high degree of defects from graphite involves high-energy ball milling through oscillatory motion in a dry environment. The precursor is flake graphite with a length between 2 and 50 µm and a thickness of less than 100 nm. The milling is carried out in a metal container coated with a material that has a Knoop hardness greater than 1000 kg/mm2 in the presence of balls of the same material. The oscillation frequency ranges between 1000 and 1500 rpm, and the mass ratio between the ball or balls and the graphite used is between 1:20 and 1:30.

The milling process is carried out at ambient temperature and pressure, in an inert atmosphere, and without additives or subsequent processing treatments. The graphite flakes are monocrystalline, and the mass ratio between the graphite and the ball or balls used is preferably 1:25. The material coating the metal container and forming the balls is tungsten carbide. The milling is carried out with a single tungsten carbide ball, which has a volume ratio between 0.8:1 and 1.2:1, preferably 1:1 with respect to the graphite used, and a volume between 1:48 and 1:52, preferably 1:50 with respect to the container.

The milling time varies between 20 and 300 minutes to obtain mesoporous structures. If the milling time varies between 20 and 100 minutes, different size distributions corresponding to the starting graphite and FLG are obtained. For milling times between 120 and 300 minutes, homogeneous size distributions corresponding to FLG are achieved. FLG with a thickness between 3 and 10 graphene layers is obtained depending on the milling time so that the number of layers decreases as the milling time increases.

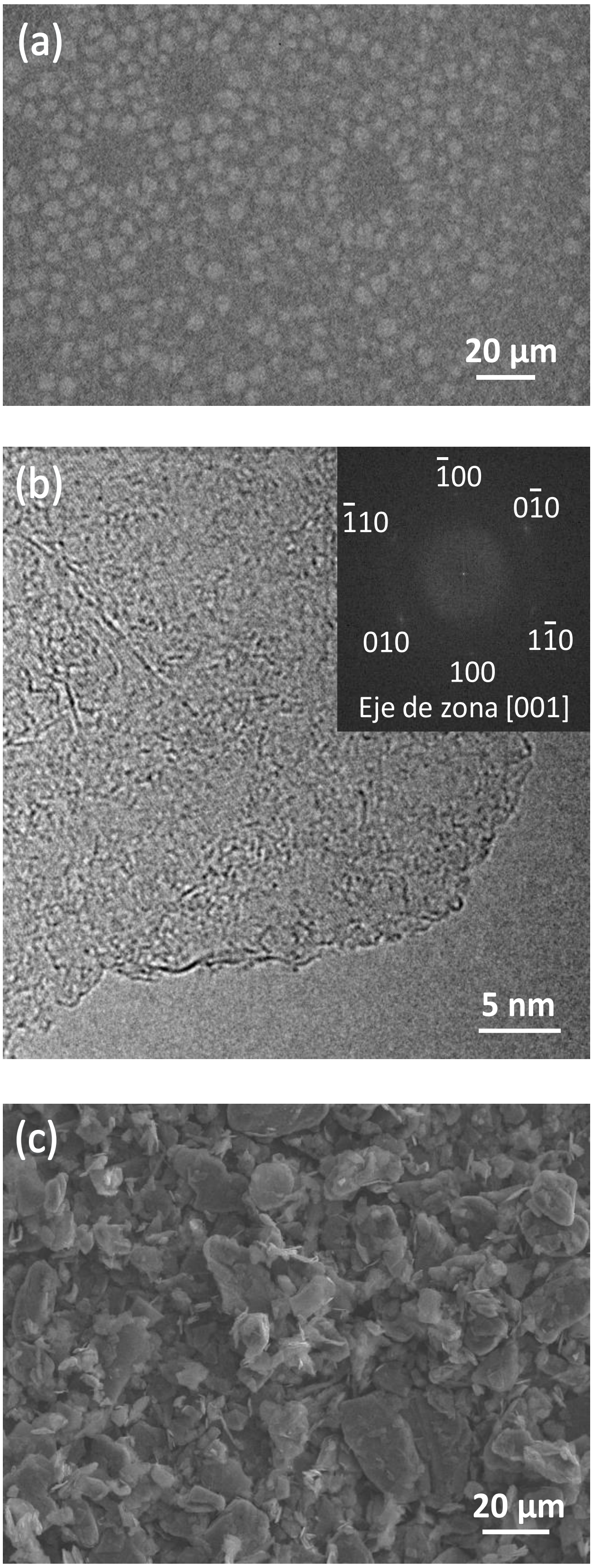

*Figure A. 4 High-quality reproduction of the Figure 1 of patent ES2779151 (B2)*

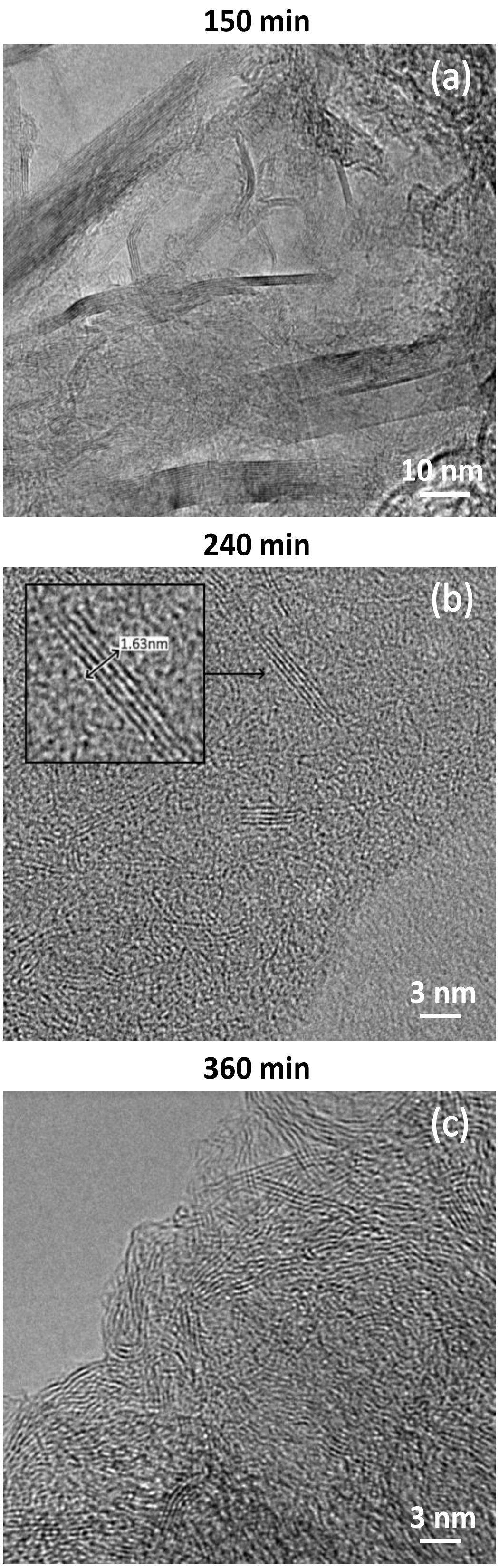

*Figure A. 5 High-quality reproduction of the Figure 3 of patent ES2779151 (B2)*

| **Title:** | **Sensor químico resistivo para la detección de muy bajas concentraciones de NO2 a temperatura ambiente basado en partículas nanoestructuradas en dominios de unas pocas capas atómicas de grafeno obtenidas por molienda mecánica de alta energía.** | | |
|---|---|---|---|
| **Authors:** | Pilar Marín; Elena Navarro; Jesús López Sánchez; **Álvaro Peña**; María del Carmen Horrillo; Daniel Matatagui | | |
| **Office:** | Spanish office, US pending | | |
| **Year:** | 2020 | Number: | ES2890726 (B2) |
| **Claims:** | 1. Un sensor químico resistivo para la detección de $NO_2$ caracterizado por que comprende:<br>• un sustrato aislante;<br>• al menos dos electrodos metálicos sobre el sustrato aislante;<br>• al menos una capa activa situada sobre el sustrato y los electrodos, que conecta dichos electrodos metálicos, donde la capa activa tiene un espesor entre 20 nm y 100 µm y está formada de partículas de un tamaño de entre 20 nm y 1000 nm aglomeradas y nanoestructuradas en nanodominios, donde dichos nanodominios son de grafeno de pocas capas atómicas no oxidado, donde dicho grafeno contiene defectos estructurales, y donde dichos defectos son vacantes de carbono en la estructura del grafeno de pocas capas atómicas.<br>2. Sensor químico según la reivindicación 1, donde el sustrato aislante está formado de un material seleccionado de entre FR4, silicio, alúmina, polímeros, plásticos, papel, arseniuro de galio, nitruro de aluminio y vidrio.<br>3. Sensor químico según cualquiera de las reivindicaciones 1 o 2, donde los electrodos son al menos dos y el material conductor que los forma es seleccionado de entre cobre, oro, platino, aluminio, cromo, titanio, plata y cualquier aleación de los anteriores, o cobre recubierto de un material seleccionado de entre oro, platino, cromo, titanio y cualquier combinación de los anteriores.<br>4. Sensor químico según la reivindicación 3, donde los electrodos son de cobre recubierto de oro y ambos electrodos se encuentran interdigitados, con un espacio entre los electrodos de entre 25 pm y 250 pm, y donde el área confinada entre los electrodos es entre 0,04 $mm^2$ de 4 $mm^2$.<br>5. Sensor químico según cualquiera de las reivindicaciones 1 a 4, donde el tamaño de las partículas de la capa activa es de entre 50 nm y 450 nm.<br>6. Sensor químico según cualquiera de las reivindicaciones 1 a 5, donde la capa activa tiene una absorbancia en la región ultravioleta-visible con un máximo de absorción localizado entre 200 y 400nm<br>7. Sensor químico según cualquiera de las reivindicaciones 1 a 6, donde la capa activa presenta un espectro de Raman que comprende 4 bandas características, donde dichas bandas son D, G, D" y 2D, donde la relación entre las amplitudes entre las bandas D y G tiene un valor comprendido entre 0,6 y 1,5, y donde la relación entre las amplitudes entre las bandas D' y G tiene un valor comprendido entre 0,1 y 0,4.<br>8. Procedimiento de obtención del sensor químico resistivo según cualquiera de las reivindicaciones de 1 a 7, caracterizado por que comprende las siguientes etapas<br>a) moler escamas de grafito mediante un método mecánico seco oscilatorio de alta energía, durante un tiempo de al menos 80 min a una velocidad de molienda de entre 1300 y 1700 rpm, y a una temperatura de entre 15 °C y 35 °C;<br>b) dispersar el polvo obtenido en la etapa (a) en un disolvente orgánico seleccionado de entre 1-metil-2-pirrolidinona, dimetilformamida o dimetilsulfóxido, y donde el rango de concentraciones varía entre 0,1 mg/mL y 10 mg/mL; y<br>c) depositar por goteo la disolución obtenida en la etapa (b) sobre un circuito integrador que comprende un sustrato aislante con al menos dos electrodos metálicos sobre dicho sustrato y evaporar el disolvente al aire; opcionalmente, durante la etapa (c) de depositar por goteo, se mide la resistencia entre los electrodos.<br>9. Procedimiento de obtención del sensor químico según la reivindicación 8, donde el tiempo de molienda de la etapa (a) es de entre 80 min y 120 min.<br>10. Procedimiento de obtención del sensor químico según cualquiera de las reivindicaciones 8 o 9, donde el disolvente orgánico es 1-metil-2-pirrolidinona. | | |

| | |
|---|---|
| | 11. Procedimiento de obtención del sensor químico según cualquiera de las reivindicaciones 8 a 10, donde se mide la resistencia entre los electrodos durante la etapa (c) de depositar por goteo, hasta que la resistencia medida entre cada gota depositada es estable, manteniéndose sin un cambio mayor al 10% de la resistencia<br>12. Uso del sensor según cualquiera de las reivindicaciones 1 a 7 para detectar la presencia de $NO_2$ en una concentración de al menos 25 ppb en aire a una temperatura de entre 15 ºC y 35 ºC<br>13. Uso según la reivindicación 12 donde el sensor está expuesto a radiación ultravioleta de entre 200 nm y 400 nm, preferiblemente de entre 250 nm y 270 nm.<br>14. Uso según cualquiera de las reivindicaciones 12 o 13, donde si en el aire hay alguno de los elementos seleccionados de la siguiente lista NH3, humedad relativa entre 0,5% y 40% y cualquier combinación de los anteriores, se detecta la presencia de $NO_2$ en una concentración de al menos 25 ppb en aire a una temperatura de entre 15 ºC y 35 ºC.<br>15. Método para detectar la presencia de $NO_2$ en aire, caracterizado por que comprende las siguientes etapas<br>a) circular aire de referencia sin $NO_2$, con un flujo de 25 mL/min a 1000 mL/min, establecido mediante un controlador de flujo, por una cámara donde se sitúa el sensor descrito según cualquiera de las reivindicaciones 1 a 7, durante un tiempo de entre 10 min y 30 min para que la resistencia del sensor tenga un valor inicial estable;<br>b) exponer el sensor de la etapa (a) al aire para detectar la presencia de $NO_2$ hasta que al menos el valor de la respuesta del sensor sea tres veces el 25 ruido del mismo, durante un tiempo máximo de 30 min; y<br>c) purgar de la cámara el aire que se encuentra en su interior durante un tiempo de entre 10 min y 30 min;<br>opcionalmente durante la etapa (a) y (b) el sensor es expuesto a una radiación ultravioleta de entre 200 nm y 400 nm, preferiblemente de entre 250 nm y 270 nm. |

This patent refers to the fabrication and conditions of the FLMG-based device use for $NO_2$ that was discussed in Chapter 4.

Once again, the original claims are presented in Spanish to avoid the possible loss of legal nuances. Briefly, the claims state the following:

The method for large-scale production of few-layer graphene (FLG) with a high degree of defects from graphite involves high-energy ball milling through oscillatory motion in a dry environment. The precursor is flake graphite with a length between 2 and 50 µm and a thickness of less than 100 nm. The milling is carried out in a metal container coated with a material that has a Knoop hardness greater than 1000 kg/mm2 in the presence of balls of the same material. The oscillation frequency ranges between 1000 and 1500 rpm, and the mass ratio between the ball or balls and the graphite used is between 1:20 and 1:30.

The milling process is carried out at ambient temperature and pressure, in an inert atmosphere, and without additives or subsequent processing treatments. The graphite flakes are monocrystalline, and the mass ratio between the graphite and the ball or balls used is preferably 1:25. The material coating the metal container and forming the balls is tungsten carbide. The milling is carried out with a single tungsten carbide ball, which has a volume ratio between 0.8:1 and 1.2:1, preferably 1:1 with respect to the graphite used, and a volume between 1:48 and 1:52, preferably 1:50 with respect to the container.

The milling time varies between 20 and 300 minutes to obtain mesoporous structures. If the milling time varies between 20 and 100 minutes, different size distributions corresponding to the starting graphite and FLG are obtained. For milling times between 120 and 300 minutes, homogeneous size distributions corresponding to FLG are achieved. FLG with a thickness between 3 and 10 graphene layers is obtained depending on the milling time so that the number of layers decreases as the milling time increases.

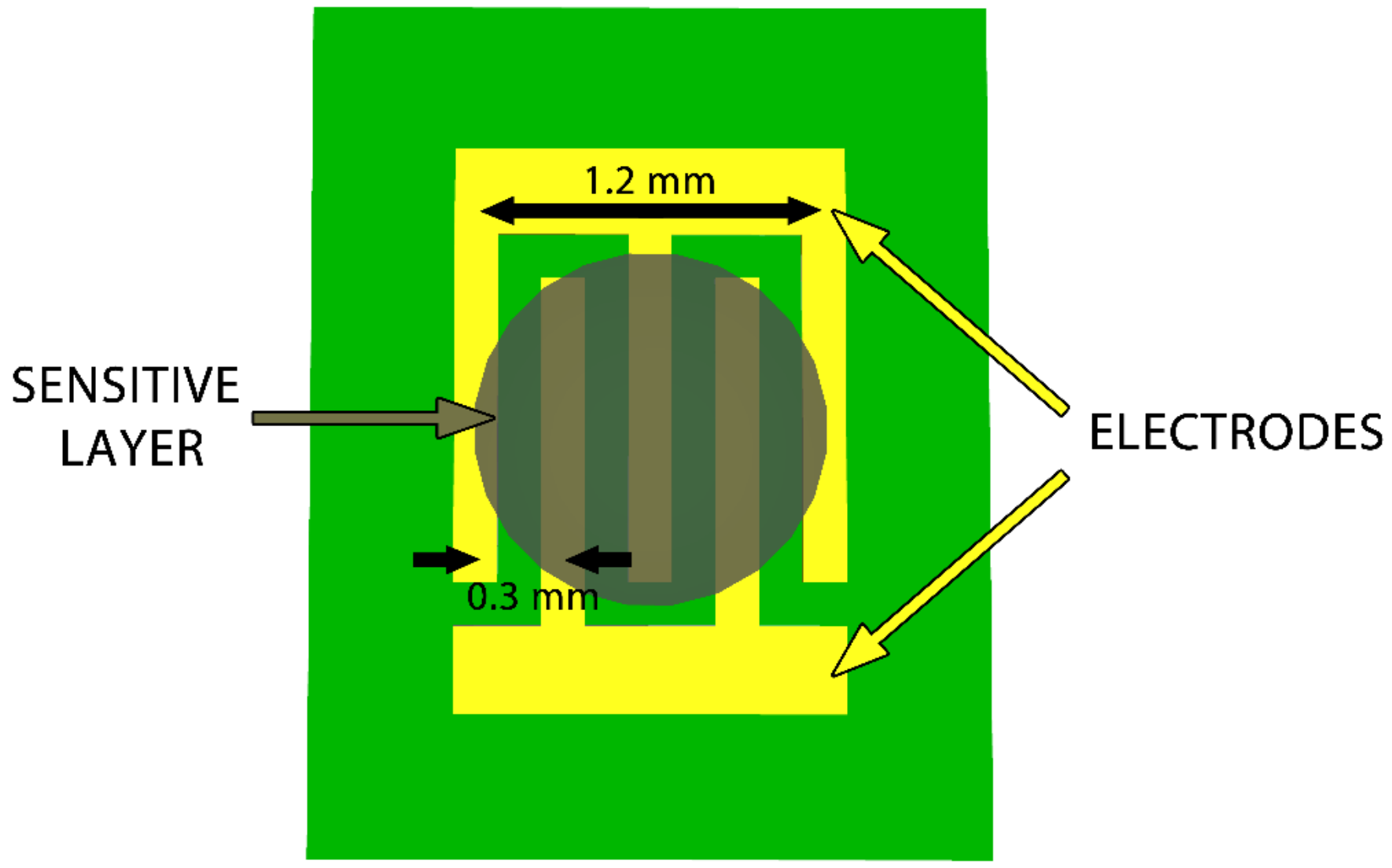

*Figure A. 6 High-quality reproduction of the Figure 7 of patent ES2890726 (B2)*

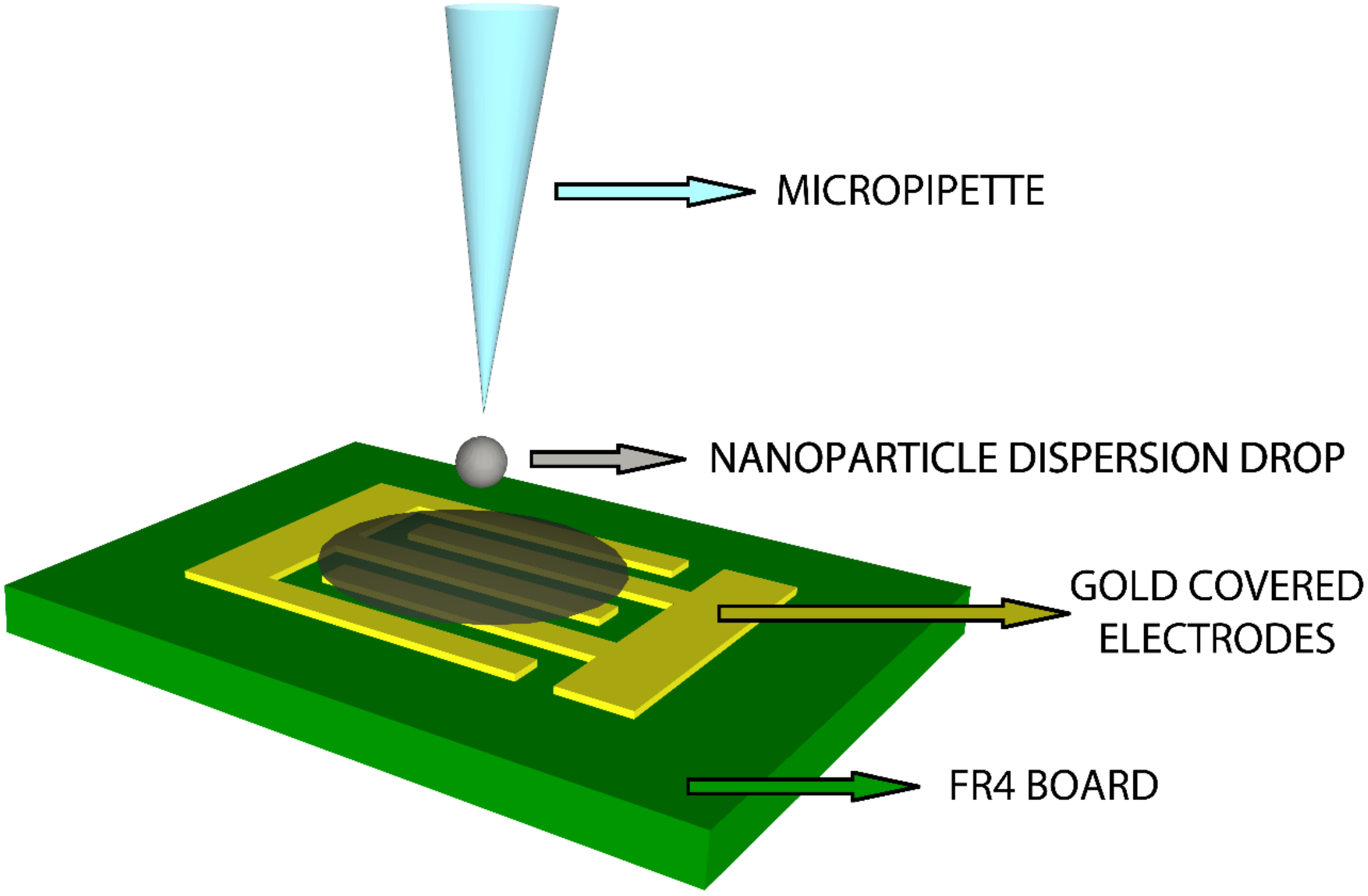

*Figure A. 7 High-quality reproduction of the Figure 8 of patent ES2890726 (B2)*

### *A.3 Contributions to Scientific Events*

#### *A.3.1 2020*

| | |
|---|---|
| **Title:** | **Tuning Microwave Absorption Property of Few-layered Graphene/ Magnetic Microwires Composite Materials for Electromagnetic Interference Shielding** |
| **Authors:** | **Álvaro Peña**; Jesús López-Sánchez; Elena Navarro; Pilar Marín |
| **Event:** | 4th Young Researchers in Magnetism |
| **Dates and venue:** | November 23, 2020 (online due to the pandemic) |
| **Type** | Poster communication |

*A.3.2 2021*

| Title: | **Few-layered mesoporous graphene obtained through high-energy dry ball-milling.** |
|---|---|
| Authors: | **Álvaro Peña**; Jesús López-Sánchez; Daniel Matatagui; Elena Navarro; María del Carmen Horrillo; Pilar Marín |
| Event: | Graphene Online & 2DM |
| Dates and venue: | April 20-21, 2021 (online due to the pandemic) |
| Type | Poster communication |

Interestingly, discussion around this topic during the conference led to the collaboration with the TU Delft researchers.

| Title: | **Study of magnetoelastic resonance for chemical sensors: Ribbons vs microwires** |
|---|---|
| Authors: | **Alvaro Peña**; Daniel Matatagui; Carlos Cruz; Patricia De La Presa; Pilar Marín; María del Carmen Horrillo. |
| Event: | 13th Spanish Conference on Electron Devices |
| Dates and venue: | June 9-11, 2021 (online due to the pandemic) |
| Type | Oral presentation |

As previously indicated, this contribution resulted in a proceeding publication under the same title.

| Title: | **Microwires optimization for magnetoelastic resonance-based sensors.** |
|---|---|
| Authors: | **Alvaro Peña**; Daniel Matatagui; Carlos Cruz; Patricia De La Presa; Pilar Marín; María del Carmen Horrillo. |
| Event: | 5$^{th}$ Young Researchers in Magnetism |
| Dates and venue: | November 10-11, 2021. Girona, Spain. |
| Type | Oral presentation and abstract committee |

This contribution consisted of a general overview of the use of magnetoelastic materials on sensing applications and optimization routes.

| Title: | **Optimization of microwires for magnetoelastic resonance-based sensors** |
|---|---|
| Authors: | **Alvaro Peña**; Daniel Matatagui; Carlos Cruz; Patricia De La Presa; Pilar Marín; María del Carmen Horrillo. |
| Event: | X Franco-Spanish Workshop CMC2 - IBERNAM-CMC2 |
| Dates and venue: | November 25-26, 2021. Arcachon, France. |
| Type | Oral presentation |

This contribution focused on the optimization of amorphous magnetic microwires with different diameters and nucleus-to-shell ratios and thermal annealing treatments, which has been covered in Chapter 5.

*A.3.3 2022*

| Title: | **Chemosresistive Graphene-Based olfatory systems for air pollutant sub-ppm detection** |
|---|---|
| Authors: | **Alvaro Peña**; Daniel Matatagui; Pilar Marín; María del Carmen Horrillo. |
| Event: | ROE Red Olfativa Española. VIII Jornadas Olfativas |
| Dates and venue: | May 3-7, 2022. Asturias, Spain. |
| Type | Oral presentation |

Interestingly, the ROE conference was focused on the neurobiology and olfactory systems. The opportunity to share the research about solid-state gas detection systems opened interesting discussions with other speakers and attendees.

| Title: | **UV-Assisted Graphene-Based Chemiresistive Gas Sensors for Sub-ppm Detection** |
|---|---|
| **Authors:** | **Alvaro Peña**; Daniel Matatagui; Pilar Marín; María del Carmen Horrillo. |
| **Event:** | ESMolNa, 15th European School on Molecular Nanoscience. |
| **Dates and venue:** | May 22-27, 2022. Asturias, Spain. |
| **Type** | Oral presentation |

This contribution occurred during the discussion phase about the results of UV irradiation optimization presented in Chapter 4 and helped to create the approach that was finally taken.

| Title: | **Gas sensors, What for?** |
|---|---|
| **Authors:** | **Alvaro Peña** |
| **Event:** | CEMAG, Club Español de Magnetismo. Curso de Verano 2022 |
| **Dates and venue:** | June 6-10, 2022. Asturias, Spain. |
| **Type** | Oral presentation |

The contribution of this magnetism-focused workshop presented the possibilities of using magnetoelastic materials in developing gas sensor devices.

| Title: | **Novel experimental setup for real-time measurements of magnetoelastic resonance-based gas sensors** |
|---|---|
| **Authors:** | **Álvaro Peña**; Daniel Matatagui; Juan Diego Aguilar; Pilar Marín; Carmen Horrillo. |
| **Event:** | EMSA22, European Magnetic Sensors and Actuators |
| **Dates and venue:** | July 5-8, 2022. Madrid, Spain. |
| **Type** | Oral presentation and organizing comittee |

This contribution anticipated the real-time magnetoelastic resonance monitoring system that has been presented in Chapter 5, generating great hype and interest among the attendees.

| Title: | **Optimization of graphene-based gas sensors by ultraviolet photoactivation.** |
|---|---|
| **Authors:** | **Alvaro Peña**; Daniel Matatagui; Filiberto Ricciardella; Leandro Sacco; Sten Vollebregt; Daniel Otero; Jesús López-Sánchez; Pilar Marín; María del Carmen Horrillo |
| **Event:** | X Jornadas de Jóvenes Investigadores ICV |
| **Dates and venue:** | October 20, 2022. Madrid, Spain. |
| **Type** | Poster presentation |

This contribution was based on the scientific paper under the same title that was being published at the time.

| Title: | **Synthesis and Applications of a novel Few-Layered Mesoporous Graphene** |
|---|---|
| **Authors:** | **Álvaro Peña.** |
| **Event:** | Patents for Innovation International Summit & Expo 2022 |
| **Dates and venue:** | October 26-27, 2022. Madrid, Spain. |
| **Type** | Oral presentation |

The patent pitch in this event was an excellent opportunity to promote the patents previously presented. During the expo, and animated video summarizing both patents could be seen at the Comunidad de Madrid – Universidad Complutense de Madrid stand.

| **Title:** | **Real-time monitoring of breath biomarkers with a magnetoelastic contactless gas sensor: a proof of concept.** |
|---|---|
| **Authors:** | **Alvaro Peña**; Juan Diego Aguilera; Daniel Matatagui; Patricia de la Presa; Carmen Horrillo; Antonio Hernando; Pilar Marín. |
| **Event:** | $6^{th}$ Young Researchers in Magnetism |
| **Dates and venue:** | November 17, 2022. Cadiz, Spain. |
| **Type** | Poster presentation and organizing comittee |

This contribution was based on the scientific paper under the same title that was being published at the time.

*A.4 Bachelor's and Master's Degree Theses.*

| Title: | Rare-earth free magnets |
|---|---|
| Author: | Guillermo Lozano |
| Type | Bachelor's degree thesis |
| University | Universidad Nebrija |
| Year | 2020-2021 |

Ball-milling and mechanical alloying of thermally annealed to crystallization of amorphous magnetic materials to develop rare-earth free magnets. My contribution to his work was providing support for all his experimental work, including operation of the mechanical mills and the furnace for thermal treatments, and DSC and magnetic characterization.

| Title: | Materiales magnéticos para aplicaciones en biosensores y en el campo biomédico |
|---|---|
| Author: | Laura Mesa |
| Type | Bachelor's degree thesis |
| University | Universidad Complutense de Madrid |
| Year | 2021-2022 |

Laura Mesa worked with Fe-based amorphous microwires, the 79.3 / 100.6 µm from Section 5.7.1, to develop magnetoelastic resonance-based sensors. In particular, she optimised the magnetoelastic resonance figures of merit with a thermal treatment based on electrical currents. My contribution to his work was providing support and samples for all his experimental work, including the setup described in Section 5.4.1.

| Title: | Sensores basados en grafeno para detectar $NO_2$ |
|---|---|
| Author: | Daniel Otero |
| Type | Master's degree thesis |
| University | Universidad Complutense de Madrid |
| Year | 2021-2022 |

Daniel participated in the experiments described in Section 4.5 Effect Of UV On Multi-Layer Graphene Chemiresistive Sensors. and co-authored "Optimization of multilayer graphene-based gas sensors by ultraviolet photoactivation" as previously indicated. My contribution to his work was providing support and samples for all his experimental work.

| Title: | Medidas de la atenuación de ondas electromagnéticas en micro y nanomateriales |
|---|---|
| Author: | Clara Gutierrez |
| Type | Bachelor's degree tesis & Internship |
| University | Universidad Complutense de Madrid |
| Year | 2021-2022 |

Clara worked on the VNA for microwave characterization of electromagnetic shielding materials and ferromagnetic resonance on nanowires. In addition, Clara participated in the experiment of metallic and plastic clips described in Section 1.3.1.1 Optimization of the FSM Technique. My contribution to his work was providing support for all her experimental work.

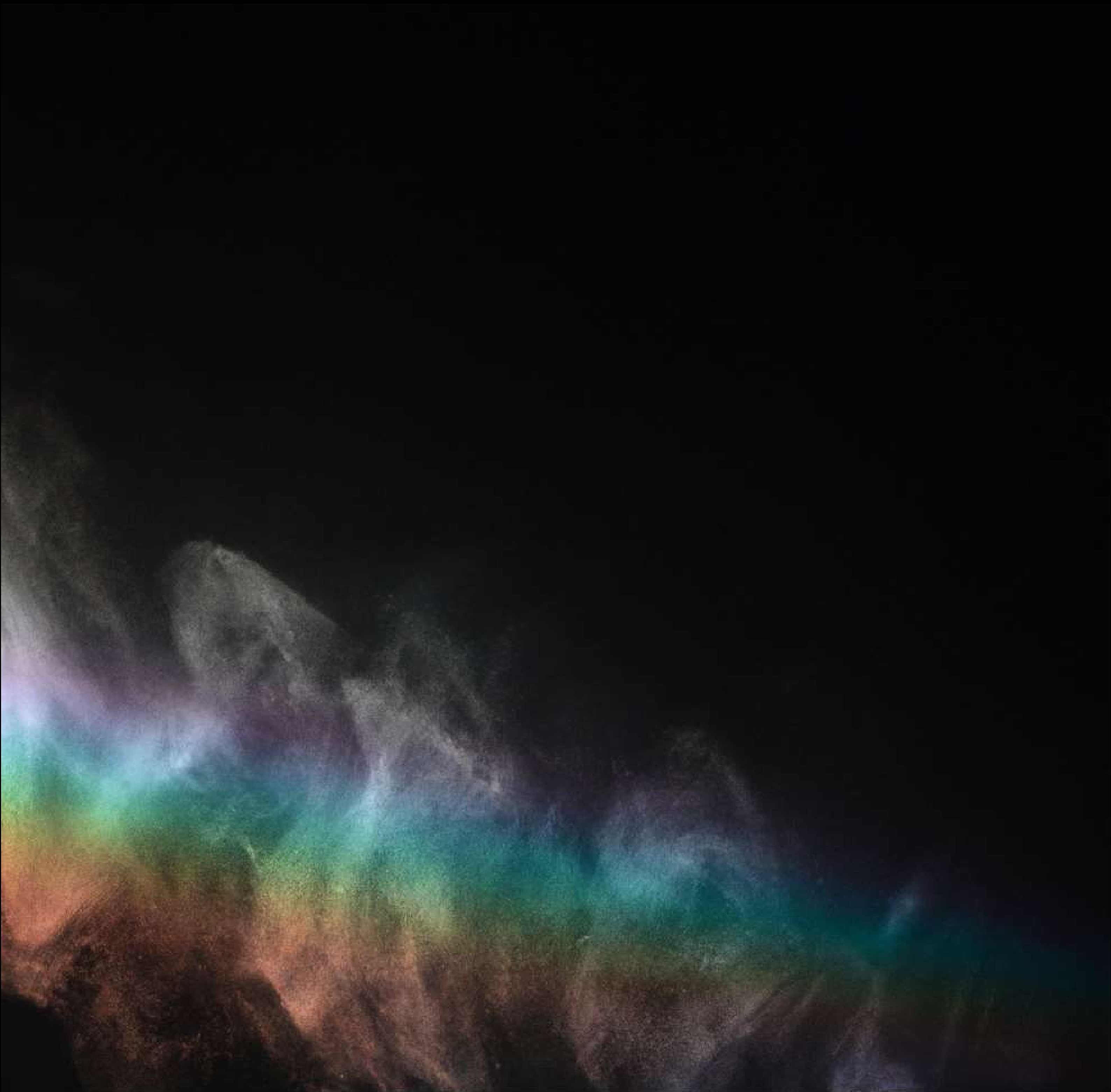